\pdfoutput=1

\documentclass[
10pt, 
a4paper, 
twoside, 
headinclude,footinclude, 
BCOR5mm, 
]{scrartcl}

\usepackage[
nochapters, 
beramono, 
pdfspacing, 
dottedtoc 
]{classicthesis} 

\usepackage{arsclassica} 

\usepackage{lmodern} 

\usepackage[T1]{fontenc} 

\usepackage[utf8]{inputenc} 

\usepackage{graphicx} 
\graphicspath{{Figures/}} 

\usepackage{enumitem} 

\usepackage{lipsum} 

\usepackage{subfig} 

\usepackage{amsmath,amssymb,amsthm} 

\usepackage{varioref} 

\usepackage{amssymb}
\usepackage{graphicx}
\usepackage{amsthm,amsmath,amsfonts}
\usepackage{hyperref}
\usepackage{bm}
\usepackage[figuresright]{rotating}
\usepackage{epstopdf}
\usepackage{xspace}
\usepackage{tabularx}
\usepackage{threeparttable}
\usepackage{lscape}
\usepackage{xcolor}
\usepackage{cite}
\usepackage{lineno}
\usepackage{multirow}

\hypersetup{
colorlinks=true, breaklinks=true, bookmarks=true,bookmarksnumbered,
urlcolor=webbrown, linkcolor=RoyalBlue, citecolor=webgreen, 
pdftitle={}, 
pdfauthor={\textcopyright}, 
pdfsubject={}, 
pdfkeywords={}, 
pdfcreator={pdfLaTeX}, 
pdfproducer={LaTeX with hyperref and ClassicThesis} 
}

\advance\topmargin by +1cm
\global\oddsidemargin=-2mm
\global\evensidemargin=-2mm

\newcommand{\pp}{\ensuremath{\rm pp}\xspace}

\newcommand{\pPb}{{p--Pb}\xspace}

\newcommand{\PbPb}{{Pb--Pb}\xspace}

\newcommand{\sqrtsNN}{\ensuremath{\sqrt{s_{\mathrm{\scriptscriptstyle NN}}}}\xspace}

\newcommand{\jpsi}{\ensuremath{\mathrm{J}/\psi}\xspace}

\newcommand{\MeV}{\ensuremath{\text{~MeV}}\xspace}
\newcommand{\GeV}{\ensuremath{\text{~GeV}}\xspace}

\newcommand{\gev}{\ensuremath{\text{GeV}}}
\newcommand{\tev}{\ensuremath{\text{TeV}}}

\newcommand{\nbinv}{\ensuremath{\text{~nb}^\text{$-$1}}\xspace}
\newcommand{\Lunits}{\rm cm^{-2}s^{-1}}

\newcommand{\dd}{\ensuremath{\mathrm{d}}}

\newcommand{\beq}{\begin{equation}}
\newcommand{\eeq}{\end{equation}}
\newcommand{\beqn}{\begin{eqnarray}}
\newcommand{\eeqn}{\end{eqnarray}}

\newcommand{\beqa}{\begin{eqnarray}}
\newcommand{\eeqa}{\end{eqnarray}}
\def\lsim{\raise0.3ex\hbox{$<$\kern-0.75em\raise-1.1ex\hbox{$\sim$}}}
\def\gsim{\raise0.3ex\hbox{$>$\kern-0.75em\raise-1.1ex\hbox{$\sim$}}}

\newcommand{\pt}{\ensuremath{p_{\mathrm{T}}}\xspace}

\graphicspath{{THEORY/Figures/}{ALICE/Figures/}{EXPINTRO/Figures/}{NA60PLUS/Figures/}{FURTHER/Figures/}}

\title{{\LARGE INFN What Next:\\ Ultra-relativistic Heavy-Ion Collisions}} 

\begin{document}


\pagestyle{scrheadings} 

\lehead{\mbox{{\small\thepage}\kern1em\color{halfgray} \vline\color{halfgray}\hspace{0.5em}\rightmark\hfil}} 
\rohead{\mbox{\color{halfgray}\rightmark \hspace{0.5em} \small\color{halfgray}  \vline \kern1em \color{black}\thepage\hfil}} 


\date{ } 

\maketitle 

\thispagestyle{empty} 

\setcounter{tocdepth}{2} 

\noindent
\mbox{A.~Dainese\,$^{6,a}$},
\mbox{E.~Scomparin\,$^{8,a}$}, 
\mbox{G.~Usai\,$^{13,a}$},
\mbox{P.~Antonioli\,$^{3,b}$},
\mbox{R.~Arnaldi\,$^{8,b}$},
\mbox{A.~Beraudo\,$^{8,b}$},          
\mbox{E.~Bruna\,$^{8,b}$},\\
\mbox{G.E.~Bruno\,$^{11,b}$},
\mbox{S.~Bufalino\,$^{10,b}$},            
\mbox{P.~Di~Nezza\,$^{1,b}$}
% ...

\vspace{0.5cm}
{\small ~~~~~~~~~~~~~~~~~~~~~~~~~~~~~~~~~~~~~~~~~~~~~~~~~~~~~~~~~~~~\today} 


\section*{Abstract} 

 This document was prepared by the community that is active in Italy,
within INFN (Istituto Nazionale di Fisica Nucleare), in the field of ultra-relativistic heavy-ion collisions.
The experimental study of the phase diagram of
strongly-interacting matter and of the Quark--Gluon Plasma (QGP) deconfined
state will proceed, in the next 10--15 years, along two directions:
the high-energy regime at RHIC and at the LHC, and the low-energy regime
at FAIR, NICA, SPS and RHIC.
The Italian community is strongly involved in the present and future
programme of the ALICE experiment, the upgrade of which
will open, in the 2020s, a new phase of high-precision
characterisation of the QGP properties at the LHC.  
As a complement of this main activity, there is a growing interest in
a possible future experiment at the SPS, which would target the 
search for the onset of deconfinement using dimuon measurements.
On a longer timescale, the community looks
with interest at the ongoing studies and discussions on a possible
fixed-target programme using the LHC ion beams and on the Future
Circular Collider.

{\let\thefootnote\relax\footnotetext{ $^a$ General editors}}
{\let\thefootnote\relax\footnotetext{ $^b$ Topic editors}}


\newpage 

\thispagestyle{empty}

\noindent
{$^1$ INFN - Laboratori Nazionali di Frascati}\\
{$^2$ INFN - Sezione di Bari}\\
{$^3$ INFN - Sezione di Bologna}\\
{$^4$ INFN - Sezione di Cagliari}\\
{$^5$ INFN - Sezione di Catania}\\
{$^6$ INFN - Sezione di Padova}\\
{$^7$ INFN - Sezione di Roma}\\
{$^8$ INFN - Sezione di Torino}\\
{$^9$ INFN - Sezione di Trieste}\\
{$^{10}$ Politecnico di Torino e INFN - Sezione di Torino}\\
{$^{11}$ Dipartimento Interateneo di Fisica di Bari e INFN - Sezione di Bari}\\
{$^{12}$ Universit\`a di Bologna e INFN - Sezione di Bologna}\\
{$^{13}$ Universit\`a di Cagliari e INFN - Sezione di Cagliari}\\
{$^{14}$ Universit\`a di Catania e INFN - Laboratori Nazionali del Sud}\\
{$^{15}$ Universit\`a di Catania e INFN - Sezione di Catania}\\
{$^{16}$ Universit\`a di Firenze e INFN - Sezione di Firenze}\\
{$^{17}$ Universit\`a di Padova e INFN - Sezione di Padova}\\
{$^{18}$ Universit\`a del Piemonte Orientale, Alessandria, e gruppo collegato INFN}\\
{$^{19}$ Universit\`a di Pisa e INFN - Sezione di Pisa}\\
{$^{20}$ Universit\`a di Salerno e gruppo collegato INFN}\\
{$^{21}$ Universit\`a di Torino e INFN - Sezione di Torino}\\
{$^{22}$ Universit\`a di Trieste e INFN - Sezione di Trieste}\\
{$^{23}$ Museo Storico della Fisica e Centro Studi e Ricerche ``E. Fermi'' e INFN - Sezione di Bologna}\\
{$^{24}$ Museo Storico della Fisica e Centro Studi e Ricerche ``E. Fermi'' e INFN - CNAF}\\
{$^{25}$ Universit\`a di Trieste e INFN - Sezione di Trieste}\\
{$^{26}$ Univerisity of Houston, USA}\\
{$^{27}$ LAL,  Universit\'e Paris-Saclay, France}\\
{$^{28}$ Massachussets Institute of Technology, USA}\\
{$^{29}$ iThemba LABS and University of Witwatersand, South Africa} \\
{$^{30}$ Lawrence Berkeley National Laboratory, USA}\\
{$^{31}$ CERN, Switzerland}\\
{$^{32}$ Wayne State University, USA}\\
{$^{33}$ University of of Chinese Academy of Sciences, Beijing, China} \\
{$^{34}$ University of Heidelberg, Germany}\\
{$^{35}$ GSI, Germany}\\
{$^{36}$ Slovak Academy of Sciencies, Slovakia}\\

\thispagestyle{empty}

\newpage

\thispagestyle{empty} 

\tableofcontents 

\cleardoublepage


\section{Introduction}
\label{sec:introduction}

In 2014 INFN (Istituto Nazionale di Fisica Nucleare) started a broad internal discussion (INFN What Next) on future physics programmes, organised around a series of plenary meetings and a number of working groups devoted to specific topics~\cite{whatnextindico}.
The mandate was the investigation of the various possible directions for the development of our research fields, and the identification of the most 
promising projects for the next decade. In the frame of this discussion, the working group on ``Standard Model precision measurements'' has addressed, among various topics, the status and future directions for precision studies of the phase diagram of strongly-interacting matter (also denoted QCD phase diagram).

This research field employs collisions of heavy ions at ultra-relativistic energies (energy per nucleon--nucleon collision in the centre-of-mass $\sqrtsNN > 1~\GeV$) and it has by now a long tradition. Starting in the 1980s with exploratory studies at fixed-target facilities at BNL and CERN, then brought to maturity in the following decade, it has reached high-precision levels with experiments at the RHIC collider and, more recently, at the LHC. Evidence for the creation of the Quark--Gluon Plasma (QGP), a state of matter where quarks and gluons are deconfined has by now been firmly reached~\cite{Arsene:2004fa,Adcox:2004mh,Back:2004je,Adams:2005dq,Muller:2012zq,Roland:2014jsa,Armesto:2015ioy}. In particular, the results from \mbox{Pb--Pb} collisions in the LHC Run-1 show that a system with an initial temperature that exceeds by more than a factor of two the critical temperature $T_{\rm c}\approx 155~\MeV$ for the phase transition from hadronic matter to QGP has been created. It has also been shown that such a system is opaque to hard probes (jets, heavy quarks) traversing it, and that quarkonium states are dissociated due to the screening of the colour charge in the QGP.  

In this situation, advances in this field can be pursued by moving towards two well defined directions.
\begin{itemize}
 \item
 First, in high-energy studies at the LHC ---which provide a QGP with the highest initial temperature, longest lifetime and largest volume--- higher-precision data and the investigation of new observables are clearly needed for a complete characterisation of the deconfined state. 
\item
Second, the phase diagram of strongly-interacting matter is still largely unexplored in the domain of high baryonic densities, which can be studied via experiments at lower collision energies. Among the highlights of these studies, the identification of the critical point of the QCD phase diagram plays a prominent role. 
\end{itemize}

A strong and motivated Italian community, devoted to these studies, exists since the very beginning of the field. Experimental physicists have played a key role both in the fixed-target experiments with Pb beams at the CERN SPS (WA97, NA50, NA57, NA60) and later on in the design, construction and operation of the ALICE experiment at the LHC. In parallel, a theory community, significantly growing in the past decade, is providing the field with high-level fundamental and phenomenological studies.
With the present document, we want to convey to high-energy nuclear and particle physicists, in the frame of the INFN What Next discussions, our view of the field and the prospects for our activities in the next years. We follow the two directions outlined above: high-energy, in the frame of the ALICE experiment, and high baryonic density, in the frame of the proposal of a new fixed-target experiment at the CERN SPS.
   
The document is structured as follows. In Chapter~2, starting from a short discussion of the phase diagram of strongly-interacting matter, we show how  ultra-relativistic heavy-ion collisions can lead to the creation of a system with the characteristics of a Quark-Gluon Plasma. Adopting a theory-driven perspective, it is then shown that the study of the spectra of soft particles produced in the collision and of their collective motions is an important tool for the determination of the global QGP properties. The use of hard probes, as jets or particles carrying heavy quarks, is also discussed, as a way to investigate the temperature of the deconfined system and the transport coefficients of the created medium. Then, electromagnetic probes, such as photons and dileptons, are reviewed. They are not sensitive to the strong interaction and represent a powerful tool to extract information about the first stages of the collision. Finally, the use of lattice calculation techniques as a theoretical tool to study the thermodynamics, phase diagram and spectral functions of QCD is discussed. 

In Chapter~3, we turn to an experimental perspective and we shortly introduce the main observables and results obtained at the RHIC and LHC machines in the study of nuclear collisions. We also outline the landscape of the experimental facilities for heavy-ion studies that are presently in operation or under construction.

Chapter~4 brings us into a discussion of the ALICE experiment, with an emphasis on the immediate prospects for the freshly started LHC Run-2, and in particular on physics plans for the next decade, during Run-3 and Run-4, when the upgraded detectors will be in operation. Areas where the contribution of Italian groups will be stronger are then discussed in detail. We start by reviewing the status of open heavy flavour studies, and we then turn to heavy quarkonia and jets. These three areas represent the ``core'' of the physics effort of the ALICE Italian community. Significant results will also be obtained on soft observables, including in particular the study of collective flow and the analysis of high-multiplicity pp and p--Pb collisions, where intriguing effects were observed in Run-1. Finally, high-energy nuclear collisions represent a copious source of nuclei and anti-nuclei, that can be used for studies of rare processes such as hypernuclei production and tests of fundamental symmetries, as CPT.

Chapter~5 presents the first feasibility studies for a future fixed-target experiment, NA60+, devoted to the investigation of electromagnetic probes and heavy quarkonia with nuclear collisions at the CERN SPS. The experiment aims at an energy scan from low to high SPS energy ($\sqrtsNN$ from a few GeV to about 20~GeV), studying muon pair production. We start by describing the relation of this observable with key concepts in the study of the QCD phase diagram, as the restoration of chiral symmetry, the onset of deconfinement, the order of the phase transition from hadronic matter to QGP and the evaluation of the QGP temperature. We then move to a first conceptual study for such an experiment, describing a possible set-up, and we analyse the foreseeable running conditions and the performance of the experiment. On the latter point, we show results of simulation studies on the detection of low-mass hadronic resonances ($\rho$, $\omega$, $\phi$) and charmonia (J/$\psi$), and on the characterisation of the dimuon invariant-mass continuum and $p_{\rm T}$ spectra.

Chapter~6 briefly presents further future projects, such as studies for fixed-target experiments using the LHC beams and prospects for measurements with heavy-ions at the Future Circular Collider (FCC), the developments of which are also followed with interest by the community. 
Finally, Chapter~7 summarises our view on the future involvement of the Italian heavy-ion community in the studies discussed in the document.
 
\section{The QCD phase diagram and heavy-ion collisions: theory}
\label{sec:theory}

\begin{figure}[!b]
\begin{center}
\includegraphics[clip,height=6cm]{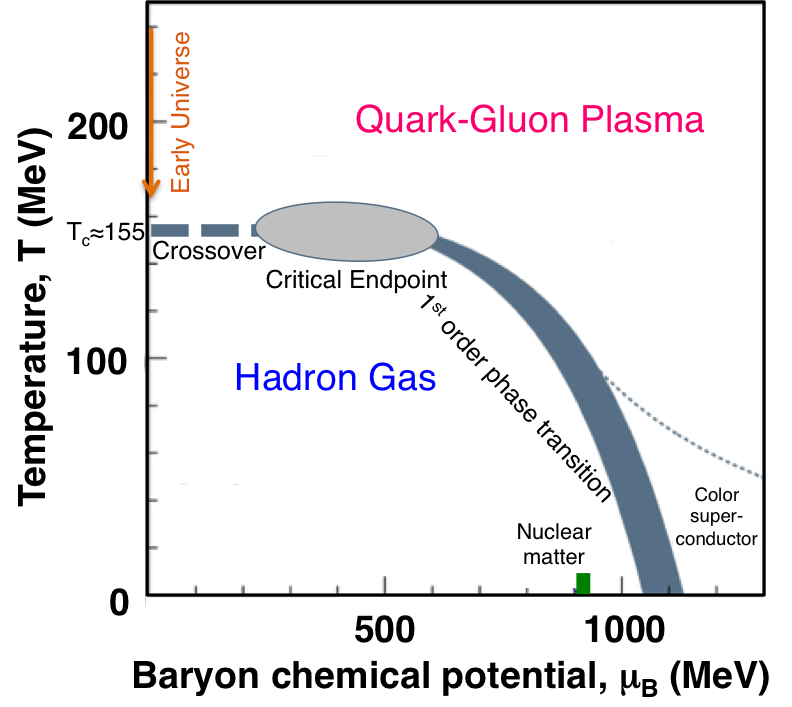}
\includegraphics[clip,height=6cm]{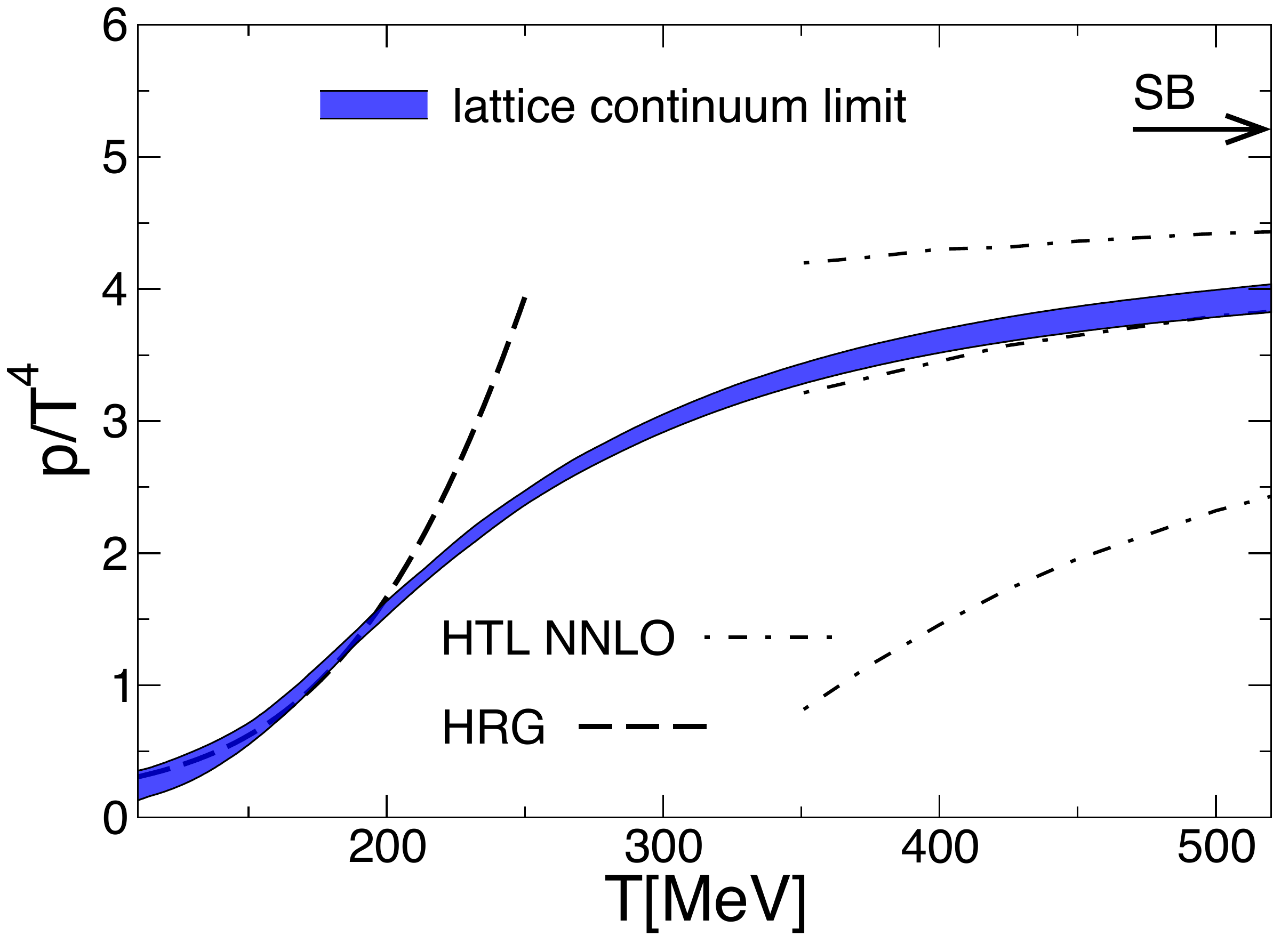}
\caption{The QCD phase diagram (left panel) and the equation of state $P(T)/T^4$ in the limit of vanishing baryon density~\cite{Borsanyi:2013bia} measured on the lattice (right panel): the latter is characterised by a rise in the effective number of active degrees of freedom, indicating the cross-over transition to a QGP.}
\label{fig:QCDPD}
\end{center}
\end{figure}

\begin{figure}[h]
\begin{center}
\includegraphics[clip,width=0.35\textwidth]{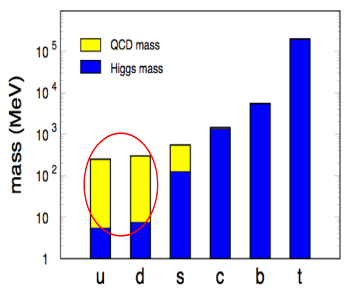}
\hfill
\includegraphics[clip,width=0.61\textwidth]{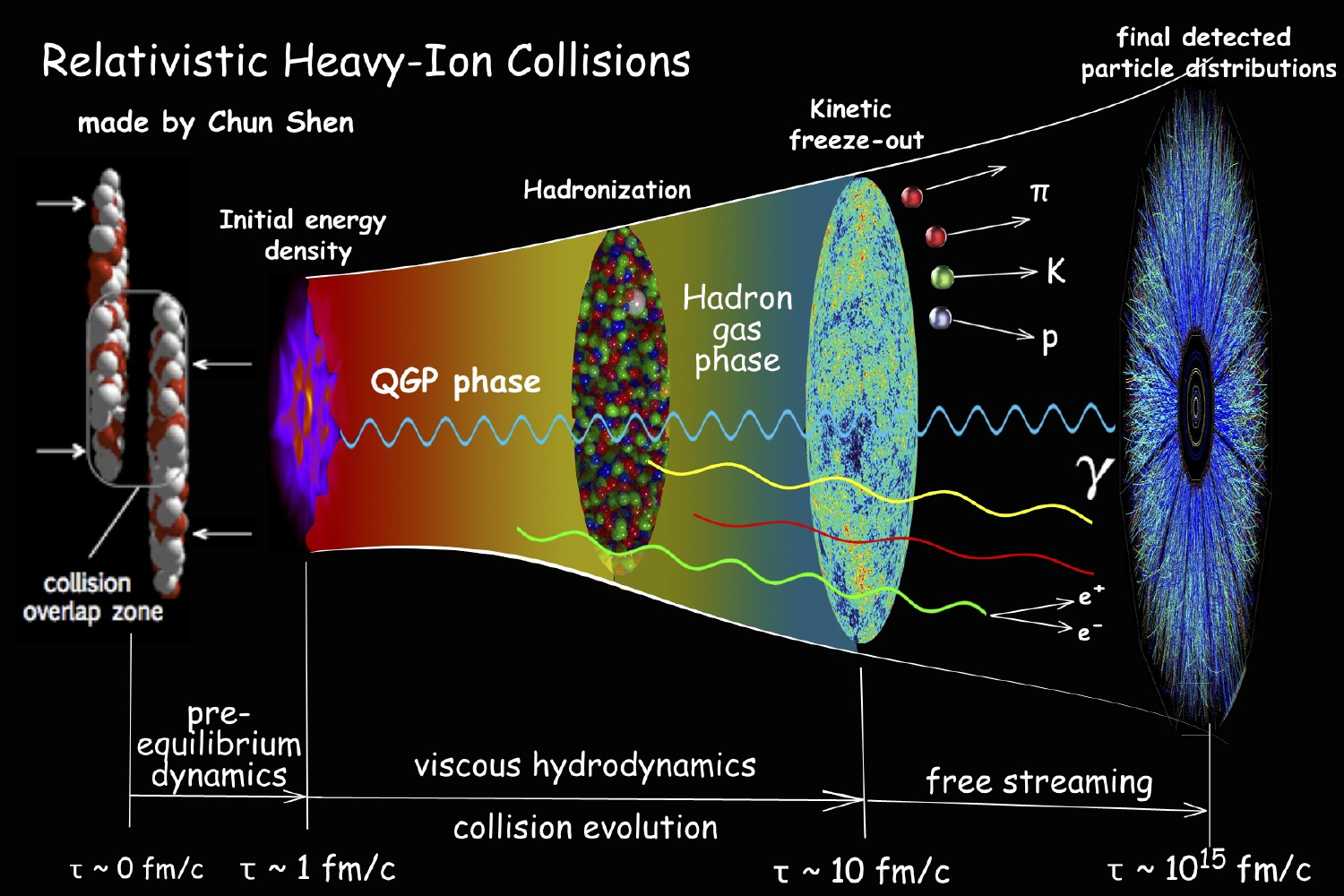}
\caption{Left: the contributions to the quark masses from the electro-weak symmetry breaking (bare quark mass, indicated as Higgs mass) and chiral symmetry breaking (constituent quark mass, indicated as QCD mass)~\cite{Muller:2004kk}. Right: schematic evolution of ultra-relativistic heavy-ion collisions.}
\label{fig:evolution}
\end{center}
\end{figure}

Quantum Chromo-Dynamics (QCD), the theory of strong interactions, is characterised by a rich phase diagram, which is schematically shown in the left panel of Fig.~\ref{fig:QCDPD} as a function of temperature $T$ and baryon chemical potential $\mu_B$. The elementary coloured degrees of freedom of the theory, quarks and gluons, under ordinary conditions are \emph{confined} in colour-neutral composed objects, mesons and baryons, and get free to propagate over distances larger than the typical size of a hadron ($\sim$1 fm) only in an extremely hot or dense environment, like the one present in the early Universe or, possibly, in the core of compact stars. From the theory point of view,  information on the QCD phase diagram comes from lattice-QCD simulations (discussed in the following) and from chiral effective Lagrangians. These calculations predict a \emph{cross-over} between hadronic matter and a Quark-Gluon Plasma (QGP) phase in the high temperature and low baryon chemical potential region. They suggest that the transition may become of first order moving towards higher values of the baryon chemical potential.
QCD calculations on the lattice provide nowadays reliable and accurate results for the thermodynamic properties of a hot strongly-interacting system, 
as we will discuss in more detail in Sec~\ref{sec:lQCD}. 
The right-hand panel of Fig.~\ref{fig:QCDPD} shows the Equation of State (EoS) that relates the pressure $P$ and the temperature $T$ in terms of $P(T)/T^4$, in the $\mu_B\!=\!0$ case. This quantity is characterised by a smooth rise in the effective number of active degrees of freedom as the temperature increases. Numerical results, in nice agreement with a Hadron-Resonance Gas at low $T$, in the deconfined phase slowly approach, due to asymptotic freedom, the limit of an ideal gas of quarks and gluons (Stefan-Boltzmann limit). The cross-over between the hadronic and QGP phase occurs around the critical temperature $T_c\!\approx\!155$~MeV. The transition from QGP to confined hadrons is also associated with the breaking of chiral symmetry. During the thermal evolution of the Universe, the chiral condensate acquired a non-vanishing expectation value $\langle q\overline{q}\rangle\ne 0$ and the baryons got most of their mass (see left panel of Fig.~\ref{fig:evolution}~\cite{Muller:2004kk}): most of the present baryonic mass in our Universe actually arises from the QCD rather than from the electro-weak phase transition.

Experimentally the deconfinement transition is studied using ultra-relativistic heavy-ion collisions. Different regions of the phase diagram can be covered by changing the centre-of-mass energy of the collision: indeed, as the energy increases, the initial temperature of the system increases and its baryo-chemical potential decreases (because the baryon number carried by the two incident nuclei has a large separation in rapidity from the central region where the hot system forms). The experiments at the LHC and at the highest RHIC energies are suited to reproduce the conditions of high-energy and low baryon-density present in the early Universe with a cross-over connecting the QGP and hadron-gas phase. Instead, the search for the Critical End-Point, where the transition is expected to become of first order, is the main motivation of the Beam-Energy Scan (BES) at RHIC and of the fixed-target experiments running at the SPS and planned at NICA and FAIR. A schematic cartoon of the evolution of the matter formed in these collisions is displayed in the right panel of Fig.~\ref{fig:evolution}, where the various stages (pre-equilibrium, QGP, hadron gas, decoupling) of its dynamics are shown: they will be discussed in the following sections, together with the probes used to get information on the properties of the QGP. An overview of experimental measurements will be provided in Chapter~\ref{sec:expintro}. 

\subsection{Space-time evolution of heavy-ion collisions}
\label{sec:evolution}

The only way to get experimental access to the QCD phase diagram in the region of deconfinement under controlled conditions is represented by ultra-relativistic heavy-ion collisions. 
The initial state of the two nuclei as well as the the very first instants following the collision can be described by Classical Yang-Mills (CYM) equations: $[\mathcal{D}_\mu,F^{\mu\nu}]=-J^\nu$. The colour-current $J^\nu$ of the fast (i.e. carrying a sizeable fraction of longitudinal momentum) valence partons of the incoming nucleons acts as a source of the gauge field $F^{\mu\nu}$  describing the small-$x$ gluons~\cite{McLerran:1993ni,McLerran:1993ka,McLerran:1994vd}. These fields turn out to be strong 
and the corresponding gluon states are characterised by large occupation numbers $\sim 1/\alpha_s$, so that a description in terms of classical fields results \emph{a posteriori} justified. The effective theory to describe the small-$x$ gluons in the wave-function of the colliding nuclei is known as Color Glass Condensate (CGC): gluons are in fact coloured particles, their fields arise from sources which, in the relevant time-scales for the collision, appear as frozen (like a glass) and they form a dense system of bosons. As a result, the transverse-momentum distribution of small-$x$ gluons turns out to be peaked around a \emph{saturation momentum} $Q_S$, which introduces a typical inverse-length scale into the system. For high energy and heavy nuclei the latter can reach values $Q_S\gg\Lambda_{\rm QCD}\!\sim\! 200$~MeV for which the running coupling $\alpha_s(Q_S)$ is sufficiently small to make perturbative calculations possible.

CYM equations can be also used to study the field configuration after the collision~\cite{Kovner:1995ja,Kovner:1995ts}. The evolution of the system is usually described in terms of the \emph{longitudinal proper-time} $\tau\equiv\sqrt{t^2-z^2}$ and the space-time rapidity $\eta_s\equiv\frac{1}{2}\ln\frac{t+z}{t-z}$. In the high-energy limit the initial conditions and the evolution of the medium do not depend on rapidity. At the very beginning, the system arising from the collision is characterised by purely longitudinal electric and magnetic colour fields, which then evolve (for some fractions of fm/$c$) according to the CYM equations, developing also transverse components. Such a state of matter, which interpolates between the initial CGC and the thermalised Quark-Gluon Plasma (QGP), is referred to in the literature as Glasma~\cite{Lappi:2006fp}.
One of the most striking peculiarities of the Glasma is given by the strong event-by-event spatial fluctuations of its initial energy density profile.

Experimental data strongly support a picture in which, after this pre-equilibrium stage,
the matter produced in the collision reaches a state close to local thermal equilibrium, thus allowing a hydrodynamic description, with the evolution of the system driven by pressure gradients. Due to the strong interactions acting in the medium, the initial anisotropy and fluctuations are then mapped into the final hadron spectra, which will be affected by the collective flow of the fireball.

The first evidence for an early development ($\tau_0\lsim 1$~fm/$c$) of an hydrodynamic expansion of the system formed in heavy-ion collisions came from the observation of the elliptic flow $v_2\equiv\langle\cos[2(\varphi-\psi_{\rm RP})]\rangle$ characterising the distribution of the azimuthal angle $\varphi$ of the produced hadrons with respect to the direction of the \emph{reaction plane} $\psi_{\rm RP}$ (the plane defined by the collision impact parameter). The elliptic flow arises from the elongated shape of the overlap region of the two nuclei in non-central collisions: the larger pressure gradients along the reaction plane give rise to a stronger flow in this direction~\cite{Ollitrault:1992bk}. In particular, since the expansion of the system would tend to dilute the initial spatial asymmetry, this observation provided strong support to the hypothesis of an early thermalization of the matter produced in the collision. 


The main theoretical tool to map the initial geometrical anisotropy into the final spectra of the hadrons is given by \emph{relativistic}  hydrodynamics (both the speed of sound $c_s$ and the velocity of the fluid are in fact quite close to $c$)~\cite{Romatschke:2009im}. 
Hydrodynamics is an effective theory to describe the propagation
of long-wavelength excitations in a medium at local thermodynamical
equilibrium with a microscopic interaction length (mean free path $\lambda_{\rm mfp}$) much smaller
than the size of the system ($L$ of the order of the nucleus radius) and the length of variation of the 
thermodynamic fields, that is temperature, velocity and chemical 
potentials.
The small values of the Knudsen number $K_n\equiv\lambda_{\rm mfp}/L$
led people to describe the plasma produced at the temperatures accessible at RHIC and at the LHC as a strongly-coupled system. In the \emph{ideal} hydrodynamic limit the evolution of the hot QCD matter is completely described by the conservation law for the energy-momentum tensor ($\epsilon$ and $P$ are the energy density and the pressure, respectively)
\beq
\partial_\mu T^{\mu\nu}_{\rm id}=0,\quad{\rm with}\quad T^{\mu\nu}_{\rm id}\equiv(\epsilon+P)u^\mu u^\nu-P g^{\mu\nu}\quad{\rm and}\quad u^2=1.
\eeq
The Equation-of-State (EoS) $P\!=\!P(\epsilon)$ closes the system and is the only place where the information on the microscopic degrees of freedom enters. The symmetries of the underlying Lagrangian also identify the conserved charges of the system and the corresponding continuity equations.

Ideal hydrodynamics provides the strongest possible response of the medium to the initial-state anisotropy. However, more systematic studies pointed out the necessity to introduce dissipative corrections in the hydrodynamic evolution of the medium, accounting for deviations from full thermal equilibrium and arising (within a kinetic description) from the finite mean-free-path of the constituents. The energy momentum tensor receives then viscous corrections (with \emph{bulk} $\Pi$ and \emph{shear} $\pi^{\mu\nu}$ components):
\beq
T^{\mu\nu}\equiv T^{\mu\nu}_{\rm id}+\Pi^{\mu\nu}\equiv \epsilon\, u^\mu u^\mu-(P+\Pi)\Delta^{\mu\nu}+\pi^{\mu\nu},\quad{\rm with}\quad \Delta^{\mu\nu}\equiv g^{\mu\nu}-u^\mu u^\nu\quad{\rm and}\quad u_\mu\pi^{\mu\nu}=\pi^{\mu}_{\mu}=0.
\eeq

One is left with the problem of providing a rigorous definition of the viscous stress-tensor $\Pi^{\mu\nu}$. A first possible approach consists in imposing the second law of thermodynamics $\partial_\mu s^\mu\ge 0$, with the entropy current given by its equilibrium expression $s^\mu\!\equiv\! s u^\mu$. This leads to ($\nabla^\mu\equiv\Delta^{\mu\alpha}\partial_\alpha$)
\beq
{\Pi=-\zeta(\partial_\mu u^\mu)}\quad{\rm and}\quad \pi^{\mu\nu}=2\eta\nabla^{<\mu}u^{\nu>}\equiv 2\eta\left[\frac{1}{2}(\nabla^{\mu}u^{\nu}\!+\!\nabla^{\nu}u^{\mu})\!-\!\frac{1}{3}\Delta^{\mu\nu}(\partial_\mu u^\mu)\right],
\eeq
where $\zeta$ is the bulk viscosity and $\eta$ the shear viscosity. Equation 3 represents the relativistic extension of the \emph{Navier-Stokes} (NS) \emph{theory} and fully accounts for all first-order terms in a gradient expansion. The latter, however, poses conceptual problems due to its breaking of causality arising from the superluminal propagation of short-wavelength modes (found for instance in the study of shear perturbations). 
Including in the entropy current corrections proportional to $\Pi^2$ and $\pi^{\mu\nu}\pi_{\mu\nu}$ is sufficient to recover a causal behaviour. 
In this case the viscous components of the stress tensor are no longer simply defined in terms of the velocity field as in the NS theory, but require the solution of evolution equations of the form 
\beq
\dot\Pi\approx-\frac{1}{\tau_\Pi}[\Pi+\zeta(\partial_\mu u^\mu)],\quad
\dot\pi_{\alpha\beta}\approx-\frac{1}{\tau_\pi}[\pi_{\alpha\beta}-2\eta\nabla_{<\alpha}u_{\beta>}],
\eeq
where $\tau_\Pi$ and $\tau_\pi$ play the role of \emph{relaxation times} necessary for the viscous tensor to approach its NS limit. The above formalism goes under the name of \emph{Israel-Stewart theory}~\cite{Israel:1979wp}. It was developed by the authors in the '70s, with the purpose of applying it to astrophysical problems, but it was essentially ignored by the scientific community for many years until when it was rediscovered in the contest of the study of heavy-ion collisions. Now its conceptual importance for the consistent inclusion of dissipative effects in a relativistic framework is finally appreciated by a broad community, going beyond heavy-ion studies and representing a nice example of cross-fertilisation between different fields. 
In this regard, part of the Italian community recently developed a relativistic viscous hydrodynamic code by adapting to heavy-ion collisions an existing code for astrophysical applications. As a result, a first version of the ECHO-QGP code~\cite{DelZanna:2013eua} was released, representing the first public numerical tool to perform (3+1)D hydrodynamic calculations with dissipative effects.  

\begin{figure}[t]
\begin{center}
\includegraphics[clip,height=5.5cm]{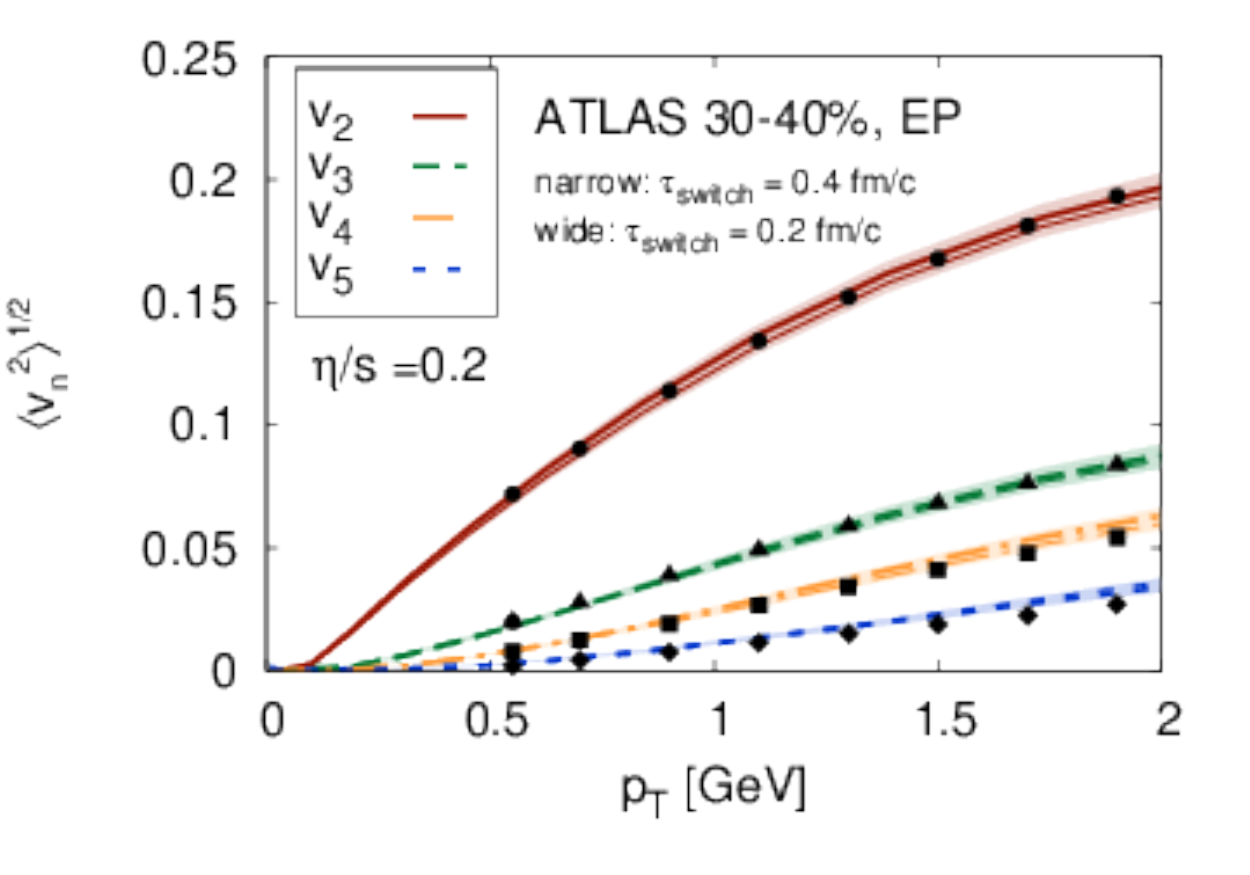}
\includegraphics[clip,height=5.5cm]{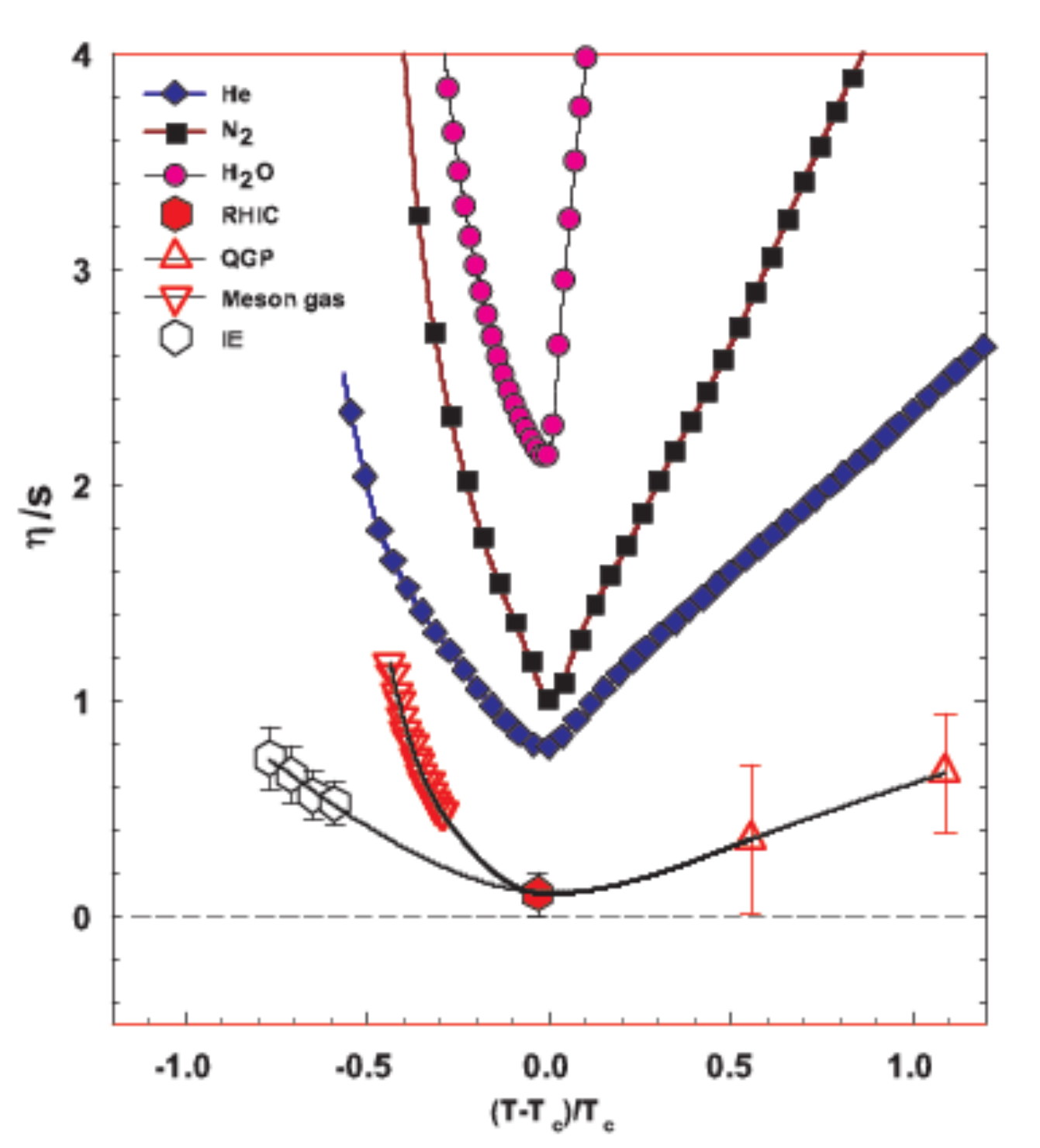}
\caption{Flow harmonics in Pb--Pb collisions at $\sqrt{s_{\rm NN}}$. Theory predictions with an initial ``Glasma'' dynamics interfaced with a subsequent viscous hydrodynamic evolution~\cite{Gale:2012rq} are compared to experimental data. Different choices of the ``Glasma''-hydro switching time are explored. In the right panel the viscosity-to-entropy ratio $\eta/s$ estimated for the QGP is compared to the one measured for other fluids around their liquid--gas phase transition~\cite{Lacey:2006bc}.}
\label{fig:IP-glasma}
\end{center}
\end{figure}

It is then necessary to go from a fluid to a particle description. A common assumption is that this transition occurs when the rate of elastic interactions of the particles 
becomes comparable with the expansion rate of the fluid or when the mean-free-path becomes of the same order of the system size. For simplicity one usually assumes that the decoupling happens suddenly, i.e. when, locally, the temperature of the fluid cell goes below the kinetic \emph{freeze-out} value $T_{\rm FO}$.
The condition $T(x)=T_{\rm FO}$ defines a three-dimensional kinetic freeze-out hyper-surface from which hadrons are emitted according to the \emph{Cooper-Frye} spectrum~\cite{Cooper:1974mv}
\beq
E\frac{\dd N}{\dd\vec{p}}=\int_{\Sigma_{\rm FO}}p^\mu d\Sigma_\mu\,f(x,p),\quad{\rm where}\quad f(x,p)=f_{\rm eq}(p\!\cdot\! u(x))+\delta f(x,p).
\eeq
In the above $f(x,p)$ is given by the sum of an equilibrium (Bose or Fermi) distribution in the local fluid rest-frame plus an off-equilibrium correction arising from presence of dissipation. The final hadrons (at variance with the case of proton-proton collisions) are then affected by the collective flow of the medium, which carries the fingerprints of the initial anisotropy and pressure gradients of the system via the Euler equation (here given in the absence of viscosity)
\beq
(\epsilon+P)\dot u^\mu=  \nabla^\mu P\,.
\eeq
Particles are accelerated outward, i.e.\, in the direction opposite to the pressure gradient.
In particular, the collective \emph{radial flow of the system} leads to harder distributions of transverse momentum ($p_{\rm T}$) of the final-state hadrons.  Several analyses are devoted to the study of more refined features of particle distributions like the observation of \emph{higher} \emph{harmonics} in the azimuthal distribution of produced particles. These are quantified by the coefficients $v_n$ of the Fourier expansion of the azimuthal distribution with respect to the estimated reaction plane $\psi_{\rm RP}$. The second harmonic coefficient $v_2$ is the \emph{elliptic flow}. It characterises non-central collisions, because of the larger pressure gradient along the reaction plane. Higher harmonics ($n>2$) arise from event-by-event fluctuations in the energy-density transverse profile in the initial state. Within the Glasma picture initial conditions can be very bumpy, with longitudinal colour fields (\emph{flux-tubes}) with a very short (even shorter than the nucleon size) correlation-length $\sim Q_S^{-1}$ in the transverse plane. Results of viscous hydrodynamic simulations with such initial conditions~\cite{Gale:2012rq} are displayed in Fig.~\ref{fig:IP-glasma}.

A systematic comparison of theory calculations with experimental results for the various flow-harmonics $v_n$~\cite{Adare:2011tg,Sorensen:2011fb,ALICE:2011ab,ATLAS:2012at,Chatrchyan:2013kba} (see Fig.~\ref{fig:IP-glasma} as an example) provides information on the initial state of the system and on important properties of the produced medium, like in particular the $\eta/s$ ratio between shear viscosity and entropy density. The above ratio turns out to be quite close to the \emph{universal lower-bound} $\eta/s=1/4\pi$ predicted by the AdS/CFT correspondence for any gauge theory with a gravity dual~\cite{Kovtun:2004de}, making the QGP produced at RHIC and LHC, when compared to other substances, the ``most perfect'' fluid~\cite{Lacey:2006bc}. 
Although QCD in the region close the deconfinement phase-transition is far from being conformal, recently such an approach was widely employed in the literature, since it is perhaps the only one able to provide predictions for \emph{real-time} quantities (like for instance transport coefficients) in a strong-coupling non-perturbative regime. First-principle lattice-QCD calculations are in fact formulated in imaginary-time and provide direct results only for equilibrium thermodynamic quantities.

\subsection{Hard probes of the QGP properties}
\label{sec:hardprobes}

The term \emph{hard probes} indicates particles (hadrons or partons) that a) are chacterized by a hard scale (mass or momentum) and are therefore produced in the first instants of the nucleus--nucleus collision in hard partonic scatterings, and b) are affected by the presence of the strongly-interacting QGP, which they cross after their production. Hard probes include: 
quarkonia (charmonia and bottomonia); heavy quarks, which are detected in the final-state as open heavy flavour hadrons or their decay products; high-momentum light quarks and gluons, which are detected in the final-state as high-momentum hadrons and jets.
The measurement of the modification of the yield and kinematic properties of hard probes is regarded as a rich source of information on their interaction 
with the QGP and on the QGP properties. In the following, we outline the main theoretical ideas on the description of the medium-induced modification of hard probes. More details will be given in Sections~\ref{sec:ohf}, \ref{sec:onia} and~\ref{sec:jets}.

The suppression of ${\rm J}/\psi$ production in heavy-ion collisions was one of the first proposed signatures of the onset of deconfinement~\cite{Matsui:1986dk}. The argument was based on the \emph{Debye-screening} of the $Q-\overline{Q}$ potential
$-\alpha/r\longrightarrow -\alpha/r\exp(-r/r_D)$ due to the large density $n\sim T^3$ of free colour charges in the plasma. The initially-produced ${\rm J}/\psi$'s have to cross a medium characterised by a Debye screening radius smaller than the size of the state $r_D<\overline{r}_{{\rm J}/\psi}$ so that the potential is no longer able to support bound states. Clearly, the smaller the binding energy (the larger the radius), the earlier the state will dissociate: a \emph{sequential suppression} scenario of the different charmonium and bottomonium states as a function of the centrality of the collisions was then proposed, with the purpose of using quarkonia as a \emph{thermometer} of the produced medium. Current measurements at the LHC allow one to test this scenario using a systematic comparison of the suppression of ${\rm J}/\psi$, $\psi'$  (for charm) and $\Upsilon$(1S), $\Upsilon$(2S), $\Upsilon$(3S) (for bottom) states~\cite{Adam:2015jsa,Chatrchyan:2012lxa}. From the theory point of view, the challenge is to make the above picture more quantitative. The first possibility is to take advantage of the numerical results of lattice-QCD simulations (described in Section~\ref{sec:lQCD}): one can get information on the free-energy of static (infinitely heavy) $Q\overline{Q}$ pairs in the QGP~\cite{Kaczmarek:2005ui} and on the in-medium quarkonium spectral functions in the different channels, looking for the survival/disappearance of bound-state peaks as the temperature increases. Analytic approaches, although facing the difficulties arising from the strongly-coupled nature of the medium at the experimentally-accessible conditions, have the advantage of providing a cleaner physical insight on the involved processes. Modern approaches, combining Effective Field Theory (EFT) techniques and thermal field-theory calculations~\cite{Laine:2006ns,Beraudo:2007ky,Brambilla:2008cx}, have displayed the complex nature of the effective potential describing the evolution of $Q\overline{Q}$ pairs in the QGP, with a real part accounting for the screening of the interaction and an imaginary part arising from the collisions with the QGP constituents.
 
The role of open heavy-flavour observables ($\rm D$ and $\rm B$ mesons and their decay products) to characterize the QGP will be discussed in detail in Section~\ref{sec:ohf}. Here we summarize the theoretical framework (for a review see for instance Refs.~\cite{Rapp:2009my,Beraudo:2014iva}). Due to their large mass $c$ and $b$ quarks are produced in initial hard processes, which can be calculated within pQCD~\cite{Alioli:2010xd}. At variance with the proton-proton case, in heavy-ion collisions the heavy quarks emerging from the hard interaction have to propagate in a hot deconfined medium. One expects then that their spectra and correlations get modified with respect to proton-proton events. In the limit in which the interaction with the medium is very strong, heavy quarks could reach thermal equilibrium with the rest of the system and the final observables like $D$-meson spectra might even display common features with light hadrons, like signatures of radial and elliptic flow.
The standard tool to study heavy-flavour particles in a hot medium is represented by \emph{transport calculations} based on the Boltzmann equation. The latter is very often replaced by its soft momentum-exchange limit, the Fokker-Planck or Langevin equation, which has easier numerical implementation. 
The comparison of experimental measurements with the results of these calculations provides information on the transport coefficients of the QGP 
and on the degree of thermalization of heavy quarks in the system.
In the high-$\pt$ regime, heavy quarks become a tool to study the mass and colour-charge dependence of parton energy-loss, which will be discussed in the next paragraph and in Section~\ref{sec:jets}.   

\begin{figure}[t]
\begin{center}
\includegraphics[clip,height=5.6cm]{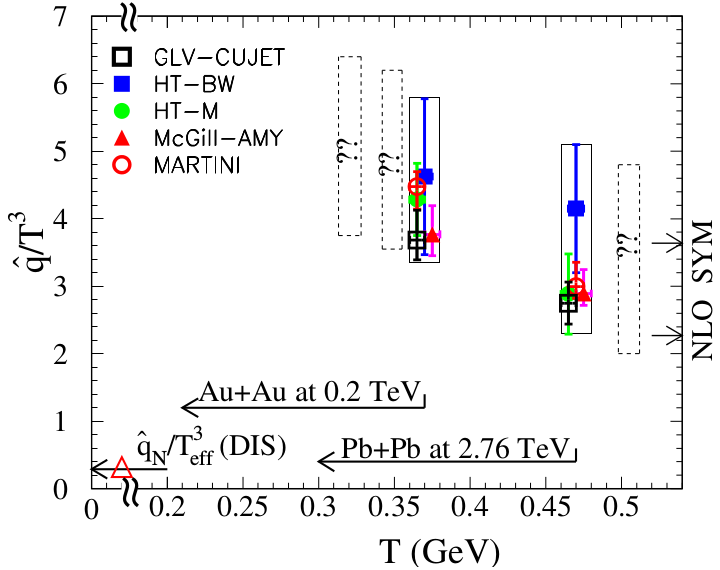}
\includegraphics[clip,height=6cm]{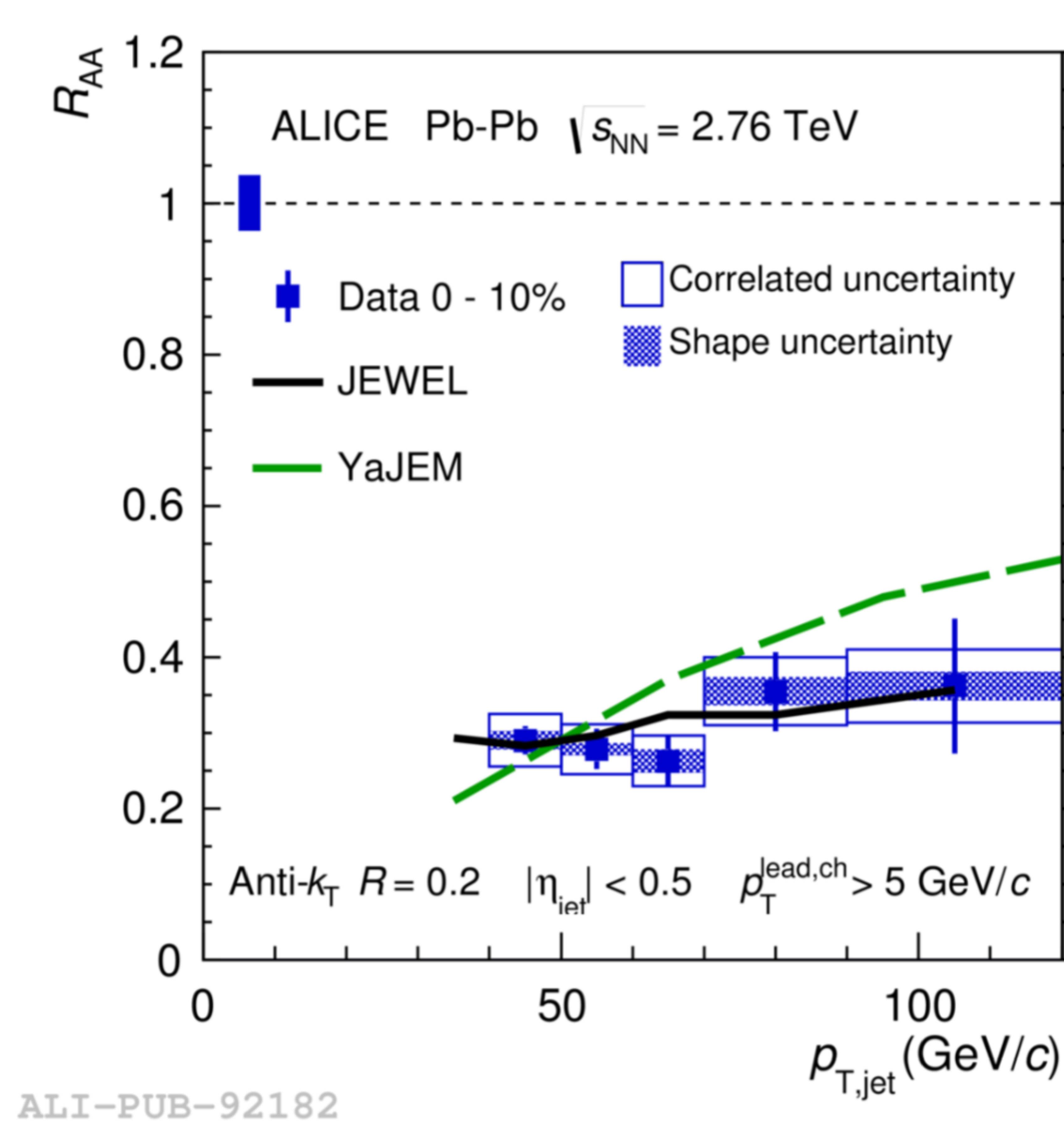}
\caption{Left panel: estimates for the value of the jet-quenching parameter $\hat{q}$ obtained from the comparison of various model calculations with RHIC  ($\sqrt{s_{\rm NN}}=0.2$ TeV) and LHC ($\sqrt{s_{\rm NN}}=2.76$ TeV) data~\cite{Burke:2013yra}. The dashed boxes indicate expected values at $\sqrt{s_{\rm NN}}=0.063$, 0.130 and 5.5 TeV. Right panel: the nuclear modification factor of jet $\pt$ spectra in central Pb--Pb collisions at the LHC. Theory predictions~\cite{Zapp:2013vla,Renk:2013rla} are compared to experimental results from the ALICE experiment~\cite{Adam:2015ewa}.}
\label{fig:jetq}
\end{center}
\end{figure} 

High-$\pt$ particles are produced in hard processes occurring in the initial stage during the crossing of the two nuclei and described by pQCD. High-energy partons cross a few fm of QGP before hadronizing. In such an environment, rich  of coloured gluons with softer momenta, the hard quarks and gluons lose part of their energy via elastic and inelastic processes. At very high energy the main role is played by medium-induced gluon radiation~\cite{Baier:1996sk,Gyulassy:2000fs}: hard partons interact with quarks and gluons of the medium, exchanging with them momentum and colour, becoming radiators of, mostly, soft and collinear gluons (for a comprehensive review see e.g.~\cite{Armesto:2011ht}). The transport coefficient $\hat{q}$ quantifies the average squared transverse-momentum exchange per unit length and is one of the parameters one aims at extracting from the data: results are shown in the left panel of Fig.~\ref{fig:jetq}. The development of a parton shower, in the vacuum simply driven by virtuality degradation, is thus modified in the presence of a medium, because the interaction increases the probability of gluon radiation. This picture has been recently implemented in numerical codes (see e.g. Ref.~\cite{Zapp:2013vla}). 
As a result of in-medium parton energy loss and of the colour-decoherence of the radiated gluons~\cite{Beraudo:2011bh,Beraudo:2012bq} the production of high-$\pt$ hadrons and jets in heavy-ion collisions is suppressed with respect to the proton-proton case. The suppression is quantified using the \emph{nuclear modification factor} $R_{\rm AA}\equiv N^{\rm AA}/\langle N_{\rm coll}\rangle N^{\rm pp}$, which is the ratio of the yields (usually differential in $\pt$ and/or $y$) in the A-A and pp case, rescaled by the average number of nucleon-nucleon collisions. This phenomenon is usually referred to as \emph{jet quenching}. The right panel of Fig.~\ref{fig:jetq} shows the comparison of a jet nuclear modification factor measurement at the LHC~\cite{Adam:2015ewa} with the predictions of jet quenching simulations~\cite{Zapp:2013vla} and of an analytical calculation of parton energy loss~\cite{Renk:2013rla}.

\subsection{Electromagnetic probes:  initial state and thermal dilepton radiation}
\label{sec:EMtheory}

In contrast to hadrons, photons and lepton pairs (dileptons) directly probe the entire space-time evolution of the expanding fireball formed in  high energy nucleus-nucleus collisions, escaping freely without final-state interactions. 

Before thermalization, dileptons might be produced from Drell-Yan, open-charm and charmonia. After thermalization, further processes contribute:

\begin{itemize} 
\item in the  QGP phase, lepton pairs can be 
produced by thermal $q\overline q$ annihilation;
\item after hadronization, the resulting strongly-interacting hadronic medium has still thermal properties and 
continues to expand while cooling down. In this phase, the main sources of thermal lepton pairs are the $\rho(770)$ meson and multi-pion processes related to $\rho-a_1$ chiral-mixing.
\end{itemize}

It was suggested that dileptons and photons might  provide 
 a benchmark for the initial stage before thermalization~\cite{Chiu:2012ij,McLerran:2014hza,Ryblewski:2015hea,Tanji:2015ata}. 
Generally speaking, both photons and dileptons are sensitive to the
distribution functions of quarks in the medium. In the early stages of the collision,
these distribution functions bring informations on the anisotropy of the system;
as a natural consequence, particle production will be affected by anisotropy,
which is quite large in the early stage.
Moreover, one has to consider that in addition to the particle production due to
quantum inelastic processes, there is also the dynamics of the classical fields,
whose evolution is determined by the expectation value of quark currents.
These currents act as sources for the electromagnetic field
\cite{Tanji:2015ata}, hence giving a further early stage contribution to the photon spectrum,
which has to be added in the theoretical computation to the single particle production due
to scattering processes.

Results presented in~\cite{Chiu:2012ij,McLerran:2014hza,Ryblewski:2015hea,Tanji:2015ata} 
show  first estimates of initial stage production
of photons and dileptons, which has to be added to the better understood production in the
quark-gluon plasma and hadron phases. It would be desirable to complete the aforementioned
studies using a single theoretical framework which encodes production of particles
both from the classical fields and from the quantum scattering processes, which consistently
follows the dynamical evolution of the system from the early stage up to the final stage of hadronization.
While this requires a considerable amount of  work, eventually it would lead to a firm 
quantitative understanding of the role of the initial stage dynamics on empirical observables.

\begin{figure}[t]
\begin{center}
\includegraphics[width=12.cm]{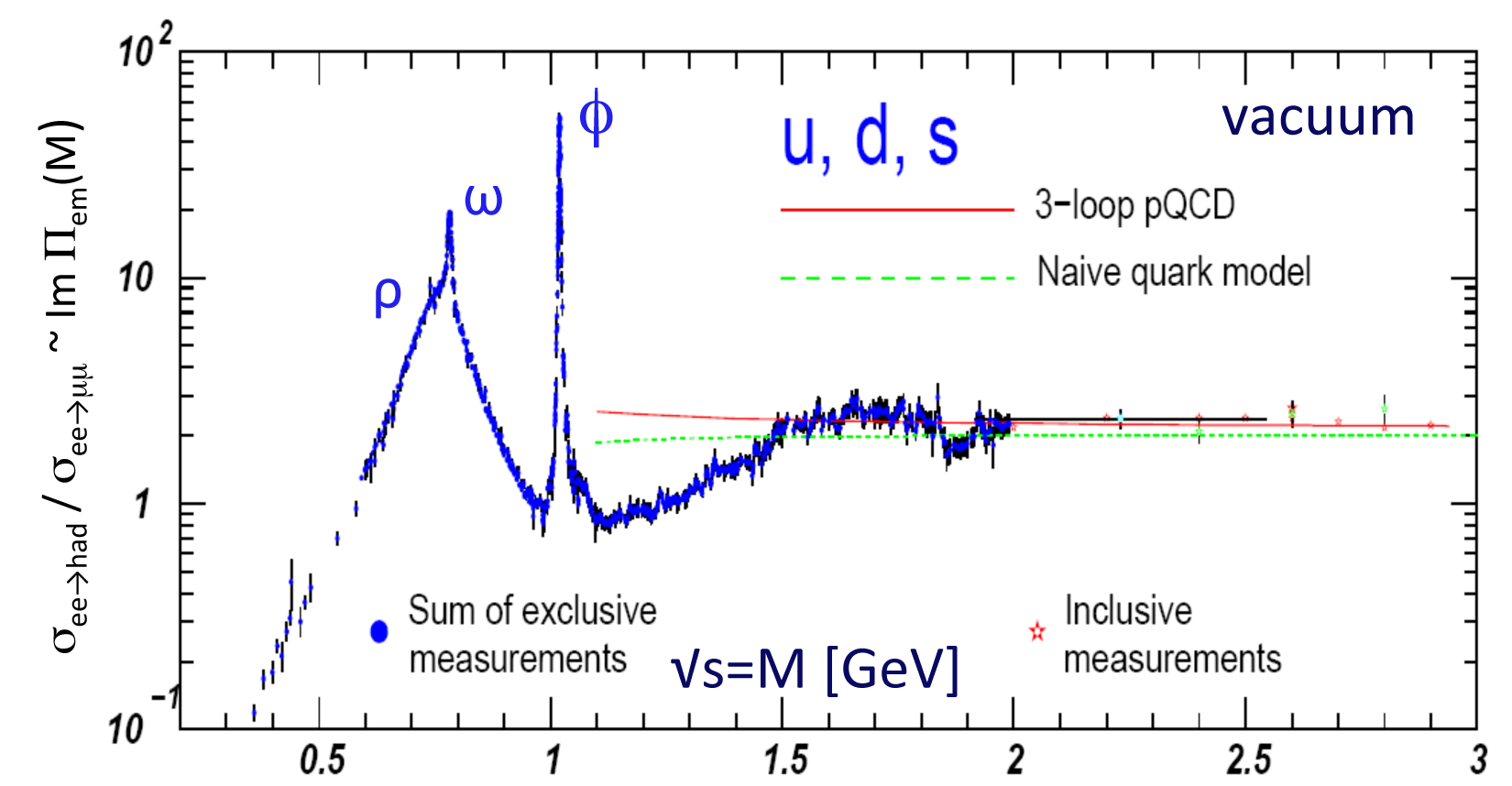}
\caption{Ratio $R$ of the $e^+e^-\to$hadrons  to
$e^+e^-\to\mu\mu$ cross-sections~\cite{Agashe:2014kda}, related to the electromagnetic spectral function in vacuum.}
\label{fig:vacuumSpectralEm}
\end{center}
\end{figure}

The thermal dilepton rate per unit space-time and 4-momentum volume can be  expressed by~\cite{McLerran:1984ay,Weldon:1990iw,Gale:1990pn}

\begin{equation}
\frac{\dd N}{\dd^4x\dd^4q}= - \frac{\alpha_{em}}{\pi^3 M}f^B(q^0,T)Im\Pi_{em}(M,q,\mu_B,T),
\label{eq:dileptonrate}
\end{equation}

where $f^B(q^0,T)$ is a Boltzmann factor and $Im\Pi_{em}(M,q,\mu_B,T)$ is the electromagnetic spectral function. In the vacuum, the electromagnetic spectral function is proportional to the ratio $R$ of the $e^+e^-\to$hadrons to $e^+e^-\to\mu^+\mu^-$ cross-sections (Fig.~\ref{fig:vacuumSpectralEm}), with $R=-\frac{12\pi}{s}Im\Pi_{em}$.

\begin{figure}[t]
\begin{center}
\includegraphics[width=9.cm]{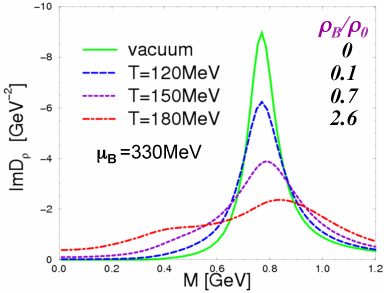}
\caption{The in-medium $\rho$ spectral function at fixed baryon chemical potential $\mu_B = 330$~MeV and
temperatures $T = 180$~MeV (corresponding to  total baryon density $\rho_B = 2.6\rho_0$), $T = 150$~MeV ($\rho_B = 0.7\rho_0$) and $T = 120$~MeV ($\rho_B = 0.1\rho_0$)~\cite{Rapp:1999us}.}
\label{fig:RhoSpectralFunction}
\end{center}
\end{figure}

For $M<1$~GeV/$c^2$ (Low Mass Region), the self-energy is dominated by the vector mesons $\rho$, $\omega$ and $\phi$. For M>1.5 GeV, several overlapping resonances lead to a continuum with a flattened spectral density corresponding to a simpler description in terms of quarks and gluons (hadron-parton duality).
Thus one has a {\it non-perturbative} regime for $M<1$~GeV/$c^2$ with

\begin{equation}
Im\Pi_{em}\sim ImD_{\rho,\omega,\phi},
\end{equation}

where $ImD_{\rho,\omega,\phi}$ are the light vector meson spectral functions. For $M>1.5$~GeV/$c^2$ the $q\overline q$ continuum leads to

\begin{equation}
\frac{Im\Pi_{em}}{M^2}\sim const\cdot\left(1+O[T^2/M^2]\right).
\end{equation}

As said above, the broad $\rho(770)$ is by far the most important among the vector mesons, due to its strong coupling 
to the $\pi^+\pi^-$ channel and its life-time of only 1.3~fm$/c$, making it subject to regeneration in the much longer-lived fireball. Indeed, for a long time the $\rho$ has been 
considered as the test particle for “in-medium modifications” of hadron properties close to the QCD phase boundary. 

In the hot hadronic matter,  the vector meson propagators  are
calculated using many-body models~\cite{Rapp:1997fs,Rapp:1999us,Urban:1998eg,Rapp:2000pe}.
The calculations show that the $\rho$ spectral function strongly broadens due to the coupling of the $\rho$ to several baryonic resonances. This is shown in Fig.~\ref{fig:RhoSpectralFunction} -  around the phase transition region the width diverges ($\rho$ 'melting').  The mass range $1<M<1.5$~GeV/$c^2$ is dominated by multi-pion annihilation (see Chap. 5 for more details) where the process  $a_1\pi^0\to\mu^+\mu^-$ plays a very important role~\cite{Dey:1990ba,Steele:1996su}.
Here, medium effects are related to the $\rho-a_1$ chiral mixing and thus may be sensitive to chiral symmetry restoration.

\begin{figure}[t]
\begin{center}
\includegraphics[width=8.cm]{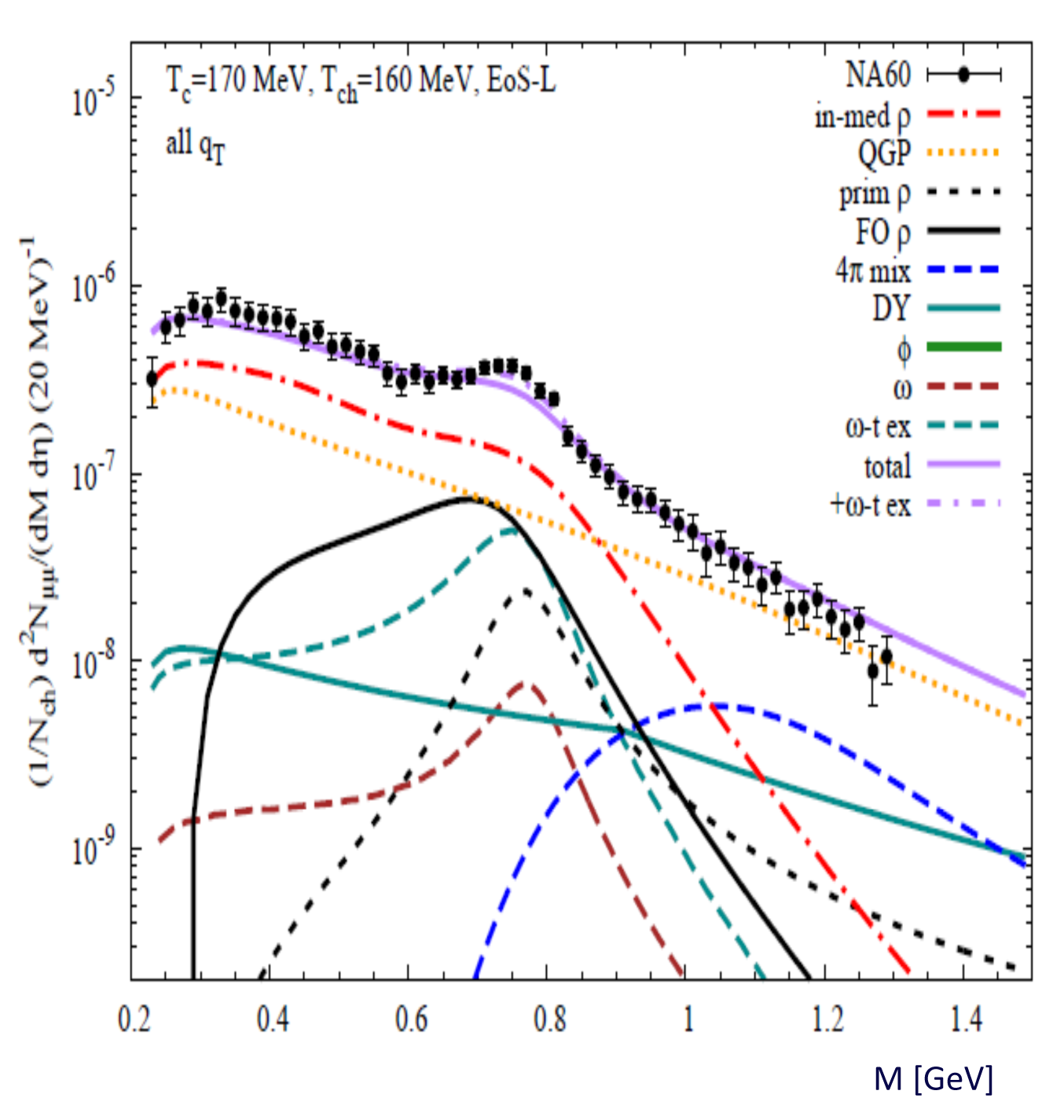}
\caption{Theoretical calculations for the thermal radiation based on the in-medium $\rho$ (dotted-dashed red line), omega spectral functions, multi-pion
annihilation with chiral mixing (dashed blue line), and QGP radiation (dotted orange line). The total yield is compared to the NA60 data~\cite{Arnaldi:2008er}.}
\label{fig:ThermalRadiation}
\end{center}
\end{figure}

Finally, theoretical rates from the QGP have been estimated based on lattice QCD or hard thermal loops calculations~\cite{Ding:2010ga,Brandt:2012jc,Ding:2013qw} or hard thermal loops~\cite{Braaten:1990wp}. 

The resulting thermal dilepton spectrum in heavy ion collisions is obtained by integrating Eq.~\ref{eq:dileptonrate} over emission volume and momenta along the fireball evolution. For $M<1.5$~GeV this leads 
to

\begin{equation}
\frac{\dd N}{\dd  M}\propto M^{3/2} \langle \exp(-M/T)\rangle \langle {\rm spectral\, function} (M) \rangle\,,
\end{equation}

while for $M > 1.5$~GeV one has

\begin{equation}
\frac{\dd  N}{\dd  M}\propto M^{3/2} \langle \exp(-M/T)\rangle.
\end{equation}

The shape of the dilepton spectrum in this mass region, directly related to the medium temperature, provides a {\it thermometer} to distinguish the partonic and hadronic origin (see Chap. 5 for more details).

The theoretical thermal dilepton yields from the QGP and hadronic phases estimated for Indium-Indium collisions at 160~GeV/nucleon in the lab system are shown in Fig.~\ref{fig:ThermalRadiation}. In the low mass region ($M<1$~GeV) the yield is dominated by the $\rho$, while
the QGP is dominant for $M>1$~GeV/$c^2$. At these energies the yield from the $a_1\pi^0\to\mu^+\mu^-$ process is  small in comparison to the QGP.
The theoretical estimate for the total thermal yield is in quantitative agreement with the measurements performed by the NA60 experiment~\cite{Arnaldi:2008er}.

\subsection{Lattice calculations: a tool to study QCD thermodynamics, phase diagram and spectral functions}
\label{sec:lQCD}

Numerical simulations of QCD discretised on an Euclidean space-time lattice
represent the best available first-principle tool to explore the thermal properties
of strongly-interacting matter
in the non-perturbative regime.
One rewrites the QCD thermal partition function in terms of an Euclidean
path integral, which is then discretised on a space-time lattice
and evaluated numerically by Monte-Carlo methods.
In this way 
we can obtain information  about
basic equilibrium thermodynamics (e.g., pressure and energy density),
equilibrium particle and conserved charge distributions, 
and other quantities which are relevant to a description of the 
phases of strongly-interacting matter.

The liberation of colour degrees of freedom
is clearly visible from the rapid change of various
thermodynamic quantities and roughly coincides 
with the restoration of chiral symmetry.
By now, it is well established that, for null baryon chemical potential $\mu_B =0$, 
such a rapid change takes place at 
a temperature $T_c \sim 155$~MeV (with an uncertainty of about 8 MeV~\cite{Wu:2006su,Cea:2007vt})and does
not correspond to a true phase transition~\cite{Aoki:2006we,Aoki:2006br,Borsanyi:2010bp,Bazavov:2011nk,Bhattacharya:2014ara},
but instead to a cross-over with no associated critical behaviour.
Our present knowledge becomes less well defined as 
we consider the extension of the QCD phase diagram 
to $\mu_B\gg 0$: this is due
to a technical problem, the complex nature of the path integral 
measure (the so-called \emph{sign problem}),
which hinders the application of standard Monte-Carlo sampling 
techniques.
Various methods exist to circumvent the
problem in the regime of small chemical potentials, 
while other methods to completely
solve it are currently under study.
 


Two issues of primary importance, related to the introduction of 
a non-null baryon chemical potential $\mu_B$, regard a) how the critical
temperature changes as a function of $\mu_B$ (pseudo-critical line) and b) whether
the cross-over turns into a first-order transition at some critical value
of $\mu_B$, corresponding to a critical endpoint. The first issue is 
interesting for a comparison with the line of 
chemical freeze-out, as deduced from heavy-ion experiments, while
the second would represent a key point in the QCD phase diagram,
with clear experimental signatures.

In the regime of small chemical potentials,
the pseudo-critical
line can be approximated by:
\begin{equation}\label{corcur}
\frac{T_c(\mu_B)}{T_c}=1-\kappa \left(\frac{\mu_B}{T_c}\right)^2\, +\, 
O(\mu_B^4)\, ,
\end{equation}
where the coefficient $\kappa$ defines the curvature of the line
$T_c (\mu_B)$ (see, for illustration, Fig~\ref{fig:QCDPD}-left). 
Information about $\kappa$ can be obtained, in lattice simulations, in 
various ways:
by suitable combinations of expectation values computed at $\mu_B  = 0$ (Taylor
expansion method~\cite{Allton:2002zi,Kaczmarek:2011zz,Endrodi:2011gv,Borsanyi:2012cr}), {by determining $T_c(\mu_B)$ for purely imaginary
values of $\mu_B$ -- for which numerical simulations are feasible -- and then 
inferring the behaviour for small and real $\mu_B$ by analytic
continuation}~\cite{deForcrand:2002ci,D'Elia:2002gd,Azcoiti:2005tv,Wu:2006su,Cea:2007vt,Cea:2009ba,Cea:2010md,Nagata:2011yf,Cea:2012ev,Laermann:2013lma,Cea:2014xva,Bonati:2014rfa}, {by 
re-weighting techniques~\cite{Fodor:2001pe,Fodor:2004nz} and, finally, 
{by a reconstruction of the canonical partition function}~\cite{Kratochvila:2005mk,Alexandru:2005ix}.

Lattice results obtained for $\kappa$ have been usually smaller
than those obtained for the curvature of the freeze-out curve
~\cite{BraunMunzinger:1994xr,Becattini:1995if,Becattini:1997rv,Becattini:2000jw,BraunMunzinger:2001ip,Andronic:2005yp,Cleymans:2005xv,Becattini:2012xb},
however recent numerical investigations, adopting the method of
analytic continuation with improved discretisations at or close to the physical
point of (2+1)-flavours QCD, have provided larger results, 
compared to previous estimates obtained by the Taylor expansion
technique~\cite{Kaczmarek:2011zz,Endrodi:2011gv,Borsanyi:2012cr}.
Such a tendency has been confirmed by very recent
updates, aiming at obtaining a full control over the continuum
extrapolation~\cite{Bonati:2015bha,Bellwied:2015rza}: a value 
$\kappa \approx 0.014$ is compatible with all these studies.

Regarding the location, and even the existence, of the critical endpoint, the situation is 
currently less well defined. Present studies
are based on re-weighting techniques, on the determination
of the canonical partition function or on estimates of the radius
of convergence of the Taylor series in $\mu_B$; however,
various limitations still do not permit a complete control 
over systematic errors.
%
A conclusion that can be drawn at the moment is that the critical endpoint,
if any, is not to be found in the small $\mu_B$ region. This reduces the chances to accurately predict its position 
before a complete solution to the sign problem is reached.
Various efforts are being pursued in this direction, including
Langevin simulations for generic complex actions~\cite{Karsch:1985cb,Aarts:2008rr},
lattice simulations on a Lefschetz thimble~\cite{Cristoforetti:2012su}, 
density of states methods~\cite{Fodor:2007vv,Langfeld:2012ah},
formulation in terms of dual variables~\cite{Gattringer:2014nxa},
tensor Renormalization Group techniques~\cite{Denbleyker:2013bea} and
effective Polyakov loop models~\cite{Greensite:2014isa,Langelage:2014vpa}.

The equation of state of QCD, as well as the fluctuations of conserved charges, are the most prominent example of observables which allow to directly relate fundamental theory and heavy-ion collision experiments. The experimental results available so far show that the hot QCD matter produced experimentally exhibits robust collective flow phenomena, which are well and consistently described by 
near-ideal relativistic hydrodynamics \cite{Teaney:2000cw,Teaney:2001av,Kolb:2003dz}. These hydrodynamical models need as an input an Equation of State (EoS) which relates the local thermodynamic quantities.  Therefore, an accurate determination of the QCD EoS is an essential ingredient to understand the nature of the matter created in heavy ion
collisions, as well as to model the behaviour of hot matter in the early Universe.
The EoS of QCD is now available, in the continuum limit, for zero and small values of the chemical potential \cite{Borsanyi:2010cj,Borsanyi:2013bia,Bazavov:2014pvz,Borsanyi:2012cr} and is displayed in Fig.~\ref{fig:QCDEOSmu}; one of the challenges for the future is to extend these results to the entire phase diagram. This will be of fundamental importance, in view of the second Beam Energy Scan at RHIC, of the continuation of the SPS programme and of the future projects at NICA and FAIR, but it will only be achieved once the sign problem is solved.

\begin{figure}[!t]
\begin{minipage}{.48\textwidth}
\parbox{6cm}{
\scalebox{.51}{
\includegraphics{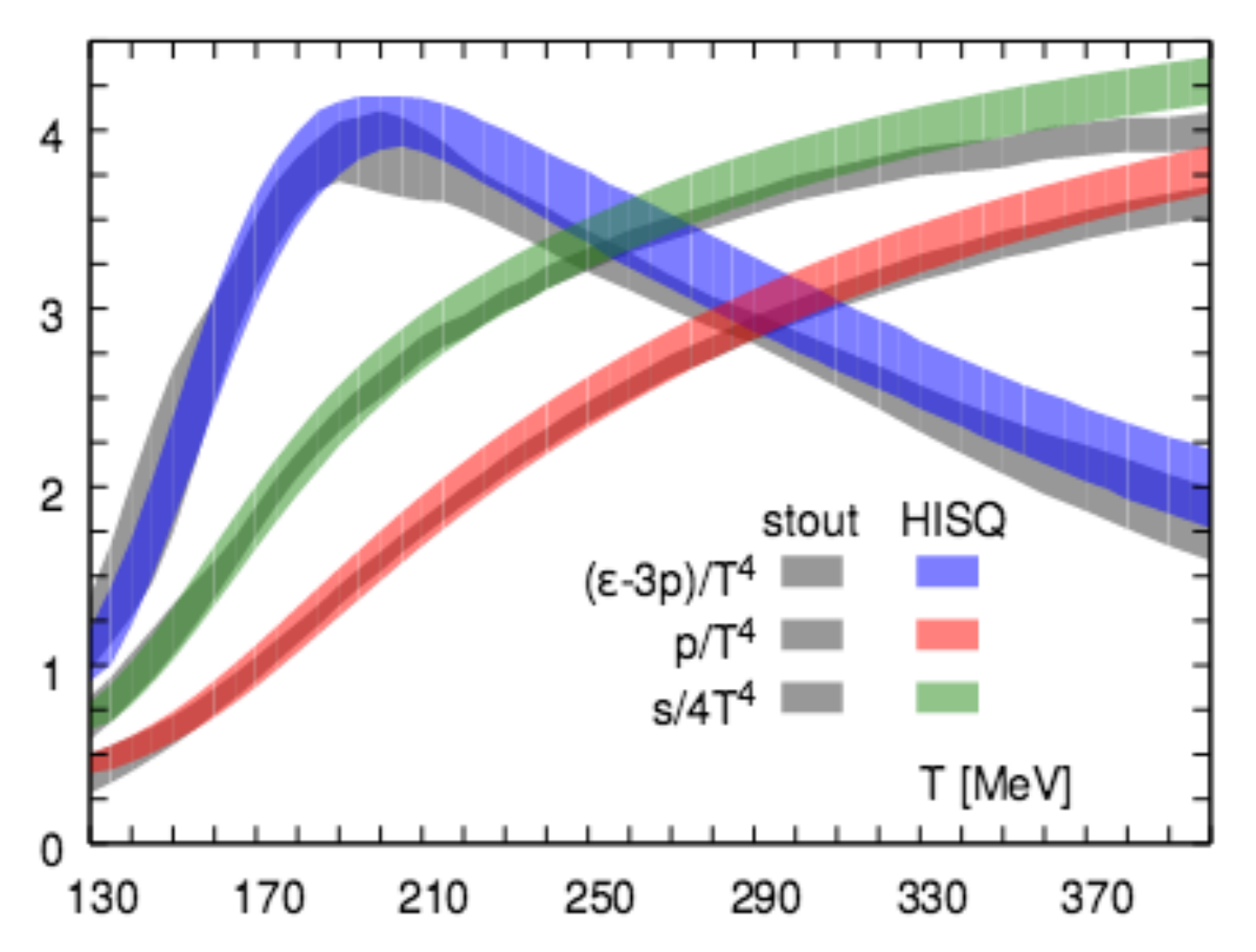}\\}}
\end{minipage}
\begin{minipage}{.48\textwidth}
\parbox{6cm}{
\scalebox{.35}{
\includegraphics{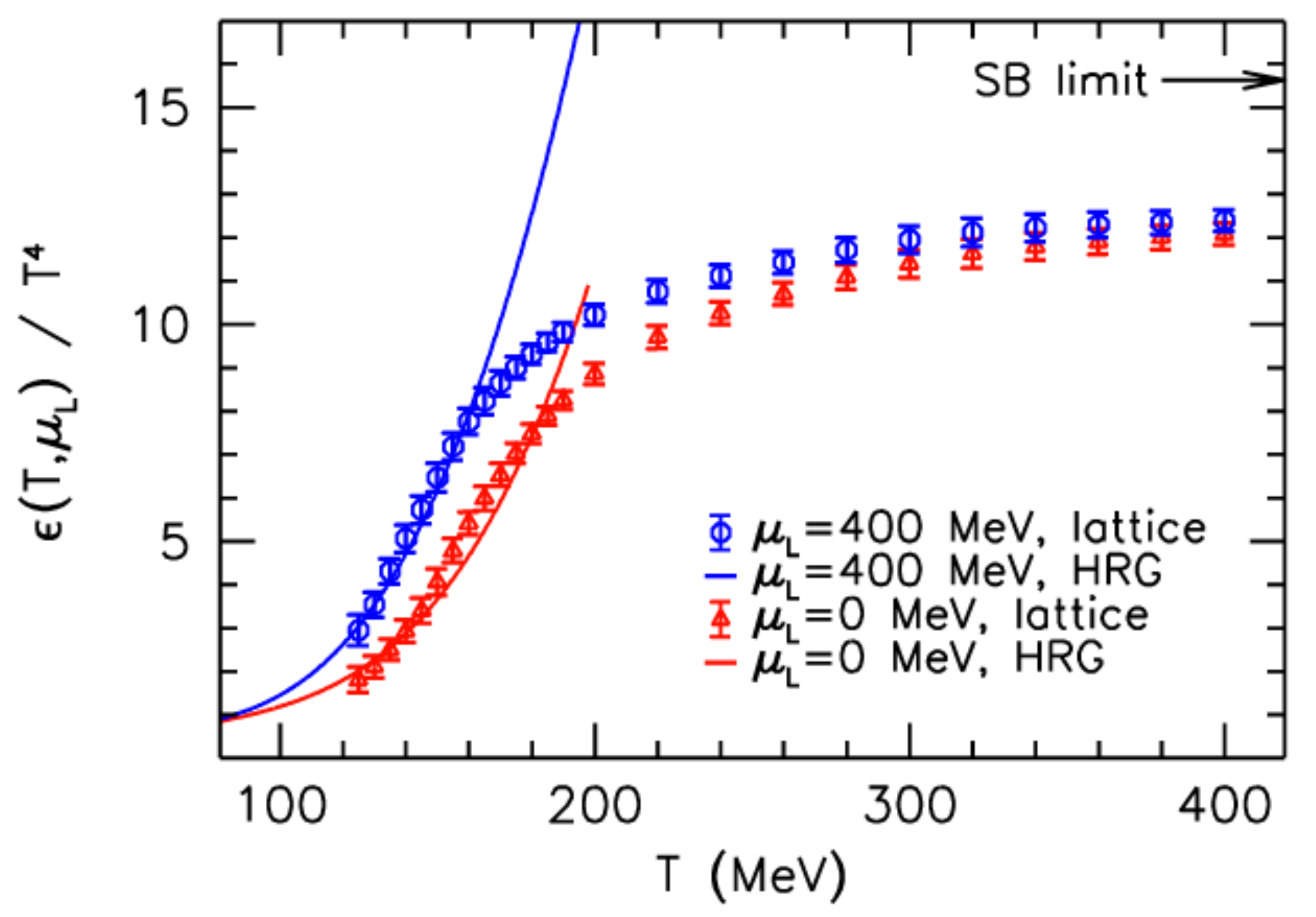}\\}}
\end{minipage}
\caption{\label{fig1} Left: equation of state of QCD at $\mu_B=0$: comparison between the results of Ref. \cite{Borsanyi:2013bia}, obtained with the stout action, and those of Ref. \cite{Bazavov:2014pvz}, obtained with the HISQ action. Right: equation of state of QCD for small values of the chemical potential \cite{Borsanyi:2012cr}.}\label{fig:QCDEOSmu}
\end{figure}

Event-by-event  fluctuations  of  the  net-electric  charge and  net-baryon  number,   which  are  conserved  charges of  QCD,  are  expected  to  become  large  near  a  critical point \cite{Berges:1998rc,Halasz:1998qr}:  for this reason,  they have been proposed as ideal observables to verify its existence and to determine its position in the QCD phase diagram \cite{Stephanov:1999zu,Gavai:2008zr}.  Experimental  results  for  these  measures  were  recently  reported  for
several collision energies \cite{Adamczyk:2013dal,Adamczyk:2014fia}.  In addition,  as a consequence of the increasing precision achieved in 
the numerical simulations, it is becoming possible to extract the chemical  freeze-out  parameters  (i.e.   freeze-out  temperature $T_{ch}$ and  corresponding  baryo-chemical  potential $\mu_{B,ch}$)
from first principles, by comparing the measured fluctuation observables to corresponding susceptibility ratios calculated  in  lattice  QCD  \cite{Karsch:2012wm,Bazavov:2012vg,Borsanyi:2013hza,Borsanyi:2014ewa}.   
An example of these calculations is shown in Fig. \ref{fig2}: the present limitations in lattice results allow at the moment to extract only an upper limit for the freeze-out temperature; also, the values of the freeze-out chemical potentials are limited to the higher collision energies. Future improvements in the lattice precision, as well as extension to larger chemical potentials, will allow to extract the freeze-out parameters in the entire range of energies addressed by the experiments. This will also test whether the non-monotonic behaviour observed in the net-proton fluctuations \cite{Adamczyk:2013dal} is a signal of the vicinity to the critical point.

\begin{figure}[!t]
\parbox{6cm}{
\scalebox{.6}{
\includegraphics{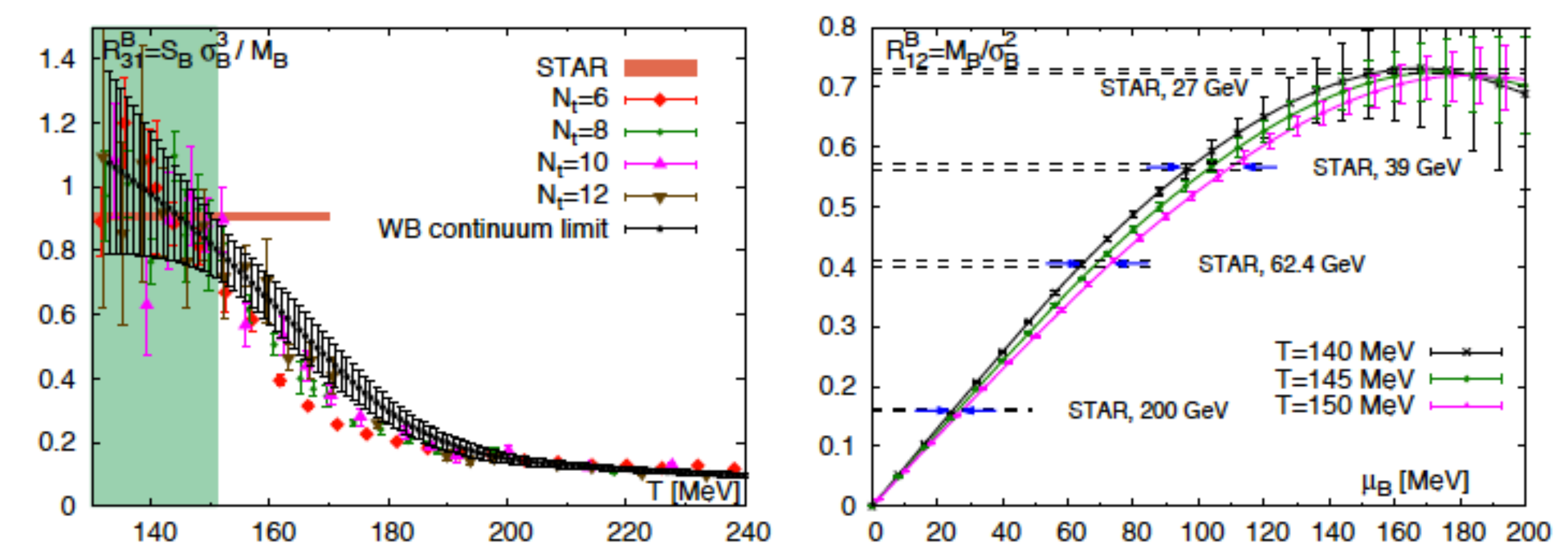}\\}}
\caption{\label{fig2} Left: Determination of the freeze-out temperature by comparing the lattice QCD simulations of a ratio of net-baryon number fluctuations to the experimental measurement~\cite{Borsanyi:2014ewa}. Right: Determination of the freeze-out chemical potential through the same procedure \cite{Borsanyi:2014ewa}. The experimental measurements are from Ref. \cite{Adamczyk:2013dal}.}
\end{figure}

Spectral functions play an important role in the study of real-time quantities, for instance in understanding how
heavy hadrons are modified in a thermal medium\cite{Andronic:2014sga}: here we will focus on the issue of quarkonium dissociation in the QGP.
In a relativistic field theory approach the  temperature $T$
is realised through (anti)periodic boundary conditions in the Euclidean time 
direction and the spectral decomposition  of an Euclidean propagator $G(\tau)$ at finite temperature  is given by
\begin{equation}
\label{eq:K}
G(\tau) = \int_{0}^\infty \frac{\dd\omega}{2\pi}\,  K(\tau,\omega)
\rho(\omega),\hskip 1truecm 0 \le \tau < \frac{1}{T}, 
\end{equation}
where $\rho(\omega)$ is the spectral function and  the kernel $K$ (in the case of a bosonic operator) is
given by
\begin{equation}
K(\tau,\omega) =
\frac{\left(e^{-\omega\tau} + e^{-\omega(1/T - \tau)}\right)}
{1 - e^{-\omega/T}}. 
\end{equation}
The $\tau$ dependence of the kernel reflects the periodicity
of the relativistic propagator in imaginary  time, as well
as its $T$ symmetry. The Bose--Einstein distribution, 
intuitively,  describes the wrapping around the periodic box which
becomes increasingly important at higher temperatures. When
the significant $\omega$ range greatly exceeds the temperature,  
$K(\tau,\omega) \simeq \left(e^{-\omega\tau} + e^{-\omega(1/T - \tau)}\right)$,
backward and forward propagations are  decoupled and the spectral
relation reduces to 
\begin{equation}
G(\tau) = \int_{\omega_0}^\infty\frac{\dd\omega'}{2\pi}\, \exp(-\omega'\tau)\rho(\omega').
\end{equation}
Note that this approximation holds true at zero temperature, and 
also in non-relativistic QCD (NRQCD):  
the interesting physics takes place around the two-quark threshold,
$\omega\sim 2M \sim 8$~GeV for $b$ quarks, which is still much
larger than our 
temperatures $T < 0.5$~GeV. 

Turning to the actual computational methodology,  
the calculation of the spectral
functions using Euclidean propagators as an input is a difficult,
ill-defined problem.  For several years
it has been  tackled by using the Maximum Entropy
Method (MEM) \cite{Asakawa:2000tr}.  
Recently, an alternative Bayesian reconstruction of the spectral
functions has been proposed in ref. \cite{Rothkopf:2011ef,Burnier:2013nla},
and applied to the analysis of HotQCD configurations \cite{Kim:2014iga}.
The spectral functions of the charmonium states have been studied  
as a function of both temperature and momentum, using as input 
relativistic propagators with two light quarks
\cite{Aarts:2007pk,Kelly:2013cpa} and, more recently, including
  the strange quark.  
Bottomonium mesons have been studied using the 
non-relativistic approximation for the bottom quark \cite{Aarts:2010ek}. 
The results~\cite{Aarts:2014cda} for the $\Upsilon$, shown in Fig.~\ref{fig:u},
demonstrate the persistence of
the fundamental state $\Upsilon$(1S) above $T_c$ as well as the suppression
of the excited states, in qualitative agreement with the experimental results that will be presented in Section~\ref{sec:onia}.  

\begin{figure}[!t]
\centering
\includegraphics[height=5.2cm,clip]{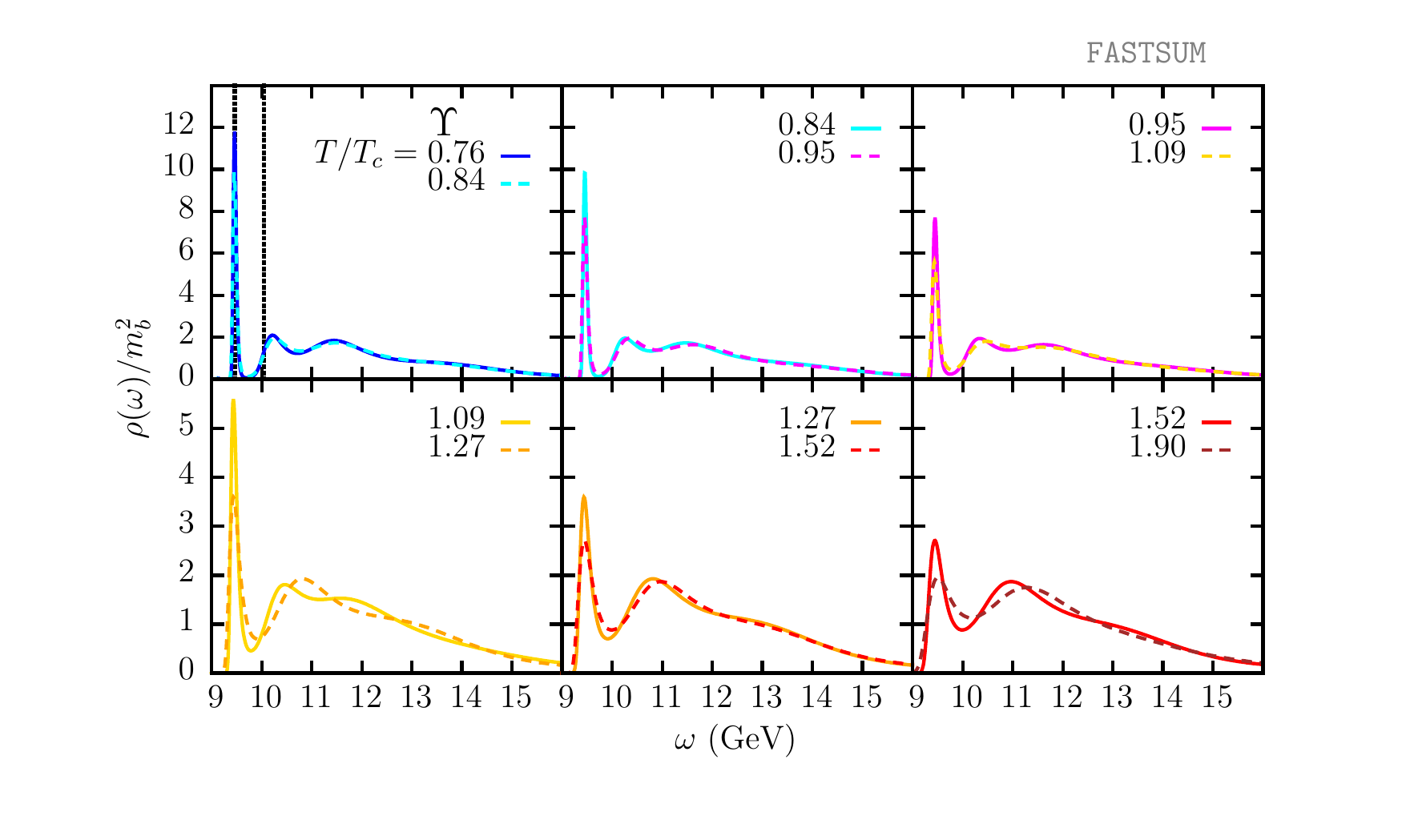}
\includegraphics[height=5cm,clip]{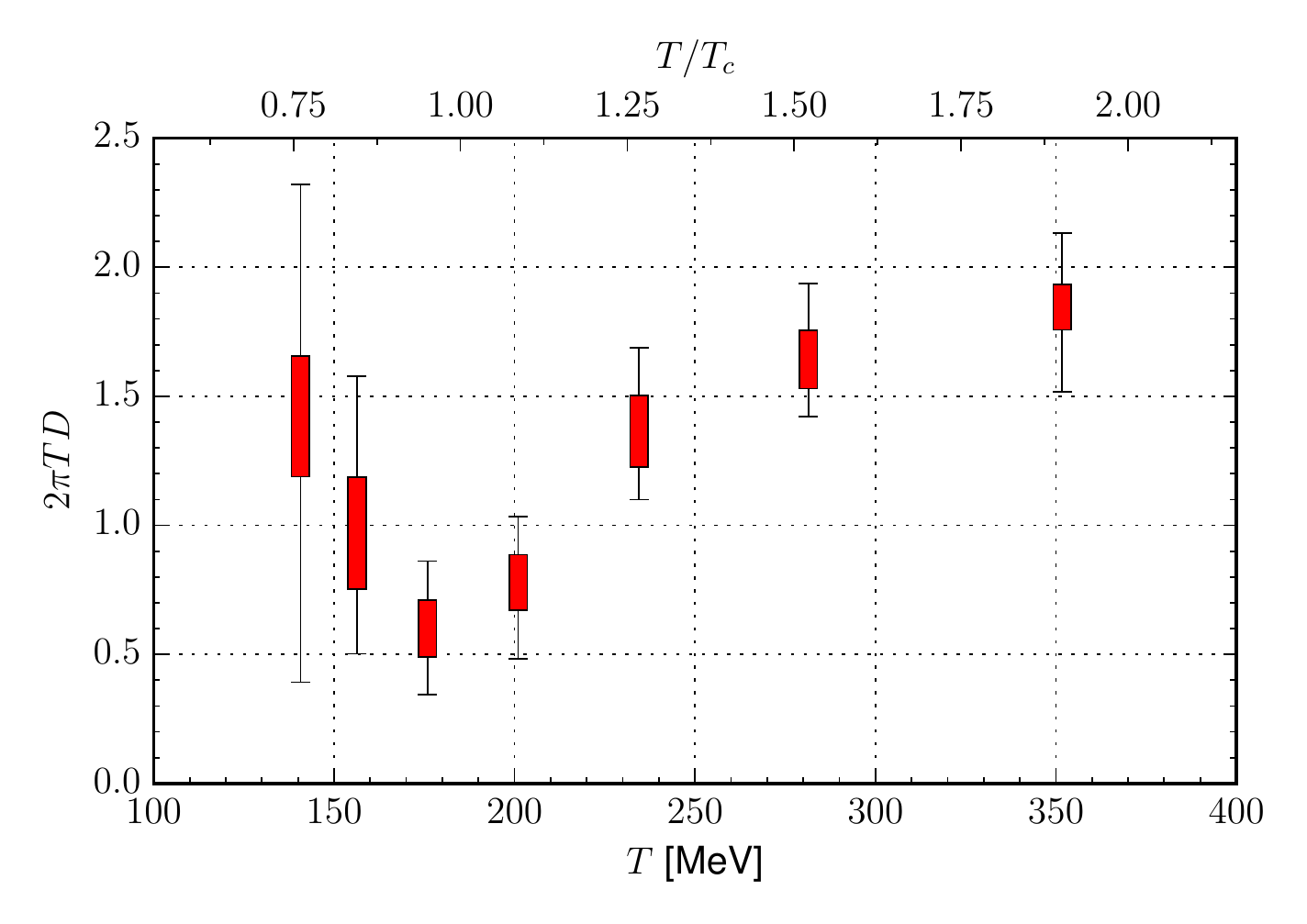}
\vskip -0.5 truecm
\caption{Examples of real-time quantities extracted from lattice Euclidean correlators. Left panel: the $\Upsilon$ spectral function at different
  temperatures, obtained using the maximum entropy method~\cite{Aarts:2014cda}. Right panel: the charge diffusion coefficient~\cite{Amato:2013naa}.}
\label{fig:u}       
\end{figure}

Spectral functions play an important role also in the extraction of transport coefficients from lattice-QCD simulations of euclidean correlators.  
Transport coefficients
describe the long real-time evolution of the medium. In principle, one can conceive a direct analytic continuation from imaginary to real time. In practice, since both real and imaginary time correlators are related by the same spectral
functions, transport-coefficient calculations are done by exploiting
appropriate Green-Kubo relations. 


The first relevant example of calculations of transport coefficients on the lattice was the estimate of the $\eta/s$ ratio~\cite{Nakamura:2004sy,Meyer:2007ic}. The results, within large
errors, were supportive of the observation of a very small value for
this ratio, close to the conjectured lower limit expected for a strongly
interacting gauge theory. 
Further analyses still support this
observation, although errors are still large.
Recently, calculations of other transport coefficients have been carried out, namely the heavy-quark trasport coefficients, the 
electrical conductivity and the charge diffusion coefficient $D$. An example of the latter is shown in the right panel of Fig.~\ref{fig:u}~\cite{Amato:2013naa}. 

Future work will focus on robust quantitative checks
of these results. Moreover these studies can equally
benefit from the methodological advances in the calculations of the spectral 
functions --indeed a controlled reconstruction of the low frequency sector
is particularly subtle and work is in progress in this direction. 
From a more theoretical viewpoint, analytically
tractable conformal and quasi-conformal models will continue providing
a useful guidance~\cite{Lombardo:2014mda,Evans:2012yi,Bellantuono:2015hxa}. 

As a final note, we mention an interesting connection between lattice QCD and the theoretical studies on axions,
which are plausible dark matter candidates~\cite{Preskill:1982cy}.
The axion mass $m_a$ is linked to the QCD topological susceptibility $\chi$ by the relation
$m_a^2(T) f_a^2 = \chi(T)$, where $T$ is the temperare and $f_a$ is the axion decay constant. 
The value of $\chi(T)$  can be computed on the lattice
as the second moment of the  distribution of the topological charge.
From the behavior of $\chi(T)$ some features of the axion evolution can be inferred, for instance the fact that the axion becomes heavier while the Universe cools down, till
the QCD transition temperature $T_c\approx 155$~MeV, then its mass ``freezes''.  
Because of this relation with $\chi(T)$ the cosmological evolution of axions is different from other dark matter candidates: 
they become more and more massive when approaching the temperature of the QCD transition from above.  
Higher momenta of the
same topological charge distribution are needed to compute the full cosmological equation
of motion of the axion. Because of the interest of these questions several groups
are involved in these calculations~\cite{Berkowitz:2015mja,Borsanyi:2015cka,Bonati:2015vqz,Trunin:2015yda}.

\section{Experimental study of the QCD phase diagram}
\label{sec:expintro}

\subsection{Experimental observables: a short summary of RHIC and LHC achievements}
\label{sec:expobs}

The experimental observation of signals related to the occurrence of deconfinement in heavy-ion collisions and sensitive to the properties of the QGP
 represents a fundamental and necessary step for the advancement of our knowledge of high-temperature QCD and, more generally, for the comprehension of
the features of the phase diagram of strongly-interacting matter~\cite{Cabibbo:1975ig,Shuryak:1978ij}. Over the years, various observables have been proposed, which often require complex detector
systems and sophisticated analysis procedures. When colliding heavy ions, typically two main kinds of difficulties arise: (i) the very high charged hadron multiplicity
${\rm d}N_{\rm ch}/{\rm d}\eta$, which exceeds 10$^3$ at mid-rapidity at the LHC~\cite{Aamodt:2010pb} and imposes the use of high resolution detectors for the measurement of inclusive observables;
(ii) the background levels for low-cross section processes, which can be very large due to combinatorial effects related to the large number of produced particles 
and have to be tackled with complex topological selections and with the help of particle identification techniques. Such issues were first met, and solved, at fixed
target experiments at the BNL AGS~\cite{Stachel:1993uh} and CERN SPS~\cite{Schmidt:1992ge} (collision energy per nucleon pair up to $\sqrtsNN\sim 20$~GeV) and then at collider experiments, currently 
in progress, at RHIC~\cite{Arsene:2004fa,Adcox:2004mh,Back:2004je,Adams:2005dq} and LHC~\cite{Muller:2012zq,Roland:2014jsa,Armesto:2015ioy}, where energies up to $\sqrtsNN=0.2$ TeV and 5 TeV are reached, respectively.

An exhaustive description of all the experimental observables studied in heavy-ion collisions is clearly beyond the scope of this document. We will therefore 
limit ourselves to a discussion of the main classes of signals and of the related physics aspects. 
Typically, 
one distinguishes between soft (low-$Q^2$) and hard (high-$Q^2$) observables. The first correspond to the bulk of the particle production processes occurring in the 
collision. 
Soft particles are mainly sensitive to what happens at hadronization, being produced when the created system cools down and crosses the confinement temperature from above. Still, a connection with the early deconfined state is in some cases possible, 
as some of the initial features of the medium survive along the history of the collision. Hard processes are connected with the study of the properties of the hot 
medium in a different way. Hard parton production occurs on a timescale ($< 1$fm/$c$) shorter than the formation time of the QGP and represents
therefore a probe of the deconfined phase. The modifications of the yields of hard processes with respect to elementary proton-proton collisions are a sensitive tool 
for quantitatively establishing many of the properties of the medium.

Soft observables, and in particular the inclusive distribution of the produced charged hadrons, are usually among the first measurements to be carried out for the 
characterisation of nucleus-nucleus collisions at the various energies. First of all, the charged particle multiplicity is directly related to the geometry of the 
interaction. For a given collision system (e.g., Pb--Pb) a larger multiplicity is connected with head-on (or central) collisions, where the impact parameter 
$b$, corresponding to the distance between the centers of the nuclei when their superposition is maximum, reaches the smallest value ($b\sim 0$). In such conditions,
the number of nucleons participating in the collisions ($N_{\rm part}$) and the total number of nucleon-nucleon collisions ($N_{\rm coll}$) is maximum. The latter 
quantities can be connected to $b$ via standard approaches as the Glauber 
model~\cite{Miller:2007ri}. It is customary
to sub-divide the event sample in so-called centrality classes, denoted as percentage ranges of the total hadronic nucleus-nucleus cross section (for example 0-10\%,
10-20\% and so on, with the lower percentages corresponding to more central events) and to study the dependence of various observables on the centrality of the 
collision.

Some of the observables connected with soft particle production are important for the determination of various physics quantities. In particular, under
rather general conditions, the charged-hadron multiplicity at mid-rapidity has a direct relation with the energy density reached at the formation time of the partons 
in the collision, via the Bjorken formula
\begin{equation}
\epsilon_{\rm Bj}=\frac{\langle m_{\rm T}\rangle\cdot {\rm d}N_{\rm ch}/{\rm d}y}{\tau_{\rm f}\cdot A}
\end{equation}
where $\langle m_{\rm T}\rangle$ is the average transverse mass of the particles, ${\rm d}N/{\rm d}y$ is the number of (charged) hadrons at central rapidity,
$\tau_{\rm f}$ the parton formation time and $A$ the transverse area of the interaction zone. With reasonable assumptions on $\tau_{\rm f}$ (from $\sim$ 1~fm/$c$ at
fixed target energies down to about one order of magnitude less at collider energies), one obtains from the measured hadronic multiplicities an estimate of the energy
density reaching $\epsilon_{\rm Bj}\sim 14$~GeV/fm$^3$ at the LHC~\cite{Chatrchyan:2012mb}. Also at fixed target energies~\cite{Margetis:1994tt} the values were found to exceed 1~GeV/fm$^3$, the order of magnitude 
of the energy density needed to reach deconfinement~\cite{Karsch:2001cy}.

On a more general level, the measurement of the charged-hadron pseudorapidity and multiplicity distributions represent a crucial tests for the theoretical
models. The main feature from the data is a power-law increase of the multiplicity at mid-rapidity with 
$\sqrtsNN$~\cite{Adam:2015ptt}. Such an increase is larger than in the corresponding pp interactions, indicating clearly that nucleus-nucleus collisions cannot be represented as a 
superposition of elementary nucleon-nucleon collisions. The charged particle production per participant nucleon, 
(${\rm d}N_{\rm ch}/{\rm d}\eta|_{\eta=0})/(N_{\rm part}/2)$, shows, at both LHC~\cite{Adam:2015ptt,Aamodt:2010cz,ATLAS:2011ag,Chatrchyan:2011pb}, and RHIC~\cite{Adler:2004zn,Bearden:2001xw,Bearden:2001qq,Back:2002uc,Adams:2004cb}, an increase by about 50\% from peripheral to central events, which is 
reasonably reproduced by theoretical models.

Another important class of observables connected with the global properties of the event is the study of the anisotropy of the angular distributions of the produced particles,
produced particles, which can be quantified by the coefficients of its Fourier expansion. 
Large values of the 2$^{\rm nd}$ harmonic coefficient, $v_2$, known as
elliptic flow~\cite{Ollitrault:1992bk}, for relatively low $p_{\rm T}$ (up to a few GeV/$c$) imply that collective 
effects develop very early in the history of the collision, when the system is in a deconfined phase, and point to a fast thermalization of such system. 
Experiments at RHIC~\cite{Ackermann:2000tr} and LHC~\cite{Aamodt:2010pa,ATLAS:2011ah,Chatrchyan:2012xq} have measured $v_2$ values that are well described by hydrodynamical calculations for an almost ideal fluid (i.e., with very small viscosity), while pure
hadronic models do not reproduce the data, implying that indeed the contribution of the QGP phase to the elliptic flow is important.
The coefficients of higher-order harmonics, as well as event-by-event flow distributions have been measured at LHC energies, thanks to the large hadronic multiplicity~\cite{ATLAS:2012at,ALICE:2011ab,Chatrchyan:2013kba}.
The results are very sensitive to fluctuations in the initial conditions of the collision and represent a very stringent test to theoretical model of the collision 
dynamics.

Collective effects are also studied using long-range correlations between pairs of produced particles. By 
correlating ``trigger'' particles in a given $p_{\rm T}$ range with the remaining charged particles with similar or lower $p_{\rm T}$, the correlation 
$\Delta\varphi$ vs $\Delta\eta$ can be studied~\cite{Dumitru:2008wn}. It shows a distinct peak at $(\Delta\varphi, \Delta\eta)\sim 0$, related to pairs of particles originating from jets, 
and an elongated structure at   
$\Delta\varphi\sim\pi$, from back-to-back correlated particles. More interestingly, a {\it ridge}-like structure emerges at $\Delta\varphi\sim 0$, over a wide 
$\Delta\eta$ region. Theoretical models relate such a structure to jet-medium interactions or to a collective behaviour of the medium itself, i.e., to a common correlation of all the particles with the reaction plane. It has been studied in detail at both RHIC~\cite{Alver:2009id,Abelev:2009af} and LHC~\cite{Aamodt:2011by,Chatrchyan:2011eka,Chatrchyan:2012wg} 
energies in nucleus-nucleus collisions. Interestingly, at the LHC, the same structure has been observed also in central \mbox{p--Pb}~\cite{CMS:2012qk,Abelev:2012ola,Aad:2012gla,Adam:2015bka} and even in pp collisions with 
large hadron multiplicity~\cite{Khachatryan:2010gv,Khachatryan:2015lva,Aad:2015gqa}, opening a lively discussion on the possibility of sizeable collective effects also in collisions of smaller systems.

Still in the domain of soft observables, interesting information can be derived from the study of the yields of identified hadrons. In particular, from the {\it chemical composition} of the system, i.e. the fraction of various particle species, one can, in the frame of statistical models~\cite{Becattini:1997ii,Andronic:2008gu} that assume thermalization of 
the QGP, 
derive the temperature $T_{\rm ch}$ at the {\it chemical freeze-out}, the moment when the particle abundances are fixed. These models also give the value of the 
baryochemical potential $\mu_B$, proportional to the net baryon density of the system. The results~\cite{Andronic:2012dm} show that, increasing $\sqrtsNN$ from SPS to RHIC to 
LHC, energy $T_{\rm ch}$ increases and then saturates, already at RHIC energy, at $T_{\rm ch}\approx 155$ MeV, a value significantly close to theory predictions for the 
deconfinement temperature. As a function of $\sqrtsNN$, $\mu_B$ decreases steadily and becomes zero at LHC energy, indicating a situation of 
{\it nuclear transparency}, where the net baryonic number of the QGP region becomes negligible. 

Further information from identified particle production studies can be obtained by comparing their $p_{\rm T}$ distributions and/or various particle ratios vs 
$p_{\rm T}$. Here, a significant hardening of the spectra can be observed when considering particles of increasing mass. This effect is visible at all energies and
it has been connected with the occurrence of a {\it radial} flow of particles, superimposed to the thermal motion in the expanding fireball~\cite{Hung:1997du}. A quantitative study in the
frame of blast-wave models gives information on the flow velocity, which increases with $\sqrtsNN$, reaching $\beta~\sim 0.7$ and on the 
temperature of the system when the interactions stop (kinetic freeze-out), which is of the order of 100 MeV at both RHIC~\cite{Abelev:2008ab} and LHC~\cite{Abelev:2012wca} for central collisions.


Moving to hard probes of the medium, several observables have been investigated. A non-exhaustive list includes the study of high-$p_{\rm T}$ hadrons, of jet 
production, and of particles containing heavy quarks. The modification of their yields in the QGP is quantified through  the nuclear modification factor $R_{\rm AA}$,
defined as the ratio between the measured yields in \mbox{A--A} and pp collisions, normalised to $N_{\rm coll}$. This quantity is usually studied as a
function of $p_{\rm T}$, and values smaller than 1 at high $p_{\rm T}$ are attributed to energy-loss of hard partons.

The study of the $R_{\rm AA}$ of unidentified high-$p_{\rm T}$ charged particles is one of the basic measurements in the sector of hard probes. A  
suppression from a few GeV/$c$ onwards was one of the main discoveries of the RHIC experiments~\cite{Adler:2003au,Adams:2003kv}. This observation has been confirmed at LHC energies, where measurements
were pushed up to $\sim 100$~GeV/$c$~\cite{Aamodt:2010jd,CMS:2012aa,Abelev:2012hxa,Aad:2015wga}. The suppression is larger than at RHIC, it reaches about a factor 7 in the range $5<p_{\rm T}<10$~GeV/$c$, and then is reduced
down to a factor $\sim 2$ at very large $p_{\rm T}$. The suppression is  understood as a consequence of the mainly radiative energy loss of fast partons. 
The energy loss is expected to be larger for gluons than for 
quarks, and for light quarks with respect to heavy quarks~\cite{Dokshitzer:2001zm}. The energy loss values that can be derived from the data are very large, up to several tens of GeV for high-energy partons in a 
central \mbox{Pb--Pb} collision, implying that the medium is essentially opaque except for partons produced near the surface. This view is confirmed by
measurements of angular correlations between high-$p_{\rm T}$ particles, which show that their back-to-back correlation is completely washed out for central 
nucleus-nucleus collisions~\cite{Adler:2002tq}.

A deeper insight into the features of the energy loss processes comes from the comparison of the suppression patterns for various hadron species. In particular,
open heavy flavour hadrons have been studied with various techniques, ranging from the detection of electrons/muons from their semi-leptonic decays to the full
reconstruction of hadronic decays (for D mesons) or the measurement of J/$\psi$ from B mesons. The first observations at RHIC showed a significant 
energy loss for heavy quarks~\cite{Adare:2006nq,Abelev:2006db}. At the LHC, much more detailed investigations have been carried out, and a factor $\sim 5$ suppression for intermediate to 
high-$p_{\rm T}$ D-mesons was established~\cite{Adam:2015nna,ALICE:2012ab}. Strong indications for a smaller energy loss for b-quarks compared to c-quarks have also been obtained recently~\cite{Chatrchyan:2012np,Adam:2015nna}. A stringent
test for the various theoretical interpretations comes from the simultaneous measurement of $R_{\rm AA}$ and $v_2$. The two quantities are correlated in the frame of heavy-quark transport models~\cite{Alberico:2011zy}, since a strong modification of the spectra implies also a strong interaction with the medium and therefore a participation to the collective expansion.
Current results in the heavy-quark sector still pose a significant challenge to the models~\cite{Abelev:2014ipa}. 

Heavy quarkonia can be dissociated in a deconfined medium, due to the screening of their attractive color interaction~\cite{Matsui:1986dk}. Various
states, corresponding to different binding energies, are expected to be suppressed at different temperatures of the medium~\cite{Digal:2001ue}. A comparative study of their $R_{\rm AA}$ 
can in principle yield information on the temperature of the QGP. A rather strong suppression of the J/$\psi$ was indeed observed at SPS~\cite{Alessandro:2004ap} and RHIC~\cite{Adare:2011yf} energy and found
to be compatible with a QGP-related effect. At the LHC, results on the various $\Upsilon$ states have been obtained, showing a hierarchy of suppression with the less 
bound 2S and 3S states exhibiting a systematically lower $R_{\rm AA}$~\cite{Chatrchyan:2011pe,Chatrchyan:2012lxa}. Interestingly, the J/$\psi$ has been found to be less suppressed than at lower energies~\cite{Abelev:2012rv,Adam:2015isa}, a new 
effect which has been interpreted in terms of a recombination of deconfined charm quarks in the medium~\cite{BraunMunzinger:2000px,Thews:2000rj}.

The measurement of high-$p_{\rm T}$ jets offers the possibility of a deeper understanding of the energy loss process, since one can study how the energy lost by the
leading parton is redistributed into softer particles and how this affects the jet shape. After exploratory studies at RHIC~\cite{Adamczyk:2013jei}, the larger jet yields at the LHC coupled 
with the excellent capabilities of the experiments have led to several interesting results for this observable. In particular, one can analyze the difference in the 
fragmentation functions, related to how the jet energy in \mbox{Pb--Pb} is redistributed in terms of particle $p_{\rm T}$, and the differential jet-shapes, related to 
how the jet energy is redistributed in radius. The results~\cite{Aad:2010bu,Chatrchyan:2011sx,Abelev:2013kqa,Aad:2014wha} show a depletion at intermediate radius, and an enhancement at larger
radius, showing that the radiated energy is redistributed at large distances from the jet axis outside the jet cone. On the contrary, the 
modification of the jet fragmentation functions, when comparing \mbox{Pb--Pb} and pp collisions, is small, showing that after traversing the medium, high-energy 
partons lose momentum but the momentum distribution among particles within the jet cone corresponds to what observed for jets fragmenting in vacuum. In addition, no significant angular decorrelation of back-to-back jets, when comparing nucleus-nucleus results to pp~\cite{Chatrchyan:2011sx}.

All the observables discussed up to now concern hadrons. A distinct class of observables, electromagnetic probes, is also extensively studied.
At high $p_{\rm T}$, real photons and vector bosons are expected to be insensitive to the presence of the QGP and represent therefore a good reference probe. In addition, they can be sensitive to initial state effects such as the  nuclear modification of the nucleon parton distribution functions (shadowing~\cite{Eskola:2009uj}) and their study helps in disentangling those effects from the ones related to the
QGP. Indeed, the $R_{\rm AA}$ of hard photons, $W$ and $Z$ bosons, once initial state effects are accounted for, are compatible with 1, i.e. with no medium effects~\cite{Chatrchyan:2012vq,Aad:2015lcb,Aad:2014bha,Chatrchyan:2014csa}.
At low $p_{\rm T}$, electromagnetic probes (real and virtual photons) produced in the early stage of the collision can carry information about the properties of the medium
at early times, since, once emitted, they are practically immune to further strong interactions that dominate at later stages. In particular, the identification of a thermal photon signal
from the medium allows the extraction of the average temperature of the QGP phase. Results from RHIC~\cite{Adamczyk:2013caa,Adare:2008ab,Adare:2015ila} and LHC~\cite{Adam:2015lda} point to temperatures between 200 and 300 MeV, well 
above the deconfinement temperature. Finally, the study of the dilepton spectrum from low to intermediate masses (below the J/$\psi$ mass) allows a complementary
determination of a thermal signal from the medium, with better accuracy and lower background with respect to real photons. Furthermore, modifications in
the spectral function of low-mass hadrons decaying to lepton pairs (in particular the $\rho$-meson) are sensitive probes of the restoration of chiral symmetry, 
expected when the system is close to deconfinement~\cite{Koch:1997ei}. Precision results at top SPS energy~\cite{Arnaldi:2006jq,Arnaldi:2007ru}, and subsequent RHIC observations~\cite{Adamczyk:2013caa,Adare:2015ila}, show intriguing features, including a significant broadening of 
the $\rho$ and the observation of a thermal dilepton signal corresponding again to temperatures well above deconfinement. At the LHC, these studies are difficult 
because of the strong increase of the background levels and are currently still in progress.

In summary, a wealth of experimental observables is available today for the study of strongly interacting matter at high temperatures. After the first observations
at the BNL AGS and CERN SPS, suggesting that deconfinement was indeed achieved, the field has reached maturity with the experiments at the RHIC and LHC collider, where
the properties of the QGP are being studied. The study of soft observables has shown that a system reaching equilibrium at early times has been produced, while hard
probe studies indicate that the medium reaches temperatures well beyond deconfinement and its large energy density induces a strong energy loss of high-$p_{\rm T}$ 
particles and jets.

\subsection{Future experimental prospects}
\label{sec:expfuture}

In the next 10--15 years the present experimental facilities for studies of ultra-relativistic heavy-ion collisions will remain in operation and 
for most of them an increase in instantaneous luminosity is planned. In addition, new machines will become available in the low-energy regime.
Figure~\ref{fig:explandscape} shows an overview of the existing and planned facilities with the corresponding centre-of-mass energies and 
expected operation schedule. 
In the following, the various facilities are briefly discussed.

\begin{figure}[!ht]
\begin{center}
\includegraphics[width=0.8\textwidth]{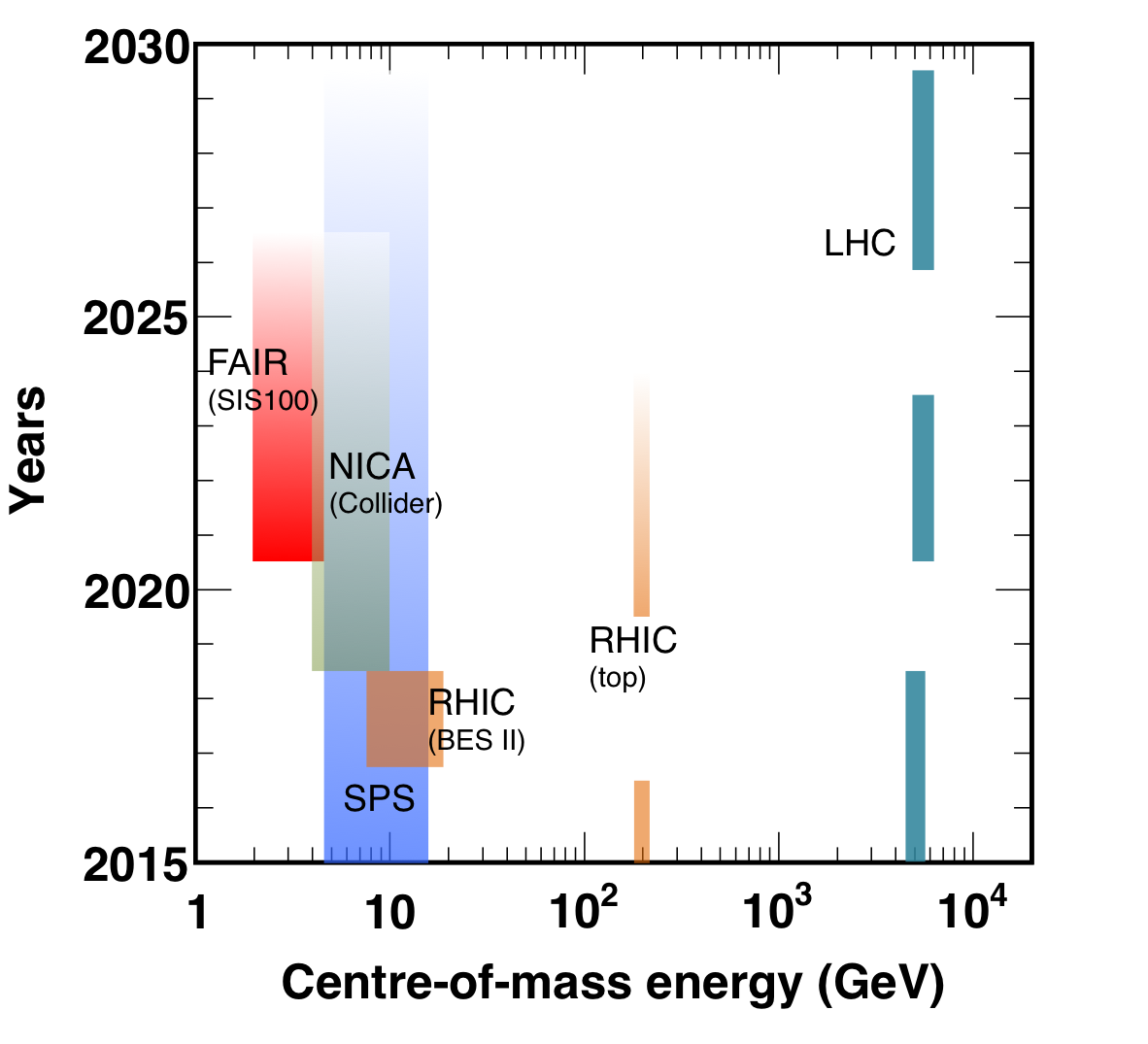}
\caption{Centre-of-mass energy coverage of the ultra-relativistic heavy-ion facilities scheduled for the next 15 years. 
The facilities that are not approved yet or do not have
a well-defined timeline are not shown.}
\label{fig:explandscape}
\end{center}
\end{figure}

\begin{description}

\item[\uppercase{LHC}] The ongoing LHC Run-2 spans the period 2015--2018 and it includes two Pb--Pb running periods at the centre-of-mass energy $\sqrtsNN=5~\tev$ and one running period with p--Pb collisions at $\sqrtsNN$ of either 5 or 8~TeV (still to be decided). For Pb--Pb collisions the instantaneous luminosity is expected to be of the order
of $10^{27}~{\rm cm^{-2}s^{-1}}$ and the luminosity integrated over the two periods of the order of $1~{\rm nb^{-1}}$, i.e.\,about 10 times larger than for the LHC Run-1. After the second long shutdown (LS2), the energy will reach the LHC design value of 5.5~TeV and the instantaneous luminosity is expected to increase by a factor 3--6. The ATLAS, ALICE and CMS experiments have requested a sample of at least $10~\nbinv$ to be delivered during Run-3 (2021--2023) and Run-4 (2026--2029). For example, the request of the ALICE Collaboration is of $13~\nbinv$~\cite{Abelevetal:2014cna}. In addition, Runs 3 and 4 will comprise data-taking periods with p--Pb collisions and pp collisions at $\sqrt s=5.5~\tev$, for reference data collection.
The possibility for limited periods with nuclei lighter than Pb, e.g.\,Ar--Ar or O--O , is considered as well. The LHCb experiment will participate in all heavy-ion runs starting from Run-2.
There is an advanced proposal for studying electron--proton and electron--ion collisions using the LHC hadron beams and electron beams provided by a new, moderate-size, electron acceleration ring~\cite{LHeC}.
Such Large Hadron electron Collider (LHeC) would enable, among others, studies of the nuclear modification of the parton distribution functions and of the possible saturation of parton densities
with unprecedented precision. This proposal, which would have a timeline starting after 2030, is now also discussed in the context of the Future Circular Collider (FCC)~\cite{FCCkickoff}.

\item[\uppercase{RHIC}] The RHIC collider at BNL, in operation since year 2000, is a dedicated heavy-ion machine and it has a large flexibility in the choice of ion species and centre-of-energy in the range $\sqrtsNN=7.5$--$200~\gev$. The future schedule of RHIC includes three campaigns~\cite{RHICfuture}. 
The one presently ongoing (2014--2016) is aimed at exploiting 
the recently-installed new inner trackers of the PHENIX and STAR detectors with pp, d--Au and Au--Au collisions at top energy of 200~GeV.  
After a shutdown in 2017 the second phase of the beam-energy scan programme (BES II) will take place in 2018--2019, with measurement in Au--Au collisions at centre-of-mass energies in the range 7.5--19~GeV. The goal is the exploration of the QCD phase diagram in the region around the expected position of the critical endpoint, with an increase of about one order of magnitude in the instantaneous luminosity (in the range $10^{25}$ to $10^{27}~\Lunits$, depending on $\sqrtsNN$). During the 2019 shutdown, the PHENIX experiment will be replaced by
sPHENIX~\cite{sPHENIX}, with focus on jet and quarkonium measurements, 
and the STAR Collaboration is considering a number of upgrades, in particular for the inner tracker and for the forward rapidity region.
The sPHENIX and STAR experiments plan a campaign of data-taking at top RHIC energy (200~GeV) in 2021--2022.
The implementation of an electron--ion collider (eRHIC) at BNL is an option that is considered for the period after 2025~\cite{eRHIC}.

\item[\uppercase{SPS}] The SPS provides Pb and lighter-ion beams for fixed-target experiments since the late 1980s. The beam energy ranges in 10--158~GeV per nucleon, 
corresponding for Pb--Pb collisions to $\sqrtsNN$ in 4.5--17.3~GeV.
At present, the NA61/SHINE experiment~\cite{NA61} is carrying out a systematic scan in beam energy and colliding system size (pp, p--Be, p--C, Be--Be, Ar--Sc, Pb--Pb, Xe--La)
with the goal of studying the onset of deconfinement and searching for the critical endpoint using hadronic observables. 
The approved programme extends to 2017, but the Collaboration is considering a detector upgrade and a proposal for an extension of the programme by a few years~\cite{NA61upgrade}.
The SPS is also used as injector for the LHC, therefore it will remain in operation well beyond 2030.
At present, the SPS Pb beam intensity leads to Pb--Pb interaction rates of a few hundred kHz with a target of 0.1 interaction lengths. 
With the implementation of a new injection scheme, interaction rates larger than 1~MHz could be reached, as discussed in detail in Chapter~\ref{sec:na60plus}.

\item[\uppercase{NICA}] The NICA facility at JINR will provide both collider and fixed-target mode heavy-ion interactions (see e.g.~\cite{NICA}). With beam energies in the interval 0.6--4.5~GeV per nucleon, the centre-of-mass energies will be in the ranges $\sqrtsNN=4$--$11~\gev$ for collider operation 
and $\sqrtsNN=1.9$--$2.4~\gev$ for fixed-target operation. For the former, where the energy enables the search of the critical endpoint, the instantaneous luminosity is expected to 
be of the order of $10^{27}~\Lunits$ (at $\sqrtsNN=9~\gev$), corresponding to interaction rates of about 6~kHz. NICA is scheduled to start fixed-target operation in 2017 
and collider-mode operation in 2019.

\item[\uppercase{FAIR}] The FAIR facility at GSI~\cite{FAIR} will provide fixed-target collisions with $\sqrtsNN$ values in the range 2--4.5~GeV in the SIS100 phase, with very high interaction rates of up to 10~MHz (essentially limited by the detector technology). This phase is presently planned to start in 2021. A SIS300 phase with $\sqrtsNN$ reaching up to 8~GeV was also in the original plan, but it is not approved at present and its operation would in any case only start well beyond 2030. The CBM experiment~\cite{CBMPhysBook} will focus on the exploration of the phase diagram with Au--Au collisions at FAIR.

\end{description}

\begin{figure}[!t]
\begin{center}
\includegraphics[width=0.8\textwidth]{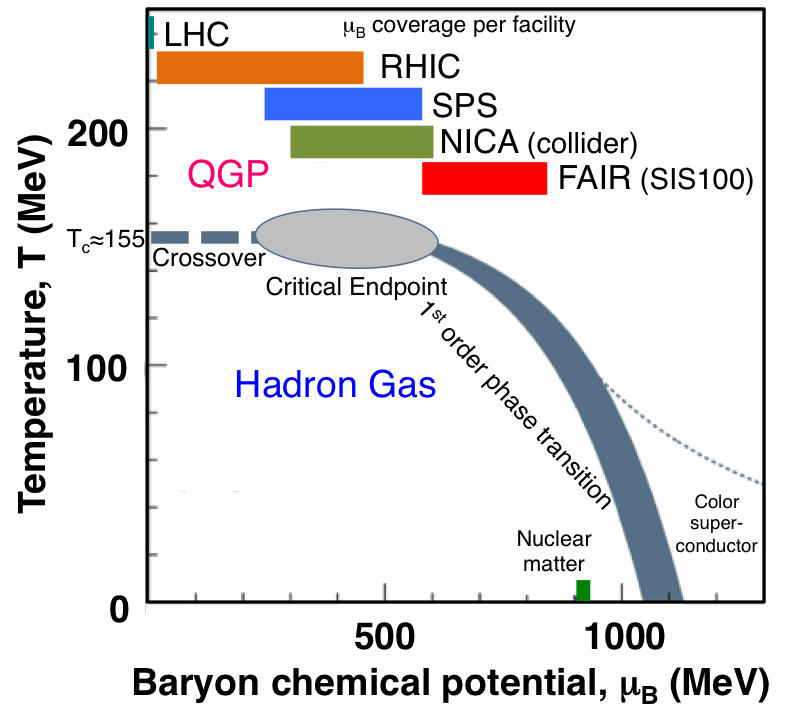}
\caption{Schematic representation of the phase diagram of strongly-interacting matter, with the approximate coverage of the various experimental facilities.
The coverage is shown in terms of a range in baryon chemical potential $\mu_B$ (the vertical position of the bands representing the facilities does not represent the coverage in temperature).
The $\mu_B$ coverages were obtained from the corresponding $\sqrtsNN$ ranges using the relation reported in Ref.~\cite{Cleymans:2014xha}, which is 
based on the statistical hadronization model.}
\label{fig:phasediagexp}
\end{center}
\end{figure}

Figure~\ref{fig:phasediagexp} shows a schematic representation of the phase diagram of strongly-interacting matter, with the approximate coverage of the various experimental facilities,
in terms of baryon chemical potential. The high-energy machines, LHC and RHIC at top energy, cover the region with vanishing baryon chemical potential,
where the best conditions are reached for the measurement of the QGP properties and their comparison with first-principle QCD calculations.
The low-energy machines, SPS, NICA-collider, RHIC-BES and FAIR-SIS100, instead cover the region with baryon chemical potential of a few hundred MeV, 
where the critical endpoint is expected to be located. Note that the SIS100 machine does not cover completely the region relevant for the 
search of the endpoint and the onset of deconfinement.

In addition to the running or planned facilities that we have described, there are a number of further possibilities that are being proposed and discussed in the community.
In the low-energy regime,
the possibility to accelerate heavy ions at the J-PARC facility at KEK is also being considered. The centre-of-mass energy interval would be $\sqrtsNN=1.9$--6.2~GeV
and the interaction rates could reach 10~MHz (essentially limited by the detector technology)~\cite{JPARC}. The timeline of this possible project is not yet defined.
In the high-energy regime, the possibilities for the longer-term future include the study of fixed-target collisions using the LHC beams ($\sqrtsNN\sim 100~\gev$, see Section~\ref{sec:after}) and 
the operation of nucleus--nucleus collisions at the Chinese SppC~\cite{SppC,SppCHI} or the CERN FCC ($\sqrtsNN\sim 30$--$40~\tev$). These high-energy opportunities will be discussed
in Section~\ref{sec:fcc}.


\section{High-energy frontier: future ALICE programme at the LHC}
\label{sec:alice}


\subsection{Timeline of LHC heavy-ion programme and ALICE upgrade}
\label{sec:upgrade}

Between $2009$ and $2013$  (Run-1) the LHC collider at CERN operated successfully providing the experiments with pp, p--Pb and Pb--Pb collisions at centre-of-mass energies of $0.9,\,2.76,\,7,\,8$~TeV,  $5.02$~TeV and $2.76$~TeV, respectively.
The LHC schedule for the coming years is shown in Fig.~\ref{LHCplan}, which emphasises the heavy-ion periods and reports the integrated luminosity requested by the ALICE experiment~\cite{Abelevetal:2014cna}.
During Run-2 (2015--2018) the LHC runs at the increased energies of  $13$~TeV and $5$~TeV for pp and Pb--Pb collisions, respectively. From $2021$ to $2023$  (Run-3) the LHC will operate at the nominal 14 TeV/5.5 TeV centre-of-mass energy for pp/Pb--Pb collisions and will also make a further step in the luminosity. The long shutdown LS3 will prepare the machine and the experiments to a jump of a factor $10$ in luminosity, with the High-Luminosity LHC entering operation in $2026$  with two runs presently foreseen (Run-4 and Run-5). Concerning Pb--Pb collisions, for Run-3 and Run-4 the experiments have requested a total integrated luminosity of more than 10~nb$^{-1}$ (e.g. $13$~nb$^{-1}$ requested by ALICE~\cite{Abelevetal:2014cna}) compared to $\sim 0.1$~nb$^{-1}$  in Run-1 and the expected  $\sim 1$~nb$^{-1}$ of Run-2. 
During Run-3 and Run-4, reference samples with pp collisions at 5.5 TeV will also be collected , as well as a sample with p--Pb collisions at 8.8 TeV. 
The possibility of extending the programme to collisions of nuclei lighter than Pb (e.g. Ar--Ar or O--O) is being discussed.

\begin{figure}[!b]
\centering
\includegraphics[width=\textwidth,angle=0.]{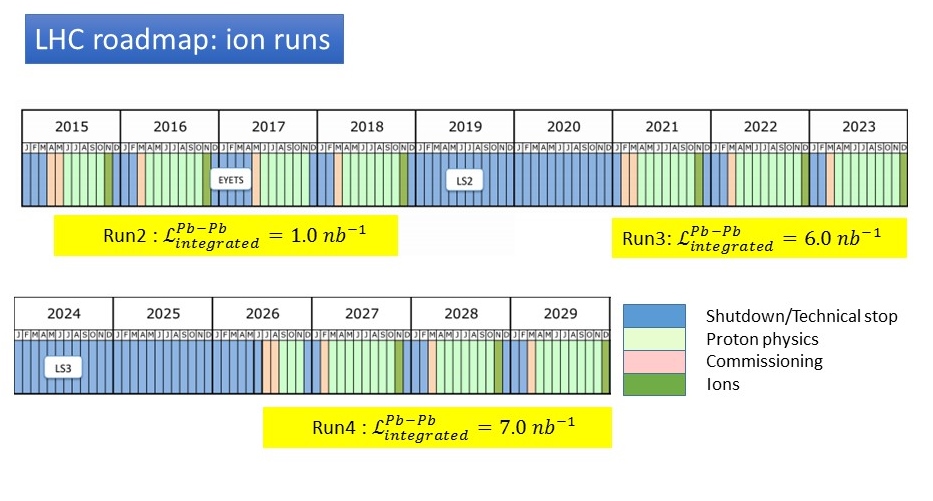} 
\caption{LHC schedule for Runs 2, 3 and 4. The integrated luminosity requested by the ALICE experiment is also reported~\cite{Abelevetal:2014cna}.}
\label{LHCplan}
\end{figure}

The  ALICE experiment~\cite{Aamodt:2008zz,Abelev:2014ffa}  was designed specifically for heavy-ion physics, with a redundant and robust tracking system 
and several systems for particle identification.
The detector  (Fig.~\ref{ALICE}) consists of a central barrel, a muon spectrometer at forward rapidity and a set of detectors for triggering and event characterisation.
The central barrel covers $|\eta|<0.9$ and is equipped with: an Inner Tracking System (ITS) made of 
three double-layer silicon detectors (pixels, drifts and strips), a large Time Projection Chamber (TPC), a Transition Radiation Detector (TRD), a Time of Flight System (TOF), a High Momentum Particle IDentification detector (HMPID), an ElectroMagnetic CALorimeter (EMCAL+DCAL) and a PHOton Spectrometer (PHOS). The last three have only partial azimuthal coverage. In the forward region the muon spectrometer (with the MTR trigger system and the MCH tracking system)  covers the range $2.5<\eta<4$. 
Two small-acceptance detectors at forward rapidity are used for the measurement of charged-particle and photon multiplicity (FMD and PMD). 
 Several detectors along the beam pipe are devoted to triggering (V0), collision time (T0) and centrality determination  (V0 and ZDC calorimeters). The ACORDE system triggers on cosmic-ray muons.

As described in the previous Chapter, the results of Run-1 have represented an important step forward in our understanding of the QGP.
Unexpected hints for collective effects have been observed also in high-multiplicity collisions of small systems (pp and p--Pb) and motivate a detailed comparison of pp, p--Pb and Pb--Pb collisions as a function of multiplicity.
Data collected during Run-2 will provide further insight on some of the questions left open by Run-1.
However, as discussed in the rest of this Chapter, a detailed characterisation of the properties of the QGP requires high-precision measurements of
heavy-flavour, quarkonium, jet and thermal dilepton production over a wide momentum range.

In order to achieve these goals, the ALICE Collaboration is preparing a major upgrade of the experimental apparatus that will operate during 
Run-3 and Run-4. The upgrade strategy was driven by the following requirements.
\begin{itemize}
\item Improvement of the track reconstruction performance, in terms of spatial precision and efficiency, in particular for low-momentum particles, in order to select more effectively the decay vertices of heavy-flavour mesons and baryons. 
\item Increase of the event readout rate up to 50~kHz for Pb--Pb collisions triggered with a minimum-bias selection that provides the highest efficiency for low-momentum processes. This increase enables recording during Run-3 and Run-4 of a sample of minimum-bias collisions two orders of magnitude larger than during Run-2.
\item Consolidation of the particle identification capabilities of the apparatus, which are crucial for the selection of heavy-flavour, quarkonium and dilepton signals at low momentum.
\end{itemize}

The ALICE upgrade programme is described in several documents~\cite{Abelevetal:2014cna,Abelevetal:2014dna,Antonioli:1603472,CERN-LHCC-2013-020,CERN-LHCC-2015-001,CERN-LHCC-2015-006}.
In summary, it entails the following changes to the apparatus:
\begin{itemize}
\item a new Inner Tracking System (ITS) with seven layers equipped with Monolithic Active Pixel Sensors (MAPS) (see Fig.~\ref{ITS}-left)~\cite{Abelevetal:2014dna}; the innermost layer will have a radius of 23~mm, to be compared with 39~mm of the present ITS; the hit resolution of the detector will be of about 5~$\mu$m and the material budget of the three innermost layers will be reduced from the present 1.1\% to 0.3\% of the radiation length; these features provide an improvement by a factor about three for the track impact parameter resolution in the transverse plane (see Fig.~\ref{ITS}-right);
\item a new Muon Forward Tracker (MFT) made of five planes of the same MAPS used in the ITS, which will provide precise tracking and secondary vertex reconstruction for muon tracks in $2.5<\eta<3.5$~\cite{CERN-LHCC-2015-001};
\item new readout chambers for the TPC based on the Gas Electron Multiplier (GEM) technology, in order to reduce the ion backflow in the drift volume 
and enable continuous readout of Pb--Pb events for an interaction rate up to 50~kHz~\cite{CERN-LHCC-2013-020};
\item a new Fast Interaction Trigger detector (FIT) based on  Cherenkov radiators and scintillator tiles at forward rapidity around the beam pipe~\cite{Antonioli:1603472};
\item an upgrade of the readout electronics of the TOF, MUON and ZDC detectors that enables recording Pb--Pb interactions at a rate of up to 50~kHz~\cite{Antonioli:1603472};
\item a new integrated Online/Offline system for data readout, compression and processing (O$^2$) to reduce the volume of data by more than one order of magnitude before shipping them to permanent storage~\cite{CERN-LHCC-2015-006}.
\end{itemize}

\begin{figure}
\centering
\includegraphics[scale=0.6,angle=90.]{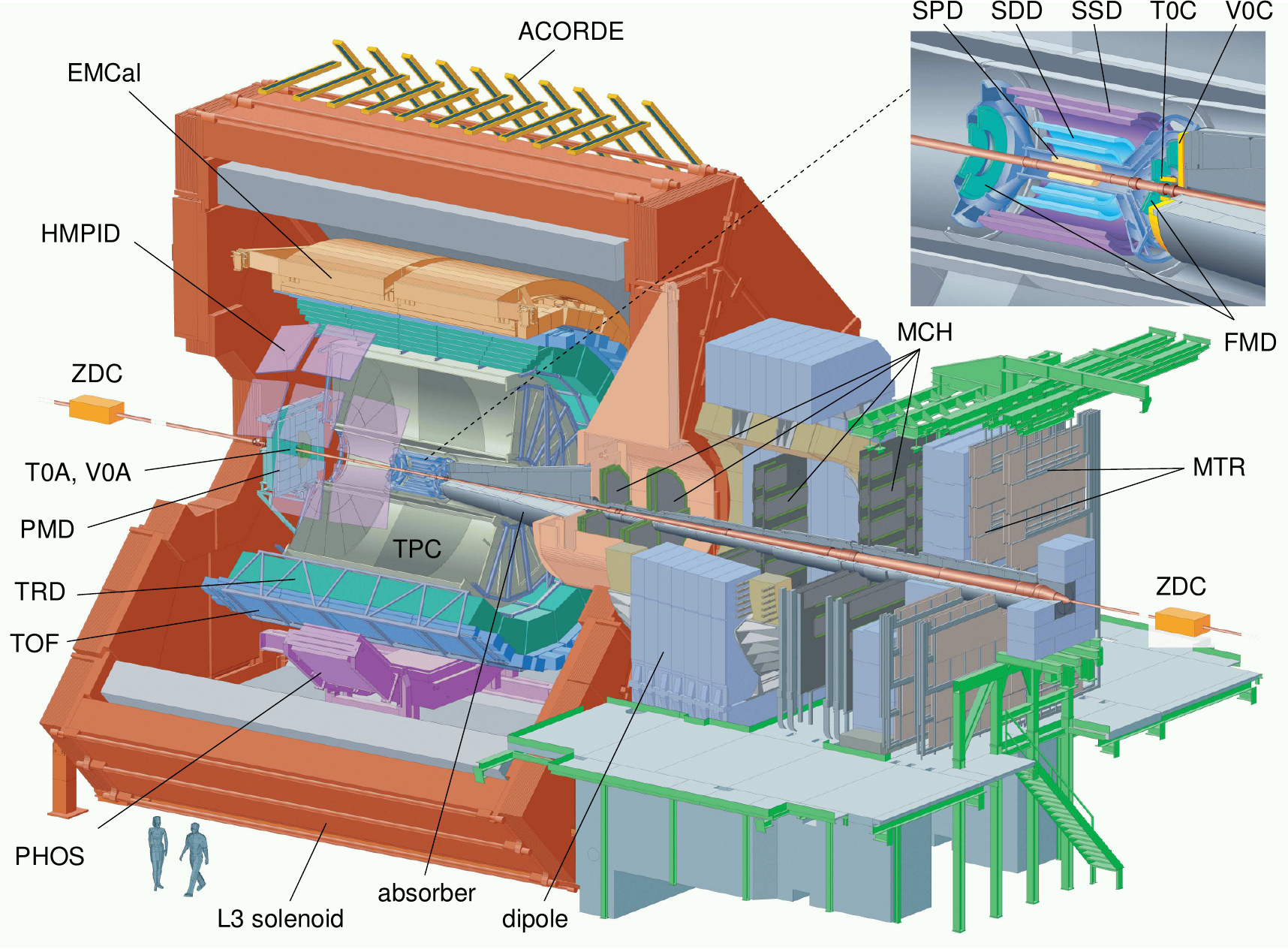} 
\caption{ALICE experimental apparatus during Run-1 and Run-2.}
\label{ALICE}
\end{figure}

\begin{figure}
\centering
\includegraphics[width=0.48\textwidth,angle=0.]{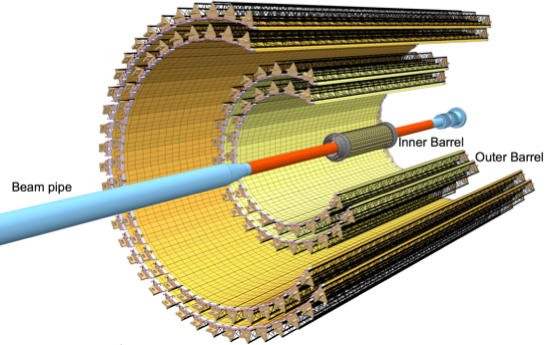} 
\includegraphics[width=0.5\textwidth,angle=0.]{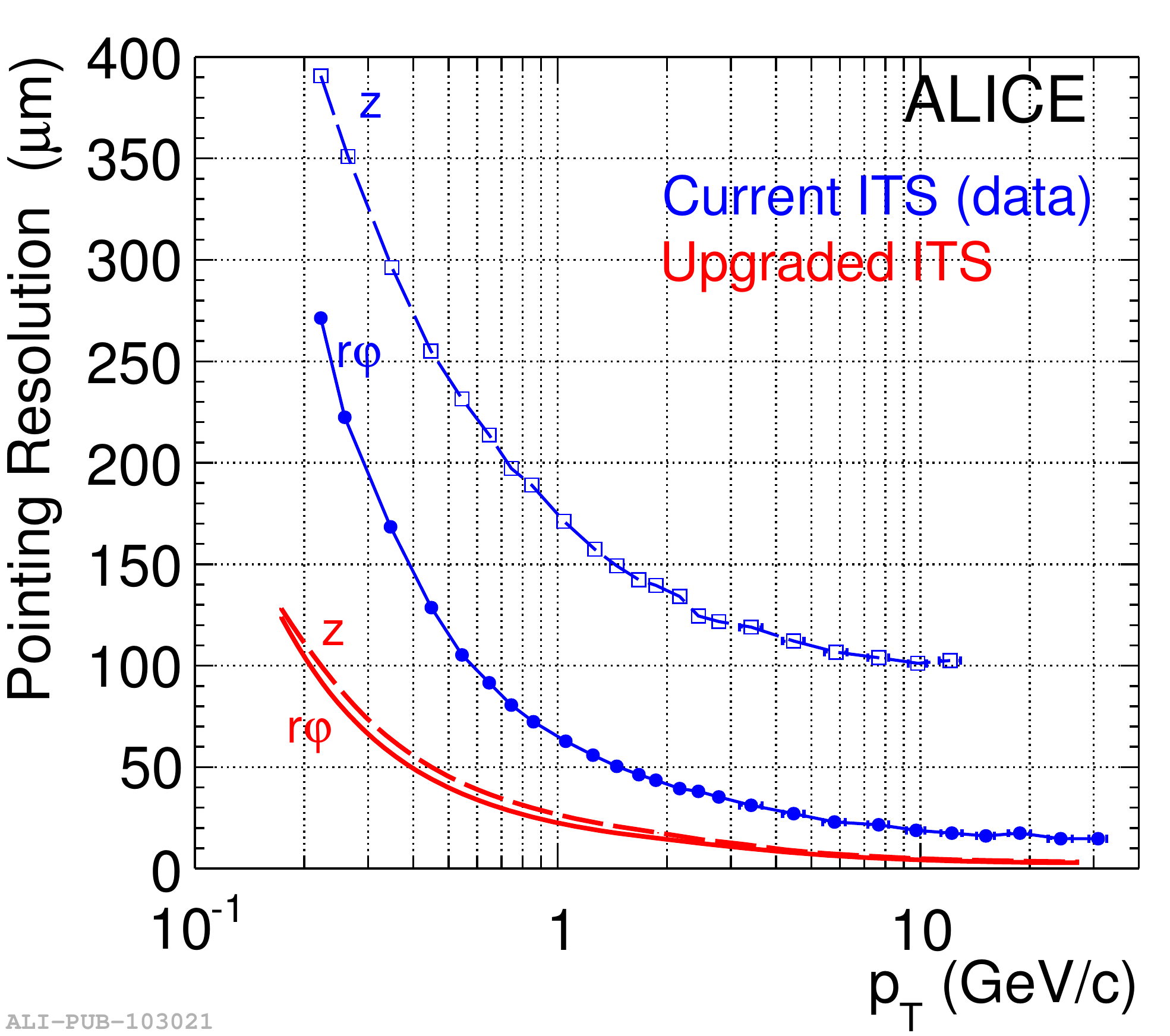} 
\caption{Left: schematic layout of the new ITS detector. Right: impact parameter resolution for primary charged pions as a function of the
transverse momentum for the present ITS (blue) and the upgraded ITS (red) in the transverse plane
(solid lines) and in the longitudinal direction (dashed lines)~\cite{Abelevetal:2014dna}. } 
\label{ITS}
\end{figure}

The installation of the new detectors and the commissioning for the upgraded ALICE experiment are scheduled for LS2 
(2019--2020), with data taking starting in 2021. Several Italian groups are involved in the new ITS project and in the upgrade of the readout electronics for TOF, Muon Tracking and Trigger detectors, and the ZDC.

\subsection{Comparison and complementarity with the other LHC experiments}
\label{sec:otherLHCexp}

The other three large LHC experiments, ATLAS, CMS and LHCb, will participate in the entire heavy-ion programme. In particular, 
the ATLAS and CMS Collaborations have requested a Pb--Pb integrated luminosity of about 10~nb$^{-1}$ for Run-3 and Run-4~\cite{ATLASKrakow,CMSKrakow,CMSFTR13025}, 
which is similar to that requested by ALICE. Both ATLAS and CMS will undergo some specific upgrade 
during LS2 and then a major upgrade during LS3
in view of the High-Luminosity LHC machine startup in Run-4. In particular, the muon systems will be 
upgraded during LS2 and new higher-precision inner trackers will be installed during LS3, 
which will strongly improve heavy-flavour hadron and jet measurements in Pb--Pb collisions.  
The LHCb experiment participated in the p--Pb run in 2013 and in the Pb--Pb run in 2015. A number of important measurements based on the p--Pb data sample
were already reported, in particular in the sector of quarkonium production~\cite{Aaij:2016eyl,Aaij:2014mza,Aaij:2013zxa}. The Collaboration expressed interest in continuing the heavy-ion programme
in Runs 2, 3 and 4, exploiting also the major detector upgrade planned for LS2~\cite{mancaQM2015}. The upgrade includes a complete replacement of the vertex detector and a faster readout system.

The data-taking strategy of the ATLAS and CMS experiments will be based mainly on highly-selective triggers on muons, jets and displaced high-$\pt$ tracks.
The Pb--Pb hadronic interaction rate of 50~kHz will be reduced by the trigger to the rate of a few kHz, which is the input to the High Level Trigger, 
and then to $\sim 100$~Hz, which is the rate of event recording to storage. This strategy is orthogonal to the ALICE approach 
of a 50~kHz recording to storage with minimum-bias trigger. Therefore, the programmes of the ATLAS/CMS and ALICE experiments will be highly complementary, 
with the former focusing on the higher $\pt$ region and high signal-to-background observables and ALICE focusing on low and intermediate $\pt$ 
and having unique access to some of the low signal-to-background probes that can not be selected with an online trigger. 
The performance of the LHCb detector in the conditions of high occupancy of central Pb--Pb events is still to be demonstrated, but in principle the experiment 
has very strong potential for the measurements of heavy-flavour production at low $\pt$ and forward rapidity.

The specificities of the various experiments are briefly summarised in the following, in the context of some of the main physics items for the LHC heavy-ion programme after LS2. 
\begin{description}
\item[\uppercase{H}eavy flavour:]  precise characterisation of the quark mass dependence of in-medium parton energy loss; study of the transport and possible thermalization of heavy quarks in the medium; study of heavy quark hadronization mechanisms in a partonic environment. These studies require measurements of the production and azimuthal anisotropy of several charm and beauty hadron species, over a broad momentum range, as well as of $b$-tagged jets.   ALICE will focus mainly on the low-momentum region, 
down to zero $p_{\rm T}$, and on reconstruction of several heavy-flavour hadron species  (including baryons). 
ATLAS and CMS will focus mainly on $b$-tagged jets and on D and B mesons at higher $\pt$. 
LHCb has a strong potential for all these measurements, pending the detector performance in central Pb--Pb collisions.
\item[\uppercase{Q}uarkonia:] study of quarkonium dissociation and possible regeneration as probes of deconfinement and of the medium temperature. ALICE will carry out precise measurements, starting from zero $p_{\rm T}$, of J$/\psi$ yields and azimuthal anisotropy, $\psi$(2S) and $\Upsilon$ yields, at both central and forward rapidity. ATLAS and CMS will carry out precise multi-differential measurements of the $\Upsilon$ states to map the dependencies of their suppression pattern. They will also complement
to high momentum the charmonium measurements.
Also in this case, LHCb has a strong potential, pending the detector performance in central Pb--Pb collisions.
\item[\uppercase{J}ets:] detailed characterisation of the in-medium parton energy loss mechanism,
that provides both a testing ground for the multi-particle aspects of QCD and a probe of the QGP density. 
The relevant observables are: jet structure and di-jet imbalance at TeV energies, $b$-tagged jets and
jet correlations with photons and $\rm Z^0$ bosons (unaffected by the presence of a QCD medium). These studies are crucial to address the flavour dependence of the parton energy loss and will be the main focus of ATLAS and CMS, which have unique high-$p_{\rm T}$ and triggering capabilities. ALICE will complement in the low-momentum region, and carry out measurements of the flavour dependence of medium-modified fragmentation functions using light flavour, strange and charm hadrons reconstructed within jets.
\item[\uppercase{L}ow-mass dileptons and thermal photons:] these observables are sensitive to the initial temperature and the equation of state of the medium, as well as to the chiral nature of the phase transition. The study will be carried out by ALICE, which will strengthen its unique very efficient electron and muon reconstruction capabilities down to almost zero $p_{\rm T}$, as well as the readout capabilities for recording a very high statistics minimum-bias sample. 
\end{description}
These and other physics items are discussed in detail in the rest of this Chapter from the point of view of the ALICE experiment.

\subsection{Open heavy flavour}
\label{sec:ohf}

\subsubsection{Current status of theory and experiment}

Heavy quarks (charm and beauty) are among the hard probes used to chacterize the properties of the hot QGP formed in ultra-relativistic heavy-ion collisions.
They are particularly well-suited for these studies, 
because of the large value of their mass ($M$) compared to the other scales involved in their production and in interaction processes in the medium.
First of all $M\gg\Lambda_{\rm QCD}$, so that their initial hard production is well described by perturbative QCD~\cite{Alioli:2010xd,Cacciari:1998it}. 
Their production processes are characterised by a large virtuality $Q^2<1/(4\,M^2)$, implying that heavy quarks are formed on a time scale smaller than about 1~fm/$c$
and they subsequently traverse the QGP interacting with its constituents.
Since $M\gg T$, their thermal production in the QGP is expected to be negligible at the temperatures $T$ reached at the LHC. 
Finally, their average thermal momentum $p_{\rm Q}^{\rm th} =\sqrt{3MT}$ is also much larger than the typical momentum exchange with the medium particles, $g\cdot T$, so that many independent collisions are necessary to significantly change the momentum value and direction of a heavy quark.
A recent review of the theoretical and experimental studies on open heavy flavour and quarkonium in heavy-ion collisions can be found in Ref.~\cite{Andronic:2015wma}.
In the following we outline the theoretical approach to which the Italian community is contributing and describe the experimental measurements carried out with the
ALICE experiment during the LHC Run-1 with important contributions of the Italian collaborators.

Depending on the momentum scale, heavy quarks are used to address different aspects of the QGP medium and of its evolution.
At high momentum (much larger than $M$), the main goal of heavy-flavour studies is gaining insight on the parton energy loss mechanism. 
 Various approaches have been developed to describe the interaction of the heavy quarks with the surrounding plasma.
In a perturbative treatment, QCD energy loss is expected to occur
via both inelastic (radiative energy loss, via medium-induced gluon radiation)~\cite{Gyulassy:1990ye,Baier:1996sk} and elastic (collisional energy loss)~\cite{Thoma:1990fm,Braaten:1991jj,Braaten:1991we}
processes. 
In QCD, quarks have a smaller colour coupling factor with respect to gluons, 
so that the energy loss for quarks is expected to be smaller than for gluons. 
In addition, the \emph{dead-cone effect} should reduce small-angle gluon radiation 
for heavy quarks with moderate 
energy-over-mass values, thus further attenuating 
the effect of the medium. This idea was first introduced in~\cite{Dokshitzer:2001zm}. 
Further theoretical studies confirmed the reduction of the total induced gluon radiation~\cite{Armesto:2003jh,Djordjevic:2003zk,Zhang:2003wk,Wicks:2007am}.
Other mechanisms such as in-medium hadron formation and dissociation~\cite{Adil:2006ra,Sharma:2009hn}, would determine a
stronger suppression effect on heavy-flavour hadrons than light-flavour hadrons, because of their smaller formation 
times.

At low momentum (of the order of $M$), an outstanding open question is whether heavy quarks take part in the collective expansion of the QGP and whether 
they approach thermal equilibrium. 
Transport calculations are used to study this problem from a theoretical viewpoint. As long as the medium admits a particle description, the Boltzmann equation is the most rigorous tool to carry out such simulations~\cite{Gossiaux:2008jv,Uphoff:2011ad,Das:2013kea}. The latter however is an integro-differential equation for the (on-shell) heavy-quark phase-space density and its numerical solution is challenging. Relying on the hypothesis of dominance of soft scatterings, the exact Boltzmann equation can be approximated with a form more suited to an easy numerical implementation: the relativistic Langevin equation. The latter enables the study of the evolution of the momentum $\vec{p}$ of each heavy quark, through the combined effect of a friction and noise term, both arising from the collisions with the QGP constituents~\cite{vanHees:2007me,Moore:2004tg,Akamatsu:2008ge,Alberico:2011zy,Alberico:2013bza,Beraudo:2014boa}:
\begin{equation}
  \frac{\Delta\vec{p}}{\Delta t}=-\eta_D(p)\vec{p}+\vec{\xi}(t)\quad{\rm with}\quad \langle\xi^i(t)\xi^j(t')\rangle=b^{ij}(\vec{p})\delta_{tt'}/\Delta t\equiv \left[\kappa_{\rm L}(p)\hat{p}^i\hat{p}^j+\kappa_{\rm T}(p)
(\delta^{ij}-\hat{p}^i\hat{p}^j)\right]\delta_{tt'}/\Delta t\,.
\label{eq:lange_r_d}
\end{equation}
 The strength of the noise is set by the momentum-diffusion coefficients $\kappa_{\rm L/T}$, reflecting the average longitudinal/transverse squared momentum exchanged with the plasma. The friction coefficient $\eta_D(p)$ has to be fixed in order to ensure the approach to thermal equilibrium through the Einstein fluctuation-dissipation relation
\begin{equation}
\eta_D(p)\equiv\frac{\kappa_{\rm L}(p)}{2TE_p}+{\rm corrections},\label{eq:friction}
\end{equation}
where the corrections, subleading by a factor ${\cal O}(T/E_p)$, depend on the discretisation scheme and are fixed in order to reproduce the correct continuum result in the $\Delta t\to 0$ limit~\cite{Beraudo:2009pe}. The advantage of the Langevin approach is that, independently of the nature of the medium, it allows one to summarise the interaction of heavy quarks with the medium into three simple transport coefficients (only two of which independent) with a clear physical interpretation. Results for the momentum-diffusion coefficients, obtained with very different approaches (resummed weak-coupling calculations and lattice-QCD simulations) are displayed in Fig.~\ref{fig:transport}. Lattice calculations are so far performed only for static quarks, i.e.\ at zero momentum, and in this case their result is much larger than the perturbative value~\cite{Francis:2015daa}. 
However, weak-coupling calculations~\cite{Alberico:2011zy,Alberico:2013bza} provide evidence of a quite strong momentum dependence of $\kappa_{\rm L/T}$ (the latter is found to be even larger in other approaches, like AdS/CFT calculations~\cite{Gubser:2006nz}), which unfortunately at present cannot be accessed on the lattice. 
This limits to low $p_{\rm T}$ the interval in which one can use experimental data to discriminate among the two theoretical results. 
In this kinematic domain, the precision of the current experimental result is limited and the interpretation of the results is complicated by
other effects like nuclear modification of the PDF and in-medium hadronization with possible changes of the heavy-flavour hadrochemistry.
On the other hand, for beauty quarks $\kappa_{\rm L}$ and $\kappa_{\rm T}$ are similar and quite constant for an extend momentum range ($p<5~\mathrm{GeV}/c$, see the right panel of Fig.~\ref{fig:transport}), opening the possibility to put tight constraints on the value of the transport coefficients with beauty measurements at low $p_{\rm T}$. 

\begin{figure}[t]
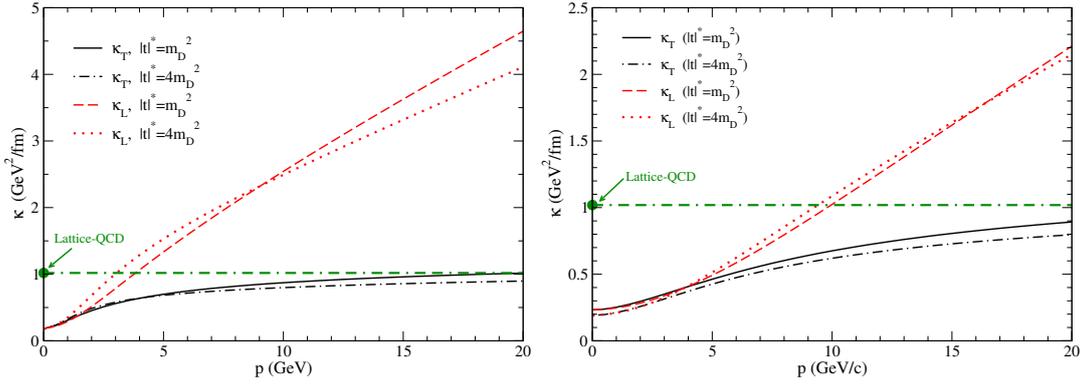

\begin{center}
\includegraphics[height=5cm]{../Figures/charmT400_run.pdf}
\includegraphics[height=5cm]{../Figures/beautyT400_run.pdf}
\caption{Charm (left panel) and beauty (right panel) momentum-diffusion coefficients $\kappa_{\rm T}$ and $\kappa_{\rm L}$  in the QGP at $T\!=\!400$ MeV. Lattice-QCD results are compared to resummed perturbative calculations for two values of $|t|^*$, which is an intermediate cutoff to separate the contributions of hard and soft collisions~\cite{Alberico:2011zy,Alberico:2013bza,Francis:2015daa}.}
\label{fig:transport}
\end{center}
\end{figure}

Model calculations are challenged to consistently describe various observables, in particular the nuclear modification factor $R_{\rm AA}$ and elliptic flow coefficient
$v_2$. At low and intermediate $\pt$ these observables provide information on the degree of thermalization of heavy quarks in the expanding QGP and on their hadronisation mechanisms, namely whether recombination with light quarks is a relevant effect. At high $\pt$, $R_{\rm AA}$ and $v_2$ are mainly sensitive to heavy-quark energy loss and to its dependence on the path length within the QGP. A comprehensive description of both observables would also put tight constraints on the temperature dependence of the drag and diffusion coefficients~\cite{Das:2015ana}. 
For the future, the challenge for theory is to provide predictions for more differential observables, including various kinds of heavy-flavour correlations, and to address the issue of possible collective effects for heavy quarks in high-multiplicity collisions of small systems, like proton--proton and proton--nucleus.

\begin{figure}
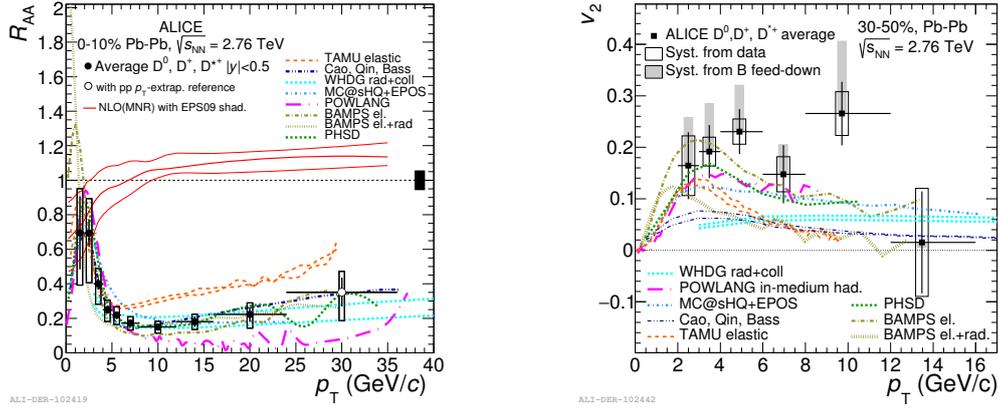

\begin{center}
\includegraphics[height=5.5 cm]{../Figures/2016-Feb-10-Comp_DAve010_240915_v2Models_shad.pdf}
\hspace{1.5 cm}
\includegraphics[height=5.5 cm]{../Figures/2016-Feb-10-v2-AverD-Models.pdf}
\caption{D-meson $R_{\rm AA}$ (left)~\cite{Adam:2015sza} and $v_2$ (right)~\cite{Abelev:2013lca,Abelev:2014ipa} as a function of $p_{\rm T}$
in Pb--Pb collisions at $\sqrtsNN=2.76$~TeV compared to model 
calculations.}
\label{fig:Dvsmodels}
\end{center}
\end{figure}

Open heavy-flavour production was measured with ALICE in pp, p--Pb and Pb--Pb collisions exploiting various 
techniques.
Charm and beauty were studied inclusively by measuring electrons (at mid-rapidity)~\cite{Abelev:2012xe,Abelev:2012sca,Abelev:2014gla,Abelev:2014hla,Adam:2015qda} and muons (at forward rapidity)~\cite{Abelev:2012pi,Abelev:2012qh,Adam:2015pga} from their semi-leptonic decays.
The beauty contribution to heavy-flavour decay electrons was estimated with two alternative methods.
The first consists in an analysis of the impact parameter distribution of the electrons to exploit 
the longer lifetime of beauty as compared to charm hadrons.
The second method is based on the azimuthal correlations between heavy-flavour decay electrons and 
charged hadrons and exploits the larger width of the near-side peak for beauty than for charm hadron 
decays.
Beauty production was also studied at mid-rapidity via measurements of 
non-prompt J/$\psi$~\cite{Abelev:2012gx,Adam:2015rba}.
Charm hadrons were fully reconstructed from their hadronic decays
$\mathrm{D}^0\rightarrow \mathrm{K}^-\pi^+$, 
$\mathrm{D}^+\rightarrow \mathrm{K}^-\pi^+\pi^+$, 
$\mathrm{D}^{*+}\rightarrow \mathrm{D}^0\pi^+$, and
$\mathrm{D_s}^+\rightarrow \phi\pi^+\rightarrow \mathrm{K}^-\mathrm{K}^+\pi^+$~\cite{ALICE:2011aa,Abelev:2012vra,Abelev:2012tca,ALICE:2012ab,Abelev:2013lca,Abelev:2014ipa,Abelev:2014hha,Adam:2015nna,Adam:2015sza,Adam:2015jda} .
Performance studies of reconstruction of jets associated to beauty hadrons 
exploiting the long lifetime and large mass 
of beauty hadrons are ongoing (see e.g. Ref.~\cite{yasserQM2015}).

The measurements with the LHC Run-1 data samples focused on the $p_{\rm T}$-differential
production yield (cross section) of heavy-flavour hadrons (or their decay products),
the nuclear modification factor $R_{\rm AA}$ ($R_{\rm pPb}$) in Pb--Pb~\cite{Abelev:2012qh,ALICE:2012ab,Adam:2015nna,Adam:2015sza,Adam:2015jda,Adam:2015rba} (p--Pb~\cite{Abelev:2014hha,Adam:2015qda}) collisions,
and the elliptic flow $v_2$~\cite{Abelev:2013lca,Abelev:2014ipa,Adam:2015qda} in Pb--Pb interactions.
In addition, first studies of angular correlations between D mesons (heavy-flavour decay electrons)
and charged hadrons were carried out on the pp and p--Pb data samples~\cite{Bjelogrlic:2014kia}.

The nuclear modification factor $R_{\rm AA}$ of D mesons (average of 
$\mathrm{D^0}$, $\mathrm{D^+}$ and $\mathrm{D^{*+}}$) as a function of
$p_{\rm T}$ in central Pb--Pb collisions
is shown in the left panel of Fig.~\ref{fig:Dvsmodels}~\cite{Adam:2015sza}.
A substantial suppression of the yield, as compared to binary-scaled pp reference, 
is observed at intermediate and high $p_{\rm T}$ (above $3$--$4~\mathrm{GeV}/c$).
Since in p--Pb collisions, the nuclear modification factor $R_{\rm pPb}$ is
observed to be compatible with unity within uncertainties~\cite{Abelev:2014hha}, the suppression
observed in Pb--Pb collisions is due to the interactions of the charm quarks
with the hot and dense medium.
In the left panel of Fig.~\ref{fig:Dvsmodels}, the data are compared
to model calculations including parton in-medium energy 
loss~\cite{Djordjevic:2014tka,Xu:2014tda,Xu:2015bbz,Wicks:2005gt,Horowitz:2011gd,Horowitz:2011wm,He:2014cla,Nahrgang:2013xaa,Alberico:2013bza,Cao:2013ita,Sharma:2009hn,Uphoff:2012gb,Uphoff:2014hza}, which can
describe the magnitude of the suppression observed at high $p_{\rm T}$.
At lower $p_{\rm T}$ (below $3~\mathrm{GeV}/c$), the trend of the
nuclear modification factor is determined by the interplay between 
parton in-medium energy loss and radial flow, and it is also expected to 
be sensitive to the hadronization mechanism (recombination vs.\ fragmentation)
as well as to initial-state effects such as $k_{\rm T}$ broadening and
nuclear shadowing of the Parton Distribution Functions.
The hadronisation mechanism was studied in particular by measuring
the production of $\mathrm{D_s}^+$ mesons: an enhancement
of $\mathrm{D_s}^+$ yield relative to that of non-strange D mesons
is expected if the dominant process for D-meson formation at 
low and intermediate momenta is in-medium hadronisation of charm quarks via 
recombination with light quarks~\cite{Andronic:2003zv,Kuznetsova:2006bh,He:2012df}.
In the left panel of Fig.~\ref{fig:PionsANDBeauty}, the nuclear 
modification factor of $\mathrm{D_s}^+$ mesons in central Pb--Pb collisions is 
compared to that of non-strange D mesons~\cite{Adam:2015jda}.
The central values of the $\mathrm{D_s}^+$-meson $R_{\rm AA}$ are higher than 
those of non-strange D mesons, although compatible within uncertainties.
From the currently available data samples, it is not possible to draw a 
conclusive statement on the expected modification of the relative abundance
of charm-hadron species due to hadronization via recombination.

The elliptic flow $v_2$ of D mesons as a function of $p_{\rm T}$ measured
in semi-central Pb--Pb collisions is reported in the right panel
of Fig.~\ref{fig:Dvsmodels}~\cite{Abelev:2013lca,Abelev:2014ipa}.
A positive $v_2$ is observed, comparable in magnitude to that of light-flavour 
particles.
This indicates that interactions with the medium constituents 
transfer to charm quarks information on the azimuthal anisotropy of the system,
suggesting that low momentum charm quarks take part in the 
collective motion of the system.
The data are compared to the same model calculations that were confronted
to the nuclear modification factor. 
It is at present challenging for models to describe simultaneously
the measured $R_{\rm AA}$ and $v_2$ of heavy-flavour hadrons,
suggesting that the data have the potential to constrain the modeling of
the medium properties and the interactions of 
charm quarks with the medium constituents.

The comparison of the nuclear modification factor of D mesons,
beauty hadrons and light flavour particles (pions) was proposed 
as a test for the predicted colour-charge and quark mass 
dependence of parton in-medium energy loss~\cite{Armesto:2005iq}.
However, other effects than the energy loss, like the initial parton $\pt$-distribution 
and fragmentation into hadrons, influence the nuclear modification factor and can
counterbalance the effect of the larger energy loss of gluons with respect to 
quarks.
The result of this comparison is shown in the left panel of 
Fig.~\ref{fig:PionsANDBeauty}~\cite{Adam:2015sza}: the D-meson $R_{\rm AA}$ is found to be 
larger than that of charged pions, even though compatible within uncertainties.
The small difference between pion and D-meson $R_{\rm AA}$ can be 
described by models including colour-charge and quark-mass dependent energy loss 
together with different production kinematics and fragmentation functions of charm quarks, 
light quarks and gluons~\cite{Djordjevic:2013xoa,Djordjevic:2014tka}.
The right panel of Fig.~\ref{fig:PionsANDBeauty} shows a comparison of 
D-meson and non-prompt J/$\psi$ (from CMS~\cite{Chatrchyan:2012np,CMS:2012vxa}) 
$R_{\rm AA}$ at high $p_{\rm T}$ as a function of the collision centrality, expressed in terms of the
number of participant nucleons $N_{\rm part}$~\cite{Adam:2015nna}.
A stronger suppression of high-$p_{\rm T}$ charm-hadron yield as compared to
beauty is observed.
This difference can be described by models including a dependence of the
in-medium energy loss on the quark mass.

\begin{figure}
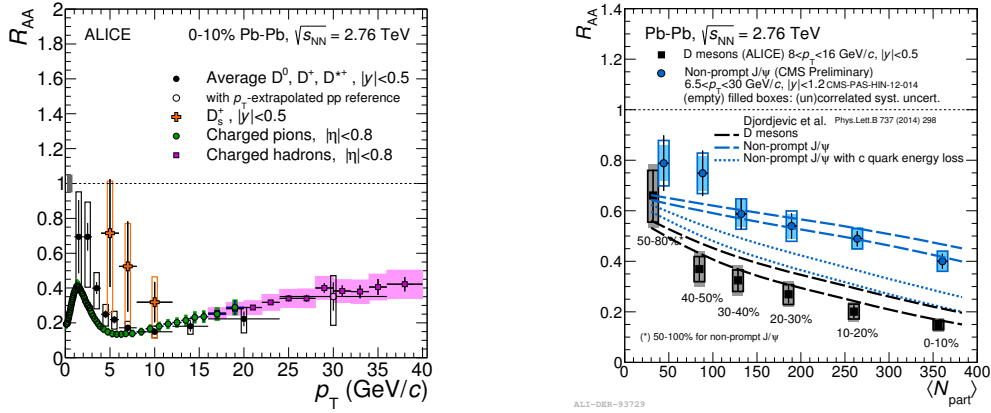

\begin{center}
\includegraphics[height=5.5 cm]{../Figures/Raa-Ds-AverD-Pions.pdf}
\hspace{1.5 cm}
\includegraphics[height=5.5 cm]{../Figures/2015-Jun-29-DmesNonPromptJpsi_8to16_CompDjordjevic_110615.pdf}
\caption{Left: D-meson $R_{\rm AA}$ as a function of $p_{\rm T}$ in central Pb--Pb collisions at $\sqrtsNN=2.76$~TeV~\cite{Adam:2015sza} compared to $\mathrm{D_s^+}$~\cite{Adam:2015jda}, charged particle~\cite{Abelev:2012hxa} and pion~\cite{Abelev:2014laa} $R_{\rm AA}$. Right: D-meson~\cite{Adam:2015nna} and non-prompt J/$\psi$~\cite{Chatrchyan:2012np,CMS:2012vxa} $R_{\rm AA}$ as a function of 
centrality in Pb--Pb collisions at the LHC.}
\label{fig:PionsANDBeauty}
\end{center}
\end{figure}

In summary, based on a data-to-theory comparison of the results from the LHC Run-1,
clear evidence emerged of substantial final-state effects due to the 
interactions of the charm and beauty quarks with the medium.
However, in order to put stronger constraints on the medium transport 
coefficients, on the possible thermalization of heavy quarks and on the role of hadronization via recombination, more precise measurements -- especially at low $p_{\rm T}$ and for different hadron species -- are needed.
This is one of the main goals to be reached with the larger data samples that will be collected in the future LHC runs.


\subsubsection{Future prospects}
\label{secHFfuture}

\begin{figure}
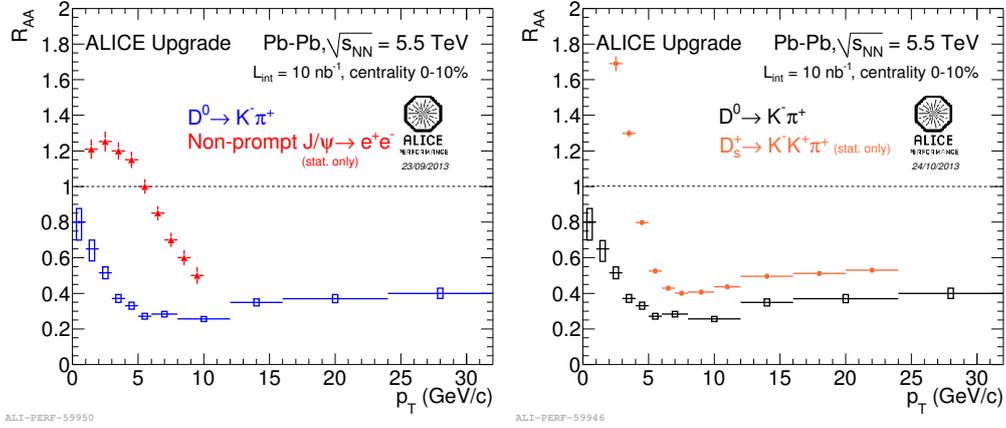

\begin{center}
\includegraphics[width=0.4\textwidth]{../Figures/2013-Oct-24-D0BJpsiRAAprompt_TDR_logo.pdf}
\includegraphics[width=0.4\textwidth]{../Figures/2013-Oct-24-D0DsRAAprompt_TDR_logo.pdf}
\caption{Comparison of prompt D-meson $R_{\rm AA}$ with $R_{\rm AA}$ of ${\rm J}/\psi$ from B-meson decay (left) and of ${\rm D^+_{s}}$ (right) as a function of $p_{\rm T}$
in the 10\% most central Pb--Pb collisions at $\sqrtsNN=5.5$~TeV. Projection of the ALICE upgrade performance with an integrated luminosity of 10~nb$^{-1}$~\cite{Abelevetal:2014dna}.}
\label{fig:PromptDmesonDsJPsifromBRAAupgradeALICE}
\end{center}
\end{figure}

\begin{figure}
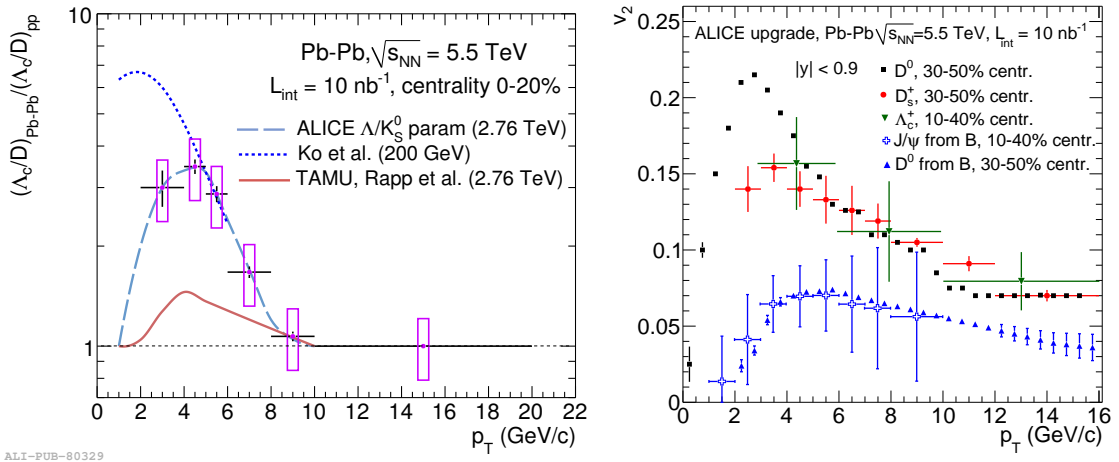

\begin{center}
\includegraphics[width=0.48\textwidth]{../Figures/2014-May-15-LambdacOverD0_DoubleRatio_10nb_TDR.pdf}
\includegraphics[width=0.4\textwidth]{../Figures/CharmBeautyMesonBaryonsCombinedv2ALICEupgrade.pdf}
\caption{Left: double ratio of $\Lambda_{\rm c}$ over prompt non-strange D-meson production in Pb--Pb and pp collisions 
  with uncertainties expected after the ALICE upgrade~\cite{Abelevetal:2014dna}. The model predictions are from~\cite{He:2012df,Oh:2009heavybaryon}. Right: comparison 
of $p_{\rm T}$ dependence of $v_2$ of prompt non-strange 
D mesons, ${\rm D^+_{s}}$, $\Lambda_{\rm c}$, and D meson and ${\rm J}/\psi$ from B-meson decay as expected to be measured 
by ALICE in Pb--Pb collisions at $\sqrtsNN=5.5$~TeV after the upgrade of the detector.}
\label{fig:LcOverDandCharmBeautyv2ALICEupgrade}
\end{center}
\end{figure}
The measurements performed with Run-1 data at the LHC represent a first step into a ``charm and beauty” era of the QGP. 
New incoming data from Run-2 and Run-3 at the LHC will allow us to 
measure with much higher precision charm and beauty nuclear modification factors and elliptic flow coefficients over a wide transverse momentum region. With 
respect to that produced at RHIC collision energies, the medium
that can be probed at the LHC is characterised by higher energy density, different temperature profile, different expansion velocity, and longer 
lifetime of the partonic phase~\cite{Muller:2012zq,Roland:2014jsa,Armesto:2015ioy}. These differences can significantly affect the interaction of heavy-quarks with the medium, in particular
in relation to charm flow, which develops at late times during the medium evolution, and to the role of hadronization via recombination,
particularly sensitive to the medium density and temperature profile. By measuring the charm and beauty nuclear modification factor 
at the two facilities, constraints can be set to the energy-density dependence of the transport coefficient of the medium. 
The measurement of the $\pt$-differential production cross-sections
of charmed hadrons at the two energies can help to constrain initial state effects, in particular nuclear shadowing, which
is expected to be significantly larger at LHC energies. Finally, at LHC the cross-section for beauty production is significantly 
larger (about a factor 70~\cite{Abelev:2014gla,Cacciari:1998it}), allowing for detailed studies of beauty quark energy loss.

With the statistics collected during Run-2 at the LHC, and considering the larger charm production cross-section at the new collision 
energy of $\sqrtsNN=5$~TeV, the ALICE experiment will improve by a factor 2--3 the statistical precision on the measurements of prompt D-meson production
in Pb--Pb collisions. The $R_{\rm AA}$ will be measured in a wider $\pt$ range, extending the current measurements
to higher as well as lower transverse momenta. For the first time, $\rm D^{0}$ mesons 
might become accessible down to $p_{\rm T}=0$ in Pb--Pb collisions, althought with limited statistical precision. This would represent an important step towards the evaluation of the total charm production 
cross section, currently the main source of 
uncertainty on the predictions of charmonium production (see Section~\ref{sec:onia}). The transverse momentum dependence of the elliptic flow of non-strange 
D mesons in semi-peripheral collisions could be assessed, allowing
for a a more quantitative comparison with pion $v_2$ in the intermediate $p_{\rm T}$ region ($2<p_{\rm T}<8~{\rm GeV}/c$) where charm flow
develops mainly with the medium collective expansion. Additional information will derive from
the measurement of ${\rm D^+_{s}}$ nuclear modification factor: with the precision expected, a first significant observation of the 
modification of charm quark hadronization could be made. 

In Run-3 the higher luminosity of the LHC machine and the upgrade of the ALICE detector~\cite{Abelevetal:2014cna,Abelevetal:2014dna,Antonioli:1603472,CERN-LHCC-2013-020,CERN-LHCC-2015-001,CERN-LHCC-2015-006} will open the precision era for charm and beauty measurements, which should bring to a 
rather complete understanding of the interaction of heavy quarks with the medium.
High-precision $p_{\rm T}$-differential measurements of nuclear modification factors and elliptic flow coefficients of charm and beauty hadrons will allow for a quantification
of the mass dependence of energy loss over a wide momentum range, of the impact of hadronization via recombination, and of 
the degree of participation of charm and beauty quarks to the collective expansion of the medium, measuring their coupling to the system
and setting constraints on the transport coefficients of the medium. In particular, thanks to the new Inner Tracking System~\cite{Abelevetal:2014dna}, ALICE will investigate
the mass dependence of energy loss at central rapidity via a precise comparison of the nuclear modification factor of prompt 
D-mesons with those of D meson and ${\rm J}/\psi$ from B-meson decay. The addition
of the Muon~Forward~Tracker (MFT~\cite{CERN-LHCC-2015-001}) will enable the measurement of the fraction of muons and ${\rm J}/\psi$ from beauty-hadron 
decay also at forward rapidity. As an example, in the left panel 
of Fig.~\ref{fig:PromptDmesonDsJPsifromBRAAupgradeALICE} the $R_{\rm AA}$ of prompt $\rm D^{0}$ meson and ${\rm J}/\psi$ from B-meson decay
at central rapidity are shown. The systematic uncertainty on the measurement of prompt $\rm D^{0}$ $R_{\rm AA}$ will be smaller than 8\% for
$p_{\rm T}>2~{\rm GeV}/c$ and around 10\% for $p_{\rm T}<2~{\rm GeV}/c$. The fraction of ${\rm J}/\psi$ from B-meson decay will be measured
with systematic uncertainty better than 8\%. Along with CMS measurement of ${\rm J}/\psi$ from B-meson decay and the measurements of 
B-meson production in central Pb--Pb collision by ALICE and CMS, these measurements will complement at higher and lower $p_{\rm T}$ the
observations made with Run-2 data and will be precise enough to track the mass dependence of energy loss over a wide momentum range. The 
new ALICE detector will give the opportunity
of accessing charm and beauty quark momenta down to $p_{\rm T}=0$ at central rapidity, a particularly appealing
region. In fact heavy quarks in this kinematic domain are the only partons probing the medium with non-relativistic 
velocities, which, among other features, implies a larger energy loss from collisional elastic scattering with the medium
constituents. Indeed, the ALICE specificity, with respect to the other LHC experiments participating to the investigation
of the QGP, will be the study of the low-$p_{\rm T}$ region for various heavy-flavour hadron species, including
both charm and beauty mesons and baryons. 

The baryon-to-meson production ratio is particularly
sensitive to recombination effects. A quantification of the role of hadronization via coalescence will be achieved by
measuring $\rm D_{\rm s}^+$ and $\Lambda_{\rm c}$ nuclear modification factors and elliptic 
flow coefficients at low and intermediate transverse momentum. Figure~\ref{fig:PromptDmesonDsJPsifromBRAAupgradeALICE} shows, in the right panel, 
the comparison between $\rm D_{\rm s}^+$ and non-strange D mesons $R_{\rm AA}$: $\rm D_{\rm s}^+$
production will be measured with statistical uncertainty smaller than 5\% down to $p_{\rm T}=2~{\rm GeV}/c$ 
in central Pb--Pb collisions. Systematic uncertainties similar to those for the measurements of prompt non-strange D mesons 
are expected. Figure~\ref{fig:LcOverDandCharmBeautyv2ALICEupgrade}
shows in the left panel the double-ratio $\rm \Lambda_c/D$ in Pb--Pb and pp collisions. The 
pseudodata points show that
the projected precision is sufficient to discriminate among predictions
from different models~\cite{He:2012df,Oh:2009heavybaryon}. 
For the first time, a possible difference
between baryon and meson nuclear modification factors could be observed also in the
beauty sector, thanks to the measurement of $\Lambda_{\rm b}$ production~\cite{Abelevetal:2014dna}. 
Figure~\ref{fig:LcOverDandCharmBeautyv2ALICEupgrade} shows the expectation for the
measurement of the elliptic flow of several charm and beauty hadrons.
With the projected precision 
the difference between charm $v_2$, accessible down to $p_{\rm T}=0$ with negligible 
uncertainties with non-strange D mesons, and beauty $v_2$, probed with statistical uncertainty smaller than 10\%
with $\rm D^{0}$ from B-meson decay down to $p_{\rm T}=2~{\rm GeV}/c$ can be quantified. 
A precise measurement of $\rm D_{\rm s}^+$ elliptic
flow as well as a first observation of $\Lambda_{\rm c}$ will also be done. Table~\ref{tablePhysicsReach} summarises the projected performance
for heavy-flavour measurements with the 
ALICE upgrade and an integrated luminosity of 10~nb$^{-1}$, compared with the expection for Run-2 and $0.1$~nb$^{-1}$ (the projected luminosity recorded with minimum-bias triggers.)

\begin{table}[t]
  \centering
  \caption{Examples of heavy-flavour performance projections with the ALICE upgrade (for a comprehensive list, see~\cite{Abelevetal:2014dna, CERN-LHCC-2015-001}): minimum accessible $\pt$ and relative statistical uncertainty in Pb--Pb collisions for an integrated luminosity of $10~{\rm nb^{-1}}$. The statistical uncertainties are given at the maximum between $\pt=2~{\rm GeV}/c$
    and $p_{\rm T}^{\rm min}$. For elliptic flow measurements, the value of $v_2$ used to calculate the relative statistical uncertainty $\sigma_{v_2}/v_2$ is given in 
    parenthesis. The case of the programme up to Long Shutdown 2, with a luminosity of
    $0.1~{\rm nb}^{-1}$ collected with minimum-bias trigger, is shown for comparison (only the statistical uncertainty is reported).}
  \label{tablePhysicsReach}
  \begin{tabular}{lccccc}
    \hline
    & \multicolumn{2}{c}{Current, $0.1~{\rm nb}^{-1}$ } & \multicolumn{3}{c}{Upgrade, $10~{\rm nb}^{-1}$} \\
     \cmidrule{2-3} \cmidrule{4-6}
    Observable & $p_{\rm T}^{\rm min}$  & \multicolumn{1}{c}{statistical} & $p_{\rm T}^{\rm min}$  & \multicolumn{2}{c}{uncertainty} \\
    & (GeV/$c$)  & \multicolumn{1}{c}{uncertainty} & (GeV/$c$)  & \multicolumn{1}{c}{statistical}  & \multicolumn{1}{c}{systematic} \\
   \hline
    D meson $R_{\rm AA}$           & 1   & 10\%  &  0 & 0.3\% & 6\% \\
    D meson from B $R_{\rm AA}$ & 3 & 30\% &  2 & 1\% & 20\%\\
    J/$\psi$ from B $R_{\rm AA}$ in 2.5<{\it y}<3.6 & \multicolumn{2}{c}{not accessible} &  0 & 2\% & 5\%\\
    $\Lambda_{\rm c}$ $R_{\rm AA}$ & \multicolumn{2}{c}{not accessible} & 2 & 15\% & 20\% \\
    $\Lambda_{\rm c}/{\rm D^0}$ ratio & \multicolumn{2}{c}{not accessible} &  2 & 15\% & 20\% \\
    D meson $v_2$ ($v_2 = 0.2$) &  1 & 10\%  &  0 & 0.2\% & -- \\
    D from B $v_2$ ($v_2 = 0.05$) & \multicolumn{2}{c}{not accessible}  &  2 & 8\% & -- \\
    $\Lambda_{\rm c}$ $v_2$ ($v_2 = 0.15$) & \multicolumn{2}{c}{not accessible}  &  3 & 20\% & -- \\
     \hline
  \end{tabular}
\end{table}

Data from Run-2 and Run-3 will open the possibility of studying 
heavy-flavour angular correlations, in particular azimuthal correlations of heavy-flavour hadron decay leptons 
or D mesons and charged particles or jets. These 
observables will provide important constraints on the dependence of the heavy quarks in-medium energy loss on the distance covered by the parton
in the medium. Moreover, the measurement of the modification of near-side correlation peak properties (associated yield and width) of
azimuthal correlations of D mesons and charged particles will enable the study of possible medium modifications to charm fragmentation and jet properties. 

A positive elliptic flow was measured for light-flavour hadrons in high-multiplicity pp and p--Pb collisions, with a $p_{\rm T}$ and particle-mass dependence resembling 
that observed in heavy-ion collisions, which is typically ascribed to the collective expansion of the medium and described by hydrodynamic models (see Section~\ref{sec:soft}). The origin of this 
effect in pp and p--Pb collisions is still debated. The study of angular correlations between heavy-flavour 
particles (D mesons or heavy-flavour hadron decay electrons) 
and charged particles produced in the event in p--Pb collisions and high-multiplicity pp collisions could reveal whether also charm and beauty show in 
``small systems'' $v_2$-like double-ridge long range angular correlations, as observed for light hadrons. These measurements
could be done with data from Run-2, though for the case of correlation between D mesons and charged particles in high multiplicity pp and p--Pb events,
a precise measurement will become possible only with the data collected with the upgraded ALICE detector in Run-3 and Run-4. 


\subsection{Quarkonia}
\label{sec:onia}

\subsubsection{Present status of theory and experiment}

In a deconfined QGP, quarkonium production is expected to be significantly suppressed with respect to the proton--proton yields scaled by the number of binary nucleon--nucleon collisions, due to a colour screening mechanism that prevents the binding of the $Q$ and $\overline{Q}$ pair~\cite{Matsui:1986dk}.
In this scenario, with increasing temperature of the system, quarkonium resonances are expected to be suppressed sequentially, according to their binding energy. 
In the charmonium sector, the loosely bound $\psi$(2S) and the $\chi_c$ states are expected to melt at a lower temperature with respect to the tightly bound J/$\psi$, while in the bottomonium sector suppression will first affect $\Upsilon$(3S) and $\Upsilon$(2S) and eventually $\Upsilon$(1S) states. 
As a consequence, the in-medium dissociation probability of the various states should be sensitive to the temperature reached in the collisions.

However, this sequential suppression scenario is complicated by other mechanisms, related to both hot or cold nuclear matter.
On the one hand, increasing the collision energy ($\sqrtsNN$), the production rates of the $Q$ and $\overline{Q}$ quarks increase. As a consequence, at high energies, a new production mechanism sets in, due to the (re)combination, either in the QGP or in the hadronization phase, of the abundantly produced $Q\overline{Q}$ pairs~\cite{Thews:2000rj,Stachel:2013zma,BraunMunzinger:2000px}. This effect enhances the charmonium yields and it might counterbalance the aforementioned suppression. Given the smaller $b\overline{b}$ production cross section, with respect to $c\overline{c}$, (re)combination is expected to play a negligible role for bottomonium states.
On the other hand, quarkonium production can also be affected by cold nuclear matter (CNM) effects, such as nuclear shadowing, parton energy loss or break-up of the $Q\overline{Q}$ pair via interactions with the nucleons. These effects are usually investigated in proton--nucleus collisions, where suppression or (re)combination, which are related to hot-matter, are not expected to affect quarkonium production.
Since CNM effects are also present in nucleus--nucleus interactions, a precise knowledge of their role is, therefore, crucial to disentangle the pure QGP effects.

In the last thirty years, quarkonium has been extensively studied by experiments at the SPS  (up to $\sqrtsNN=17$~GeV, see, for example, ~\cite{Abreu:1997jh,Abreu:2000ni,Abreu:1999qw,Alessandro:2004ap,Arnaldi:2007zz}) and RHIC (up to $\sqrtsNN=200$~GeV, see, for example, ~\cite{Adler:2003rc,Yan:2006ve,Adare:2011yf,Adamczyk:2013tvk,Adamczyk:2012pw}) facilities, and finally at the LHC.
In LHC Run-1, \mbox{Pb--Pb} collisions were collected at $\sqrtsNN$= 2.76~TeV (see, for example, ~\cite{Adam:2015isa,Adam:2015rba,Abelev:2013ila,Abelev:2014nua,Aad:2010aa,Chatrchyan:2012np,Chatrchyan:2011pe,Chatrchyan:2012lxa}) and \mbox{p--Pb} collisions at $\sqrtsNN$= 5.02~TeV (see, for example, \cite{Abelev:2013yxa,Adam:2015iga,Aaij:2013zxa}).
The quarkonium in-medium modification is usually quantified through the nuclear modification factor $ R_{\rm {AA}}$, defined as the ratio of the quarkonium yields in nucleus--nucleus interactions and a reference value obtained scaling the corresponding values in proton--proton interactions by the number of nucleon--nucleon collisions. 
In spite of the factor of ten difference in the centre-of-mass energies, the J/$\psi$ nuclear modification factors,  measured by the SPS and RHIC experiments, show a rather similar centrality-dependent suppression.
This observation suggests that the (re)combination process might set in already at RHIC energies compensating for some of the quarkonium suppression due to screening in the QGP.

\begin{figure}[t]
  \centering
  \includegraphics[width=0.45\textwidth]{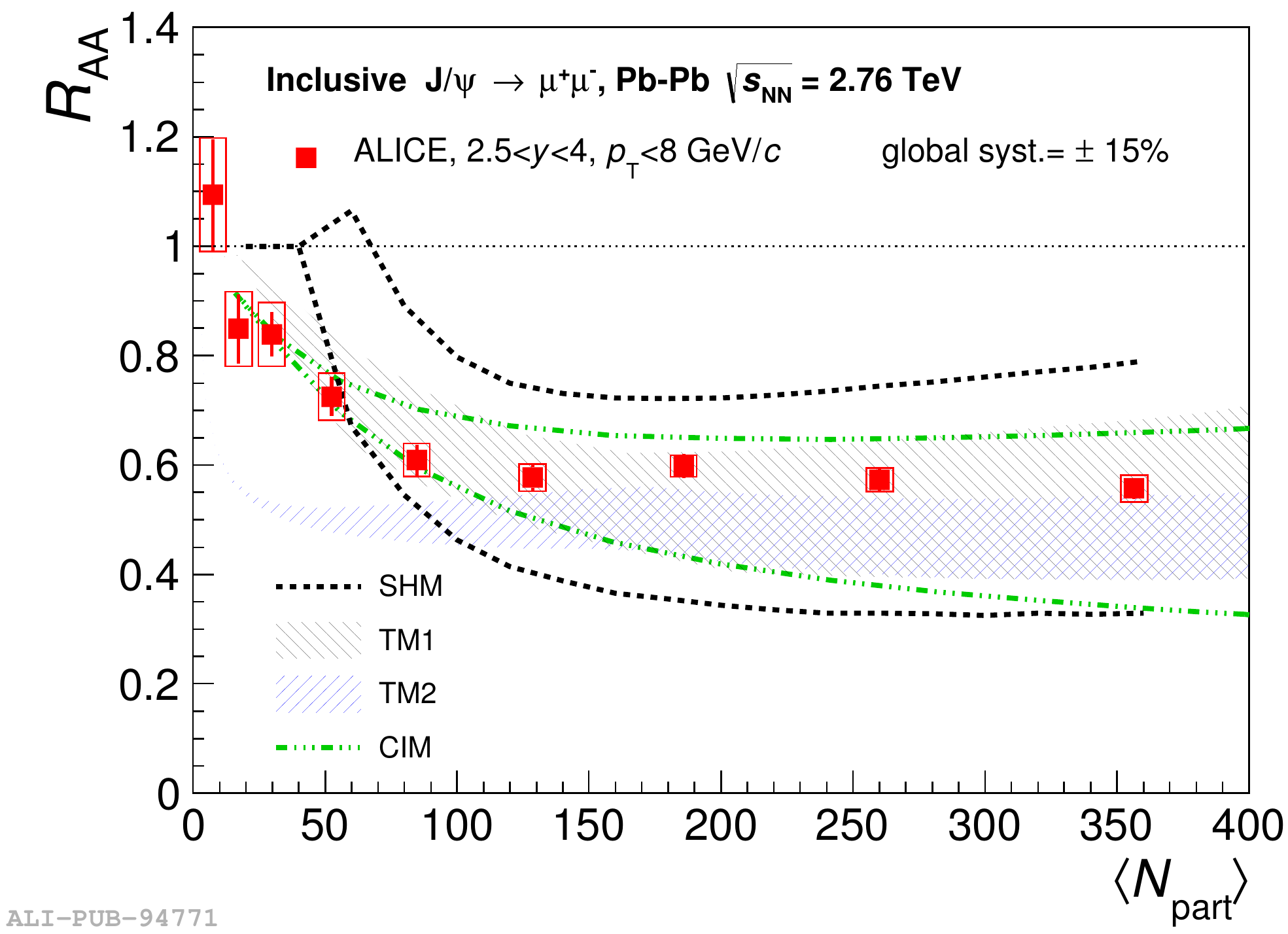}
  \includegraphics[width=0.45\textwidth]{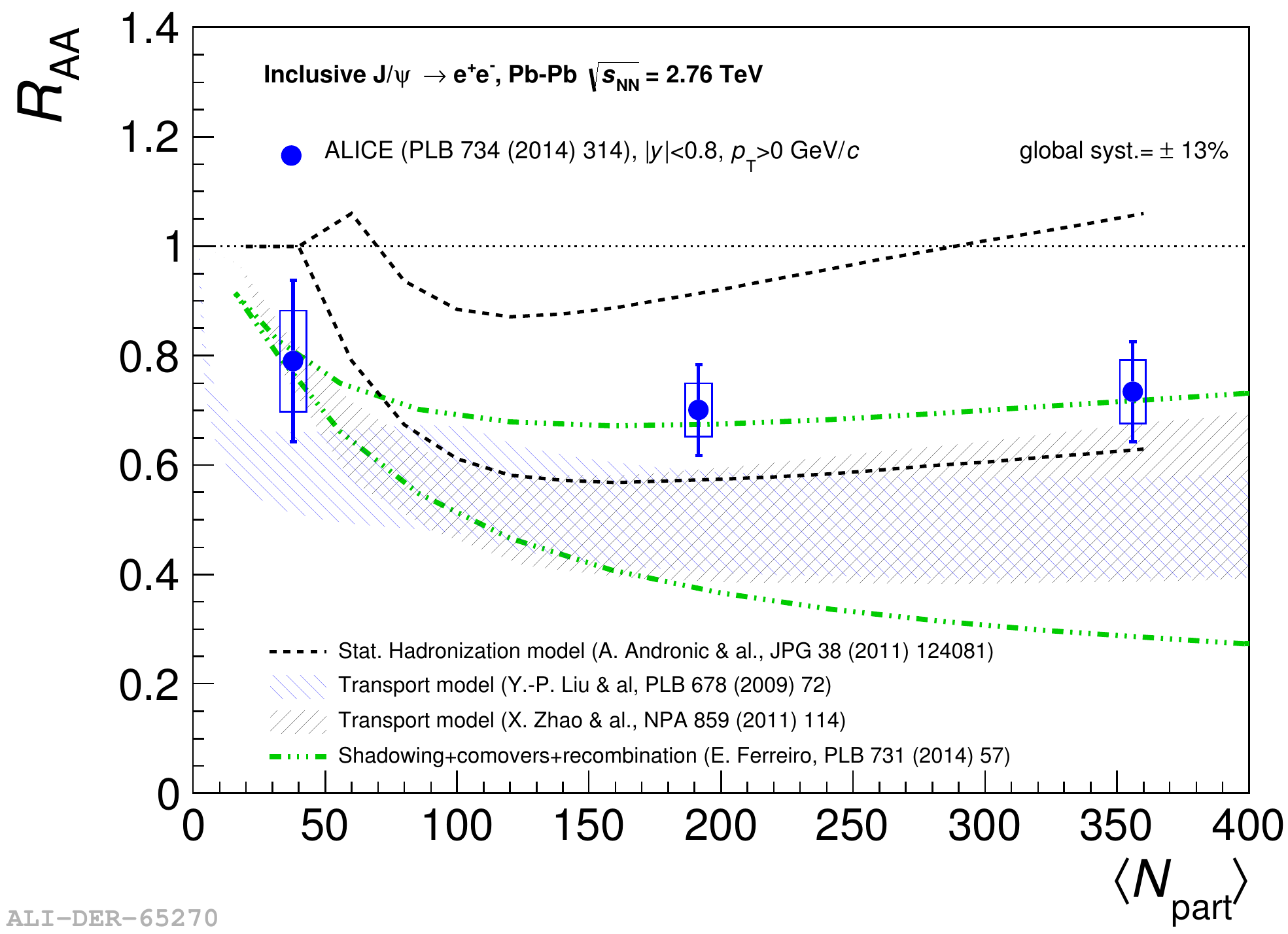}
  \caption{Comparison of the ALICE J/$\psi$ $R_{\rm{AA}}$ at forward rapidity~\cite{Adam:2015isa} (left) and mid-rapidity~\cite{Abelev:2013ila}  (right) with theory predictions from the statistical hadronization model (SHM~\cite{Andronic:2011yq}), from transport models (TM1~\cite{Liu:2009nb}, TM2~\cite{Zhao:2011cv}) and a model including shadowing, recombination and J/$\psi$ interactions with comoving hadrons (CIM~\cite{Ferreiro:2012rq}).}
  \label{fig:RAA_models_NPart}
\end{figure}

The (re)combination probability increases with increasing number of $c\overline c$ pairs in the QGP, thus with the charm production cross section. Therefore, it is expected to increase with $\sqrtsNN$ and to play an important role at LHC energies. This expectation was confirmed by the $R_{\rm{AA}}$ results obtained by the ALICE experiment for low-$p_{\rm T}$ J/$\psi$, both at forward and mid-rapidity~\cite{Adam:2015isa,Adam:2015rba,Abelev:2013ila}. The ALICE J/$\psi$ $R_{\rm{AA}}$ shows, in fact, a reduced suppression with respect to the one measured at RHIC~\cite{Adam:2015rba,Abelev:2013ila,Abelev:2012rv}, in spite of the larger energy density reached in LHC collisions. Theoretical models that include $\sim$50\% of the low-$p_{\rm T}$ J/$\psi$s produced via (re)combination are in fair agreement with the data, as shown in Fig.~\ref{fig:RAA_models_NPart}. The larger $R_{\rm AA}$ at the LHC with respect to RHIC would be due to the larger $c\overline{c}$ pair multiplicity, which compensates the suppression from colour screening in the deconfined phase. The large uncertainties associated to the theoretical calculations prevent, for the moment, a more detailed comparison with the experimental data. The dominant sources of uncertainty in the models are the total charm production cross section and the CNM effects on quarkonium production. Therefore, the 
models would strongly  benefit from a measurement of the charm production cross section down to zero $\pt$ and a precise assessment of CNM effects.

For kinematic reasons, (re)combination contributes mainly to low transverse momentum J/$\psi$, while it vanishes at high $\pt$. This is confirmed by the observation that high-$\pt$ J/$\psi$ measured by the CMS experiment show a stronger suppression with respect to the RHIC result~\cite{Chatrchyan:2012np}. 

The study of the J/$\psi$ production in p-A collisions, provided information on the role of CNM effects. As shown in Fig.~\ref{fig:CNM} (left) the J/$\psi$ nuclear modification factor $R_{\rm{pA}}$ shows a clear rapidity dependence, with a stronger suppression at forward $y$~\cite{Adam:2015iga,Abelev:2013yxa}. The comparison of the J/$\psi$ $R_{\rm{pA}}$ with theoretical calculations indicates that nuclear shadowing and/or coherent energy loss~\cite{Abelev:2013yxa,Adam:2015iga} are the main CNM effects at LHC energies.
Therefore, assuming that these effects on the two colliding nuclei in Pb--Pb collisions can be factorised, an estimate of CNM effects in Pb--Pb collisions can be carried out.  
This extrapolation is shown in Fig.~\ref{fig:CNM} (right), where it is compared with the J/$\psi$ $R_{\rm{AA}}$.
At high $\pt$ CNM effects do not account for the charmonium suppression observed  in \mbox{Pb--Pb} collisions, thus confirming that hot matter mechanisms are taking place in \mbox{Pb--Pb} collisions at $\sqrtsNN$=2.76~TeV~\cite{Adam:2015iga}. At $\pt$ close to zero, under these assumptions, the J/$\psi$ $R_{\rm{AA}}$, corrected for the CNM extrapolation, would be closer or even larger than unity, consistent with the presence of a contribution related to the (re)combination of $c\overline{c}$ pairs.

\begin{figure}[t]
  \centering
  \includegraphics[width=0.45\textwidth]{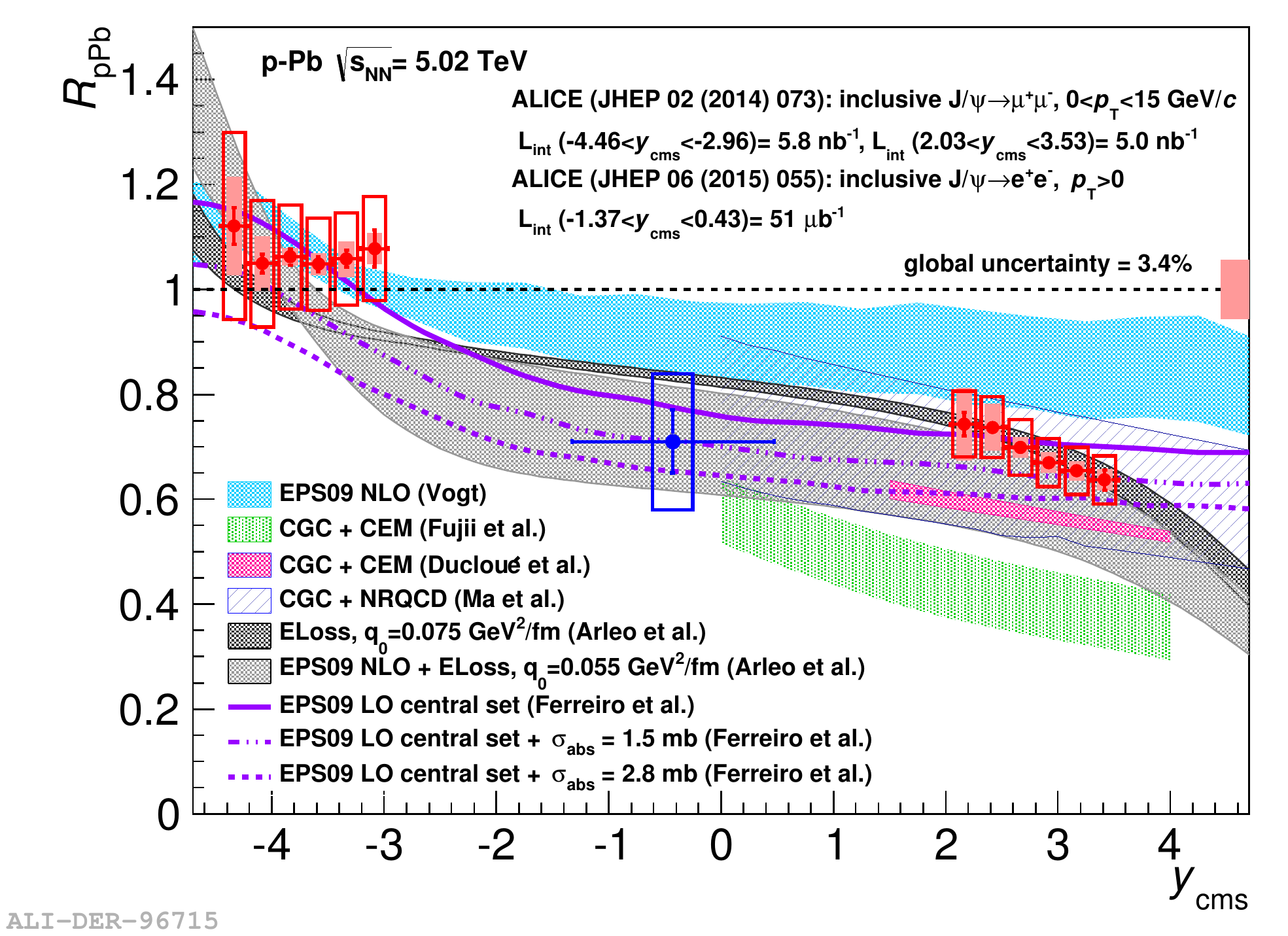}
  \includegraphics[width=0.45\textwidth]{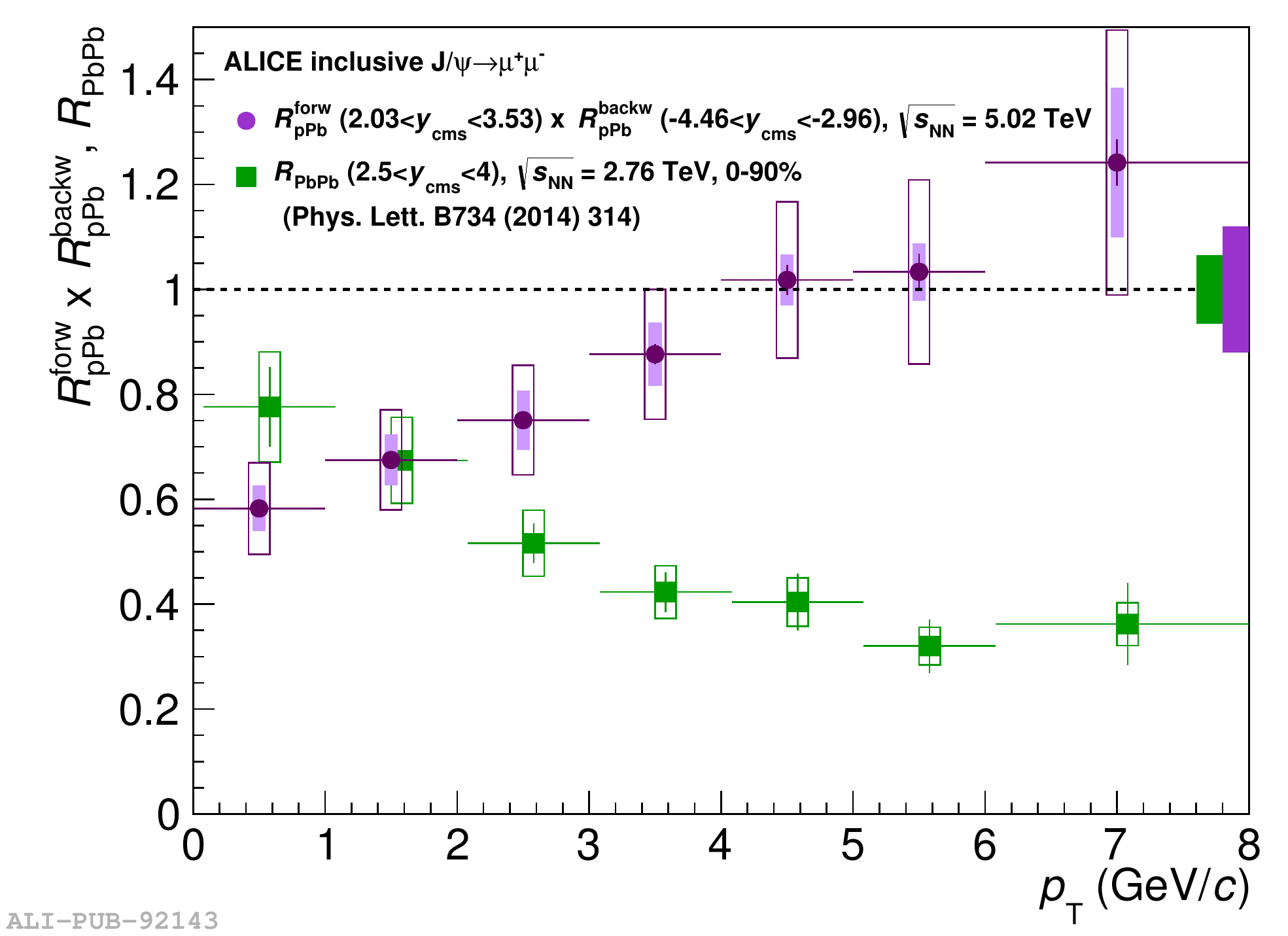}
  \caption{Left: J/$\psi$ nuclear modification factor $R_{\rm{pA}}$ as a function of rapidity compared to theoretical calculations (see~\cite{Abelev:2013yxa} and references therein). Right: Estimate of cold nuclear matter effects in \mbox{Pb--Pb}, computed as  $R_{\rm{pA}} \times R_{\rm{Ap}}$ (violet points). These CNM effects are compared with the J/$\psi$ $R_{\rm{AA}}$ (green points) measured in \mbox{Pb--Pb} collisions in a similar $y$ range~\cite{Adam:2015iga}. }
  \label{fig:CNM}
\end{figure}

The information provided by the J/$\psi$ $R_{\rm{AA}}$ can be complemented by the study of the quarkonium azimuthal distribution with respect to the reaction plane, defined by the beam axis and the impact parameter vector of the colliding nuclei. The second coefficient of the Fourier expansion describing the final state particle azimuthal distribution, $v_{2}$, is denoted elliptic flow. J/$\psi$ produced through a (re)combination mechanism, should inherit the elliptic flow of the charm quarks in the QGP, acquiring, therefore, a positive $v_{2}$ (see Section~\ref{sec:ohf}). With the Run-1 data ALICE measured a positive $v_{2}$ with a significance of 2.7$\sigma$~\cite{ALICE:2013xna}.

The measurement of the production of the excited charmonium state $\psi$(2S) with respect to the ground state, in \mbox{A--A} collisions, is an interesting tool to disentangle the mechanisms at play, but it still represents a challenging measurement. The limited $\psi$(2S) statistics, collected during LHC Run-1,  is not enough to address the sequential suppression picture in the charmonium sector. First results from the CMS~\cite{Khachatryan:2014bva} and ALICE~\cite{Adam:2015iga} experiments in \mbox{Pb--Pb} collisions show a strong dependence on $\pt$ and $y$, but the experimental precision does not allow for a clear conclusion on the $\psi$(2S) behaviour, as shown in Fig.~\ref{fig:PsiP_Upsilon} (left). 

\begin{figure}[t]
  \centering
  \includegraphics[width=0.44\textwidth]{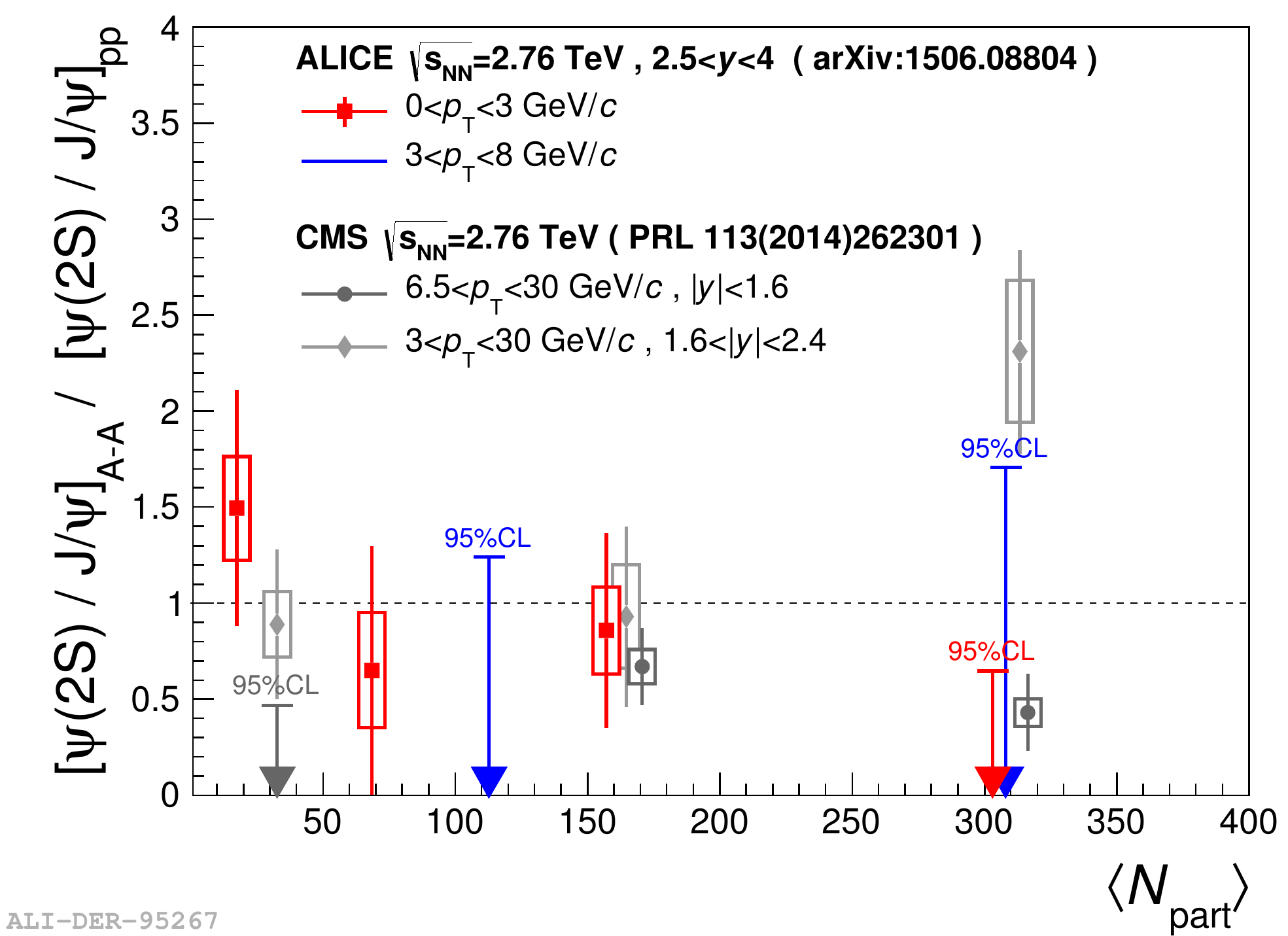}
  \includegraphics[width=0.34\textwidth]{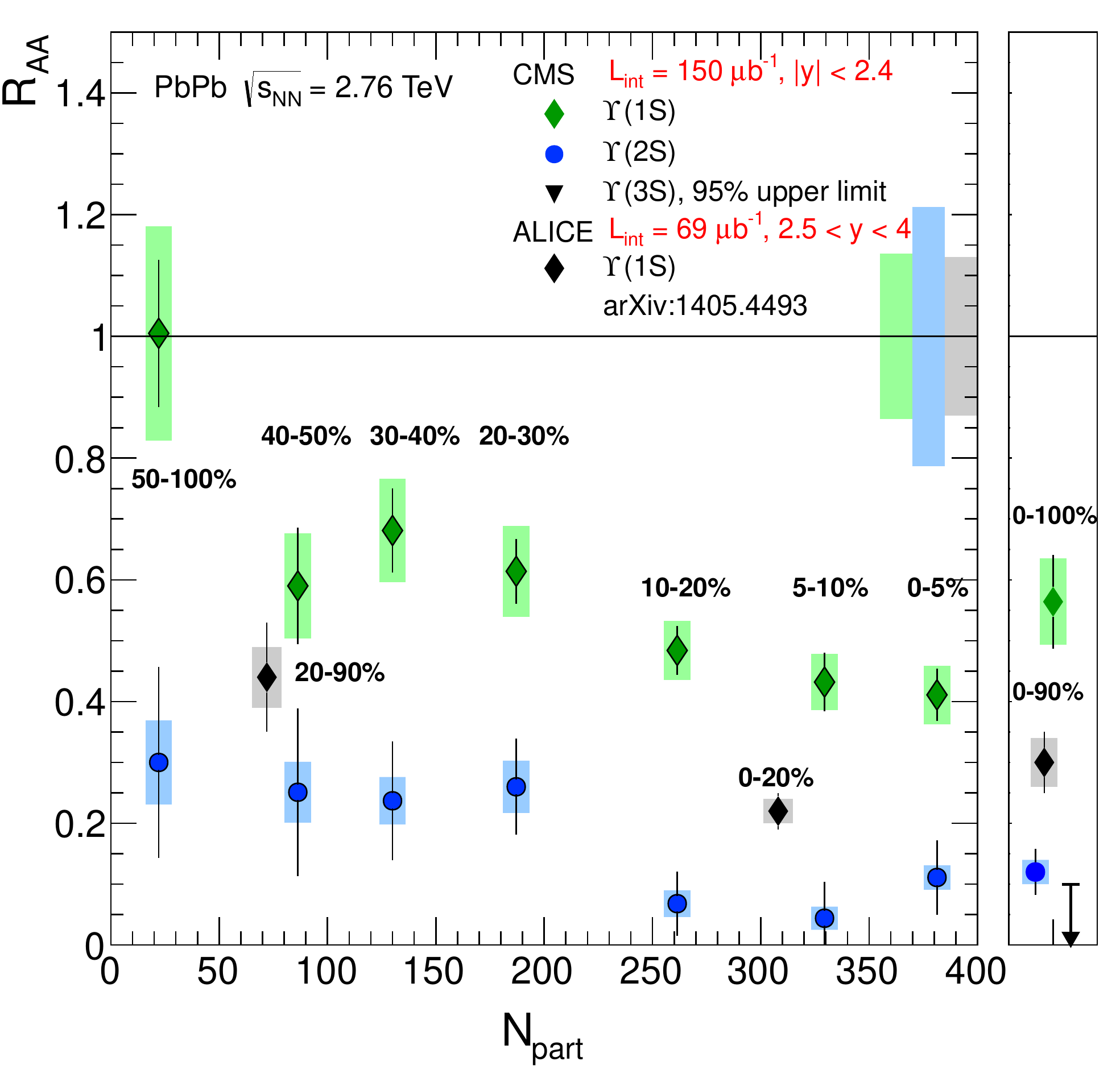}
  \caption{Left: $[\psi(2{\rm S})/{\rm J}/\psi]_{\rm PbPb}/[\psi(2{\rm S})/{\rm J}/\psi]_{\rm pp}$ ratio as a function of centrality as measured by the CMS~\cite{Khachatryan:2014bva} and ALICE~\cite{Adam:2015iga} experiments in \mbox{Pb--Pb} collisions at $\sqrtsNN=2.76$~TeV. Right: $\Upsilon$ $R_{\rm{AA}}$ as a function of centrality measured by CMS in
   $|y|<2.4$~\cite{Chatrchyan:2011pe,Chatrchyan:2012lxa} and ALICE in
    $2.5<y<4$~\cite{Abelev:2014nua} in \mbox{Pb--Pb} collisions at $\sqrtsNN=2.76$~TeV.}
  \label{fig:PsiP_Upsilon}
\end{figure}

At LHC energies the large production cross section of $b\overline b$ pairs enabled for the first time a study of the three bottomonium states $\Upsilon$(1S), $\Upsilon$(2S)  and $\Upsilon$(3S). This represents a cleaner tool to investigate the sequential suppression mechanism in the QGP, because 
bottomonium states are essentially not affected by (re)combination effects, given that the total number of $b\overline b$ pairs in the QGP at LHC energies is 
of only a few units. 
As shown in Fig.~\ref{fig:PsiP_Upsilon} (right), an order in the suppression of the $\Upsilon$ states, as predicted by the sequential dissociation picture, is clearly visible~\cite{Abelev:2014nua,Chatrchyan:2011pe,Chatrchyan:2012lxa}.
While the $\Upsilon$(2S) and $\Upsilon$(3S) are almost completely suppressed in central \mbox{Pb--Pb} collisions, a more precise knowledge of the feed-down contributions, from the higher resonances into the $\Upsilon$(1S), is required to assess whether directly produced $\Upsilon$(1S) are suppressed.

Since the $\Upsilon$ is less sensitive to recombination,  its dissociation pattern
might be more directly related with the spectral functions at
equilibrium which can be computed on the lattice~\cite{Andronic:2015wma}, as discussed also in Section~\ref{sec:lQCD}.
The bottomonium spectral functions are relatively easy to compute
on the lattice, since the non-relativistic QCD approximation, which
simplifies the  numerical analysis, is valid for beauty quarks.
The suppression pattern observed in experiments is
indeed compatible with the behaviour of spectral functions, see Fig.~\ref{fig:u}
(left) \cite{Aarts:2014cda}.
That diagram  depicts the $\Upsilon$  spectral functions at temperatures from $0.76\,T_c$ up to 
 $1.90\,T_c$, where $T_c$ is the critical temperature for the QGP formation at $\mu_B=0$.
The ground state peak is clearly visible and persists at all accessible temperatures. Current estimates of resonance width in the lattice approach would be suggestive of a very short survival time of the $\Upsilon$(1S) and it can be speculated that this is related with the apparent suppression. However, at this stage the computed width is only an upper bound and any phenomenological implication has to be treated with great care.
The second peak which may be identified with the first excited state clearly dissolves above $T_c$, and in this case the analogy with the experimental results is straightforward.

\subsubsection{Future prospects}

With the second run of the LHC an integrated luminosity of about 1~nb$^{-1}$ for \mbox{Pb--Pb} collisions at $\sqrt{s_{\rm NN}}=5$~TeV will be delivered to the experiments, an order of magnitude higher than in the first run. This increase in statistics will certainly allow for a step forward in the understanding of quarkonium production in heavy-ion collisions at LHC energies, clarifying the issues left open at the end of Run-1. 
The first physics objective will be the confirmation of the role of regeneration in the production of charmonium at LHC energies: at low $p_{\rm T}$ ALICE should be able to observe a  
dependence of the J/$\psi$ production on the energy of the collision ($\sqrt{s_{\rm NN}}=2.76$~TeV versus $\sqrt{s_{\rm NN}}=5$~TeV), which in the regeneration scenario should results in a larger 
 $R_{\rm AA}$ at the higher energy. On the contrary, at high $p_{\rm T}$ one would expect a lower $R_{\rm AA}$ at the higher energy. CMS will perform a precise measurement in the high $p_{\rm T}$ region.  
Then, the observation of a non-zero J/$\psi$ $v_2$ coefficient,   
assuming the same value for this quantity at the new energy as  the central value of the ALICE determination at $\sqrt{s_{\rm NN}}=2.76$~TeV,  
will become possible at a 5\,$\sigma$ level or better. 
Finally, at central rapidity a measurement of the $R_{\rm AA}$ of the J/$\psi$ down to $p_{\rm T}=0$, with also the possibility to separate the non-prompt component, i.e. J$/\psi$ produced from B decay, for $p_{\rm T} > 1.5$~GeV/$c$, will be 
feasible with a precision of about 10\% for the $p_{\rm T}$-integrated  $R_{\rm AA}$, and also as a function of $p_{\rm T}$.  

With the increase of integrated luminosity in Run-2, new and more precise results are expected for the very rare signals. 
Quantities that are presently measured with large statistical uncertainties or integrating over transverse momentum,   
rapidity 
or centrality, will be studied differentially with respect to these variables. 
For instance, the $\psi{\rm (2S)} / {\rm J}/\psi$  ratio should be measured with about 20\% statistical uncertainty in central Pb--Pb collisions in a few $p_{\rm T}$ bins at forward rapidity, assuming a similar value as for pp collisions. 
As a second example, the study of the $\Upsilon$(2S)/$\Upsilon$(1S) ratio versus $p_{\rm T}$, which is expected to be very sensitive to the quarkonium dissociation mechanism, will be studied with a 
much finer binning in $p_{\rm T}$ with respect to the one adopted for Run-1 results~\cite{Krouppa:2015yoa}.
Other differential studies of the $\Upsilon$ family, in particular the elliptic flow, should also become feasible for the first time. 
  
However, it is with the high luminosity phase of the LHC heavy-ion programme  after LS2 
 that quarkonium studies will enter in the precision era.  
%
As an example, ALICE will be able to determine precisely the elliptic flow of the prompt J/$\psi$ (as well as that 
of the non-prompt J/$\psi$ as discussed in the Section~\ref{sec:ohf}) as a function of $p_{\rm T}$ and $y$. 
Figure~\ref{fig:v2_Jpsi_upgrade} shows the expected statistical uncertainties on the $v_2$ coefficient as a function of the transverse momentum. 
The absolute error 
in the $p_{\rm T}$ bin 1--2~GeV/$c$ (6--7~GeV/$c$) will be 0.005 (0.008)  at forward rapidity $2.5<y<4$ and 
0.007 (0.020)  at central rapidity $|y|<0.9$.
 
\begin{figure}[t]
  \centering
  \includegraphics[width=0.4\textwidth]{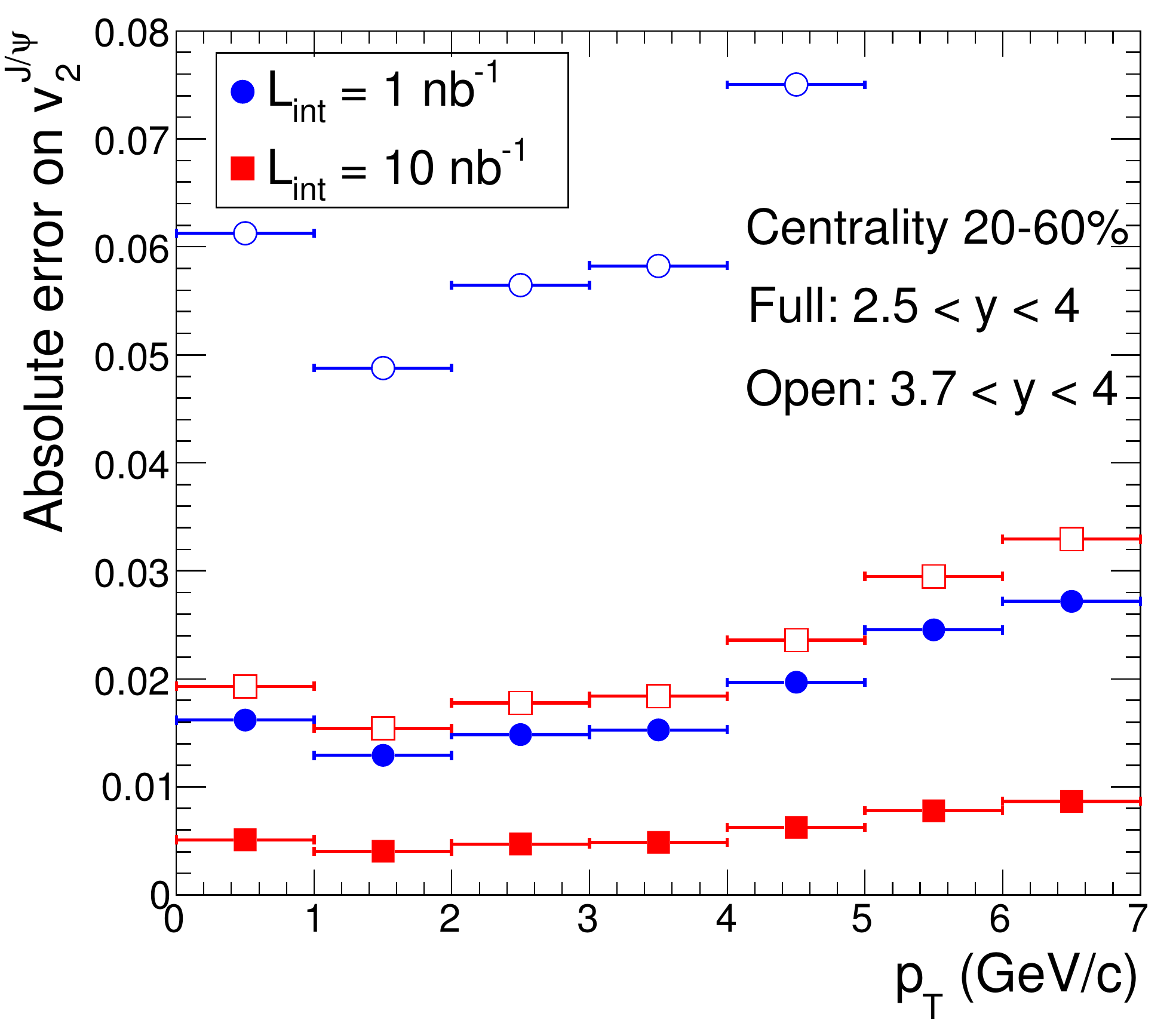}
  \includegraphics[width=0.4\textwidth]{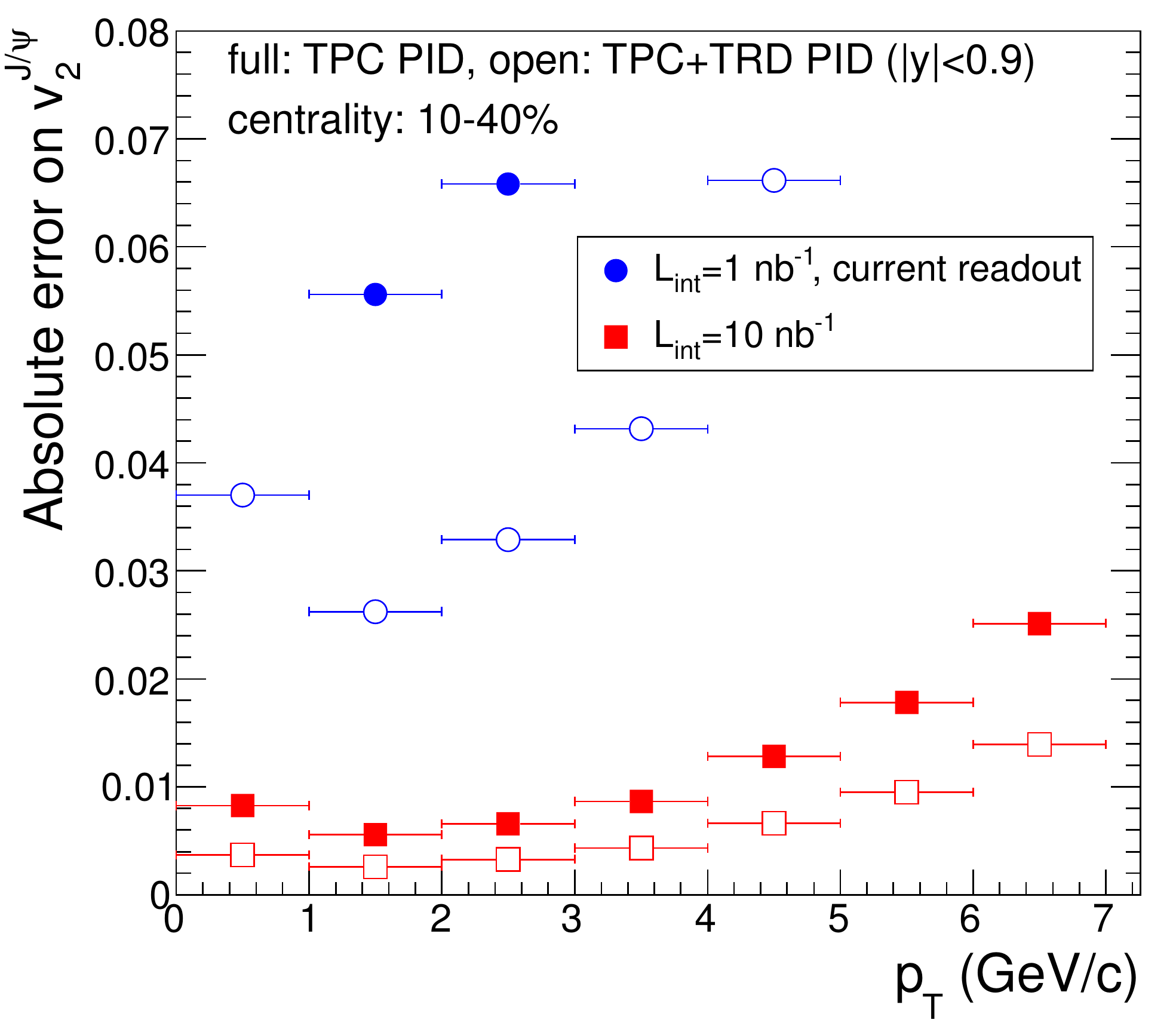}
  \caption{Projected performance for the measurement of the J/$\psi$ $v_2$ with the ALICE upgrade and an integrated luminosity of 10~nb${-1}$.
         Absolute statistical error on the $v_2$ parameter of J/$\psi$ as a function of 
           transverse momentum for the measurement at forward (left panel, centrality range 20--60\%) and 
           at central rapidity (right panel, centrality range 10--40\%)~\cite{Abelevetal:2014cna}.}
  \label{fig:v2_Jpsi_upgrade}
\end{figure}

\begin{figure}[t]
  \centering
  \includegraphics[width=0.4\textwidth]{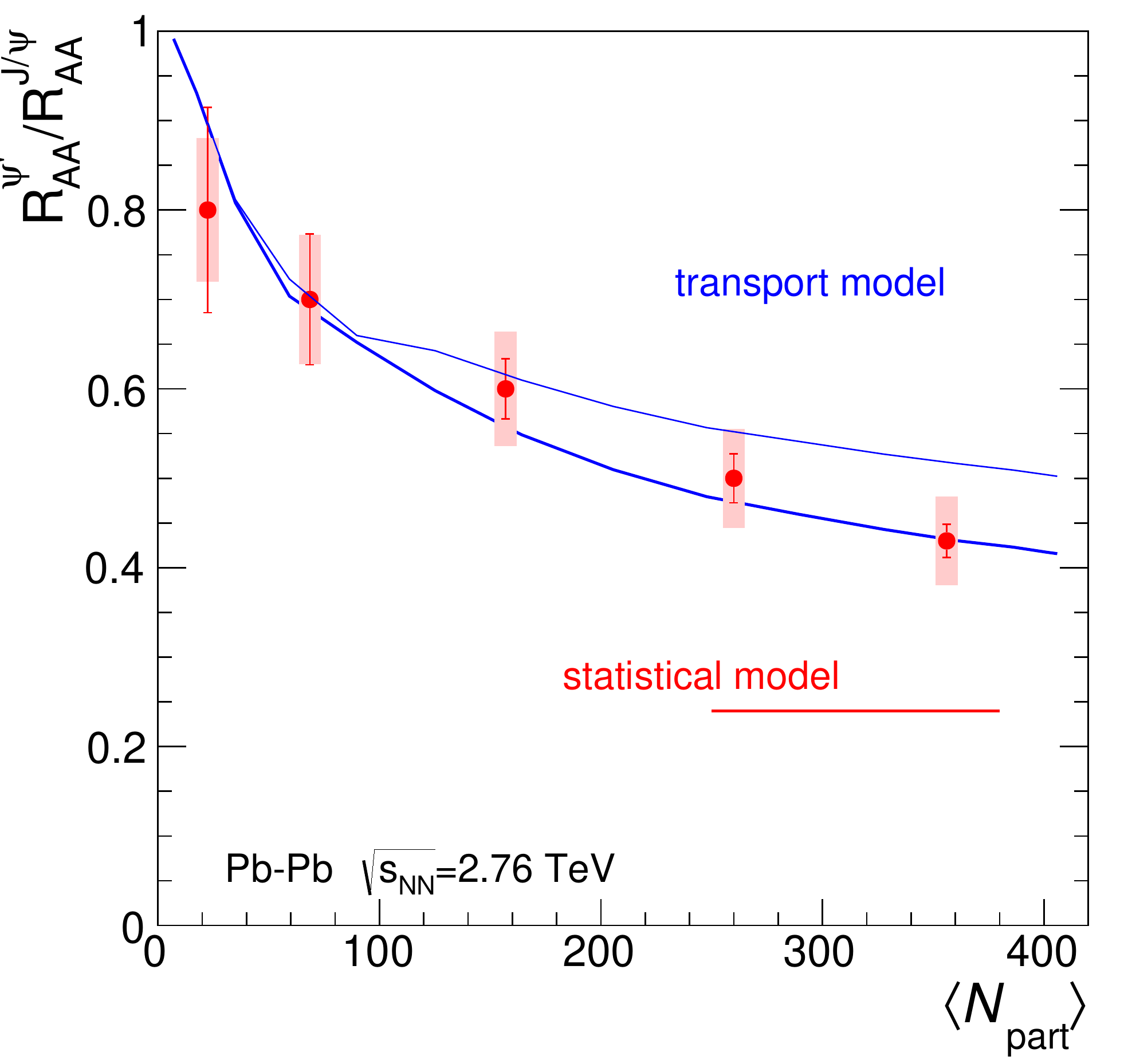}
  \caption{Projected performance for the measurement of the ratio of the nuclear modification 
           factors of $\psi$(2S) and J/$\psi$ at forward rapidity ($2.5<y<3.5$) and $\pt>0$ with the ALICE upgrade and an integrated luminosity of 10~nb$^{-1}$~\cite{CERN-LHCC-2015-001}, compared with the QGP-effect predicted by two theoretical calculations~\cite{Zhao:2011cv,Liu:2009nb,Andronic:2011yq}. 
           }
  \label{fig:DoubleRatio_upgrade}
\end{figure}

\begin{table}[!h]
\begin{center}
\caption{Projected physics performance for quarkonium studies for Run-2 ($L_{\rm int}=1$~nb$^{-1}$) and for Run-3 and Run-4 with the detector upgrade ($L_{\rm int}=10$~nb$^{-1}$). The forward rapidity region corresponds to $2.5<y<4$ in the current scenario and to $2.5<y<3.6$ in the upgrade one (because of the reduced acceptance of the MFT with respect to the Muon Spectrometer). The quoted uncertainties include both statistical and systematic contributions, unless differently specified. The uncertainties correspond either to the values reported in Refs.~\cite{Abelevetal:2014cna,CERN-LHCC-2015-001} or they were obtained by scaling those of Run-1 to the expected Run-2 luminosity. 
For the mid-rapidity inclusive J/$\psi$ $R_{\rm{AA}}$ uncertainty in Run-3 and Run-4, a reduction by a factor two of the current systematic uncertainty was assumed.
}
\small
\begin{tabular}{lcccc}
\hline
    & \multicolumn{2}{c}{Current, $1~{\rm nb}^{-1}$ } & \multicolumn{2}{c}{Upgrade, $10~{\rm nb}^{-1}$} \\
     \cmidrule{2-3} \cmidrule{4-5}
Observable      & $p_{\rm T}^{\rm min}$  &  Uncertainty     & $p_{\rm T}^{\rm min}$   &  Uncertainty \\
      & (GeV/$c$)   &       & (GeV/$c$)  &  \\
\hline
     \multicolumn{5}{c}{mid-rapidity (|$y|<0.9$)} \\
\hline    
inclusive J/$\psi$ $R_{\rm{AA}}$   & 0 & 10\% at 1~GeV/${c}$ &  0  &  5\% at 1~GeV/${c}$ \\ 
$\psi$(2S) $R_{\rm{AA}}$           & 0 & at limit &  0  &  20\% \\ 
prompt J/$\psi$ $R_{\rm{AA}}$      & 1.5 & 10\% in 1.5--4~GeV/$c$&  1  &  5\% at 1~GeV/${c}$ \\ 
J/$\psi$ from B hadrons            & 1.5 & 30\% in 1.5--4~GeV/$c$ &  1  &  10\% at 1~GeV/${c}$ \\ 
J/$\psi$ elliptic flow             & 0 & 0.05 abs. stat. err. at 1.5~GeV/${c}$&  0  &  0.007 abs. stat. err. at 1.5~GeV/${c}$\\
\hline
\multicolumn{5}{c}{forward rapidity} \\
\hline
inclusive J/$\psi$ $R_{\rm{AA}}$   & 0 & 5\% at 1~GeV/$c$ &  0  &  5\% at 1~GeV/$c$ \\ 
$\psi$(2S) $R_{\rm{AA}}$           & 0 & 30\% &  0  &  10\% \\ 
prompt J/$\psi$ $R_{\rm{AA}}$      & 0 &not accessible &  0  &  5\% at 1~GeV/${c}$ \\ 
J/$\psi$ from B hadrons            & 0 & not accessible&  0  &  6\% at 1~GeV/${c}$ \\ 
J/$\psi$ elliptic flow             & 0 & 0.015 abs. stat. err. at 1.5~GeV/${c}$& 0  &  0.005 abs. stat. err. at 1.5~GeV/${c}$\\
J/$\psi$ polarization ($\lambda_{\theta}$)  & 2 & 0.05 abs. stat. err. &  2  & 0.02 abs. stat. err.\\
J/$\psi$ polarization ($\lambda_{\varphi}$)  & 2 & 0.035 abs. stat. err. &  2  & 0.01 abs. stat. err.\\

\hline

\end{tabular}
\label{tab:onia_run3_run4}
\end{center}
\end{table}

This high-statistics data sample will allow, for the first time, to measure the J/$\psi$ polarization in \mbox{Pb--Pb} at LHC energies. This measurement will be particularly interesting since, as of today, no unique theoretical predictions on how polarization might be influenced by hot medium effects are available.
Furthermore, it has been suggested that the measurement of the J/$\psi$ and $\Upsilon$(1S) polarisation  can be a sensitive instrument to study the 
suppression of $\chi_c$ and $\chi_b$ in heavy-ion collisions, where a direct determination of signal yields 
involving the identification of low-energy photons is very difficult~\cite{Faccioli:2012kp}.  
CMS will be in an excellent position to measure the polarization of the J/$\psi$ and $\Upsilon$(1S) at high $p_{\rm T}$. 
The measurement of the J/$\psi$ polarization parameters in Pb--Pb collisions at low-$p_{\rm T}$  is a
challenging measurement that ALICE will perform with the envisaged luminosity of 10 nb$^{-1}$. It is
expected that statistical errors, e.g., on $\lambda_\theta$ --- one of the three parameters describing 
the spin state of the J/$\psi$ in a given reference frame ---  
of about 0.02 will be reached at forward rapidity (Muon
Spectrometer) for such integrated luminosities. A comparable, albeit slightly worse, accuracy is
expected at mid-rapidity (Central Barrel).

The addition of the Muon Forward Tracker~\cite{CERN-LHCC-2015-001} in the ALICE apparatus will enable a measurement of the $\psi$(2S) signal 
with uncertainties as low as  10\% down to zero $p_{\rm T}$.  
Also at central rapidity in the dielectron channel, where the measurement is more challenging, with the target integrated luminosity of 
10~nb$^{-1}$, a measurement with a precision smaller that 20\% is expected\footnote{Both at forward and mid-rapidity the most unfavourable 
scenario of the $\psi$(2S) production yield was assumed for these estimates, corresponding to the predictions from the Statistical Hadronization Model~\cite{Andronic:2011yq}.}.

The precise measurement of the $\psi$(2S),
combined with the one of prompt J/$\psi$ production, will offer an important tool to discriminate
between different models of charmonium regeneration in the QGP, as shown in Fig.~\ref{fig:DoubleRatio_upgrade}.  

ALICE has recently measured, in peripheral \mbox{Pb--Pb} collisions, an excess of the J/$\psi$ yields at very low $p_{\rm T}$, i.e. below 300~MeV/$c$~\cite{Adam:2015gba}. This excess can be interpreted as due to the coherent J/$\psi$ photo-production measured for the first time in nuclear collisions. Also in this case, increase in statistics and the capabilities of the Muon Forward Tracker will allow a more detailed study of the origin of this excess.



A summary of the physics reach for the quarkonium benchmark measurements is presented in Table~\ref{tab:onia_run3_run4} for Run-2 with present ALICE apparatus, and for Run-3 and Run-4 with the ALICE upgrade.


\subsection{Jets}
\label{sec:jets}

Jets are experimentally defined as sprays of collimated particles coming from the fragmentation and hadronization of the initial partons produced from hard scatterings in the early stages of high-energy
collisions.
Their production cross section in hadronic collisions is calculable using perturbative QCD (pQCD) and the contribution from the non-perturbative hadronization can be well calibrated via measurements in pp collisions. 
In nucleus--nucleus collisions jet fragmentation is modified relative to proton--proton collisions, 
as a consequence of the interactions of the high-$\pt$ partons with the QGP via radiative and collisional processes~\cite{Gyulassy:1990ye,Baier:1994bd}. 

The measurement of single hadrons constrains the parton kinematics very loosely. Conversely, jet reconstruction gives access to the kinematics of the original parton that produced the jet, providing insights into energy loss mechanisms and their effects on the jet structure. 
Therefore, jets are considered among the golden probes to perform QGP medium tomography.
The interaction with the medium can result in a broadening of the jet shape and a softening of the jet fragmentation~\cite{Salgado:2003rv}, leading to an increase of out-of-cone gluon radiation~\cite{Vitev:2005yg} with respect to jets reconstructed in pp collisions~\cite{Chatrchyan:2011sx}. Because of this effect, for a given initial parton energy and a jet resolution parameter $R$, the jet transverse momenta in heavy-ion collisions are expected to be smaller than those in pp collisions.

Several theoretical energy-loss models are trying to explain RHIC and LHC results on jet quenching. Most of them are based on pQCD calculations of in-medium energy loss, possibly implemented in Monte Carlo event generators, allowing the study of different jet quenching observables. Several microscopic generators (i.e. JEWEL, YaJEM, Q-PYTHIA) exploit various implementations of the parton--medium interaction to describe the evolution of parton showers~\cite{Zapp:2013vla,Renk:2008pp,Armesto:2009fj}. 


Jet reconstruction in heavy-ion collisions is challenging because the particles of the underlying event represent a large background for the definition of the jet area and energy. Dedicated jet finding algorithms and background subtraction techniques have been optimised to reconstruct all the particles resulting from the hadronization of the original hard parton (see e.g. Refs.~\cite{Cacciari:2008gp,Abelev:2012ej}). 
 At the LHC, fully-reconstructed jets measured over a wide $\pt$ range at $\sqrtsNN= 2.76$~TeV, confirm and extend the suppression pattern
observed for charged particles~\cite{Adam:2015ewa, CMS:2012rba}, as shown in the left panel of Fig.~\ref{fig:jetraa}. 
This measurement suggests that the initial parton energy is not recovered within the jet cone radius in the case of a hot and dense medium.
Additional indication can be reached by studying the jet $R_{\rm CP}$~\cite{Aad:2012vca}, as shown
in the right panel of Fig.~\ref{fig:jetraa}. The central-to-peripheral ratio $R_{\rm CP}$ is defined as the ratio of the per-event jet yields divided by the number of nucleon-nucleon collisions in a given centrality  class  to  the  same  quantity  in  a  peripheral  centrality class. For $\pt < 100$~GeV/$c$, the quantity
$R_{{\rm CP}}^R/R_{{\rm CP}}^{{R=0.2}}$, for both $R = 0.4$ and $0.5$, differs from unity beyond
the statistical and systematic uncertainties, indicating a clear jet broadening.
However, for $R \leq 0.4$ at $\pt > 100$~GeV/$c$,
the ratio is consistent with jet production in vacuum over all centralities.
This may be interpreted as an indication that the jet core remains intact
with no significant jet broadening observed within the jet cone resolution.
Other measurements~\cite{Chatrchyan:2011sx} showed that a large contribution of the radiated energy is carried by low-$\pt$ particles at large radial distance, $\Delta R > 0.8$ relative to the jet axis.

\begin{figure}[t]
\begin{center}
\includegraphics[width=0.41\textwidth]{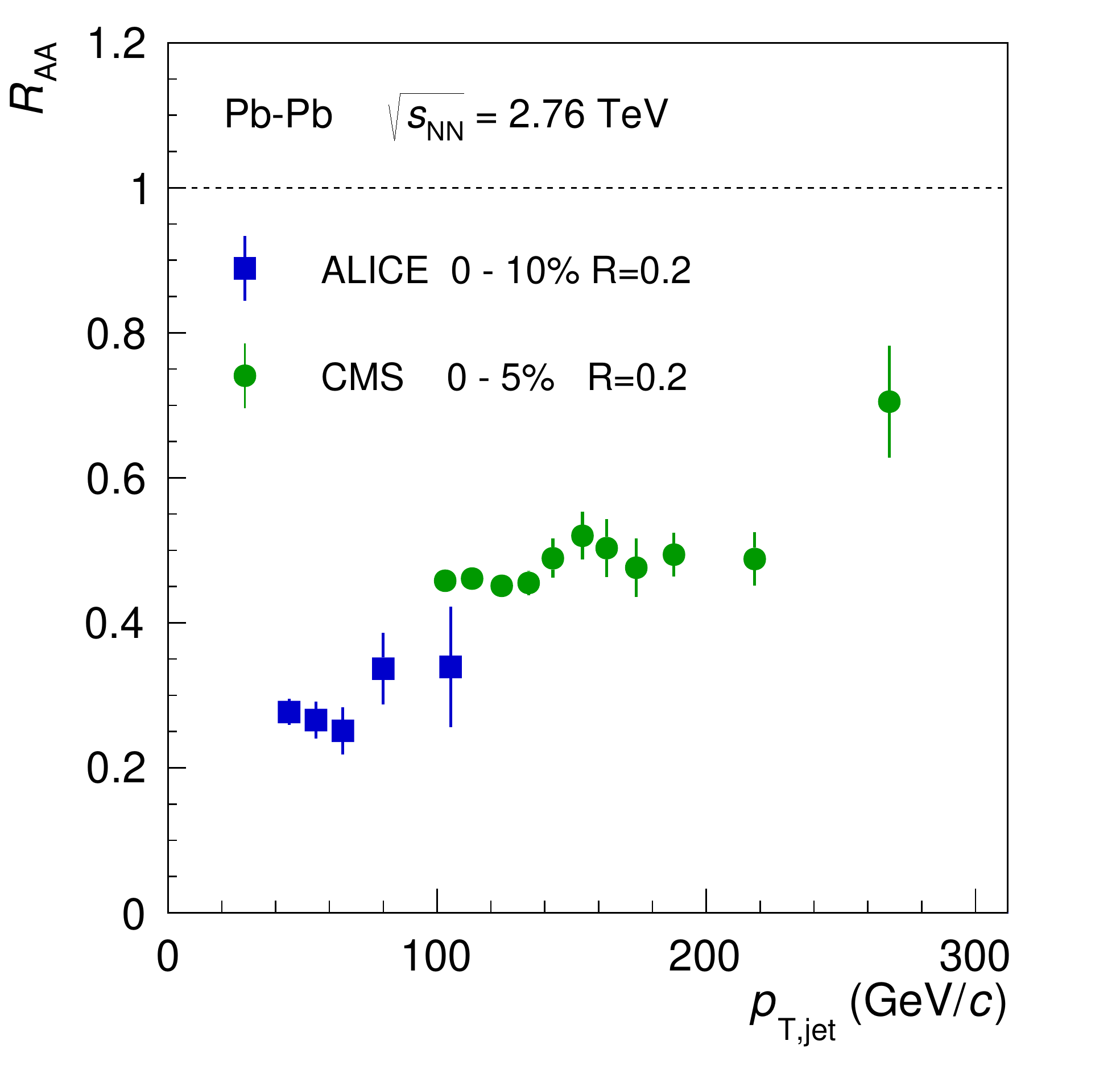}
\includegraphics[width=0.51\textwidth]{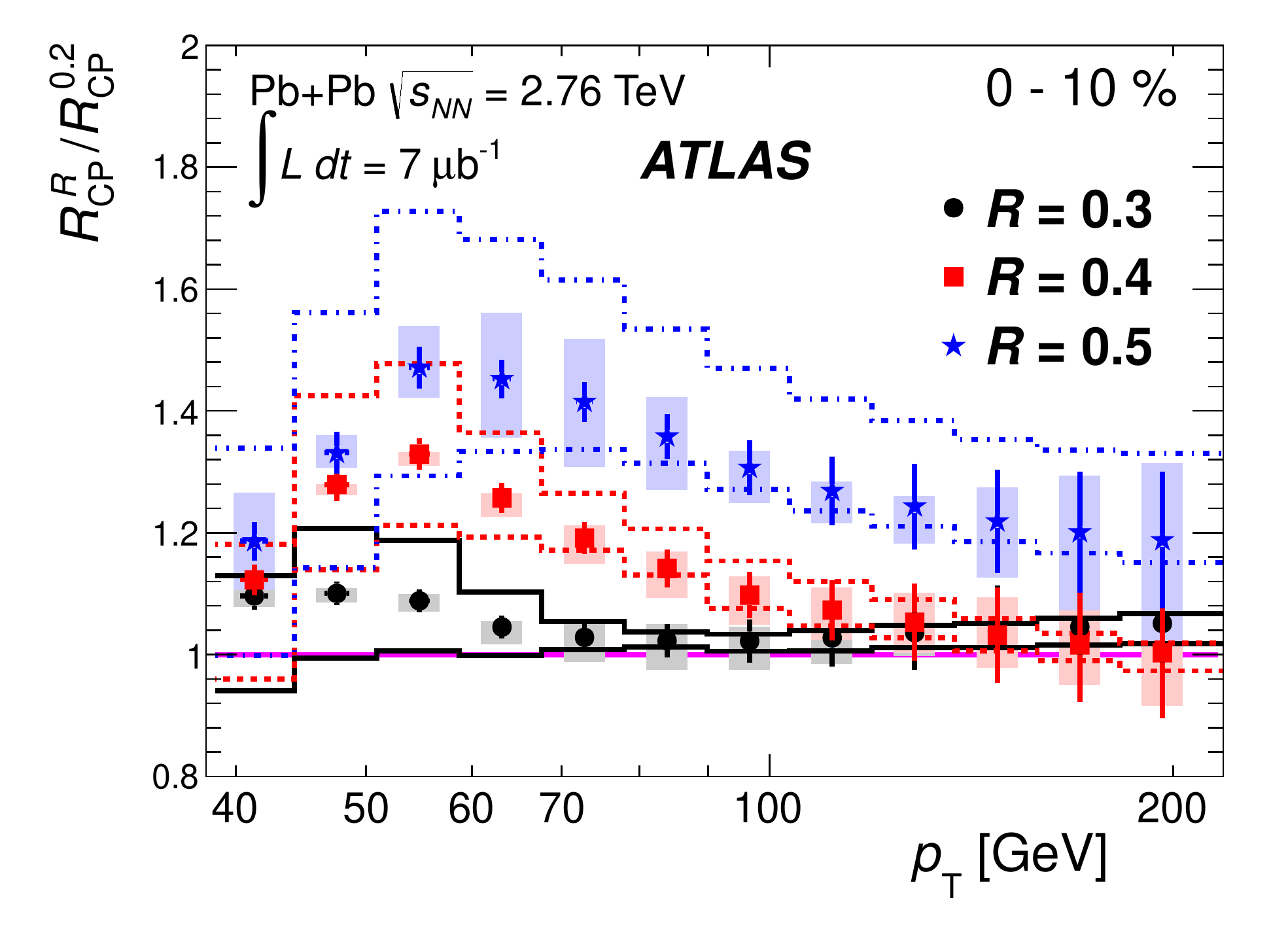}
\caption{Left: jet $R_{\rm AA}$ as a function of jet $\pt$ measured by ALICE and CMS~\cite{Adam:2015ewa, CMS:2012rba}. Right: ratios $R_{\rm CP}^R/R_{\rm CP}^{R=0.2}$ for $R=0.3$, $0.4$ and $0.5$ as a function of jet $\pt$ in the 0--10\% centrality bin. The bars show the statistical uncertainties, the lines indicate fully correlated uncertainties and the shaded boxes represent partially correlated uncertainties between different $\pt$ values~\cite{Aad:2012vca}.}
\label{fig:jetraa}
\end{center}
\end{figure}

ALICE made further steps by identifying pions, kaons and protons, providing information on the hadrochemical composition of particles in jets in pp, central and peripheral Pb--Pb collisions~\cite{Abelev:2014laa}. The nuclear modification factor $R_{\rm AA}$, reported in Fig.~\ref{fig:pid} (left), indicates that high-$\pt$ pions, kaons and protons are equally suppressed, suggesting that the chemical compositions of leading particles from jets in the medium and in vacuum are similar. These results establish strong constraints on theoretical modeling for fragmentation and energy loss mechanisms. In particular, the current data already rule out ideas in which the large energy loss leading to the suppression is associated with strong mass ordering or large fragmentation differences between baryons and mesons. In addition, ALICE measurements of $\Lambda$ and $\rm K^0_S$  in reconstructed jets in heavy-ion collisions provide important insights into the interplay of various hadronisation processes~\cite{VitKucera}.

New observables are being developed to study jet quenching in central Pb--Pb collisions based on semi-inclusive rate of jets recoiling from a high-$\pt$ charged hadron trigger. This approach enables collinear-safe jet measurements with low infrared cutoff in heavy-ion collisions~\cite{Cunqueiro:2012vga}. These measurements are directly comparable to theoretical calculations because they utilize hadrons as trigger particles, semi-inclusive jets and background suppression techniques that do not require modelisation of the underlying background.

\begin{figure}[t]
\begin{center}
\includegraphics[width=0.60\textwidth]{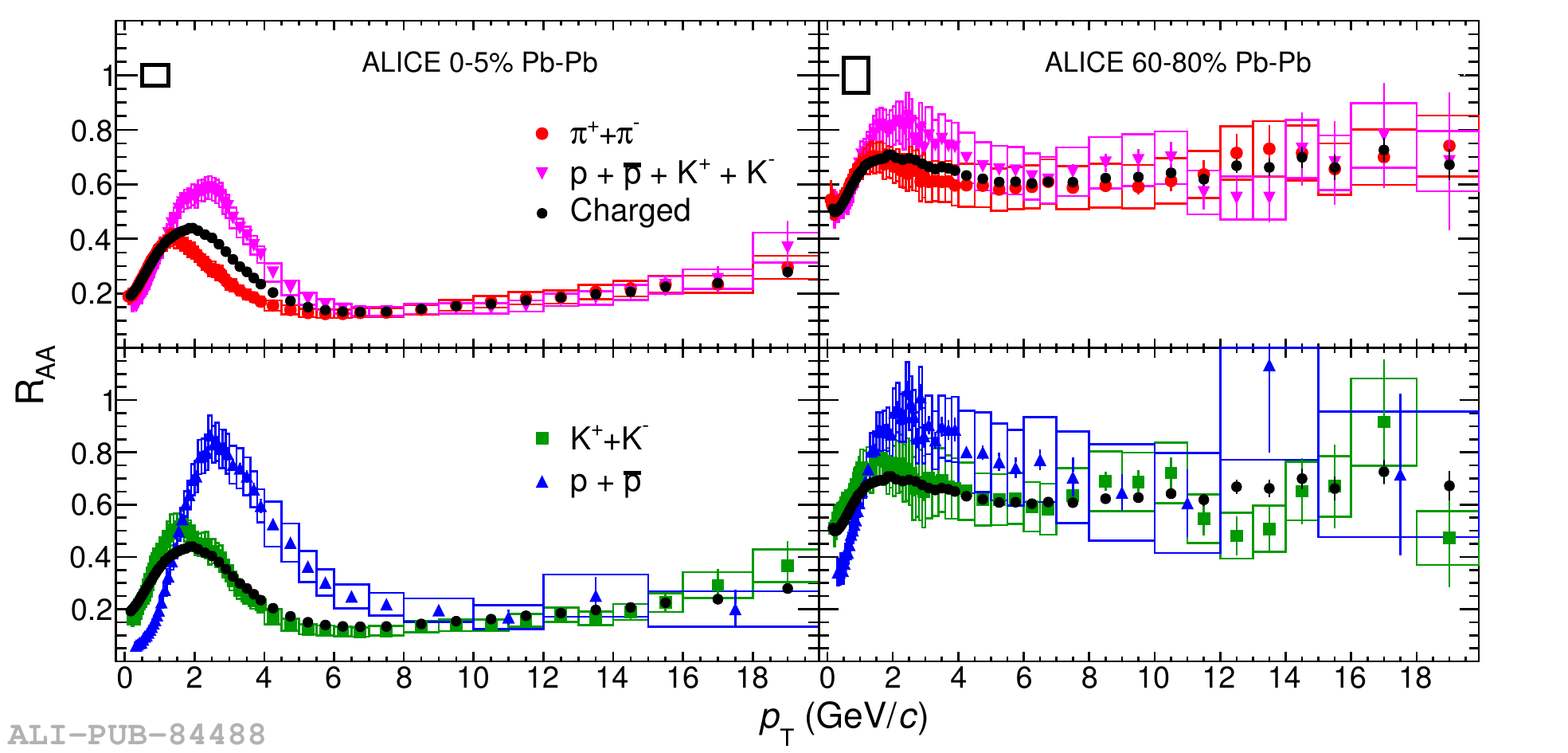}
\includegraphics[width=0.39\textwidth]{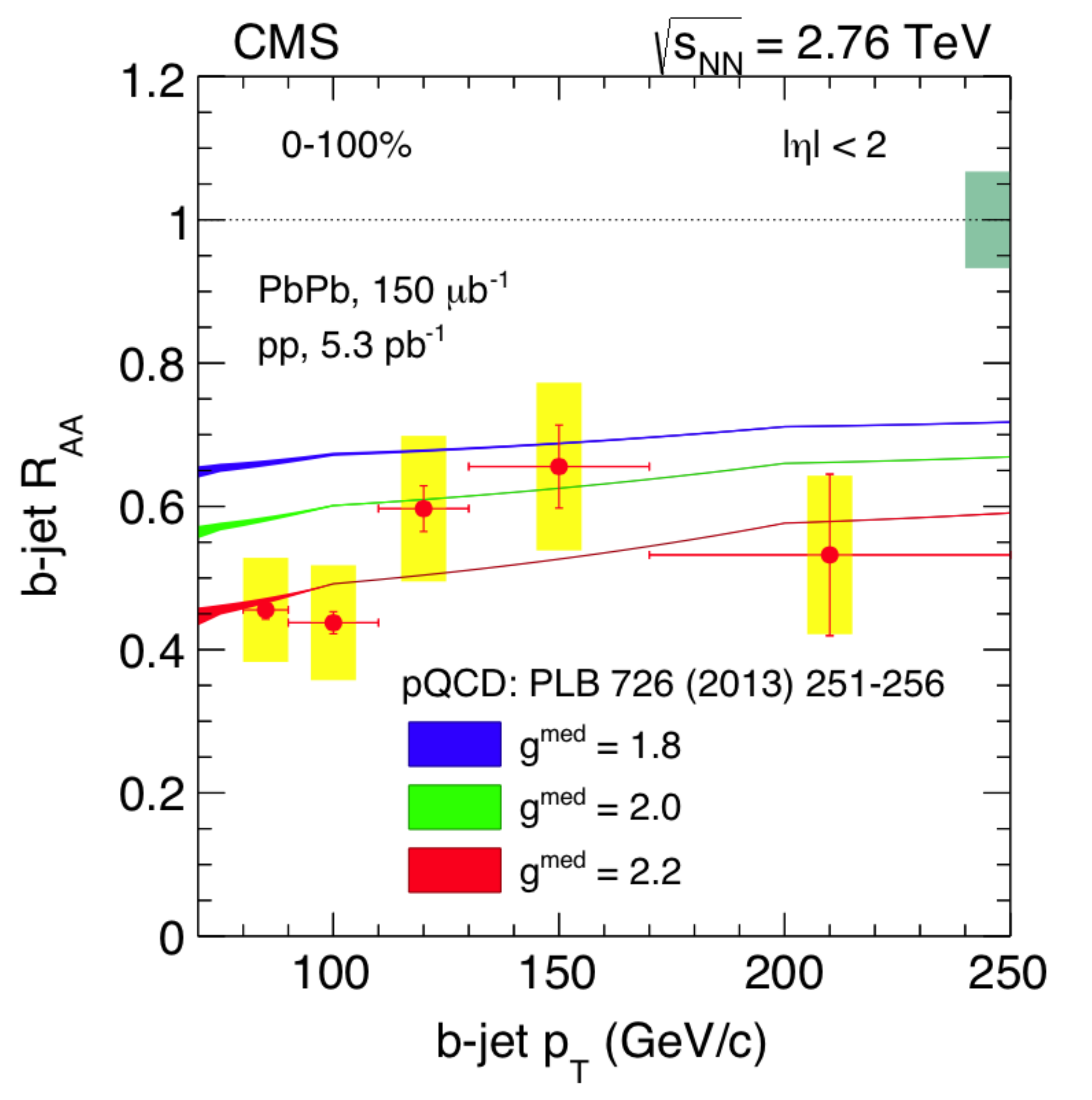}
\caption{Left: $R_{\rm AA}$ as a function of $\pt$ for different particle species, for 0--5\%
(left) and 60--80\% (right) collision centrality classes~\cite{Abelev:2014laa}. Right: $b$-jet $R_{\rm AA}$   as a function of $\pt$.  Data are compared to pQCD-based calculations~\cite{Chatrchyan:2013exa}.}
\label{fig:pid}
\end{center}
\end{figure}

The quenching of jets in heavy-ion collisions is expected to depend on the flavour of the fragmenting parton. For example, jets initiated by heavy quarks are expected to radiate less than light ones, due to the so-called dead cone effect (see Section 4.2).    
Recent data on single-particle production of B mesons (via non-prompt J/$\psi$)~\cite{Chatrchyan:2012np} show a smaller suppression compared to D mesons~\cite{ALICE:2012ab}.
These results provide a strong motivation to perform a measurement using fully-reconstructed jets, enabling a direct comparison of heavy-flavour energy loss to that of light flavours. In particular, it will be interesting to have such comparison in a wide jet $\pt$ range, where the mass-dependence of the energy loss can be probed in detail.
From an experimental point of view, jets formed from heavy-quark fragmentation can be tagged by the presence of displaced vertices, either by direct reconstruction of these vertices or by the impact parameter of tracks originating from secondary vertices~\cite{CMS:2012gik}. 
The production of jets associated to $b$ quarks was measured for the first time in heavy-ion collisions by CMS using fully-reconstructed jets~\cite{Chatrchyan:2013exa}.                                          
The $b$-jet suppression, observed in the $R_{\rm AA}$ as a function of $\pt$ (Fig.~\ref{fig:pid}, right), is found to be qualitatively consistent with that of inclusive jets~\cite{Aad:2012vca}. Although a sizeable fraction of $b$-tagged jets comes from gluon splitting, a large mass and/or flavour dependence for parton energy loss can be excluded for jets with $\pt > 80$~GeV/$c$. Models based on strong coupling (via the AdS/CFT correspondence)~\cite{Horowitz:2007su}, in which mass effects could persist to large $\pt$, would be incompatible with the current data, in contrast to a perturbative model in which mass effects are expected to be small at large $\pt$~\cite{Huang:2013vaa}.     

The advanced background-subtraction techniques developed for inclusive jet measurements in ALICE will be exploited to perform measurements of heavy-flavour tagged jets down to lower transverse momenta. Ongoing studies suggest that, with the expected luminosity of the LHC Run-2, $b$-jets could be reconstructed with ALICE in central Pb--Pb collisions in the transverse momentum range of 30--100~GeV/$c$~\cite{yasserQM2015}. These measurements will open the possibility to study the mass effect
in the energy loss of heavy and light quarks with jet observables at lower momenta, extending the
current $b$-jet cross section measurements~\cite{Chatrchyan:2013exa} that are restricted to the high end of the jet energy spectrum
(80--100~GeV/$c$), where the effect of the quark mass becomes negligible. 


Besides jet cross section observables described above for inclusive and heavy-flavour jets, more differential observables will be needed to quantitatively constrain energy loss mechanisms.
It is possible to explore jet substructure or shapes in order to probe different aspects of the jet fragmentation. Jet shapes are intra-jet observables like fragmentation functions~\cite{Chatrchyan:2014ava}, angularities and planar flow~\cite{angularities}. The idea is to consider well-defined jet shapes, i.e. calculable in pQCD without uncontrolled bias,  and measure them in heavy-ion collisions.
This kind of study is very promising to exploit different fragmentation patterns of quarks and gluons to investigate the colour-charge dependence of the energy loss by separating jets
originating from quarks and gluons.
 Using new jet shape background-subtraction methods together with unfolding techniques, the aim is to correct those shapes to particle level in order to measure different aspects of intrajet modifications relative to vacuum fragmentation. With the large sample expected in Run-3, it will be possible not only to clarify issues such as the vacuum-like fragmentation of jets, but also issues connected to the differences between parton and jet momentum and the relative logarithmic resummation terms. 

It was observed that prompt photons, produced directly in hard sub-processes, as well as vector bosons, do not strongly interact with the QGP medium~\cite{Chatrchyan:2012vq}. 
At leading order these photons are produced back-to-back with an associated parton (i.e. jet), thus having, with a good approximation, the same initial transverse momentum. Therefore, the production of a jet with an associated photon back-to-back in azimuth is considered as the golden channel to investigate energy loss of partons in the medium~\cite{Chatrchyan:2012gt}. With the expected statistics for the ALICE upgrades, it will be possible to characterise modifications of jet properties as a function of centrality using isolated-photon + jet events in Pb--Pb collisions exploiting the ALICE EMCal and DCal calorimeters.

The mass of a jet, as measured by jet reconstruction algorithms, constrains its virtuality, which in turn has a considerable effect on observables like fragmentation functions and jet shapes~\cite{jetmass}. The leading parton, propagating through a dense medium, experiences substantial virtuality (or mass) degradation along with energy loss. Having access to the virtuality evolution
via jet mass measurements adds a new, not yet experimentally accessed, dimension to jet quenching measurements by constraining both of the relevant quantities, energy and virtuality, and is expected to provide tests for models of in-medium shower evolution.


New differential measurements in the heavy-flavour sector, i.e. angular correlations of $b$-jets and $c$-jets with hadrons and of $b$-dijets ($b$-jet--$\overline b$-jet) in Pb--Pb collisions, will constitute a new terrain to study the redistribution of the jet energy after the interaction with the medium and to compare the results for light and heavy quarks. Such measurements for heavy-flavour jet observables are expected to carry information on the contribution of radiative and collisional energy loss, providing new tools to study the medium response to heavy quarks. 

The very large jet samples that will be available with the ALICE and LHC upgrades will provide deeper insight into the mechanisms of jet-shower in-medium modification and of energy flow from the jet to the medium~\cite{Qin:2009uh,Renk:2013pua}.
ALICE will carry out complementary measurements with respect to the ATLAS and CMS experiments, in particular by extending the jet spectrum to lower momenta, 
also for heavy-flavour jets using the high-precision tracking system,
and by performing jet hadrochemistry studies via charged-hadron identification over a wide momentum range.


\subsection{Low-mass dileptons}
\label{sec:dilept}

Electromagnetic radiation is produced at all stages of heavy-ion
collisions and, since photons and electrons (or muons) do not interact
strongly with the surrounding medium, their kinematic spectrum retains information of the entire system
evolution. The fundamental questions that can be addressed by a comprehensive measurement of
thermal dileptons in heavy-ion collisions at the energies of the RHIC and LHC colliders are the following:
\begin{itemize}
\item The generation of hadron masses, which is driven by the spontaneous breaking of QCD
chiral symmetry in the vacuum. Chiral restoration leads to substantial modications of
the vector and axial-vector spectral functions. Such modications, in particular of the $\rho$
meson, can be inferred from low-mass dilepton spectra. The theoretical
interpretation of the NA60 measurements at SPS described in Section~\ref{sec:EMtheory}
suggests that the $\rho$ spectral function 
strongly broadens, melting around the phase transition region. This broadening is driven by the total 
baryon density. 
At collider energies,
this  can be verified in an environment where the net-baryon density
(baryons $-$ anti-baryons) is much smaller
than at SPS energies, while the total baryon density is still large due to the significant
number of anti-baryons in the central rapidity region.
\item The temperature of the emitting medium. The invariant mass of
  thermal dileptons is not subject to blue-shift in collectively expanding systems and therefore is
most directly related to the source temperature. The study of low-mass dileptons allows the real direct photon 
production to be estimated.
\item The space-time evolution of the system. The QGP fireball lifetime can be extracted from low-mass
dilepton measurements. The potential to disentangle early from late contributions
gives access to the evolution of collectivity and, thus, to fundamental properties such as
transport coeffcients, viscosity, and the equation of state.
\end{itemize}

The PHENIX~\cite{Adare:2015ila} and
STAR~\cite{Adamczyk:2013caa,Adamczyk:2015lme} experiments at RHIC
measured $e^+e^-$ production in  Au--Au collisions at
$\sqrtsNN=200$~GeV. The dielectron spectrum measured in minimum-bias
collisions is shown in Fig.~\ref{fig:star-dielectrons}-left (STAR) and
right (PHENIX). The spectra are compared to the estimate of thermal
radiation from the QGP and hadronic phases of Ref.~\cite{Rapp:2000pe},
which successfully describe the SPS data. 
The data are both consistent with the calculation, confirming the
strong broadening of the $\rho$ spectral function.

\begin{figure*}[t]
\begin{center}
\includegraphics[width=13.cm]{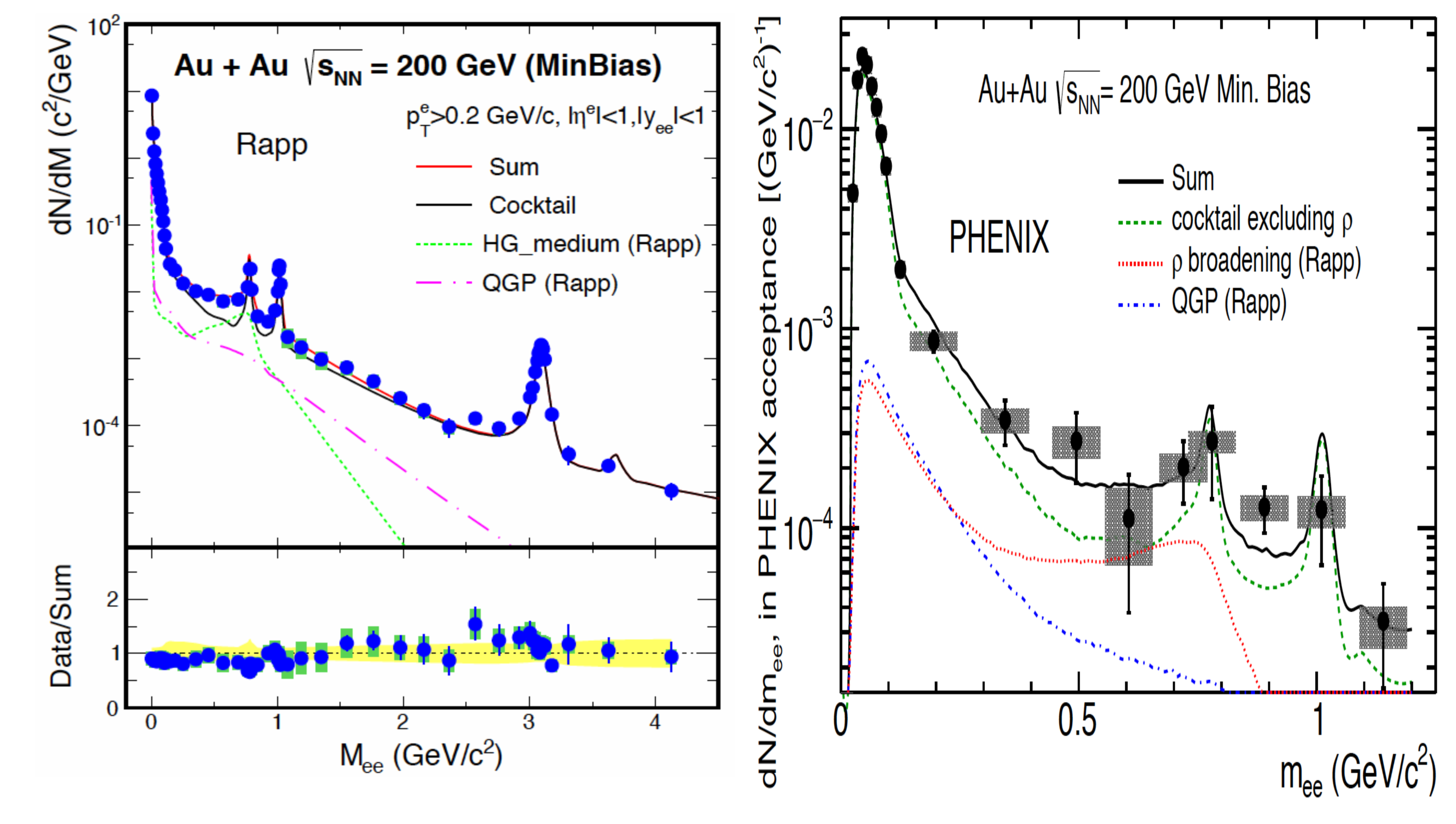}
\caption{Left: STAR dielectron invariant-mass spectrum in minimum-bias Au--Au
  collisions at $\sqrtsNN=200$~GeV~\cite{Adamczyk:2013caa,Adamczyk:2015lme}  compared to the
hadron cocktail plus the hadronic medium and partonic QGP
contributions calculated in Ref.~\cite{Rapp:2000pe} (upper left panel). Yellow bands in the bottom panels depict systematic uncertainties on the
cocktail. Right: PHENIX dielectron mass spectrum in minimum-bias
Au--Au collisions at the same energy~\cite{Adare:2015ila}  compared to the same model calculations.
The main contributions, the in-medium  broadening (dotted line), the QGP thermal radiation (dot-dashed line) and
the hadron cocktail excluding the (dashed line) are also shown.}
\label{fig:star-dielectrons}
\end{center}
\end{figure*}

\begin{figure*}[t]
\begin{center}
\includegraphics[width=0.49\textwidth]{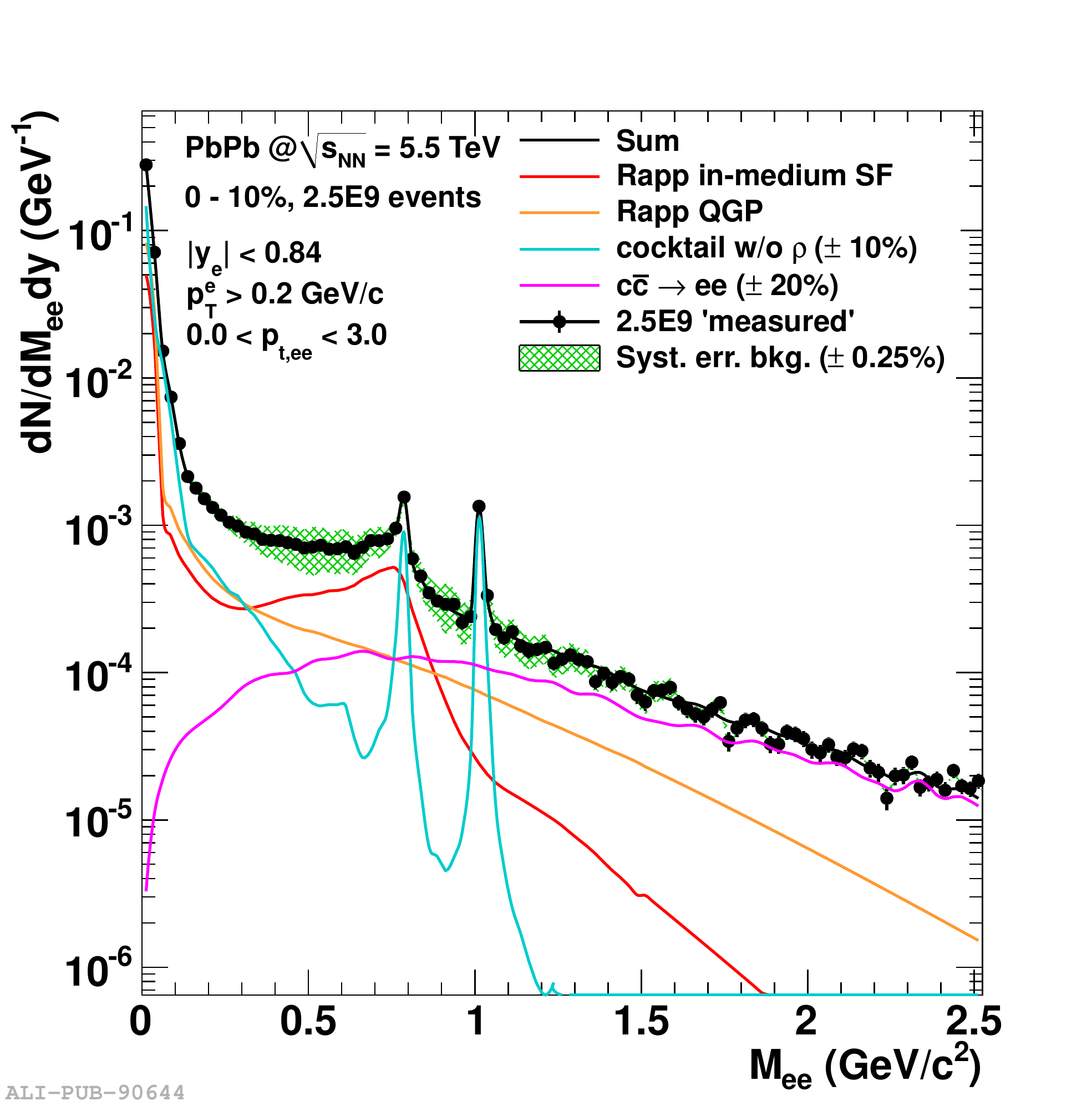}
\includegraphics[width=0.49\textwidth]{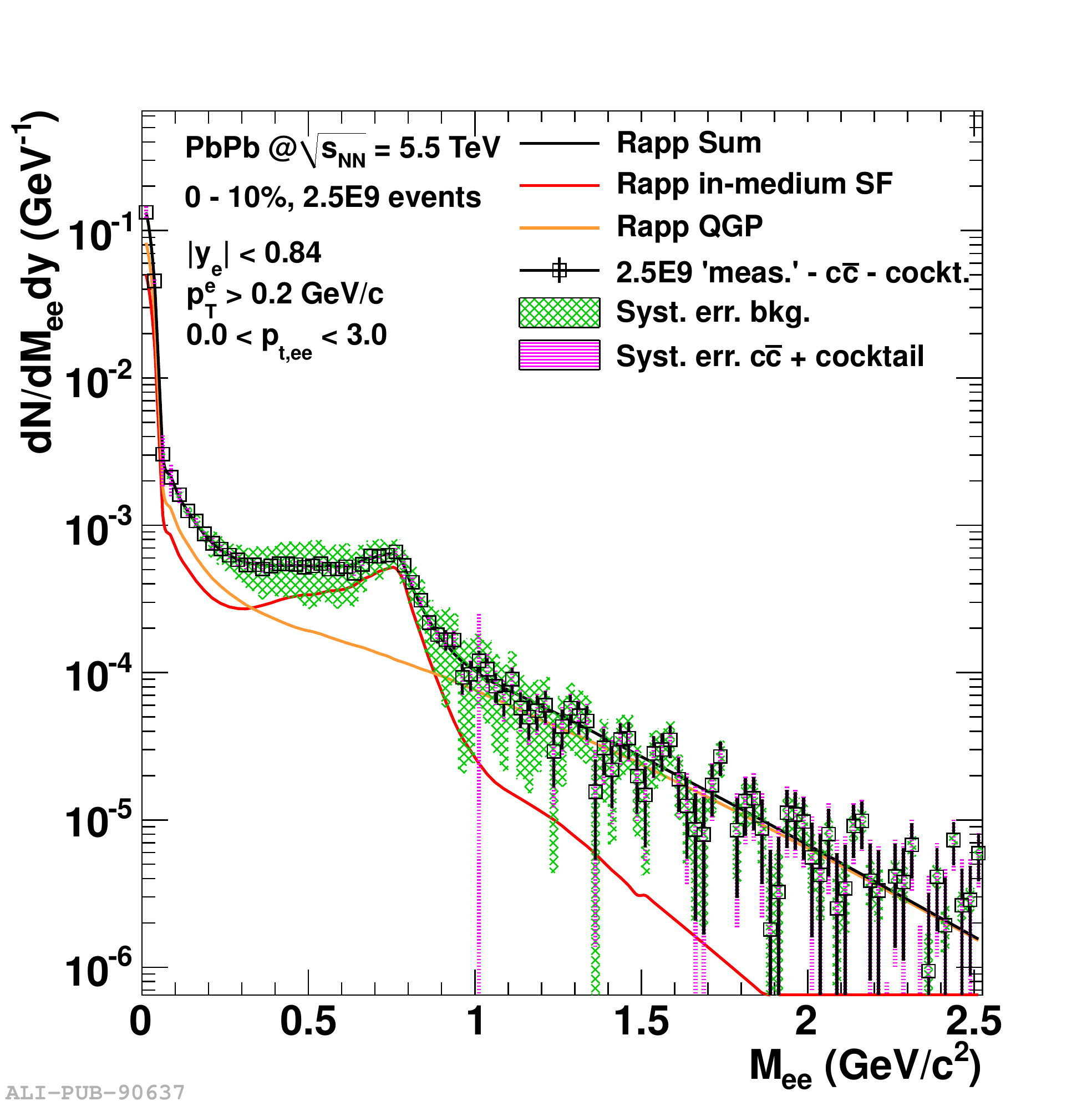}
\caption{Projected performance of the ALICE upgrade~\cite{Abelevetal:2014dna}: inclusive $e^+e^-$ invariant mass spectrum (left) and excess spectrum (right) in
10\% most central Pb--Pb collisions at $\sqrtsNN = 5.5$~TeV,  for an
integrated luminosity of 3~nb$^{-1}$. The green boxes show the systematic uncertainties from
the combinatorial background subtraction, the magenta boxes indicate systematic uncertainties
related to the subtraction of the hadron cocktail and charm contribution.}
\label{fig:alice-dielectrons}
\end{center}
\end{figure*}

At LHC energies, the measurement of low-mass $e^+e^-$ pairs in Pb--Pb collisions  poses major
experimental challenges because of the significant increase of the background in comparison
to RHIC. A very good electron identification is mandatory for the suppression of combinatorial
background from hadronic contamination. Moreover, electrons from  Dalitz decays and photon
conversions (mainly from $\pi^0\to\gamma\gamma$) form a substantial combinatorial background. This demands
a very low material budget before the  first active detector layer and
refined strategies to detect $e^+e^-$
pairs from photon conversions and Dalitz pairs for further rejection. The large combinatorial
background prevents also a straightforward online trigger scheme. Therefore, the analysis can
be carried out only using minimum-bias samples.
Finally, the measurement requires acceptance for dilepton pairs at invariant masses
and transverse momenta as low as $M_{ee}\sim  \pt\sim  T\sim 150$~MeV. This implies electron detection
down to $\pt = 100$--200~MeV/$c$.

These critical aspects have so far prevented a quantitive measurement of dielectron production at
LHC energies. The ALICE detector upgrades will make the measurement feasible in the LHC
Run-3~\cite{Abelevetal:2014cna,Abelevetal:2014dna}.

The enhanced low-$\pt$ tracking capability of the new ITS allows the electrons tracks
to be measured down to $\pt\approx 50$~MeV/$c$, improving the reconstruction efficiency of photon conversions and
Dalitz pairs for combinatorial background suppression. The better impact parameter resolution
will also enable efficient tagging of electrons from semi-leptonic charm decays,
which can be separated from prompt dileptons. In order to optimize the low-$\pt$ acceptance for
electron identication with the TPC and TOF detectors, the measurement will be carried out
with a value of 0.2~T for the magnetic field in the ALICE
central barrel, lower with respect to the default (0.5 ~T). 
With the readout upgrade of the experiment that enables
Pb--Pb collisions recording at 50~kHz, the goal is collect $L_{\rm int}=3$~nb$^{-1}$
in one month of running with this field configuration.

Figure~\ref{fig:alice-dielectrons} shows
the projected performance for the measurement of the inclusive
$e^+e^-$ invariant-mass spectrum (left) 
and the thermal spectrum after subtraction of the hadronic components
(right). 
The in-medium $\rho$ spectral function
will be measured with good precision.
Furthermore, information on the early temperature of the system can be derived from the invariant-mass
dependence of the dilepton yield at masses $M_{ee} > 1.1$~GeV/$c^2$, where the yield is completely
dominated by the thermal radiation from the QGP. In order to quantify the sensitivity of
this measurement, an exponential fit to the simulated spectra in the invariant-mass
region $1.1 < M_{ee} < 2.5$~GeV/$c^2$ was performed with the function 
$\dd N/\dd M_{ee}\sim M_{ee}^{3/2} \exp(-M_{ee}/T)$.  The temperature
of the source that emits the thermal
dileptons can be measured with a statistical precision of about $10\%$ and a systematic uncertainty
of about 20\%~\cite{Abelevetal:2014dna}. 

Finally, after the ALICE upgrade low-mass dimuons will be studied also at forward rapidity ($2.5<y<3.5$)
with improved precision using the MFT, the new silicon tracker placed before the hadron absorber
of the ALICE muon spectrometer. This measurement complements the one at central rapidity described
above, though with somewhat more restrictive momentum cuts. For more details see~\cite{CERN-LHCC-2015-001}.


\subsection{Bulk hadron production and correlations}
\label{sec:soft}

In ultra-relativistic heavy-ion collisions low-$\pt$ particles are generally used to access the
properties of the produced medium in terms of chemical composition and
global kinematic characteristics.
The chemical composition is accessible using the abundances of
various hadron species with different quark contents, while
the kinematic properties of the medium are studied by identifying collective phenomena
that involve all the particles in the same event.
In this section we will present the role of such effects in \PbPb
collisions, where they are well understood in terms of theoretical models,
and then we will move to smaller colliding systems presenting
the most recent intriguing results from the LHC Run-1.

\subsubsection{Light-flavour hadron production in Pb--Pb collisions}

The chemical composition of the particles produced in the medium was extensively
investigated so far both at RHIC and LHC. In particular, the particle
identification capability of the STAR and the ALICE experiments
enables the comparison of
the yields of several hadron species with the predictions of
thermal and statistical hadronisation models (introduced in Section~\ref{sec:evolution}).
The observations are consistent with a decreasing of the
baryonic chemical potential when increasing the energy and with a
temperature close to the one expected for the QCD phase transition
(see e.g.~\cite{Cleymans:2014xha}). 

Strangeness production plays an important role in such models
because it is expected to be favoured in high energy density systems~\cite{Rafelski:1982pu}.
An enhancement of strange baryon production in heavy-ion collisions with respect to \pp collisions
was measured by several experiments from SPS to LHC energies~\cite{Andersen:1999ym,Afanasiev:2002he,Antinori:2004ee,Abelev:2007xp,ABELEV:2013zaa}.
Recently, the ALICE experiment measured an increase of strange baryon
yields relative to pions also in high-multiplicity p--Pb events 
(see Fig.~\ref{figStrangeness}, left and central panels)~\cite{Adam:2015vsf}.
The observed trend is consistent with the prediction of some thermal models that
expect a saturation to the grand canonical value when the size of the system
is larger than a few~fm.
The increase of the centre-of-mass energy and of the projected ALICE
data samples in Run-2 and Run-3 should enable measurements in 
p--Pb multiplicity classes that overlap with the Pb--Pb ones, in order
to assess whether these ratios saturate in p--Pb at the same values as
in Pb--Pb collisions.

\begin{figure}[t]
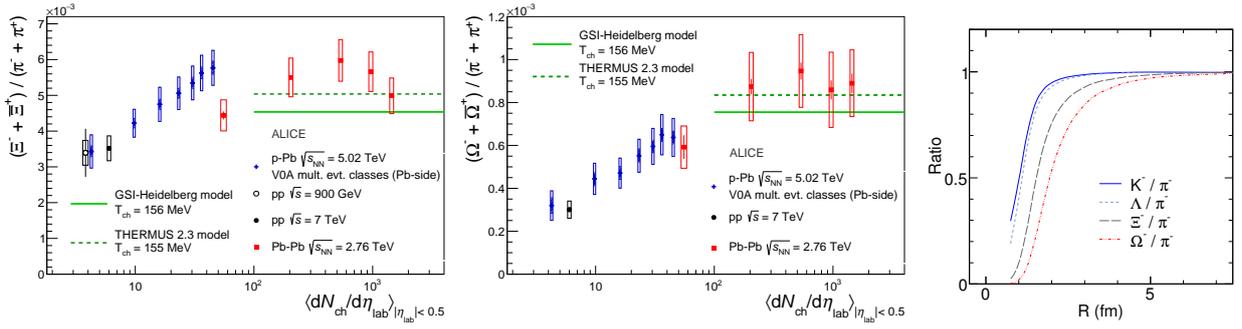

 \centering
 \includegraphics[width=0.36\textwidth]{../Figures/XiToPiRatio-pPb_final-14410.pdf}
 \includegraphics[width=0.36\textwidth]{../Figures/OmegaToPiRatio-pPb_final-14411.pdf}
 \includegraphics[width=0.25\textwidth]{../Figures/volumeRad.pdf}
\caption{$\Xi/\pi$ (left) and $\Omega/\pi$ (centre) ratios vs.
  charged-particle multiplicity in \pp, \pPb and \PbPb collisions at the LHC, measured
  by ALICE~\cite{Adam:2015vsf}.  In the right-hand panel the same
  observable is obtained from
  theoretical calculations within a thermal model with
$T = 170~{\rm MeV}$ and $\mu_{B} = 1~ {\rm MeV}$~\cite{Kraus:2008fh}; the ratio are normalised to the grand-canonical value and shown as 
  functions of the radius $R$ of the fireball.}
\label{figStrangeness}
\end{figure} 

In addition, recent lattice-QCD calculations suggest that
the freeze-out temperature (and hyper-surface) of the QGP may be
different for quarks of different
flavours. Figure~\ref{figStrangenessTemp} shows a higher critical
temperature for strange quarks than for u/d
quarks~\cite{Borsanyi:2011bm},
which could be reflected in an earlier freeze-out (at higher
temperature) for strange baryons.
The experimental search for such an effect demands very high
statistical precision, as can be achieved with the future ALICE data-taking.

\begin{figure}[t]
 \centering
 \includegraphics[width=0.49\textwidth]{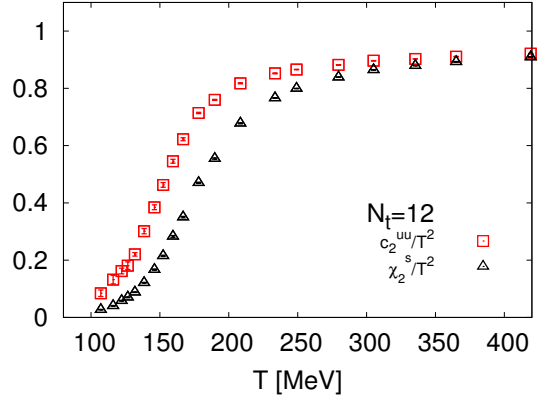}
\caption{Comparison between the lattice results for up/down and strange-quark number susceptibilities~\cite{Borsanyi:2011bm}.}
\label{figStrangenessTemp}
\end{figure} 




\subsubsection{Indications of collective effects in small colliding systems}

The evidence for collective behaviour in small systems has unexpectedly emerged from
high-multiplicity p--Pb collisions and pp collisions at LHC energies. 

Angular correlations between charged ``trigger'' and ``associated'' particles are a powerful tool
to explore the mechanisms of particle production in collisions of hadrons and nuclei at high
energy. Such correlations involve the measurement of the distributions of relative angles $\Delta
\varphi$ and $\Delta \eta$ between a trigger particle within a certain range in transverse momentum
$p_{\rm T, trig}$ and an associated particle in a $p_{\rm T, assoc}$ transverse momentum range (where
$\Delta \varphi$ and $\Delta \eta$ are the differences in azimuthal angle $\varphi$ and pseudorapidity
$\eta$ between the two particles). The  ``near side'' peak ($\Delta \varphi \sim 0$) includes
particles associated with the leading particle, while the ``away side'' peak ($\Delta \varphi
\sim \pi$) is formed by particles associated with the recoil jet. 
In pp collisions, the away-side structure is elongated along $\Delta
\eta$ because of the
longitudinal momentum distribution of partons in the colliding protons. In
nucleus--nucleus collisions, the jet-related correlations are modified and additional
structures emerge, which persist over a long range in $\Delta \eta$ both on the near
side and on the away side. These long-range correlations in A--A collisions are
commonly attributed to the formation of a QGP with a collective flow. 

In p--Pb collisions at $\sqrtsNN = 5.02$~TeV, two-particle angular correlations
show a long-range ridge structure on the near side in events characterised by higher
than average multiplicity
(see Fig.~\ref{figsoft1})~\cite{Abelev:2012ola,Aad:2012gla,CMS:2012qk}. In the low-multiplicity class, the correlation in the near side
peak for pairs of particles originating from the same jet, as well as the elongated
structure at $\Delta \varphi \sim \pi$ for pairs of particles back-to-back in azimuth are
visible. In the high-multiplicity class, the same features with higher yields can be
observed.
To remove the  contribution due to jet fragmentation and isolate the
ridge structure, 
low-multiplicity events  are subtracted from high-multiplicity correlations~\cite{Abelev:2012ola}. 
This subtraction method is based on the observation  that in p--Pb the near-side peak yield does
not depend on event  multiplicity~\cite{Abelev:2014mva} (see Fig.~\ref{figsoft1} bottom-right). 
The resulting distribution in $\Delta \varphi$ and $\Delta \eta$ for the high-multiplicity event
class is shown in Fig.~\ref{figsoft1}~(bottom-left). After this subtraction, a double ridge
excess  structure in the correlation is observed. 

\begin{figure}[!t]
 \centering
 \includegraphics[width=0.94\textwidth]{../Figures/softfig1.pdf}
 \includegraphics[width=0.42\textwidth]{../Figures/softfig2.pdf}
 \includegraphics[width=0.52\textwidth]{../Figures/softfig3.pdf}
\caption{Top: associated yield per trigger particle for pairs of
  charged particles with $2<p_{\rm T,trig}<4$~GeV/$c$ and $1<p_{\rm T,assoc}<2$~GeV/$c$ in p--Pb collisions at $\sqrtsNN=5.02$~TeV for low (left) and high (right) multiplicity event classes. 
Bottom left: associated yield per trigger particle for pairs of charged particles in p--Pb collisions at $\sqrtsNN = 5.02$~TeV for the high-multiplicity class, after subtraction of the jet contribution. Bottom right: yield in the near-side peak as a function
of event-multiplicity~\cite{Abelev:2014mva}.}
\label{figsoft1}
\end{figure} 

The away-side ridge has also been observed in pp collisions, in events with very large
multiplicity~\cite{Khachatryan:2010gv}.  However, in this case the near-side jet yield
significantly depends on multiplicity and, therefore, the subtraction method is not fully
justified.

In A--A the near-side ridge is commonly attributed to the formation of a dense medium with a
collective flow.
The Fourier decomposition of the correlations observed in p--Pb
collisions indicates that the double-ridge is dominated by the
second-order azimuthal anisotropy harmonic $v_{2}$. Both the second ($v_{2}$) and third ($v_{3}$)
order coefficients are comparable to the values measured in Pb--Pb collisions. These values are 
described by hydrodynamical model calculations that
assume a flowing medium~\cite{Bozek:2012gr, Romatschke:2015gxa}. 

Another striking evidence of collectivity in p--Pb collisions is provided by multi-particle
azimuthal correlations (between four, six and all particles), for which the contribution from 
jet fragmentation is expected to be strongly suppressed.  These correlations show consistent
$v_{2}$ values for four or more particle cumulants,  supporting the collective nature of the
observed correlations  (see Fig.~\ref{figsoft3}) ~\cite{CMS:2014bza}. 

\begin{figure}[t]
 \centering
 \includegraphics[width=0.75\textwidth]{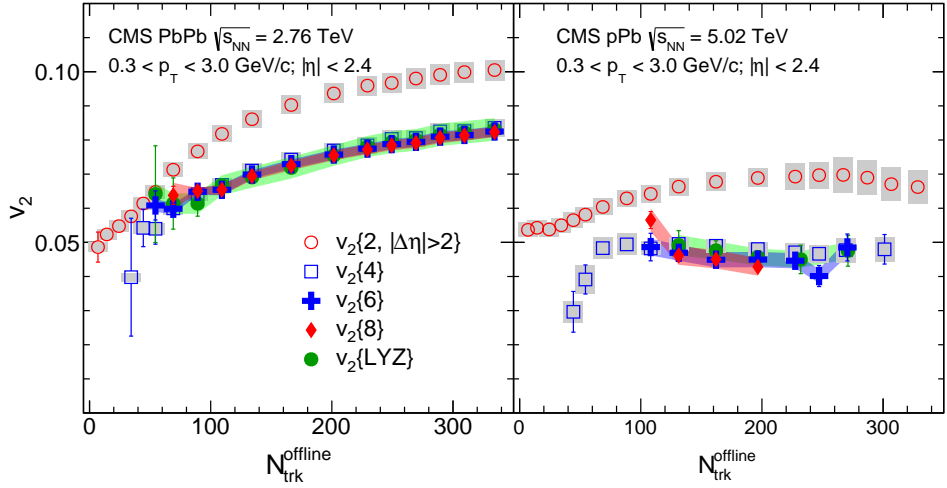}
\caption{$v_{2}$ values as a function of the number of tracks measured
  by the CMS experiment in p--Pb collisions at $\sqrtsNN=5.02$~TeV~\cite{CMS:2014bza}. Solid data points are obtained from multi-particle correlations to remove the correlation due to jet fragmentation products.}
\label{figsoft3}
\end{figure} 
%
%

Another unexpected observation that has emerged for identified hadrons is the mass ordering of
$\pt$ spectral slopes and of $v_{n}(\pt)$ after subtraction of non-flow contributions, hinting to a
common radial flow of the system.   Furthermore, hints of  scaling
with the number of constituent quarks are observed in p--Pb
collisions at $\sqrtsNN=5.02$~TeV~\cite{Khachatryan:2014jra} suggesting that the development of flow could occur at partonic level.

These observations open a very interesting field of investigation as they pose many questions
to our current knowledge.  A hot and expanding medium was not expected
to be formed in pp or in p--A collisions. 
It will be crucial to further study small systems to
understand what drives the onset of collectivity and what is the
minimum size of a system that exhibits a
fluid-like behaviour.
The mechanisms that produce the ridge in A--A, p--A and in pp collisions need to be further 
investigated.
In particular, for pp collisions it will be important to investigate the origin 
of the (double) ridge structure, after establishing a robust and reliable method to subtract the jet 
contribution.

The models currently proposed to explain the double ridge, can be divided into two
phenomenological classes (see e.g.~\cite{Torrieri:2013aqa} for a review). The first class states that the correlations
are established in the initial high gluon density state through local partonic interactions, while
in the second one they are assumed to arise due to the hydrodynamic flow that collimates local 
anisotropies in the final states. 
It is worth to note that, even if many observables in p--Pb and in pp are in satisfactory
agreement with hydrodynamic models (such as the shape of $v_{n}(\pt)$ and the mass ordering),
some basics assumptions  of hydrodynamics are at the edge of validity in small systems. In fact,
as the mean-free-path of the matter approaches the characteristic system size, the viscosity
increases, the Knudsen number becomes larger ($K_{n} \geq 1$) and the hydrodynamic description
is no more valid.
Experimentally the challenge is to understand the origin of the collective phenomena and to
discriminate between these two classes of models. 

There are many other aspects that need to be further clarified in
p--Pb collisions.
Jets are not suppressed in p--Pb
collisions~\cite{CMS:2014qca,ATLAS:2014cpa,Adam:2015hoa}, no dijet asymmetry is
observed~\cite{Chatrchyan:2014hqa} and also di-hadron yields are 
unmodified~\cite{Abelev:2012ola}. Therefore no in-medium modifications or energy loss for high
$\pt$  particles is observed, suggesting that no medium is formed in p--Pb collisions.  However,
this seems to clash with the observation of positive $v_{2}$ values at high $\pt$. A theory able to reproduce simultaneously the modification factor and the $v_{2}$ measurements at high momenta is still missing.

For many of these studies the particle identification capabilities provided by the ALICE experiment will play a crucial role.
In addition, the upgrade of the Inner Tracking System
and the new Forward Muon Tracker, together with the increase in luminosity, will provide the opportunity to extend identified two-particle correlation studies to larger $\eta$ ranges.

%
Furthermore, two-particle correlations with heavy-flavour mesons could
shed light on the roles of flow and of initial-state correlations. 
In fact, while correlations due to high initial gluon density are
expected to equally affect light and heavy quarks, in a hydrodynamic
system the effective Knudsen number depends on the mass and should
show different effects on the final correlations for light- and
heavy-flavour hadrons.


All these studies could benefit from a run at the LHC with lighter
ion--ion (such as Ar--Ar or O--O) and/or proton--light-ion collisions to
assess the system size dependence and possibly understand the variable
driving the onset of the collective behaviour. 
On the basis of the results from the p--Pb data sample of Run-2, the experiments may consider to request a light-ion run.


\subsection{Nuclei, hypernuclei and exotic hadrons}
\label{sec:intro}

The high-partonic density environment created in ultra-relativistic heavy-ion collisions  is uniquely suitable for the
production of both light (anti-)nuclei and (anti-)hypernuclei.
A hypernucleus~\cite{Botta:2012xi} contains at least one hyperon, namely a baryon 
with one or more strange quarks, in addition to protons and nucleons.
Since hypernuclei are weakly-bound nuclear states, they
are sensitive probes of the final stages of the evolution of the fireball formed in the heavy-ion collisions.
One of the striking features of particle production at high energies is the near equal abundance of matter and antimatter in the central rapidity region~\cite{Abelev:2008ab, Aamodt:2010pa}.
Although strange hadrons are abundantly produced, their strong interactions are not well understood. This interaction is not
only important for the description of the hadronic phase of a heavy-ion collision, but it also plays an important role in the description of dense
hadronic matter, as for instance in neutron stars.
Depending on the strength of the hyperon--nucleon interaction, the
collapsed stellar core could be composed of hyperons, of strange quark matter or of a kaon
condensate. 
In this context, hyperon interactions are
crucial to understand the phase structure of QCD at large baryonic densities (i.e. large $\mu_B$) and low temperatures.
Because of the presence of hyperons, (anti-)hypernuclei
provide an ideal environment to learn about the hyperon--hyperon and hyperon--nucleon interaction, responsible in part for
the binding of hypernuclei and lifetime, which is of fundamental interest in nuclear physics and
nuclear astrophysics.

The production of light (anti-)nuclei has attracted attention
already in ``low-energy'' heavy-ion collisions at BNL-AGS~\cite{Armstrong:2000gd}, CERN-SPS~\cite{SimonGillo:1995dh,Anticic:2004yj}, and BNL-RHIC~\cite{Adler:2001uy}. It has been argued that the production mechanism may depend on collision energy, e.g. with spectator fragmentation at lower energies and a production via parton or hadron coalescence at higher energies.

Measurements for the production of (anti-)nuclei in pp and Pb--Pb collisions at the LHC 
were recently reported by the ALICE Collaboration~\cite{Adam:2015vda}: light (anti-)nuclei show  the same behaviour as
 non-composite light flavour hadrons, which are governed by a common chemical freeze-out and a subsequent hydrodynamic expansion. 
Results on the production of the (anti-)hypertriton published by ALICE~\cite{Adam:2015yta} show a value of the lifetime comparable with the one measured in Au--Au collisions by the STAR Collaboration.

Besides the nuclei containing one hyperon and observation of anti-matter nuclei, more exotic forms of deeply-bound states with strangeness have been proposed, either consisting of baryons or quarks. One of these states is the H dibaryon predicted in 1977~\cite{Jaffe:1976yi}.
Later, many more bound dibaryon states with strangeness were proposed using quark potentials~\cite{Goldman:1987ma,Goldman:1998jd}
or the Skyrme model~\cite{Schwesinger:1994vd}.  However, the non-observation of multi-quark bags, like for instance strangelets,
is still one of the open problems of intermediate and high energy physics.
The HAL QCD Collaboration~\cite{Inoue:2011ai} performed phase space studies on the lattice which clearly show that the H dibaryon is not bound, while recent lattice studies report that there could be strange dibaryon systems including  $\Xi$ hyperons that can be bound~\cite{Beane:2011iw}.
An experimental confirmation of such a state would, therefore, be an enormous step forward in the understanding of the hyperon interaction.

The data-taking programme with the upgraded ALICE detector after LS2 has a strong pontential for measurements of nuclei, hypernuclei and the search for exotic 
states, because these studies require large event samples with a minimum-bias trigger, as well as high tracking precision for the separation of secondary vertices and
charged-hadron (light nucleus) identification. 

\subsubsection{Expected yields for the ALICE-upgrade programme}
\label{sec:yield}

The signal yields for the projected integrated luminosity of 10~nb$^{-1}$ were estimated for the (hyper)nuclei (d, $^{3}$He, $^{4}$He, $^{3}_{\Lambda}$H, $^{4}_{\Lambda}$H, $^{4}_{\Lambda}$He) and for the exotic bound states (${\Lambda\Lambda}$ and  $\Lambda$n), and their antiparticles.

The theoretical production yields predicted by the the statistical hadronisation model~\cite{Andronic:2010qu} for central (0--10\%) Pb--Pb collisions at 
$\sqrtsNN = 2.76$~TeV were considered. The yields per unit of rapidity at mid-rapidity are reported in Fig.~\ref{InvMass} (left) for 
two values of the chemical freeze-out temperature that were obtained from fits to the light-flavour hadron abundances at RHIC and LHC energies.
The predicted yields at $\sqrtsNN = 5.5$~TeV (the actual energy after LS2) are very close to those at 2.76~TeV~\cite{Andronic:2010qu}. 

Table~\ref{tab:yield} reports the expected yields of the (hyper)nuclei and the exotic states that could be observed in the ALICE detector 
for $L_{\rm int} = 10$~nb$^{-1}$~\cite{Abelevetal:2014dna}.  For example, the signal-to-background ratio and the significance of the $^{3}_{\Lambda}$H signal for $p_{\rm{T}} > 2$~GeV/$c$ are expected to be of about 0.1 and 60, respectively.
Figure~\ref{InvMass} (right) shows the projected $^{3}_{\Lambda}$H invariant-mass distribution with the ALICE upgrade and an integrated luminosity of 10~nb$^{-1}$.

\begin{figure}[!t]
\begin{tabular}{ccc}
\begin{minipage}{.45\textwidth}
\centerline{\includegraphics[width=1\textwidth]{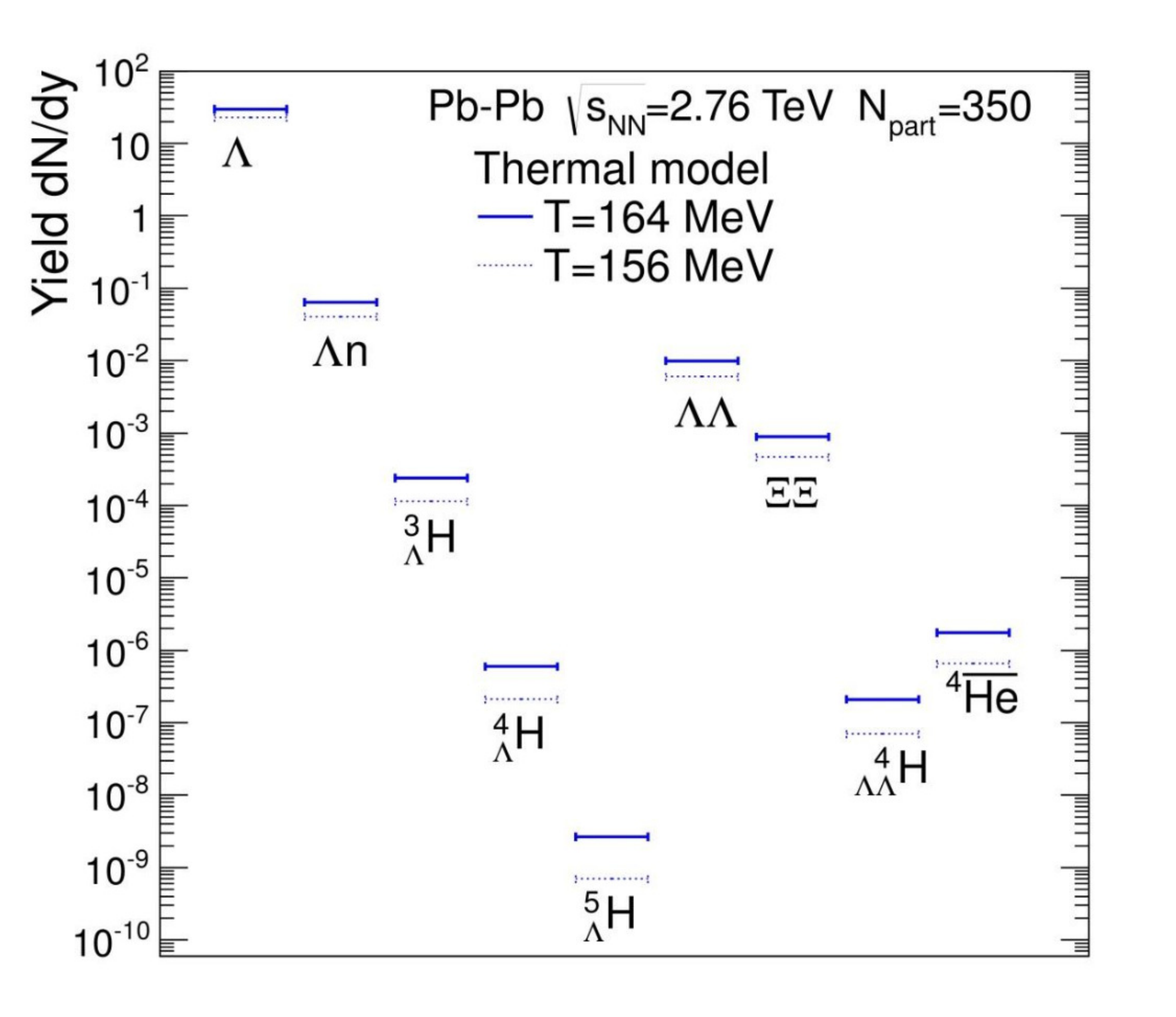}}
\end{minipage} &
\begin{minipage}{.45\textwidth}
\centerline{\includegraphics[width=1\textwidth]{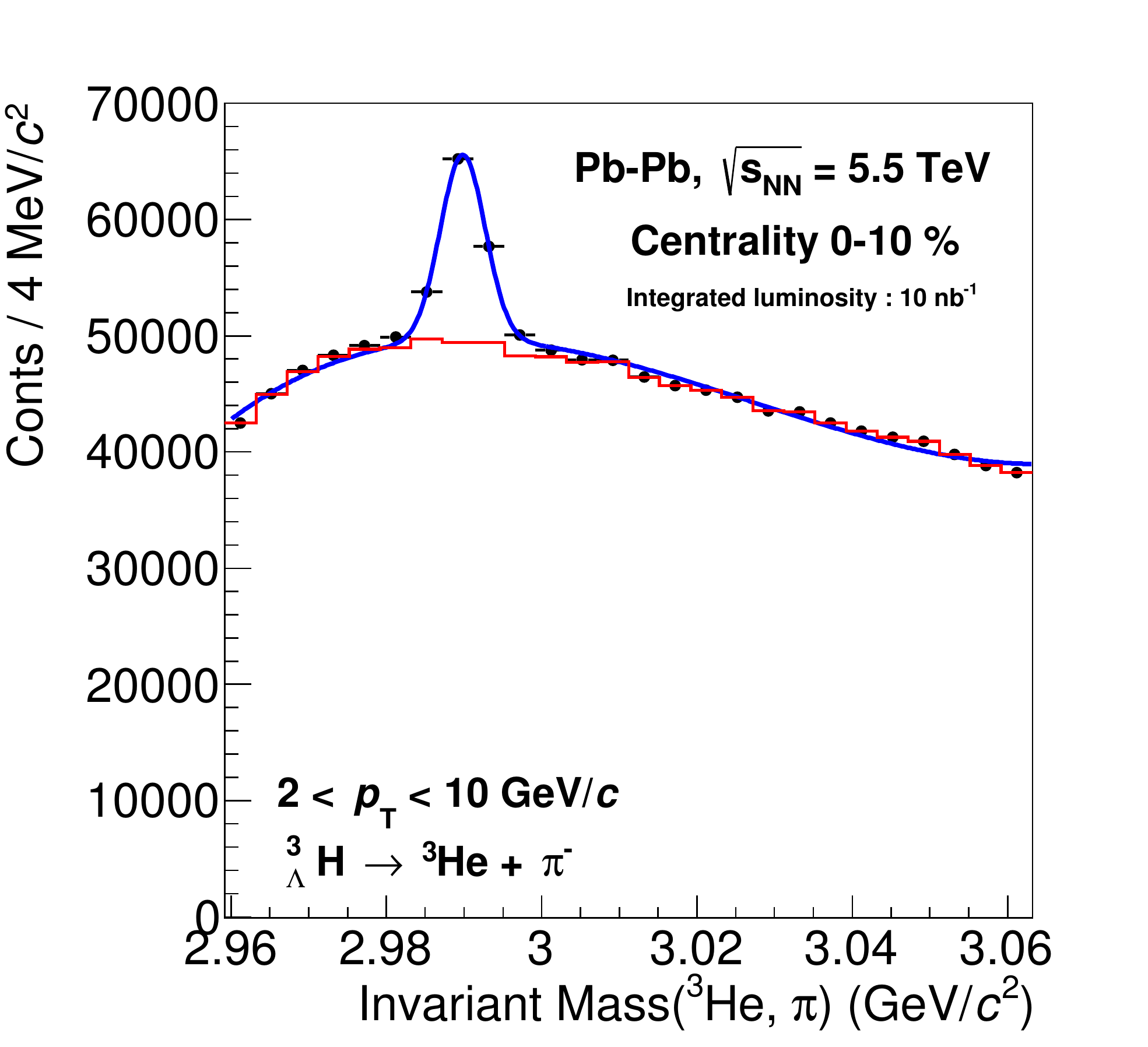}}
\end{minipage}
\end{tabular}
\caption{Left: Yields (d$N$/d$y$) calculated for strange particles and light (hyper) nuclei using the thermal model
and assuming two different temperatures ($T=164$~MeV, which corresponds to the expected temperature from RHIC,
and $T=156$~MeV, which corresponds to the best estimation using LHC data)~\cite{Andronic:2010qu}. The values were calculated for Pb--Pb
collisions at $\sqrtsNN = 2.76$~TeV in the 0--10\% centrality class.
Right: projected $^{3}_{\Lambda}$H invariant-mass distribution with the ALICE upgrade for a Pb--Pb integrated luminosity of 10~nb$^{-1}$~\cite{Abelevetal:2014dna}.}
\label{InvMass}
\end{figure}

\begin{table}[!h]
\begin{center}
\caption{Expected yields for light (hyper)nuclear states (and their antiparticles)
for central Pb--Pb collisions (0--10\%) at $\sqrtsNN = 5.5$~TeV.
From left to right: (hyper)nuclear species, production yield from the statistical hadronization model~\cite{Andronic:2010qu}, branching
ratio (only for hypernuclei and exotica states), rapidity interval, and number of expected reconstructed particles for $L_{\rm int} = 10$~nb$^{-1}$~\cite{Abelevetal:2014dna} and reference for the estimation of the average acceptance-times-efficiency $\langle Acc. \times \epsilon\rangle$ for $p_\mathrm{{T}} > 0$.}
\label{tab:yield}
\begin{tabular}{lccccc}
\hline
State              & d$N$/d$y$                & B.R.   & $|y|<$ &  Yield                & Ref.\\
\hline
d (TPC)            & 5  $\times$10$^{-2}$  & -- &      0.5        &  3.1$\times$10$^{8}$ &\cite{Adam:2015vda}\\
d (TPC+TOF)        & 5  $\times$10$^{-2}$  & -- &      0.5        &  1.4$\times$10$^{8}$ &\cite{Adam:2015vda}\\
$^{3}$He (TOF)     & 3.5$\times$10$^{-4}$  & -- &      0.5        &  2.2$\times$10$^{6}$ &\cite{Adam:2015vda}\\
$^{4}$He (TPC+TOF) & 7.0$\times$10$^{-7}$  & -- &      0.5        &  1.5$\times$10$^{3}$ &\cite{Adam:2015vda}\\
$^{3}_{\Lambda}$H  & 1.0$\times$10$^{-4}$  & 0.25 &      1          &  4.4$\times$10$^{3}$ &\cite{Andronic:2010qu}\\
$^{4}_{\Lambda}$H  & 2.0$\times$10$^{-7}$  & 0.50 &      1          &  1.1$\times$10$^{2}$ &\cite{Andronic:2010qu}\\
$^{4}_{\Lambda}$He & 2.0$\times$10$^{-7}$  & 0.54 &      1          &  1.3$\times$10$^{2}$ &\cite{Andronic:2010qu}\\
${\Lambda}$n       & 3.0$\times$10$^{-2}$  & 0.35 &      1          &  2.9$\times$10$^{7}$ &\cite{Adam:2015nca}\\
${\Lambda\Lambda}$ & 5.0$\times$10$^{-3}$  & 0.064 &      1          &  1.9$\times$10$^{5}$ &\cite{Adam:2015nca}\\
${\Lambda\Lambda}$ & 5.0$\times$10$^{-3}$  & 0.41 &      1          &  1.2$\times$10$^{6}$ &\cite{Adam:2015nca}\\
\hline
\end{tabular}
\end{center}
\end{table}

\subsubsection{Shedding light on the the XYZ states using high-$p_\mathrm{T}$ deuteron measurements}
\label{sec:deuto}


Recently, the search of exotic forms of deeply-bound states has attracted a large attention with the unexpected observation at electron--positron colliders of the new X, Y and Z states with
masses around 4~GeV$/c^2$~\cite{Abashian:2000cg,Bai:1994zm}. 
These heavy particles show very unusual properties, whose theoretical interpretation is entirely open. A number of new states have been recently discovered by BaBar, Belle and CLEO~\cite{Chen:2013wva,Agashe:2014kda}. One of the most well-established
among these is the narrow X(3872) state with a width of about 2.3~MeV$/c^2$~\cite{Choi:2003ue}. Its main hadronic decay modes are $\pi^{+}\pi^{-}$J$/\psi$ and
$\mathrm{D^{0}}\overline{\mathrm{D}}^{0}\pi^{0}$. As many other newly discovered charmonium-like hadrons, it does not seem to fit into the conventional $c\overline{c}$ spectrum~\cite{Godfrey:2008nc, Esposito:2015fsa}. The very close vicinity of the $\mathrm{D^{0}}\overline{\mathrm{D}^{0*}}$ threshold favours a molecular interpretation with these
constituents~\cite{Guerrieri:2014gfa}. Ref.~\cite{Guerrieri:2014gfa} reports a prediction for the X state $\pt$-differential cross section up to $25$~GeV/$c$ and proposes a comparison with the production cross section of the anti-deuteron, which may be regarded as a baryonic analogous of the X state. Because of the lack of experimental measurements of anti-deuterons at high $\pt$, the debate on this topic is still open and it would be interesting to measure the anti-deuteron spectrum at least up to $p_{\mathrm{T}} = 10$~GeV/$c$ in pp collisions (rather than in Pb--Pb to avoid the complications of possible QGP effects). 
This measurement could be within reach with the large minimum-bias sample of pp collisions that is planned for the ALICE data-taking after LS2, namely $L_{\rm int}\sim 5$~pb$^{-1}$~\cite{Abelevetal:2014cna}.

\subsubsection{CPT invariance tests with light (anti-)nuclei}
\label{sec:ctp}

\begin{figure}[t]
\begin{center}
\includegraphics[width=0.75\textwidth]{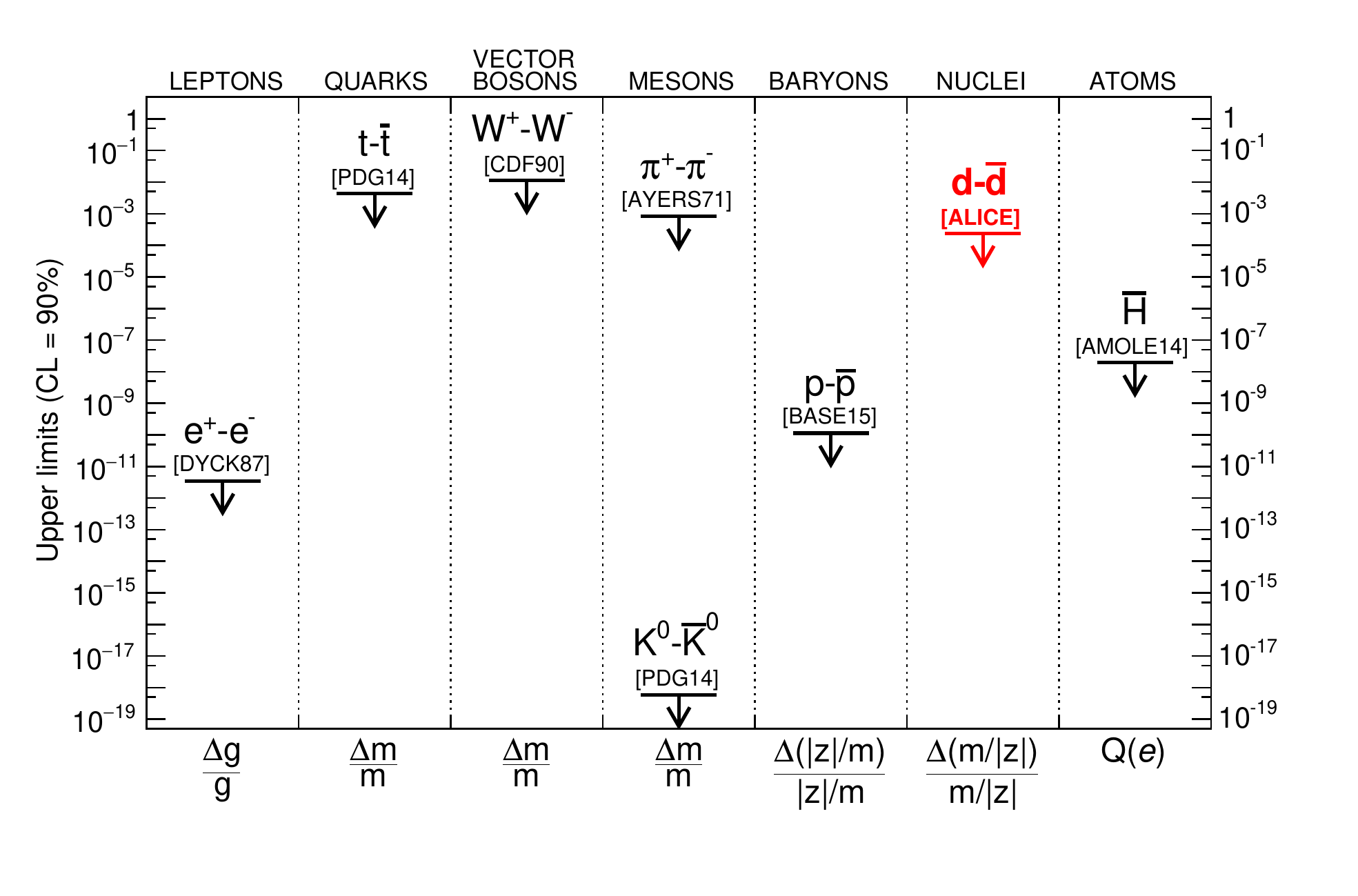}
\caption{Experimental limits for CPT invariance $(\mathrm{CL}=90\%)$ for particles, nuclei and atoms.
From left to right: measurement of $g$-factor for the electron and positron (DYCK87~\cite{VanDyck:1987ay}),
mass difference between top and anti-top (PDG average~\cite{Agashe:2014kda}),
$\mathrm{W}^{+}$--$\mathrm{W}^{-}$ (CDF90~\cite{Abe:1990pp}),
$\pi^{+}$--$\pi^{-}$ (AYERS71~\cite{Ayers:1971kz}),
$\mathrm{K}^{0}$--$\mathrm{\overline{K^0}}$ (PDG average~\cite{Agashe:2014kda}),
the mass-to-charge ratio difference between proton and anti-proton (BASE15~\cite{Ulmer:2015jra}),
the mass-to-charge ratio difference between deuteron and anti-deuteron (ALICE~\cite{Adam:2015pna,ALICE-PUBLIC-2015-002})
and the charge of anti-hydrogen (AMOLE14~\cite{Amole:2014rla}).}  
\label{TestsCpt}
\end{center}
\end{figure}

The large and similar production of matter and anti-matter particles observed in high-energy heavy-ion collisions
enables the study of their properties, such as the mass and electric charge.
The comparison of these physical quantities between a particle species and the corresponding anti-matter counterpart represents an interesting investigation of the CPT symmetry, one of the most important laws of nature.
Currently, it is tested
experimentally for elementary fermions and bosons, for QED and
QCD systems, with different levels of precision~\cite{Agashe:2014kda,Amole:2014rla}.
Such measurements can be also used to constrain
the parameters of effective field theories that add explicit CPT-violating terms to the Standard Model Lagrangian,
such as the Standard Model Extension\cite{Kostelecky:2008ts}.

The extension of such measurement
from (anti-)baryons to (anti-)nuclei allows one to probe any difference in the interactions between nucleons and
anti-nucleons encoded in the (anti-)nuclei masses.
The nuclear force is a remnant of the underlying strong interaction among quarks and gluons that can be described by
effective theories~\cite{vanKolck:1999mw}, but not yet directly derived from QCD.

Experimental limits on the mass-to-charge ratio differences between nuclei and anti-nuclei
were reported by the ALICE experiment based on the Run-1 data sample for the deuteron and anti-deuteron,
and $^{3}\mathrm{He}$ and $^{3}\mathrm{\overline{He}}$ nucleus: they are, respectively, $2.4 \times 10^{-4}$ and $2.1 \times 10^{-3}$ at a confidence level of $90\%$~\cite{Adam:2015pna,ALICE-PUBLIC-2015-002}.
The limit for  the (anti-)deuteron is reported in Fig.~\ref{TestsCpt},
which shows the best CPT invariance tests obtained to date for particles, nuclei and atoms.

The large increase of integrated luminosity expected for the LHC Run-3 and Run-4 data-taking
will push the precision of the
measurements potentially down to $10^{-6}$ (deuteron) and $10^{-5}$ ($^{3}\mathrm{He}$).
The increase in luminosity should enable this measurement also for the triton and, for the first time, for the $^{4}\mathrm{He}$ nucleus.

\clearpage

\subsection{Future ALICE programme: summary}
\label{sec:alicesummary}

Heavy-ion collisions at the LHC and at top RHIC energy provide access
to the region of the QCD phase diagram at high temperature and
vanishing baryon chemical potential. 
The results from the RHIC programme and the 
LHC Run-1 have revealed the QGP as a strongly-coupled (liquid-like),
high-density and low-viscosity
medium in which colour charge is deconfined.
Experimental research is now moving 
towards high-precision measurements, in order to constrain the
properties of the QGP and determine its equation of state and
characteristic parameters ---namely, the temperature, the
shear-viscosity-to-entropy-density ratio and the transport
coefficients, as well as their time-dependence during the collision
evolution.

The LHC Run-2 is the first step in this direction, with
a precision improvement granted by the $\sqrtsNN$
increase by about a factor two, to 5~TeV, and by a similar 
increase in instantaneous luminosity.
 In the scope of the ALICE
experiment, we have discussed
how this improvement will 
allow turning a number of, partly-qualitative, observations,
some of which were made for the first time at the LHC,
into quantitative measurements of QGP effects. Examples of these
observations include the elliptic flow of D mesons
and of J/$\psi$'s, the production of $\rm D_s^+$ and $\psi$(2S), 
and the measurements of the attenuation and structure of jets.

To provide a strong boost towards high-precision studies of
the QGP properties, the ALICE Collaboration is preparing a major
detector upgrade. This will be implemented
during the Long Shutdown 2 in 2019--20 with the goal of collecting
several samples of Pb--Pb collisions at $\sqrtsNN=5.5~\tev$ during the LHC Runs 3 and 4 
for a total integrated luminosity of 13~nb$^{-1}$, in addition to
proton--proton reference samples and data with p--Pb and, possibly,
lighter ion collisions.
The ALICE upgrade strategy entails two
main items:
\begin{itemize}
\item the upgrade of the readout of most of the detectors and a new
  online-offline system for data compression will enable recording all Pb--Pb interactions with a minimum-bias
  trigger up to the maximum projected interaction rate of 50~kHz; this will provide 
an increase by two orders of magnitude of the sample of minimum-bias
events with respect to that expected in Run-2;  
\item a large improvement of the track reconstruction precision and
  efficiency; 
  the new Inner Tracking System at central rapidity provides a
  precision improvement by a factor more than three for the
  reconstruction of heavy-flavour decay vertices, and the Muon Forward
  Tracker adds such reconstruction capability for muon-based
  measurements at forward rapidity.
\end{itemize}

In this chapter we have discussed the prospects for the physics topics
that guided the upgrade design (mainly, heavy flavour,
charmonium and low-mass dilepton studies) and for those that will 
substantially benefit of the improved detector capabilities. 
In summary:
\begin{itemize}
\item Measurements with unprecedented precision of the production and azimuthal anisotropy of
  heavy-flavour mesons and
  baryons over a broad momentum range, as well as of $b$-jets, will enable a detailed
  characterisation of the colour-charge and mass dependence of
  in-medium energy loss, of the transport and possible thermalization
  of heavy quarks in the QGP, and of their hadronization mechanism in a
  partonic environment.
\item High-precision measurements, starting from zero $\pt$ both at central and forward rapidity, of the production and
  azimuthal anisotropy of the J/$\psi$, $\psi$(2S) and $\Upsilon$
  states will provide stringent tests of quarkonium dissociation
  and regeneration as probes of deconfinement and of the QGP
  temperature.
\item First precise measurements of low-mass dileptons at LHC
  energies will carry information on the QGP temperature 
and equation of state, as well as on the chiral nature of the
phase transition in the vanishing baryon chemical potential regime.
\item Multi-differential measurements of jet production and
  properties, including flavour-dependent fragmentation functions,
  will add unique information on the parton energy loss mechanism.
\item The large increase in minimum-bias integrated luminosity 
and the extended rapidity coverage with the new inner trackers will
enhance the precision and the spectrum of multi-particle 
correlation measurements and their sensitivity to the initial
conditions in small and large colliding systems and to the medium response (in terms of, e.g., shear-viscosity).
\item The production of light nuclei, antinuclei and hypernuclei
 will be measured with unprecedented precision, enabling, among others, 
searches for yet-unobserved states and new tests of the CPT invariance.
\end{itemize}


\section{Low-energy frontier: a new dimuon experiment at the SPS}
\label{sec:na60plus}

\subsection{Exploring the structure of the phase diagram of strongly-interacting matter: chiral-symmetry restoration and the onset of deconfinement}


As discussed in previous Sections, our quantitative understanding of the QCD phase diagram is largely restricted to the region of low baryochemical potential $\mu_B$. For $\mu_B\sim0$, lattice QCD provides quantitative results (see Fig.~\ref{fig:QCDEOSmu}): a fast increase of $\epsilon/T^4$ ($\epsilon$ = energy density) 
occurs around  a critical temperature $T_c\approx 155$~MeV. In this regime, the phase transition is a cross-over~\cite{Borsanyi:2013bia}.

Also the extensive experimental campaigns conducted at the CERN SPS, RHIC and LHC accelerators have mostly explored so far this region of the phase diagram at low $\mu_B$, showing 
that a deconfined state of matter is produced in heavy ion collisions at high energies, with properties consistent with the predictions of lattice QCD. 

However, there are basic 
aspects of the phase diagram structure not yet understood:

\begin{itemize}
\item  In vacuum, the light hadron masses are largely due to the spontaneous breaking of QCD chiral symmetry rather than to the Higgs boson coupling. 
This is illustrated in the left panel of Fig.~\ref{fig:fig4}.
At the hadron-quark gluon plasma phase boundary chiral 
symmetry should be restored. This will imply a change in the hadron mass spectrum, but how this is realised is not known.
\item For moderate temperatures and high baryon densities the existence of a first order transition with co-existence of a  mixed-phase was suggested. The first order transition 
line should end with a second order critical point. No measurement has confirmed this scenario yet. 
\end{itemize}

At a theoretical level, the location  of the critical point is not well defined. Negative experimental evidence at the maximum CERN SPS energy of 160~GeV/nucleon suggests that it should be found for $\mu_B>250$--300~MeV. This is qualitatively shown in Fig.~\ref{fig:phasediagexp} of Section~\ref{sec:expfuture}.
Thus, nucleus--nucleus collisions at energies below the maximum SPS energy (160~GeV/nucleon) should be able to reach the 
baryochemical region of potential interest allowing for the exploration of chiral symmetry restoration and the first-order phase transition.

This Section outlines an experimental strategy to address these open points by performing novel high precision measurements
of muon pair production with a fixed target experiment operated at different energies below 160~GeV/nucleon at the  CERN SPS (low beam energy  scan). 
By  increasing the collision energy, initial states of matter can be produced at different points in the diagram finally crossing the phase transition line with the onset of  deconfinement. 
  
The pillars of the strategy are:

\begin{enumerate}
\item Chiral symmetry restoration:
\begin{itemize}
\item The doublet of the vector meson $\rho$ and its axial vector partner $a_1$, split in vacuum due to chiral symmetry breaking, should
become degenerate at chiral restoration. We propose to measure the $a_1$ production and spectral function via the process
$\pi^0+a_1\to \mu^+\mu^-$, which occurs in the hadronic medium and populates the dilepton invariant mass region $1 < M < 1.5$~GeV. This
would be the first measurement ever related to the $a_1$ to address chiral symmetry restoration.
\item Measurement of the modification of the production of open charm D mesons. At chiral restoration the threshold for
production of a $\mathrm{D}\mathrm{\overline D}$ pair may be reduced, leading to an enhancement of their production by a large factor.
\end{itemize}
\item Onset of deconfinement:
\begin{itemize}
\item Measurement of the strongly interacting matter caloric curve: temperature vs. energy density. This is performed  by measuring the temperature from fits of the dimuon mass spectrum above 1.5~GeV at different collision energies. A caloric curve to study the QCD phase diagram was never determined so far.
\item Measurement of QGP yield at different collision energies:
 performed by measuring the effective temperature extracted from dimuon transverse momentum spectra at different collision energies. This measurement, performed only at full SPS energy (160~GeV/nucleon), would be extremely sensitive to the onset of deconfinement.
\item Measurement of the fireball lifetime: performed by a precise measurement of the $\rho$ yield. An anomalous behaviour of the lifetime vs. hadronic transverse momentum spectra  would be very helpful to corroborate findings from other measurements.
\item Measurement of the charmonium states (J$/\psi$, $\psi(2\mathrm{S})$, $\chi_c$), as a function of collision energy and centrality. This measurement
is sensitive to the melting of charmonia, occurring at different temperatures for the different states, and is due to colour
screening in a deconfined phase. In addition, the ratio of J$/\psi$ and open charm yields might exhibit a drop at the onset of
deconfinement.
\end{itemize}

\end{enumerate}

\subsection{Chiral symmetry restoration}

\subsubsection{ $\rho$--$a_1$ mixing} 
As discussed above, chiral symmetry is spontaneously broken in the hadronic world. The contribution to the light quark masses due to Higgs coupling and to chiral symmetry 
breaking in QCD is shown in Fig.~\ref{fig:fig4}-left. Chiral symmetry breaking leads to the observed meson mass spectrum, removing in particular the degeneracy between the 
vector $\rho$ and axial-vector $a_1$ mesons (see Fig.~\ref{fig:fig4}  central). Fig.~\ref{fig:fig4}-right shows the  Aleph measurements of the vector and axial-vector 
spectral functions in vacuum~\cite{Barate:1998uf}.

\begin{figure}[t]
\begin{center}
\includegraphics[width=16.5cm]{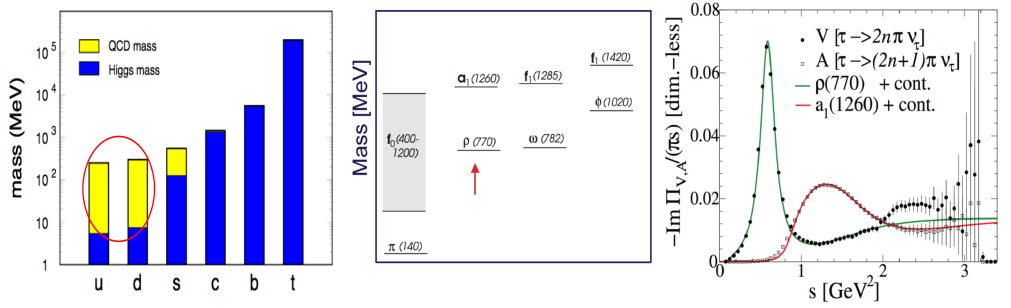}
\caption{Left: contribution to the light quark masses due to Higgs coupling and to chiral symmetry breaking in QCD; middle: the vector axial-vector meson mass spectrum; right: vector and axial-vector spectral functions in vacuum~\cite{Barate:1998uf}.}
\label{fig:fig4}
\end{center}
\end{figure}

At the phase transition boundary, it is expected that chiral symmetry is restored. The order parameter for chiral symmetry breaking is the quark-antiquark condensate (chiral 
condensate), which is non-zero in vacuum. Lattice QCD calculations for $\mu_B=0$ show that at the phase transition boundary the chiral condensate steeply decreases 
around $T_c$, indicating that chiral symmetry is restored (see Fig.~\ref{fig:fig5}-left)~\cite{Borsanyi:2010bp}. The Weinberg sum-rule~\cite{Weinberg:1967kj}, which remains valid in the medium, relates the difference of the vector and axial-vector spectral 
function to the chiral condensate: 

\begin{equation}
\int{\frac{ds}{\pi}(\rho_V-\rho_A)} = - m_q \langle \overline qq \rangle.
\end{equation}

Thus, the melting of the chiral condensate must imply that the vector meson $\rho$ and its axial-vector partner $a_1$ must become degenerate. 
Fig.~\ref{fig:fig5} displays qualitatively two possibilities, known as dropping mass and melting resonance scenarios.

\begin{figure}[t]
\begin{center}
\includegraphics[width=12.cm]{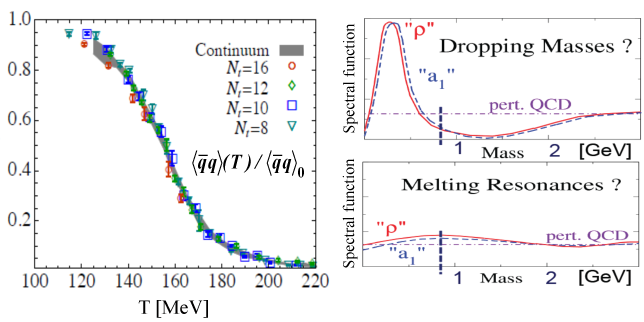}
\caption{Left:  evolution of the chiral condensate as a function of the medium temperature for $\mu_B=0$; right: possible scenarios for vector and axial-vector spectral functions when chiral symmetry is restored.}
\label{fig:fig5}
\end{center}
%
\begin{center}
\includegraphics[width=6.5cm]{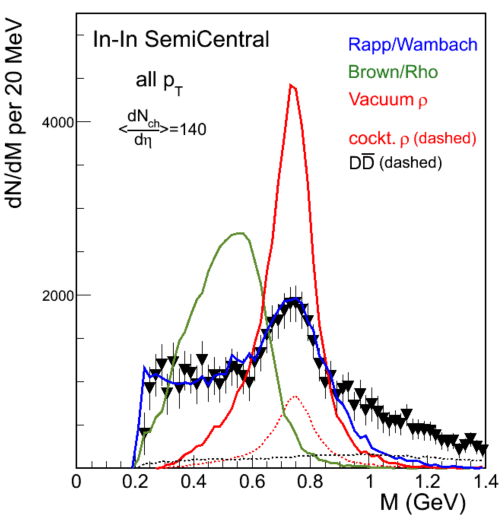}
\includegraphics[width=6.85cm]{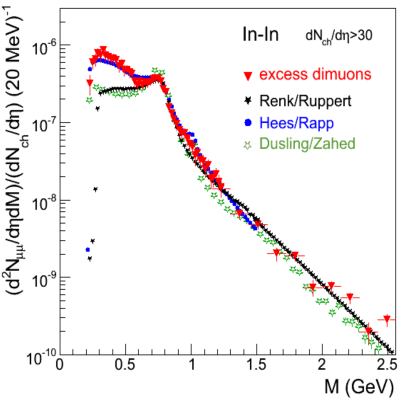}
\caption{Left: Thermal dimuon mass spectrum in In--In collisions compared to theoretical predictions, renormalised to the data in the mass interval $M<0.9$~GeV~\cite{Arnaldi:2006jq}. No acceptance correction applied. Right: Thermal dimuon mass spectrum in In--In collisions at full SPS energy~\cite{Arnaldi:2008er,Specht:2010xu}. Data corrected for acceptance and all background sources subtracted; see text for details.}
\label{fig:fig6}
\end{center}
\end{figure}

In the fireball medium produced in a ultra-relativistic heavy-ion collision, the broad $\rho(770)$ is by far the most important among the vector mesons, due to its strong coupling 
to the $\pi^+\pi^-$ channel and its life time of only 1.3~fm, making it subject to regeneration in the much longer-lived fireball. 
Changes both in width and in mass were suggested as a signature of the chiral 
transition.

The $\rho$ spectral function has been precisely measured in in Indium-Indium collisions at 160~GeV/nucleon by the NA60 experiment at the CERN SPS.
The dilepton mass spectrum, after subtraction of the $\eta$, $\omega$ and $\phi$ contributions and before acceptance correction, is shown in the left panel of Fig.~\ref{fig:fig6} ~\cite{Arnaldi:2006jq}. For $M<1$~GeV (Low Mass Region), this spectrum is dominated by the $\rho$, directly reflecting  the space-time averaged $\rho$ spectral function.
The data are compared to the two main theoretical scenarios developed historically for the
in-medium spectral properties of the $\rho$, dropping mass~\cite{Brown:1991kk,Li:1995qm,Brown:2001nh} and broadening without mass shift~\cite{Rapp:1999us,vanHees:2007th}. The dropping mass - which related the mass shift directly to the decrease of the chiral condensate - 
is ruled-out by the data. The broadening scenario, based on the hadronic many body 
model (see Section~\ref{sec:EMtheory}), successfully describe the data. This broadening is driven by the total baryon density. Recent theoretical investigation found that this broadening is 
consistent with chiral symmetry restoration (``melting of the $\rho$'')~\cite{Hohler:2013eba}.

On the other hand, no measurements exist for the axial-vector partner $a_1$ in high energy nuclear collisions.
Referring to Section~\ref{sec:EMtheory}, the dilepton mass range $1.1<M<3$~GeV (Intermediate Mass Region - IMR) is populated by radiation emitted in the quark-gluon plasma  via quark-antiquark 
annihilation and/or by radiation emitted at a later stage from the hadronic medium due to multi-pion processes. More specifically, the dilepton radiation emitted from the 
hadronic phase in the mass range $1.1<M<1.5$~GeV is dominated by the process $a_1\pi^0\to\mu^+\mu^-$, reflecting $\rho$--$a_1$ mixing which provides a direct link to chiral symmetry 
restoration~\cite{Dey:1990ba,Steele:1996su}. It is not possible to study this process in the data of Fig.~\ref{fig:fig6} - at this energy, the radiation emitted from the QGP phase overwhelms the $\pi^0 a_1$ yield (see Fig.~\ref{fig:ThermalRadiation} of Section~\ref{sec:EMtheory} for more details).
Thus, this measurement is best performed at a collision energy where the initial state is close to the phase-boundary to maximize the sensitivity to chiral restoration,
decreasing the contribution from the QGP at a negligible level.
This is expected to occur at beam energies below full SPS (see also next section), and the decrease of the background sources from the Drell–Yan and open-charm processes which also populate 
this mass region would further facilitate this delicate measurement. It is important to notice that the large QGP and open charm yields prevent completely any study of the $\rho$--$a_1$ mixing at topmost RHIC or LHC energies.

\subsubsection{Open charm production}
When chiral symmetry is
restored, the melting of the $\langle q\overline q\rangle $  condensate might also shift the threshold for $\mathrm{D}\mathrm{\overline D}$ production from about 3.73~GeV in
vacuum to about 3~GeV in the chirally-symmetric medium. This reduction is predicted to lead to an enhancement of open
charm meson yields by a factor of up to 7, with respect to binary scaling of the production in p--A collisions. Thus, a large
enhancement of D meson production is regarded as a strong medium effect on the charmed hadrons, and as a signature for
a chirally-symmetric phase~\cite{Friman:2011zz}.

\subsection{Onset of deconfinement and the nature of phase transition}

\subsubsection{Measurement of the strongly interacting matter caloric curve}
\label{sec:HeatingCurve}

Heavy-ion physics at the Fermi energy scale  explored the phase
diagram of nuclear matter (Fig.~\ref{fig:fig7}-left) to establish the properties of the phase transition from
the liquid self-bound ground state to a gas of free nucleons~\cite{D'Agostino:2005qj}.
The experimental measurement of a caloric curve, indicating that the chemical (isotopic) temperature saturates over
a broad range of excitation energies (Fig.~\ref{fig:fig7}-middle) was presented as evidence for
a phase transition in nuclei~\cite{D'Agostino:2005qj}.
In a given phase, the temperature 
increases in a monotonic way. When the system goes across the phase transition, a mixed phase occurs where the temperature remains constant during the phase transformation.

\begin{figure}[t]
\begin{center}
\includegraphics[width=16.5cm]{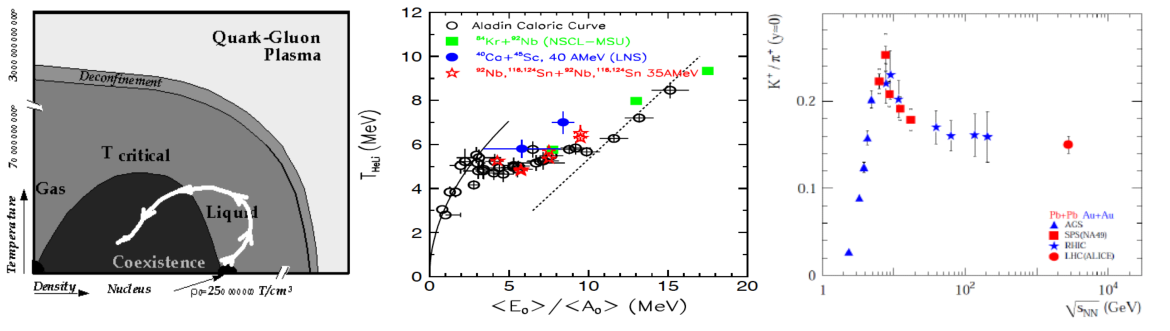}
\caption{Left: Zoom of phase diagram of nuclear matter in the liquid-gas phase transition region; middle: chemical (isotopic) temperature temperature-excitation energy correlation (caloric curve)~\cite{D'Agostino:2005qj}; right:  particle ratio $\mathrm{K}^+/\pi^+$ in central lead--lead collisions as a function of energy measured by NA49~\cite{Alt:2007aa}.}
\label{fig:fig7}
\end{center}
\end{figure}

In the exploration of the first order QCD phase transition between the hadron gas and the QGP, a similar approach based on the measurement of a caloric curve, which requires a  precise temperature measurement, could not be followed so far.

The strategy followed up to now to search for the onset of deconfinement was based
 on the study of the evolution of hadronic observables as a function of beam energy.  
Anomalies in the energy dependence of hadron production properties might be related to the transition between different phases of the strongly interacting matter 
created at the early stage of collisions. Along these line, the NA49  experiment at the CERN SPS studied the particle ratio $\mathrm{K}^+/\pi^+$ in central Pb--Pb collisions as a function of energy 
(Fig.~\ref{fig:fig7}-right)~\cite{Alt:2007aa}. The ratio passes through a sharp maximum followed by a plateau not observed in proton-proton collisions. Within a statistical model the maximum and 
the plateau were interpreted as the decrease in the ratio of strange to non-strange number of degrees of freedom when deconfinement sets in~\cite{Gazdzicki:1995ze,Gazdzicki:1998vd}.	
At present, there is no consensus on this interpretation, being model dependent. Nevertheless, though still controversial, this measurement is intriguing and the energy interval across the maximum is worth to be investigated in detail.

A new horizon for identifying where the phase transition occurs in the phase diagram would be set if the medium temperature could be measured as a function of the 
energy density, thus truly measuring a caloric curve. A structure in the caloric curve like a plateau would be revealing of the order of the phase transition. 
In the following  we outline our proposed  method for measuring $T$.

The acceptance corrected thermal dimuon spectrum measured at full SPS energy is shown in Fig.~\ref{fig:fig6}-right ~\cite{Arnaldi:2008er,Specht:2010xu}. For masses above 1.5~GeV, the
continuum of overlapping resonances leads to a flattened spectral density corresponding to a simpler description in terms of
quarks and gluons (hadron-parton duality). The space–time averaged mass spectrum is then approximately 
described by $\dd N/\dd M\propto M^{3/2}\exp(-M/T)$ (see Section~\ref{sec:EMtheory}), and the average temperature of the emitting sources can directly be extracted by a fit of the mass spectrum~\cite{Specht:2010xu}. Since 
mass is by construction a Lorentz-invariant, the mass spectrum is immune to any motion of the emitting sources, unlike transverse momentum spectra. 
The parameter $T$ in the spectral shape of the mass spectrum is therefore the true thermal temperature. 
At the full SPS energy of 160~GeV/nucleon, the fit of the spectrum of Fig.~\ref{fig:fig6} gives $T=205\pm12$~MeV\cite{Arnaldi:2008er,Specht:2010xu}. This is above $T_c$, thus showing that the QGP is already 
produced at this collision energy.
 
This new experimental program  proposes to locate where the phase transition occurs in the phase diagram, with accurate 
temperature measurements obtained from fits of the mass spectrum performed at different collision energies below the full SPS energy of 160~GeV. The temperature will be 
correlated with the energy density, which is experimentally inferred from the charged particle multiplicity density through Bjorken scaling or more advanced models~\cite{Csorgo:2008pe}.

\subsubsection{Disentangling hadronic and partonic emission}

While historically the interest has largely focused on dilepton mass $M$, transverse momentum $\pt$ or transverse mass $m_{\rm T}=(\pt^2+M^2)^{1/2}$ contain not only 
contributions from the spectral function, but encode the key properties of the expanding fireball, temperature and in particular transverse (radial) flow. In the description of 
hadron $\pt$ spectra, the study of collective flow has contributed decisively to the present understanding of the fireball dynamics in nuclear collisions. However, while hadrons 
always receive the full asymptotic flow reached at the moment of decoupling from the flowing medium, lepton pairs are continuously emitted during the evolution, sensing small 
flow and high temperature at early times, and increasingly larger flow and smaller temperatures at later times. The resulting space-time folding over the temperature-flow history 
can be turned into a diagnostic tool: the measurement of $\pt$ spectra of lepton pairs potentially offers access to their emission region and may thereby differentiate between a 
hadronic and a partonic nature of the emitting source.

Thermal transverse mass spectra for different dimuon mass bins in In--In collisions at full SPS energy are shown in Fig.~\ref{fig:fig10}-left~\cite{Arnaldi:2007ru}. The spectra have a thermal form 
$(1/m_{\rm T}){\rm d}N/{\rm d}m_{\rm T}\sim\exp(-m_{\rm T}/T_{\rm eff})$,

%
%
where the parameter $T_{\rm eff}$ can roughly be described by a superposition of a thermal and a flow part in the form $T_{\rm eff}=T+M<v^2_R>$, $v_R$ being the radial velocity of 
the collective motion. The extracted values of $T_{\rm eff}$ vs. pair mass are summarized in Fig.~\ref{fig:fig10}-right, supplemented by a set of further fit values from narrow slices 
in $M$. Here, the slope parameter $T_{\rm eff}$ rises nearly linearly with mass up to about 270~MeV at the pole position of the $\rho$, followed by a sudden decline to values of 190 – 
200~MeV for masses > 1~GeV. The increase up to $M\sim$1~GeV is a strong evidence for radial flow in the region of thermal dilepton emission dominated by the $\rho$ meson, 
which is maximally coupled to radial flow through pions. 

\begin{figure}[t]
\begin{center}
\includegraphics[width=12.5cm]{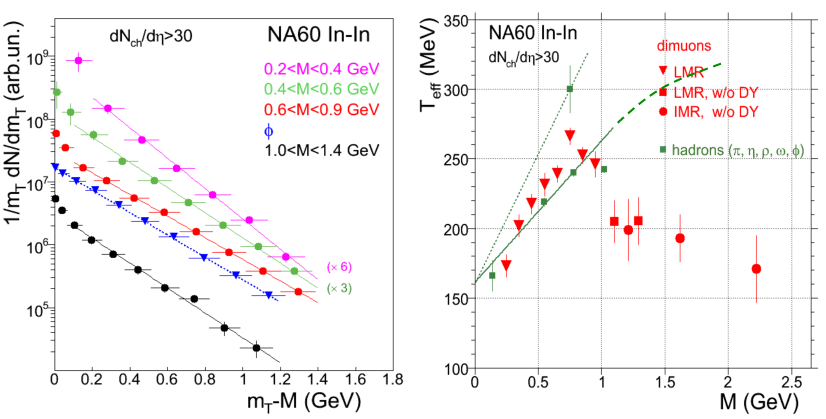}
\caption{Left: Transverse mass spectra of thermal dileptons in In--In collisions at full SPS energy~\cite{Arnaldi:2007ru} for different mass windows summed over centralities (excluding the peripheral bin), in comparison to the $\phi$; Right: $T_{\rm eff}$ of $m_{\rm T}$ spectra vs. $M$.  Dashed green line for $M$>1~GeV: qualitative expected trend of $T_{\rm eff}$ with no QGP below the onset of deconfinement.}
\label{fig:fig10}
\end{center}
\end{figure}

The sudden decline of $T_{\rm eff}$ at masses > 1~GeV is the other most remarkable feature of this data. Extrapolating the lower-mass trend to beyond 1~GeV, a jump by about 50 
MeV down to a low-flow situation is extremely hard to reconcile with emission sources which continue to be of dominantly hadronic origin in this region. If the rise is due to flow, 
the sudden loss of flow is most naturally explained as a transition to a qualitatively different source, implying dominantly early, i.e. partonic processes like quark-antiquark 
annihilation into a muon pair for which flow has not yet built up~\cite{Arnaldi:2007ru}. 

As indicated by the dashed green line in Fig.~\ref{fig:fig10}-right, based on theoretical modeling, the evolution of the pattern of $T_{\rm eff}$ vs. $M$ from low energies below the 
onset of deconfinement towards higher beam energies will be most revealing: thermal radiation from multi-pion processes should exhibit a monotonic increase of $T_{\rm eff}$ vs. 
$M$ so that the onset of deconfinement can be determined with great sensitivity by the appearance of the drop.


\subsubsection{Measurement of the fireball lifetime}

Precise thermal
dilepton measurements are sensitive to the fireball lifetime,
via the total  $\rho$ yield. Fig.~\ref{fig:RhoClock}
shows the total $\rho$ yield normalised to the expected $\rho$ yield
in elementary collisions (bound to the $\omega$ yield by
$\sigma_{\rho}/\sigma_{\omega}=1$) as a function of centrality for
In--In at 160~GeV/nucleon~\cite{Arnaldi:2008fw}. From this, it is
possible to directly extract the so-called $\rho-$clock ``ticking'',
namely the number of $\rho$ generations (production and decay)
produced during the fireball expansion. This clock ticking allowed
the fireball lifetime to be constrained with unprecedented precision:
 $\tau_{FB}=(7\pm1)$~fm/$c$ for In--In collisions
at 160~GeV/nucleon~\cite{Rapp:2014hha}.  

Such a measurement would be important in confirming the presence of a soft mixed phase: 
for increasing collision energy, an increase in $\tau_{FB}$ with identical final-state hadron transverse momentum spectra
(i.e., in terms of radial-flow) would necessarily imply a lifetime extension
without extra collective flow, i.e. a soft phase. This underlines the importance
of a parallel program of hadron measurements, which could be performed by the NA61 experiment at the
CERN SPS and by the RHIC beam energy scan. 

\begin{figure}[!htb]
\centering
\includegraphics[width=0.5\columnwidth]{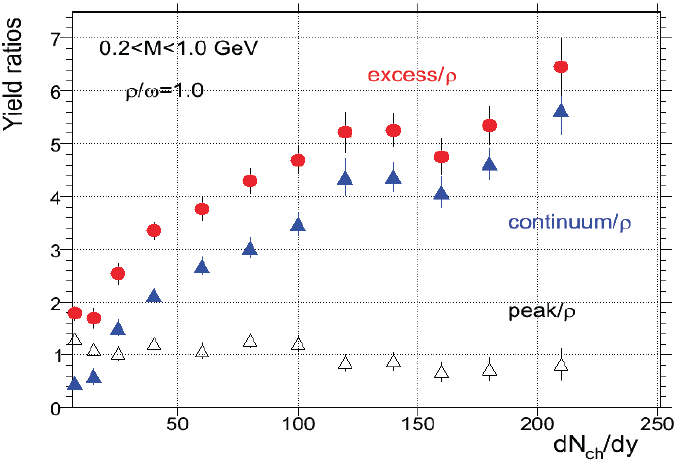}
\caption{ $\rho$ yield in In--In at 160~GeV/nucleon normalized to $\rho$ expected in elementary collisions~\cite{Arnaldi:2008fw}.
Red: total yield ratio. Black: ratio considering only the peak part of the $\rho$ spectrum. Red: ratio considering only the continuum part of the $\rho$ spectrum. The $\rho$ spectrum is displayed in Fig.~\ref{fig:fig6} .}
\label{fig:RhoClock}
\end{figure}

\subsubsection{Charmonium studies}

Charmonium production was the first hard probe to be studied in the frame of ultra-relativistic heavy-ion collisions. The seminal paper of 1986 by Matsui and Satz~\cite{Matsui:1986dk} predicted the 
suppression of the J/$\psi$ meson as an unambiguous signature of the formation of a Quark-Gluon Plasma, due to the screening of the colour binding in a deconfined medium. 

At top SPS energy, after a pioneering phase (O- and S-induced collisions) in the frame of the NA38 experiment~\cite{Baglin:1990iv,Abreu:1998if}, detailed experimental results on charmonium in \mbox{Pb--Pb} 
and \mbox{In--In} collisions were obtained by the NA50 and NA60 experiments, respectively~\cite{Alessandro:2004ap,Arnaldi:2007zz}. Concerning the J/$\psi$, a suppression of this meson was established in 
\mbox{Pb--Pb} collisions with respect to dimuon production from the Drell-Yan process, which, being electromagnetic, is not affected by QGP formation. This suppression was 
also found to exceed the expected effects of J/$\psi$ dissociation in cold nuclear matter, which were investigated via the study of \mbox{p-A} collisions at the same energy and 
in the same kinematic region of \mbox{Pb--Pb} data~\cite{Arnaldi:2010ky}. More in detail, the results (see Fig.~\ref{fig:na60psi1}(left)) showed that for $N_{\rm part}\gtrsim 200$, in the rapidity region 
$0<y_{\rm cms}<1$, a suppression of the J/$\psi$ yield with respect to expectations from suppression in cold nuclear matter, was present. Similar studies in \mbox{In--In} 
collisions, where the maximum $N_{\rm part}$ is limited to $\sim 200$, did not show a significant suppression in such region, consistently with \mbox{Pb--Pb} observations. One 
must note that the maximum effect observed ($\sim$30\%) qualitatively corresponds to the fraction of produced J/$\psi$ which comes from decays of the less bound $\chi_{\rm 
c}$ and $\psi(2\mathrm{S})$ states. Lattice-based calculations show that the latter states are expected to melt close to the critical temperature $T_{\rm c}$, contrary to the J/$\psi$ which 
should survive at least up to $T\sim 1.5 T_{\rm c}$~\cite{Adare:2014hje}. Therefore, the observations at top SPS energy suggest that the medium is hot enough to dissociate the $\chi_{\rm c}$ and $
\psi(2\mathrm{S})$ states, but not the more tightly bound J/$\psi$. 

Results on the more loosely bound $\psi(2\mathrm{S})$ state were also obtained by NA50~\cite{Alessandro:2006ju}, showing the ratio between the $\psi(2\mathrm{S})$ and J/$\psi$ yields to steadily decrease, by a factor 
$\sim2.5$ from peripheral to central \mbox{Pb--Pb} collisions (see Fig.~\ref{fig:na60psi1}(right)). Such a larger suppression for the $\psi(2\mathrm{S})$ can indeed be expected 
considering its weak binding energy (the state lies only $\sim$50~MeV below the threshold for open charm production).
 
\begin{figure}[t]
\centering
\resizebox{0.44\textwidth}{!}
{\includegraphics{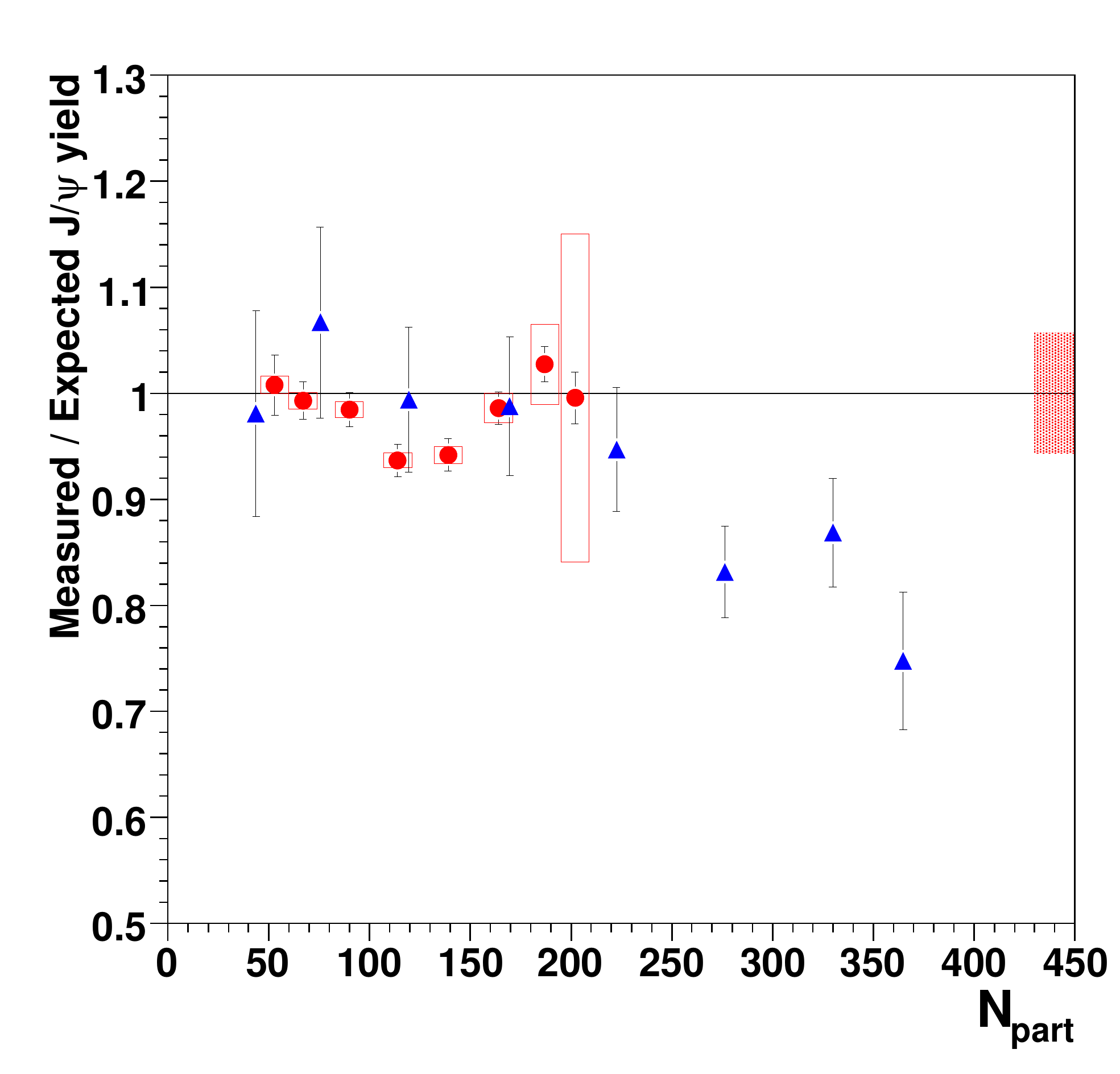}}
\resizebox{0.45\textwidth}{!}
{\includegraphics{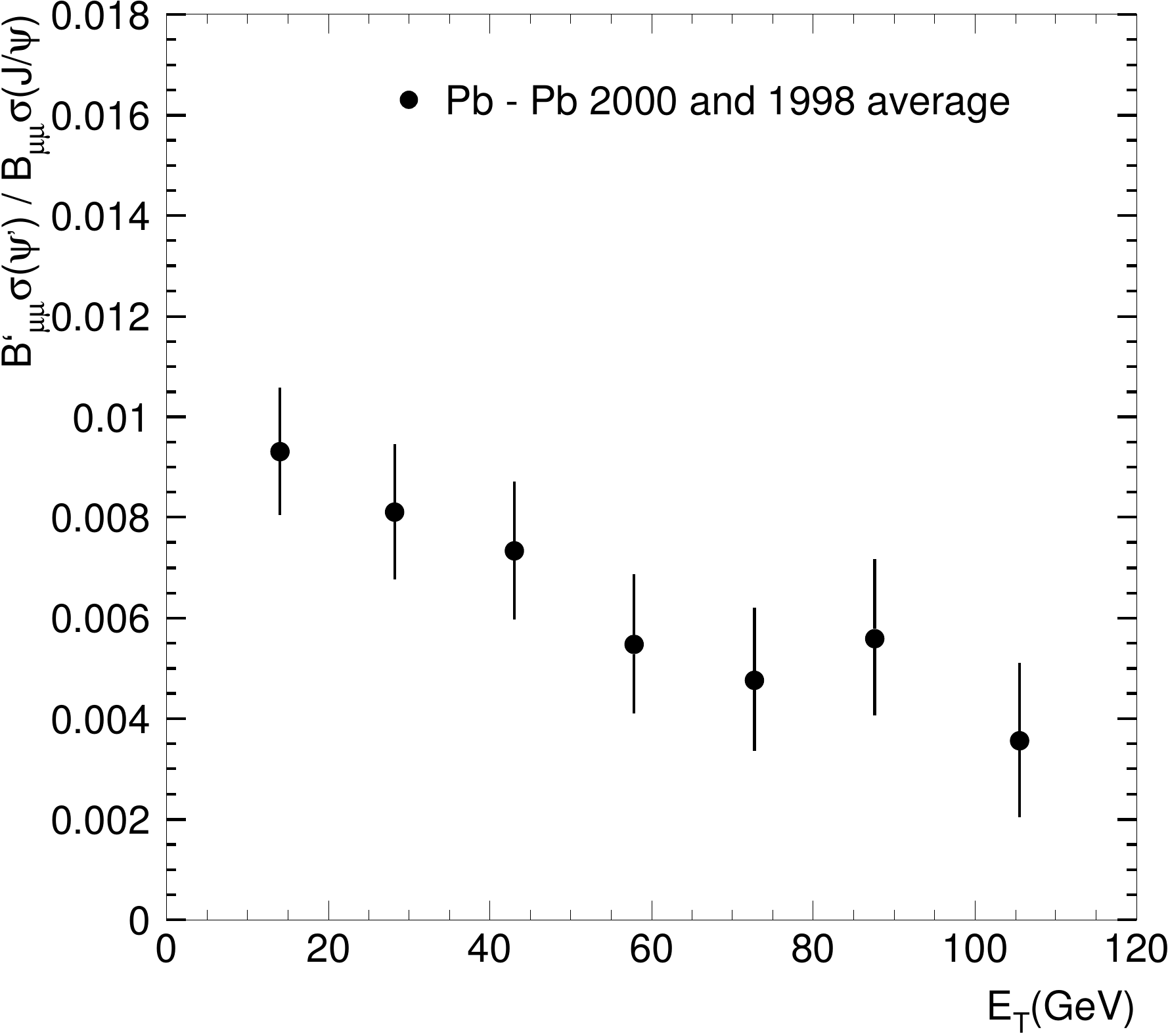}}
\caption{Left: anomalous \jpsi\ suppression in In--In (circles), measured by the NA60 experiment, and Pb--Pb collisions (triangles), measured by the NA50 experiment, as a function of $N_{\rm part}$, from~\cite{Scomparin:2009tg}. The boxes around the In--In points represent 
correlated systematic errors. The filled box on the right corresponds to the uncertainty in the absolute normalization of the In--In {\it points}. A 12\% global error, due to the 
uncertainty on cold nuclear matter effects, is not shown.  Right:$B^{\psi(2\mathrm{S})}_{\mu\mu}\sigma(\psi(2\mathrm{S}))/B^{{\rm J}/\psi}_{\mu\mu}\sigma({\rm J}/\psi)$ as a function of the 
transverse energy $E_{\rm T}$, combining the Pb--Pb 1998 and 2000 data samples., from~\cite{Alessandro:2006ju} Errors are the quadratic sum of statistical and systematic uncertainties.}
\label{fig:na60psi1}
\end{figure}

Charmonium production in heavy-ion collision has never been studied below top SPS energies (160~GeV/nucleon Pb beams, corresponding to $\sqrtsNN=17.3$~GeV). 
The extension of the charmonium studies towards lower SPS energies (down to $\sim$40~GeV/nucleon, corresponding to $\sqrtsNN\sim 8$~GeV) is very attractive, as it 
offers the possibility of investigating the onset of the so-called anomalous J/$\psi$ suppression. Such a terminology was introduced by the NA50 collaboration, referring to the 
fraction of the observed suppression in nuclear collisions that exceeds cold nuclear matter effects. 
A clear observation of the onset of the anomalous suppression in nucleus-nucleus collisions at low/intermediate SPS energy would represent a strong indication for the onset 
of deconfinement itself, as at least the $\chi_{\rm c}$ and $\psi(2\mathrm{S})$ states are expected to melt at $T\sim T_{\rm c}$. 
Clearly, once the effect of the anomalous suppression becomes small, it is mandatory to have a good control of other suppression mechanisms that may affect charmonium 
states. One of the most important suppression effects is expected to come from the break-up of charmonium states by the projectile and target nucleons. At SPS energy this 
effect is much more important than at collider energies since, at least around mid-rapidity, the $c\overline c$ pair is expected to spend a significant amount of time, and therefore 
to form the final-state resonance, inside the nucleus itself~\cite{Lourenco:2008sk}. A comparison of the J/$\psi$ production cross sections measured by NA60 in \mbox{p-A} collisions at 400 and 158 
GeV incident proton momentum, shown in Fig.~\ref{fig:na60psi2}, indicates that the data are compatible with a break-up cross section of the observed J/$\psi$ which increases 
from $\sigma^{\rm abs}_{{\rm J}/\psi}{\rm (400~GeV)}= 4.3\pm 0.8{\rm (stat)}\pm 0.6{\rm (syst)}$ mb to $\sigma^{\rm abs}_{{\rm J}/\psi}{\rm (158~GeV)}= 7.6\pm 0.7{\rm (stat)}\pm 
0.6{\rm (syst)}$ mb.

\begin{figure}[t]
\centering
\resizebox{0.44\textwidth}{!}
{\includegraphics{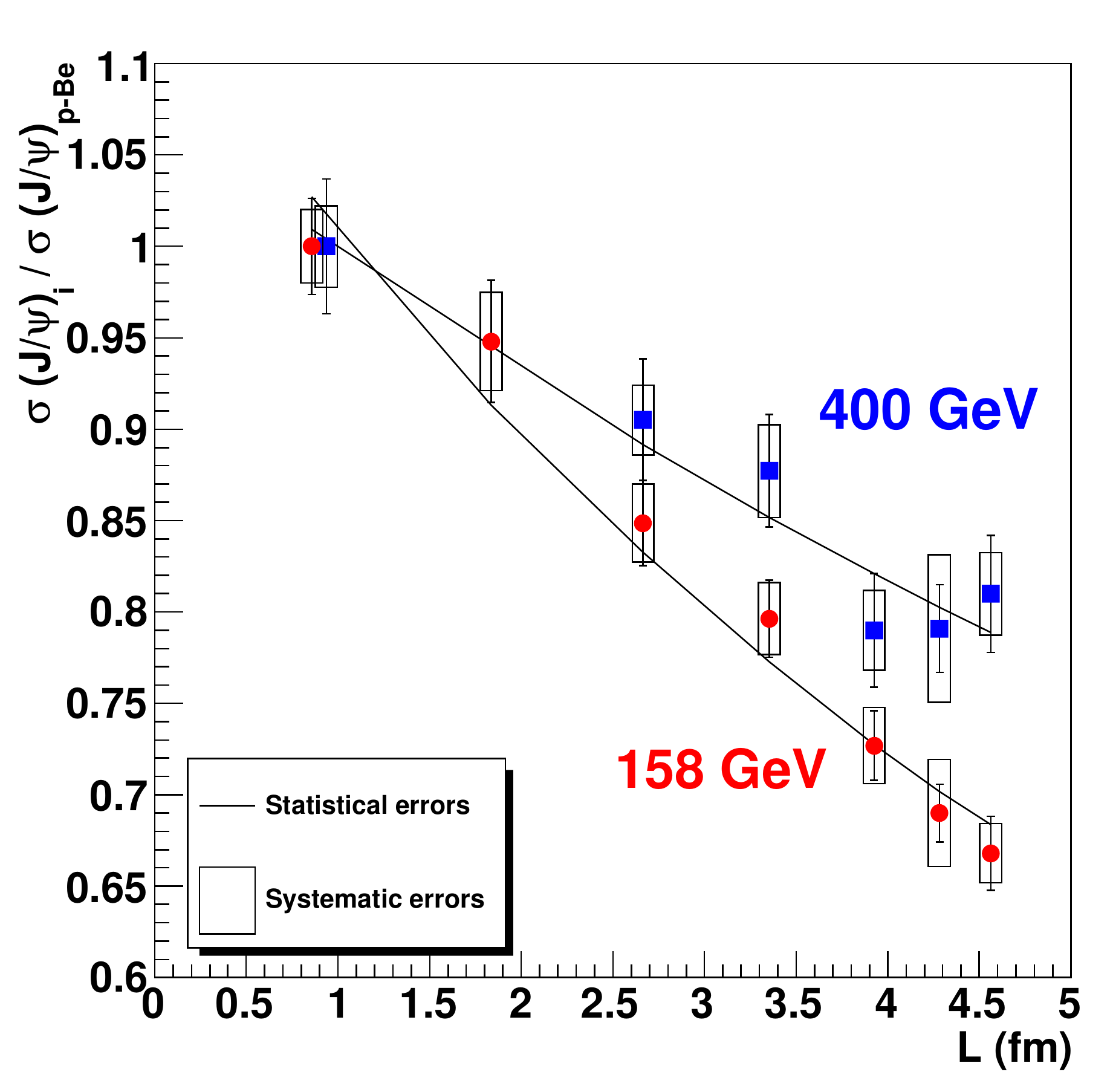}}
\caption{Cross sections measured by NA60 for J/$\psi$ production in \mbox{p-A} collisions, normalized to the \mbox{p-Be} J/$\psi$ cross section, from~\cite{Arnaldi:2010ky}. Data are plotted as a function 
of $L$, the mean path of nuclear matter crossed by the $c\overline c$ pair, which is calculated using the Glauber model of the collision geometry~\cite{Miller:2007ri}.}
\label{fig:na60psi2}
\end{figure}

Already at 160~GeV/nucleon, the nuclear break-up effects and the anomalous suppression were of the same order of magnitude. Since at lower energies the anomalous effects are 
expected to decrease, while nuclear breakup is likely to become stronger, a \mbox{p-A} data taking campaign represents a necessary part of the experimental program for 
charmonium studies. In any case, it must be underlined that the investigation of cold nuclear matter effects on charmonium production has a physics interest in itself, beyond its 
use for the calibration of the same effects in nucleus-nucleus collisions. In addition to nuclear break-up, also other mechanisms as parton shadowing in the nucleus~\cite{Eskola:2009uj} and initial/final state energy loss~\cite{Arleo:2012hn} are expected to contribute to the overall cold nuclear matter effects, and the availability of a set of data in a still unexplored energy range may give 
important constraints to the theoretical descriptions.

Another suppression effect, not related to colour screening, but which may become important at low SPS energy, is the charmonium dissociation in the hadronic phase, i.e., after 
the temperature of the system has gone below $T_{\rm c}$. Studies performed for top SPS energy~\cite{Maiani:2004qj} implied that the effect would only be important for the very loosely bound $\psi(2\mathrm{S})$, but relatively less important with respect to nuclear break-up for J/$\psi$ and $\chi_{\rm c}$ state. It is fair to say that such an issue was not definitely settled and 
further input from  theory will be necessary on this specific point. In particular, when the collision energy decreases, the contribution of baryonic matter becomes increasingly 
important, and an evaluation of baryonic suppression in the hadronic phase has to be included in the calculations.

Finally, in the proximity of $T_{\rm c}$, the presence of a strong coupling regime may lead to enhanced suppression effects that would represent a very interesting observation. 
Although this topic remains for the moment speculative, a fine enough energy scan at intermediate/low SPS energy represents an excellent way to test effects related to the 
onset of deconfinement in the quarkonium sector. 

\begin{figure}[t]
\centering
\resizebox{0.48\textwidth}{!}
{\includegraphics{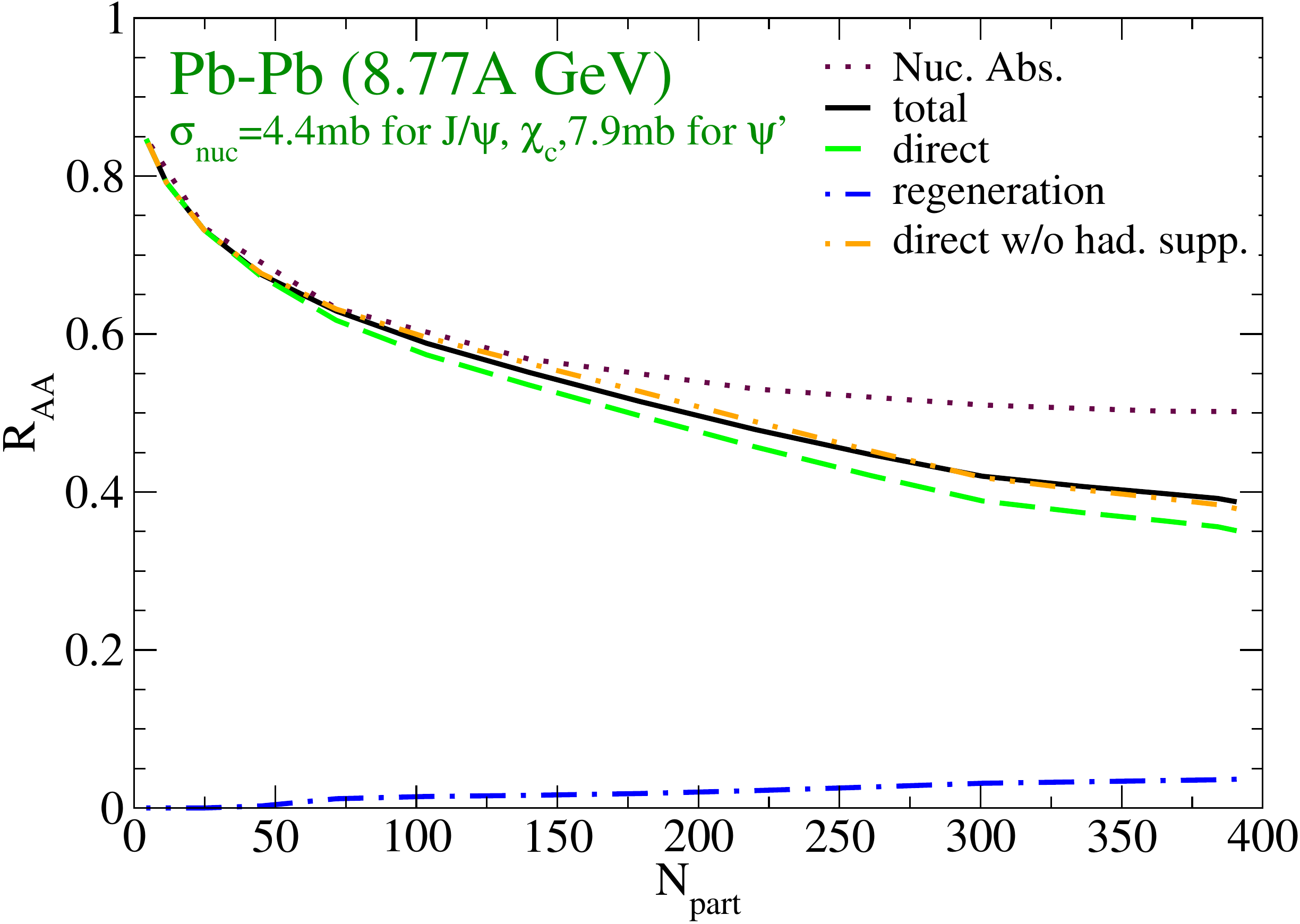}}
\resizebox{0.48\textwidth}{!}
{\includegraphics{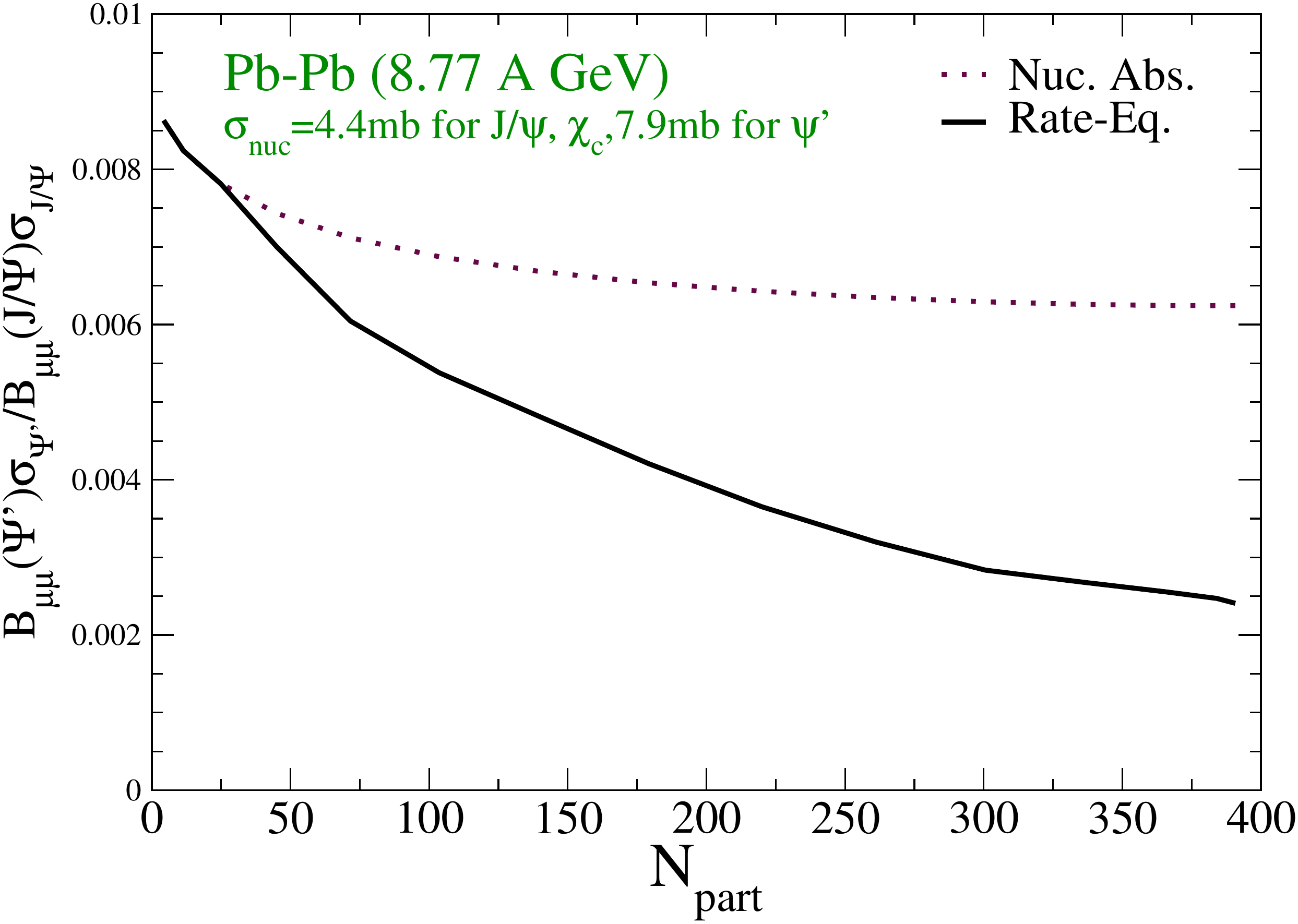}}
\caption{Left: the J/$\psi$ nuclear modification factor, calculated for Au-Au collisions at $\sqrtsNN=8.77$~GeV.  Right: the ratio between the $\psi(2\mathrm{S})$ and J/$\psi$ yields, for the same system and energy. From R.~Rapp, priv. comm. and ref.~\cite{Zhao:2010ti}.}
\label{fig:na60psi3}
\end{figure}

Few quantitative calculations exist for this energy domain. As an example we plot in Fig.~\ref{fig:na60psi3} the results of a calculation carried out for \mbox{Au-Au} collisions at 
$\sqrtsNN=8.77$~GeV (corresponding to $E_{\rm lab}=40$~GeV)~\cite{Zhao:2010ti}. For the J/$\psi$, the ``total'' nuclear modification factor was obtained taking into account the J/$\psi$ 
break-up effect in nuclear matter, the suppression both in the QGP and in the hadronic phase and also a (small) contribution from J/$\psi$ regeneration in the medium. Although 
the choice of some of the input parameters  should be considered as tentative (as an example the value $\sigma^{\rm abs}_{{\rm J}/\psi}=4.4$ mb is possibly small, seen the 
larger break-up cross section observed at $E_{\rm lab}=158$~GeV), two main features emerge, namely the dominance of the suppression in the QGP with respect to the 
hadronic phase, and the relevance of the cold nuclear matter effects in the overall suppression. The ratio between the $\psi(2\mathrm{S})$ and J/$\psi$ yields, at the same energy, is also 
shown in Fig.~\ref{fig:na60psi3}, and a much stronger effect of the medium on $\psi(2\mathrm{S})$, increasing with centrality, is clearly visible.

Finally, the measurement of the ratio J$/\psi / (\mathrm{D}+\mathrm{\overline D})$  as a function of beam energy might be a further probe of the onset of deconfinement~\cite{Friman:2011zz}. This ratio is expected to exhibit a drop at the onset of deconfinement, as a consequence of the J$/\psi$ melting in the QGP, on the
one hand, and of the possible enhancement of $\mathrm{D}-\mathrm{\overline D}$ production in the chirally-symmetric medium, on the other
hand. Therefore, a measurement of the ratio at various collision energies can provide insight on the threshold for
deconfinement, which is complementary to the study of J$/\psi$ suppression described above.

\subsection{Experiment}
\subsubsection{Running conditions and accelerator complex}
The experimental program requires to  collect data at different energies (beam energy scan) lower than full SPS energy of 160~GeV/nucleon ($\sqrtsNN=17$~GeV). As discussed in section~\ref{sec:HeatingCurve}, the NA49 measurements, though controversial, suggest that one should scan down to 10-20~GeV/nucleon at least ($\sqrtsNN=4.5-6$~GeV).

In order to reach a break-through in physics, we have to push forward the apparatus performances and high beam luminosities are needed.
The thermal spectrum from the hadronic phase should be isolated for masses  up to 2~GeV and the measurement of temperature should be performed with an accuracy at the 
level of few~MeV.  J$/\psi$ production must be studied accurately as a function of centrality. 
This requires to collect:
\begin{itemize}
\item
$\sim5\cdot10^7$ reconstructed thermal muon pairs per energy point. The statistics increase at each energy point is a factor 100 over NA60  and $\sim10^5$ over RHIC and LHC experiments. 
\item
 $\sim2$-$3\cdot 10^4$ reconstructed J$/\psi$ events per energy point.
\end{itemize}

No already approved experiment is able to cover such a large energy interval and to collect at the same time the large statistics required for truly quantitative measurements.

In the near future several accelerator complexes might deliver ion beams to perform low energy beam scans. In general terms, since dilepton radiation is produced by rare 
processes,  precision measurements require very large interaction rates. This is in general best accomplished with a fixed target experiment. Among the different accelerators, the CERN SPS appears to be the most suited:
\begin{itemize}
\item CERN SPS. This facility is continuously running since many years. Thanks to the new injection scheme it would be able to deliver  intense ion beams leading to interaction rates exceeding 1 MHz. In addition,
the facility is able to cover a very wide 
energy range $\sqrtsNN\sim$4.5 ($\sim$ 10~GeV/nucleon in the lab system) up to $\sqrtsNN=17.3$~GeV (160~GeV/nucleon in the lab system). An ion beam could be delivered to a fixed target experiment while the SPS is used as ion injector for LHC. Considering the present LHC running conditions with ions, this means that ions would be available for $\sim4$ weeks per year.
Presently the NA61 experiment, with a complementary experimental program, is running at the CERN SPS.

\item
RHIC. This accelerator provides good energy coverage spanning the interval $\sqrtsNN\sim$7.5 - 200~GeV, but the luminosity decreases significantly at smaller energies which leads to an interaction rate of $\sim 1$~kHz,   smaller by  three orders of magnitude or more with respect to the one that can be reached at the  SPS.
 In such conditions,
the STAR experiment, during the phase 2 beam energy scan program,  will target  a minimum bias statistics ranging from $4\times 10^8$ events at $\sqrtsNN
=19.6$~GeV to
$8\times 10^7$ events at $\sqrtsNN=7.7$~GeV. This minimum bias statistics is $\sim3$ orders of magnitude smaller than  the original NA60 dimuon triggered sample, so that there is no possibility to reach the needed precision required for the novel measurements here described. Within STAR, a fixed-target operation mode was also proposed.
However, this would exploit the beam halo and would suffer from the intrinsically slow detectors of the STAR apparatus. For this reason it does not look competitive unless an 
entirely new ad-hoc experiment for dileptons is built.

\item FAIR. The SIS/100 facility is presently approved and should become operational around 2023. It will deliver intense ion beams to the fixed target experiment CBM with interaction rates exceeding 1 MHz.
However, it has a rather limited energy coverage, with a maximum energy of $\sqrtsNN=4.5$~GeV ($\sim$ 10~GeV/nucleon in the lab system) so that it is not sufficient to address properly the experimental program 
outlined in the previous sections. For charmonium, SIS/100 can access a region where the charmonium cross section is very low and even close to 
the kinematical threshold. 
The SIS/300 facility, originally proposed to extend the energy coverage up to $\sqrtsNN=8$~GeV ($\sim$ 35~GeV/nucleon in the lab system), is not approved at present and its operation would in any case only start well beyond 2030. 

\item NICA. JINR aims at building a collider facility reaching $\sqrtsNN=11$~GeV for \mbox{Au-Au} collisions (corresponding to beam energies of 8-70~GeV/nucleon in a fixed target environment). Although the energy domain would have a larger overlap with the SPS, the maximum 
foreseen interaction rate is $\sim 6$ kHz,  remaining anyway smaller by  two orders of magnitude or more with respect to the one that can be reached at the 
SPS.

\end{itemize}

Table~\ref{tab:tab1} summarises the energy coverage and interaction rates at different facilities.

\begin{table}[t]
  \begin{center}
    \begin{tabular}{l l l}
      Facility & Energy range ($\sqrtsNN$)& Interaction rate\\
      \hline
      CERN SPS    & $\sim$4.5-17.3~GeV & > MHz\\
      RHIC               & 7.5-200~GeV & $\sim 1$ kHz \\
      FAIR SIS100  & 2-4.5~GeV & > MHz\\
      NICA JINR     & 4-11~GeV & $\sim6$ kHz
    \end{tabular}
  \end{center}
    \caption{Comparison of different accelerator facilities in terms of energy coverage and interaction rate.}
    \label{tab:tab1}

\end{table}

The proposed experimental program is based on a beam energy scan at the CERN SPS with periods of data-taking at several beam energies in the interval 20 to 160~GeV/nucleon. Tentatively, 
data should be collected at 20-30-40-80-120-160~GeV/nucleon with Pb ions. 
Further energy points might be required depending on the first results. 
The goal is to collect the required statistics at a given energy in a run having 
a duration of 10-15 days. The experimental program requires also to collect p-Pb data at a few energy points.
This beam scan program might be accomplished in $\sim5$ years of data-taking. 

\begin{figure}[t]
\begin{center}
\includegraphics[width=15.5cm]{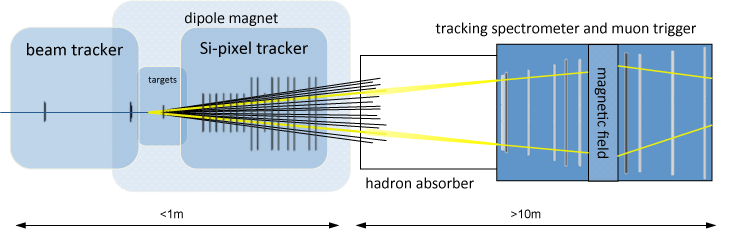}
\caption{Principle of high precision measurement of muons. The uncertainty of the muon track kinematics measured by the muon spectrometer (yellow bands) is drastically reduced by the matching to the tracks measured in the vertex region.}
\label{fig:fig14}
\end{center}
\end{figure}

\subsubsection{Apparatus layout}
\label{sec:NA60applayout}
Traditionally, muons are measured by a magnetic spectrometer placed after a hadron absorber (see Fig.~\ref{fig:fig14}  for an illustration of the muon measurement principle). 
Tracks reaching the muon tracker after the hadron absorber are muon candidates. Thus, the hadron absorber provides the muon identification, but at the cost of degrading the 
kinematics of the muons, because of energy loss fluctuations and multiple scattering. This problem is overcome by measuring particle tracks also before the hadron absorber 
with a silicon tracker, which is the key element for a precision measurement of muons. Muon tracks are then matched to the tracks measured in the silicon vertex spectrometer 
in coordinate and momentum space. In addition, the silicon tracker provides also the measurement of charged-particle multiplicity density, which 
can be used to estimate the energy density of the system via the Bjorken estimate or more advanced models~\cite{Csorgo:2008pe}.

\begin{figure}[]
\begin{center}
\includegraphics[width=0.8\textwidth]{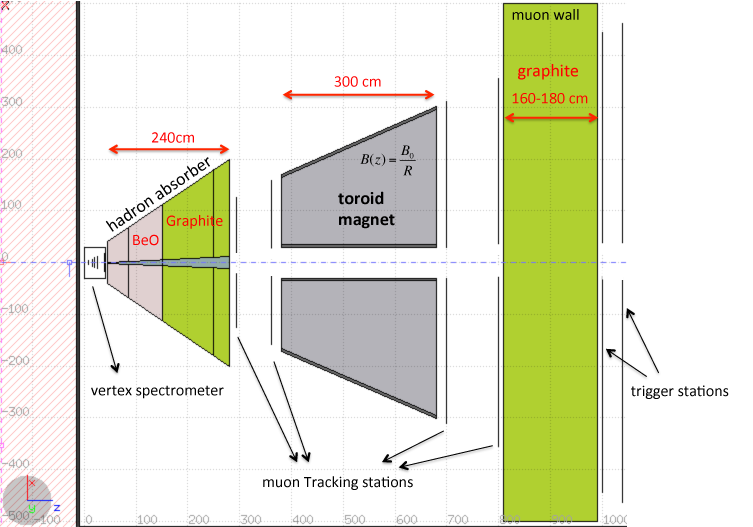}
\includegraphics[width=0.8\textwidth]{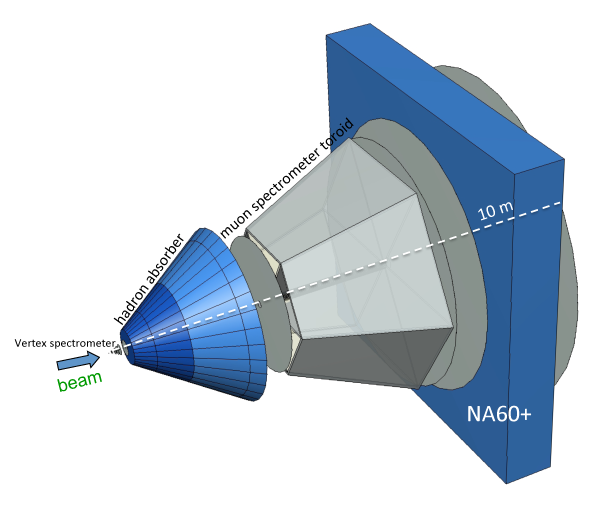}
\caption{Top: geometry of the proposed experimental apparatus. Bottom: prospective view.}
\label{fig:fig15}
\end{center}
\end{figure}

The NA60 apparatus has been  state-of-the-art for this kind of measurements in the past decade.
To reach a break-through in physics, we have to push forward the performances of the apparatus as compared to NA60:
\begin{itemize}
\item
Interaction and trigger rates. 
In order to fullfil the statistics 
requirement, the data taking must  be optimised profiting as much as possible of the machine luminosity. Assuming interaction rates of $\sim$
1 MHz or even more, the readout system must cope with dimuon trigger rates of several tens of kHz, a factor 10-20 larger than NA60.
\item
Geometric acceptance (rapidity and transverse momentum). The goal is to design a fixed target experiment covering the forward rapidity hemisphere.
The experimental set-up will be adapted
to the varying beam energy by contracting/extending the elements of the muon spectrometer in such a way to keep a reasonably good acceptance close to mid-rapidity. 

The
NA60 experiment, which took data at 160~GeV/nucleon, was optimised to cover the 
muon pseudo-rapidity range $\eta\sim2.9-4$ in the lab system, with $\eta=2.9$ corresponding to mid-rapidity at that energy.
In order to measure muons in the lab system at beam energies from 160 down to 20 
GeV or so, the apparatus must cover approximately the interval 
1.6<$\eta$<4. Moreover, the apparatus must guarantee a good coverage down to low transverse momenta in particular for low masses and low pseudo-rapidities: for M< 0.5 
GeV, the goal is to improve the acceptance at very low transverse momenta (<200~MeV) by a factor >10 with respect to NA60.
\item
Signal-to-background ratio. The main source of background comes from the muons due to pion and kaon decays. The experiment must retain a signal-to-background ratio in Pb--Pb central collisions not smaller than 1/20-1/30 in the intermediate mass region, comparable to the one reached by NA60.

\item
Mass resolution. The subtraction of the freeze-out processes was mastered in NA60 with a data driven technique thanks to the good mass resolution (20~MeV at the $\omega$ mass). The new experiment should push the mass resolution down to $\sim$10~MeV. 
\end{itemize}

The general aspects of the apparatus layout were investigated by using a fast simulation based on a semi-analytical tracking algorithm and Fluka~\cite{Usai:2014yia}.  A sketch of the apparatus layout is shown in Fig.~\ref{fig:fig15}. It is formed by the following sub-systems:
\begin{itemize}
\item
{\bf Muon spectrometer} This part of the apparatus reconstructs the muon tracks. It is composed of 4 tracking stations placed after the hadron absorber. An option which is being 
investigated  is to use GEM detectors with $\sim200$ $\mu$m space resolution. A toroid magnet placed in the middle of the tracking stations provides a field integral of 
0.75 Tm at R=1 m – approximately the one provided by the toroid magnet ACM in NA60. Simulations for the lowest energies (20 and 40~GeV/nucleon) were performed also assuming a 
reduced field integral of 0.3 Tm at R=1 m. The apparatus angular coverage allows muons to be measured down to $\eta\sim$1.6 at 20~GeV/nucleon.
\item
{\bf Hadron absorber} 
The signal-to-background ratio should be similar at all energies. This requires to study an absorber system with a scalable thickness.
At present an absorber was investigated for measurements at 20-40~GeV/nucleon.
It consists of BeO-graphite sections compromising in the best way between hadron absorption and multiple scattering. A graphite wall is placed at the 
end  before the trigger stations. The total absorber thickness for measurements at 20~GeV/nucleon corresponds to 7.3 $\lambda_I$, 14.7 $X_0$.
\item
{\bf Muon trigger} Dilepton radiation is a relatively rare processes. This requires high interaction rates and a system to select only events where muon pairs are produced. The 
candidate muons are those tracks that are detected by the stations placed after the the muon wall shown in Fig.~\ref{fig:fig15} (trigger stations). A trigger algorithm selects tracks originating from the vertex region with a real time track reconstruction in the trigger and muon stations. This system must have a sufficient band-width in order not to waste beam luminosity. The possibility to use information provided by detectors in the vertex region might be further
considered. Additionally, the possibility of a more selective trigger based on a muon transverse momentum threshold might
be interesting for the J$/\psi$ and open charm studies where beam intensities (hence interaction rates) larger than those needed
for thermal radiation might be needed.
\begin{figure}[t]
\begin{center}
\includegraphics[width=12.5cm]{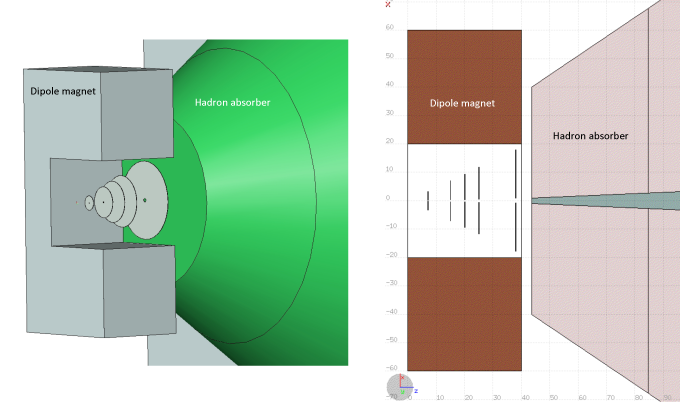}
\caption{Left: prospective view of the silicon tracker. Right: geometry details of the silicon tracker.}
\label{fig:fig16}
\end{center}
\end{figure}
\item
{\bf Silicon vertex tracker} This is the fundamental detector which allows muon kinematics to be measured with very high precision. A preliminary sketch of the silicon tracker, embedded in a dipole field is shown in Fig.~\ref{fig:fig16}~\cite{Usai:2014yia}. The 
spectrometer consists of 5 silicon tracking planes immersed in a 1.2 Tm dipole field. The tracking stations have an angular coverage that allows muons to be measured down to 
a pseudo-rapidity $\eta\sim1.6$. The required physics performance places stringent requirements for the pixel detectors in terms of temporal response, radiation 
hardness and material budget:
\begin{itemize}
\item
{\bf Temporal response and radiation hardness}. As mentioned above, since ion beams are usually circulated in the SPS only for a few weeks per year, the goal is to collect the required statistics at a given energy in a run having a 
duration of 10-15 days. Dilepton radiation is produced by rare processes. A simulation with a trigger system designed to select the events with muon pairs showed 
that an interaction rate at least at the level of $\sim$1 MHz is needed. The pixel sensors are exposed to all interactions and for this reason must have a fast temporal response. 
In addition, this requires that the pixel sensors should be working up to fluencies exceeding $10^{14}$ equivalent neutrons/cm$^2$. 
\item
{\bf Material budget}. A limiting factor for the precision of the measurement is the background of muons from pion and kaon decays before the absorber. Pairs of background 
muons constitute the so-called combinatorial background, whose subtraction is an important limit to the accuracy of the measurement of the signal spectrum. If a decay occurs 
before the pixel station closest to the interaction vertex, the decay muon will appear in the silicon tracker as a track with an offset with respect the interaction vertex. If the decay 
occurs inside the pixel tracker, the mother particle and the decay muon tracks will form a kink. The possibility to reject these tracks requires high space resolution and minimal - 
if not negligible - multiple scattering. 

\end{itemize}

The use of hybrid pixels was first explored, with a material budget per plane of $\sim1\%$ $X_0$~\cite{Usai:2014yia} . The results presented in the following are based on this layout. A
drawback of this technology is the cost and the thickness of the sensors+readout chip, that typically exceeds several
hundred microns and the pixel pitch is limited to 50 $\mu$m in state-of-the-art sensors.

A very interesting alternative is the new generation of monolithic active pixel sensors~\cite{Abelevetal:2014dna}. While these sensors have not yet
the required readout speed, new developments might lead to competitive sensors in the near future. These sensors have a
pitch of 20-30 $\mu$m and the material budget per pixel tracking plane might be reduced by a factor 5-10 with respect to hybrid
pixels – down to 0.1-0.3\% $X_0$. In this case the offset and angular resolution might improve at a level to allow practically all
the background muons from kaon and pion decays to be rejected. This might imply a  more efficient use of the beam time since 
the effective statistics – being 
proportional to the interaction rate multiplied by signal-to-background ratio – would be increased.


\end{itemize}

\subsection{Performance studies: thermal radiation}
 First performance studies with the apparatus outlined in the previous section were carried out for Pb--Pb 0-5$\%$ central collisions at 40~GeV/nucleon ($\sqrtsNN=8.8$~GeV, $<\dd N_{ch}/\dd y>=$265). The thermal $\mu^+\mu^-$ differential  spectra $\dd N/\dd M \pt \dd\pt \dd y$ are
 based on the in-medium $\rho$, $\omega$ and  $4\pi$ spectral functions and the expanding thermal fireball model of~\cite{Rapp:1999us}, with subsequent improvements according to~\cite{vanHees:2007th}. 
The QGP spectrum is calculated using a lattice-QCD constrained rate based on the equation of
state of~\cite{He:2011zx} with  $T_c=$163~MeV.  The chemical freeze-out temperature is $T_{ch}=$ 148~MeV.
The hadronic and QGP contributions to this thermal spectrum are shown in Fig.~\ref{fig:fig20}.
Finally, the hadron cocktail generator for the 2-body decays of $\eta$, $\omega$,  and $\phi$ and the Dalitz decays  $\eta\to\gamma\mu^+\mu^-$ and $\omega\to\pi^0\mu^+\mu^-$ is based on the NA60 one and on 
the statistical model of~\cite{Becattini:2005xt}. Drell-Yan and open charm are simulated with the Pythia event generator.

\begin{figure*}[t]
\begin{center}
\includegraphics[width=7.5cm]{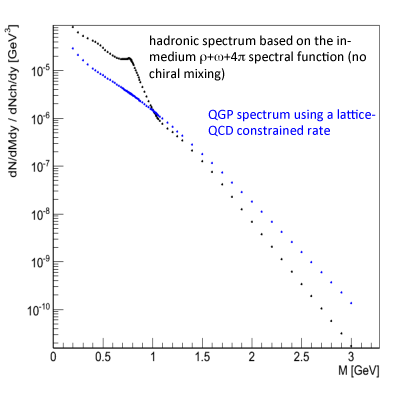}
\caption{Thermal dilepton spectrum predicted by Rapp et al. for Pb--Pb central collisions at 40~GeV/nucleon~\cite{Rapp:1999us,vanHees:2007th}. Black: thermal radiation produced in the hadronic phase; blue: thermal radiation produced in the partonic phase.}
\label{fig:fig20}
\end{center}
\end{figure*}

A fast-simulation framework with a semi-analytical tracking algorithm was developed. The particle tracks are propagated through the apparatus, defined as a sequence of active or 
passive (absorbers) layers, each with defined geometric dimensions and material properties (including magnetic field). 
Tracks are reconstructed starting from the trigger stations towards the interaction point with a Kalman filter which adds hits in muon stations and vertex detector to the candidate tracks.
 
\begin{figure*}[t]
\begin{center}
\includegraphics[width=13.5cm]{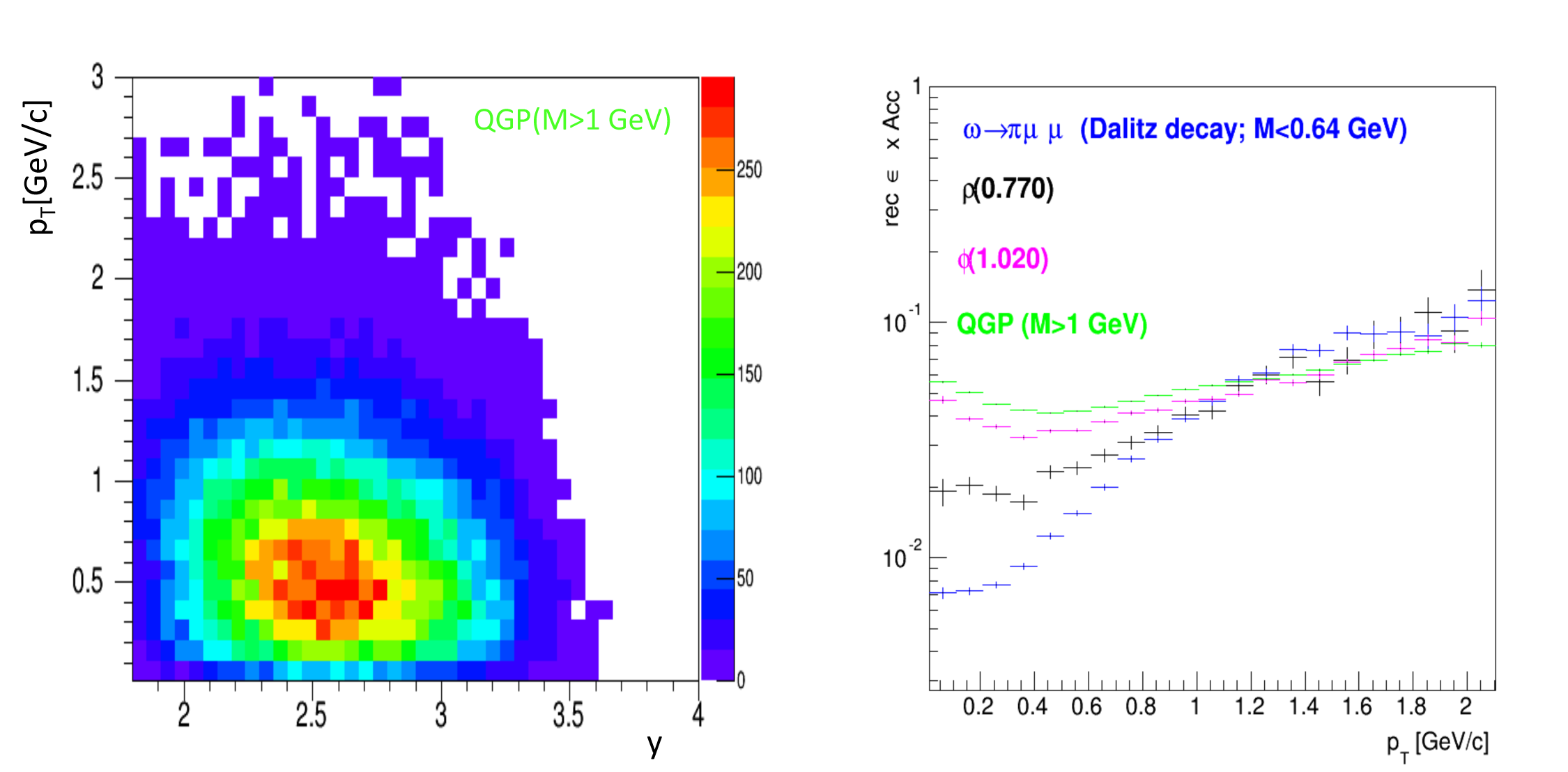}
\caption{Left: transverse momentum vs. rapidity coverage for reconstructed dimuons with $M>1$~GeV produced from the QGP phase.
Right: acceptance times reconstruction efficiency vs. transverse momentum for processes in different mass ranges.}
\label{fig:fig17}
\end{center}
\end{figure*}


As an illustration, the transverse momentum-rapidity coverage is shown in the left panel of Fig.~\ref{fig:fig17} for reconstructed dimuons with $M>1$~GeV produced from the QGP phase. In this mass region, the apparatus has a good coverage down to mid-rapidity (2.2 in the lab system) and zero transverse momentum.
Fig.~\ref{fig:fig17} - right  shows the 
 the pair reconstruction efficiency for processes in different mass ranges: this varies from  $\sim$1\% at low masses and very small $\pt$ to $\sim5-10\%$ for $M>1$~GeV at any $\pt$. 

The dilepton spectrum is dominated by a combinatorial background  arising from muons produced by decays of primary or secondary hadrons. Additionally, also punch-through of primary or secondary hadrons produced in the absorbers may occur.
In order to study this background, the Fluka package has been used to simulate in detail the full hadronic shower development in the absorber. 

Furthermore, in the signal reconstruction, it is possible that hadronic hits in silicon pixel planes are associated to a muon track . This potential contamination ({\it fake matches}) was taken into account at reconstruction level including the hadronic hits in the silicon stations according to the pion, kaon and proton multiplicities
as measured at 40~GeV/nucleon by the NA49  experiment~\cite{Afanasiev:2002mx}. 



\subsubsection{Low and intermediate mass dileptons: measurement of source temperature and sensitivity to $\rho$--$a_1$ chiral mixing}
In this section we present the results for a sample of $2\cdot10^7$ reconstructed pairs in central collisions, corresponding to a total
sample of $\sim5\cdot10^7$ integrated in centrality.

\begin{figure}[]
\begin{center}
\includegraphics[width=0.9\textwidth]{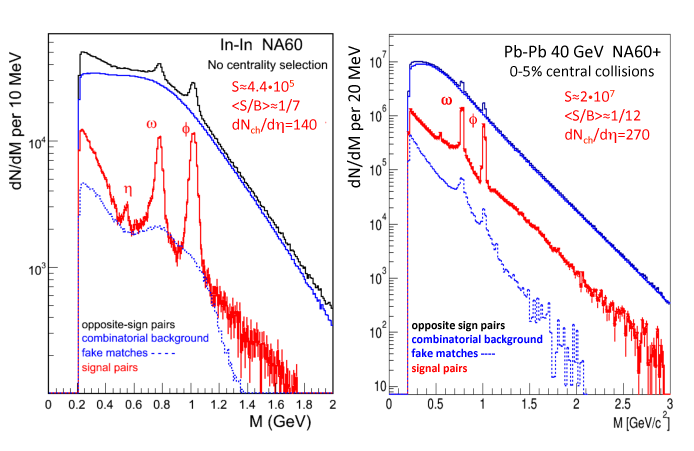}
\includegraphics[width=0.9\textwidth]{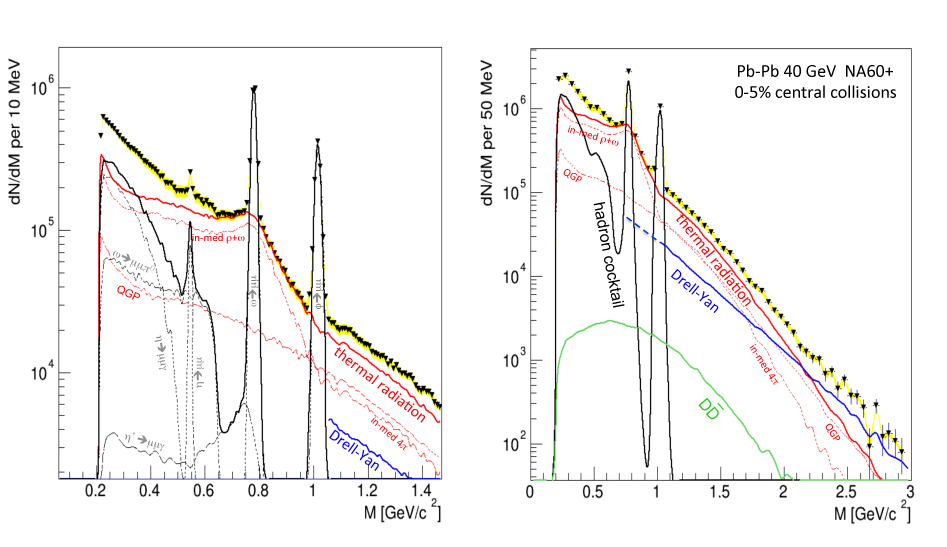}
\caption{Top left: total data sample integrated in centrality measured by NA60 in In--In collisions at 160~GeV/nucleon. Top right: expected data sample in Pb--Pb 0-5$\%$ central collision at 40~GeV/nucleon. Bottom left: expected reconstructed signal mass spectrum after subtraction of combinatorial and fake matches backgrounds - focus in the LMR. Bottom right: expected reconstructed signal mass spectrum after subtraction of combinatorial and fake matches backgrounds in LMR and IMR.}
\label{fig:fig21}
\end{center}
\end{figure}



  We first focus on the mass spectra. Fig.~\ref{fig:fig21}-top-right  shows the total reconstructed mass spectrum (black).  
The combinatorial background (blue) is subtracted assuming a $0.5\%$ systematic uncertainty, based on  the very  conservative  $1\%$ value estimated in NA60. 
The  net signal after subtraction of the combinatorial background and fake matches is shown in red. The average signal-to-background ratio is 1/12.
Assuming a lower field integral of 0.3 Tm at 1 m for the toroid magnet, the  performance remains good: the signal-to-background ratio increases by 30-40$\%$, still allowing the measurement to be performed with very  good precision.
For comparison, the dilepton spectrum integrated in centrality measured by NA60 in In--In collisions at 160~GeV/nucleon is shown in Fig.~\ref{fig:fig21}-top-left.
For what concerns minimum bias collisions, the progress in statistics over NA60 is a factor $\sim 100$, retaining a similar background ratio and a better mass resolution.

Fig.~\ref{fig:fig21}-bottom shows all the components of the signal spectrum 
after subtraction of the combinatorial background (the  uncertainty from the  background subtraction is shown as a yellow band). The left panel shows a zoom of the LMR. In this region, the thermal radiation yield is dominated by the in-medium $\rho$+$\omega$, while the QGP contribution is almost an order of magnitude smaller.
The $\omega$ and $\phi$ peaks are well resolved with a resolution of $\sim$10~MeV at the $\omega$ mass. This allows the in-medium thermal radiation to be precisely
isolated by subtracting the hadron cocktail contributions with the data-driven technique mastered by NA60~\cite{Arnaldi:2006jq}. 
The signal-to-background ratio at $M=600$~MeV is $\sim$1/20, which leads to a 10$\%$ systematic uncertainty. In the lower field configuration, the signal-to-background ratio
is $\sim1/30$, which leads to a 18$\%$ systematic uncertainty.
The right panel shows the total spectrum.
The  thermal spectrum is measurable up to 2.5-3~GeV. According to the theoretical model considered, the
QGP yield is still significant at  40~GeV/nucleon. The Drell-Yan exceeds the QGP above 2.5~GeV, while the  open charm yield  is negligible.

\begin{figure*}[t]
\begin{center}
\includegraphics[width=14.cm]{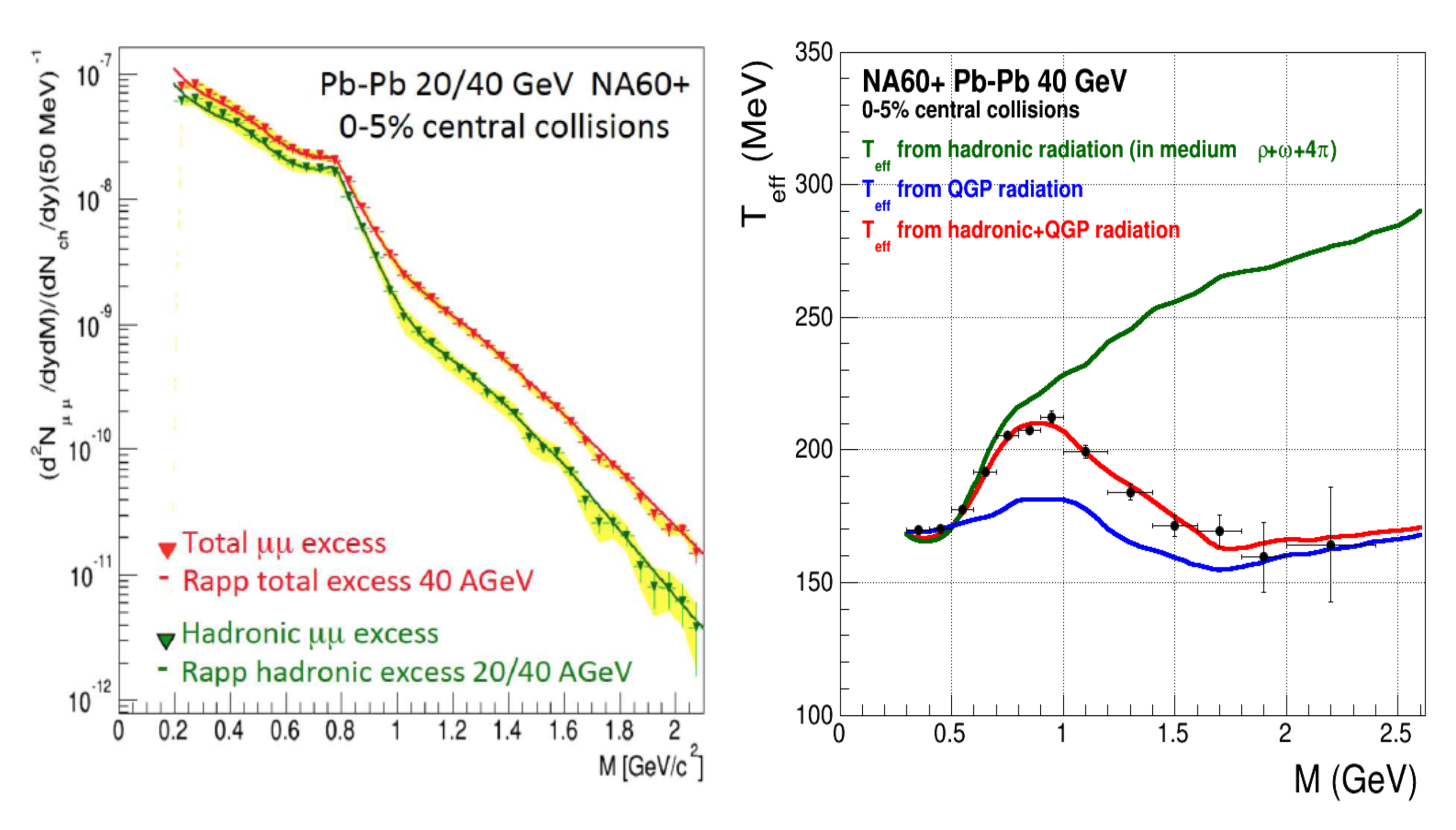}
\caption{Left: Acceptance-corrected inclusive thermal dilepton spectrum  (red) and acceptance-corrected spectrum from hadronic radiation only (green). Right: Measurement of $T_{\rm eff}$ vs. $M$ of total excess from the analysis
of $m_{\rm T}$ spectra.  }
\label{fig:fig23}
\end{center}
\end{figure*}

The most significant results  are summarized in Fig.~\ref{fig:fig23}-left, which shows the thermal dilepton spectrum obtained after subtraction of the hadronic cocktail, Drell-Yan and open charm contributions  (the latter being negligible at this energy). 
The resulting total thermal excess that would be measured at  40~GeV/nucleon is shown in red. The performance for the measurement of temperature was assessed by fitting the distribution in the range
1.5-2.5~GeV, with $\dd N/\dd M\propto M^{3/2}\exp(-M/T)$. In this way we find $T=163\pm4\pm1$~MeV, in perfect agreement with the input value from the theoretical generator of 160~MeV. This shows that a high precision measurement is feasible.

The performance for the measurement of the thermal spectrum at the onset of deconfinement has been estimated considering a scenario without QGP and a measured thermal dilepton yield at the level of the in-medium $\rho+\omega+4\pi$ processes of Fig.~\ref{fig:fig20}. Moreover, a combinatorial background yield at the  level of Pb--Pb central collisions at 20/40~GeV/nucleon was assumed. The resulting spectrum in shown in green in Fig.~\ref{fig:fig23}-left. This shows that a study of the hadronic excess up to $M \sim$ 2~GeV is  possible, with a very good sensitivity to   $\rho$-$a_1$ chiral mixing. 

\subsubsection{Low and intermediate mass dileptons: measurement of effective temperature from transverse momentum spectra}

The theoretical model employed to describe the thermal radiation provides the $\pt$ spectra in different mass intervals separately for the QGP and hadronic radiation.
These spectra were fitted with the formula $\dd N/m_{\rm T}\dd m_{\rm T}=\exp(-m_{\rm T}/T_{\rm eff})$ to determine the evolution of the effective temperature vs. mass for the two contributions.
The green line shown in the right panel of Fig.~\ref{fig:fig23} shows the evolution of $T_{\rm eff}$ for the hadronic radiation. $T_{\rm eff}$ has a monotonous rise consistent with the radial flow of a hadronic source:  $\pi^+\pi^- \to \rho\to \mu^+\mu^-$  in the LMR and $4\pi$ processes in the IMR.
The blue line shows the evolution of $T_{\rm eff}$ for the QGP radiation. In this case, $T_{\rm eff}$ is almost flat, consistent with an early source with low radial flow.
The red line shows the $T_{\rm eff}$ behavior when the QGP and hadronic spectra are convoluted into the total spectrum. 

To assess the sensitivity to the measurement of $T_{\rm eff}$, the $\pt$ spectra in different mass bins of the thermal spectrum extracted from Fig.~\ref{fig:fig23}-left (red distribution) 
were fitted with the exponential formula. The resulting values with the corresponding errors are shown in the plot.
Experimentally, $T_{\rm eff}$ can be extracted in several mass bins up to $\sim$2.5~GeV, measuring the mixture of partonic and hadronic processes 
  with a high sensitivity to even a small contribution of QGP just above the onset, signalled by the appearance of the drop above $M\sim$ 1~GeV.

\subsubsection{Running conditions}

The required statistic at each energy should be collected in a reasonably short time.
The rare electromagnetic processes impose the use of a high intensity SPS beam line, equivalent to the one available in the ECN3 experimental hall at the SPS North Area (now used by the NA62 experiment). Such a line was delivering, for NA50/NA60, beam intensities of the order of 10$^7$ Pb (or In) ions/s, with a structure consisting of 5~s long bursts, repeated 3 times in a minute. The choice of such a beam intensity was not dictated by the intrinsic limitation of the accelerator, but mainly from the maximum triggering rate on muon pair production of the experiment. This burst structure amounts to an effective continuous beam of $2.5\cdot10^6$~ions/s.

A trigger condition was simulated requiring a pair of tracklets in the trigger stations matched to tracklets in the muon tracking stations which loosely point to the vertex region.
This assures that no signal is lost while at the same time rejects all tracks not coming from the vertex region. Assuming a field integral of  0.75~Tm at $R=1$~m for the toroid magnet,
the  trigger rate  -  dominated by background - is $\sim 15$--$20$~kHz and we assume that the experiment will have  the band-width to cope with it (4--5 times more 
than NA60).  

Assuming an effective continuous beam intensity of $\sim 2.5\cdot10^6$ ions/s, a target thickness of $\lambda_i=0.15$ and the trigger logics described above,  a statistics of $\sim1$-$2\cdot10^7$ reconstructed pairs in central collisions can be collected in a  period of $\sim10$--$15$ days. With such a beam intensity, pile-up effects are very small, even considering a conservative hypothesis that the vertex detector has a 100~ns strobe (see the charmonia section for more details).

\subsection{Charmonium studies}

A measurement of charmonium production at low SPS energy (in the approximate range $40<E/A<160$~GeV) can in principle be performed with an experimental apparatus similar to the one used by the NA50/NA60 experiments. The main issues to be investigated are (i) the possibility of collecting a sizeable statistics in a reasonable amount of time (ii) the possibility of adapting the angular  coverage of the experiment to the varying kinematic conditions as a function of the energy of the collision.

Charmonium production cross sections strongly decrease when moving down from
top SPS energy. A reasonable quantitative estimate can be obtained using the so-called Schuler parameterization for the J/$\psi$ production cross section~\cite{Schuler:1994hy,Vogt:1999cu}, based on a fit of low energy pp results:

\begin{equation}
\sigma^{{\rm pp}\rightarrow {\rm J}/\psi}
(\sqrt{s},x_{\rm F}>0)=\sigma_{0}\left( 1-\frac{m_{{\rm J}/\psi}}{\sqrt{s}}\right)^{n}
\end{equation}

with $\sigma_{0}=638\pm 104$ nb and $n=12.0\pm0.9$. The cross section values for \mbox{Pb--Pb} collisions can be obtained (assuming for the moment no nuclear effects for the J/$\psi$) by multiplying the pp cross section by the Pb-ion mass number squared $A^2_{Pb}$, and are shown as a black line in Fig.~\ref{fig:na60psicross}. Assuming for the measurement a rapidity coverage $0<y_{\rm cms}<1$, as it was the case for NA50/NA60, taking into account the branching ratio for the decay channel to muon pairs BR(J/$\psi\rightarrow\mu\mu$)$=(5.96\pm0.03)$\% and applying (again as in NA50/NA60) a cut on the decay angle of the muons in the Collins-Soper reference frame $|\cos\theta_{\rm CS}|<0.5$~\cite{Collins:1977iv}, the production cross section (red line in Fig.~\ref{fig:na60psi3}) is of the order of 40 $\mu$b at $E/A\sim 100$~GeV and decreases by more than one order of magnitude at $E/A=30$~GeV. The estimate of the fraction of the J/$\psi$ yield in the $0<y_{\rm cms}<1$ is carried out assuming an empirical distribution also taken from~\cite{Schuler:1994hy,Vogt:1999cu}. Assuming no polarization for the J/$\psi$ and therefore a flat $\cos\theta_{\rm CS}$ distribution, the cut on this variable effectively removes 50\% of the mesons.
 
\begin{figure}[t]
\centering
\resizebox{0.5\textwidth}{!}
{\includegraphics{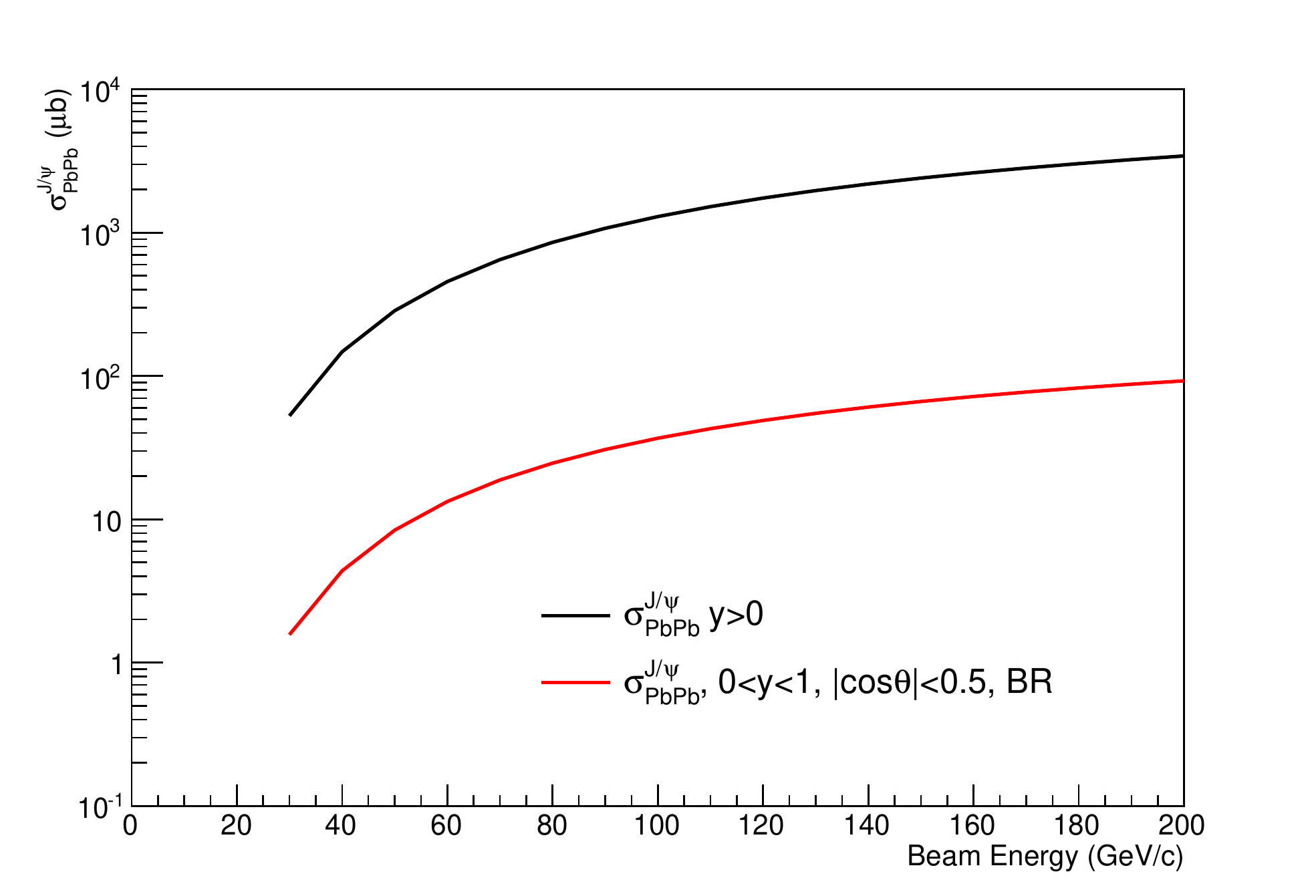}}
\caption{The J/$\psi$ production cross section as a function of the incident beam energy. The black curve represents the integrated cross section in the forward hemisphere, while the red curve takes into account the branching ratio to dimuons, a $0<y_{\rm cms}<1$ acceptance and a $\cos\theta_{\rm CS}<0.5$ cut on the muon decay angle.}
\label{fig:na60psicross}
\end{figure}

Assuming for the moment the same beam intensity of NA50/NA60 the number of Pb-ions to be delivered to the experiment in order to
collect 10$^4$ J/$\psi$ is shown in Fig.~\ref{fig:na60psi4}(left). The estimate assumes, for \mbox{Pb--Pb} collisions, a 4 mm thick target and a 15\% acceptance (which includes detector/reconstruction efficiency, and corresponds to the NA50 value). It is also assumed that a factor 3 suppression for J/$\psi$, similar to the one observed at top SPS energy, is present. The result shows that, down to $E/A\sim 50$~GeV, at most about 10$^{12}$ delivered Pb ions are necessary.  For \mbox{In--In} collisions, using the corresponding target thickness, acceptance and suppression values measured by NA60, one gets a very similar number of delivered ions to collect 10$^4$ J/$\psi$.

In a complementary way, and in order to better establish the feasibility of a charmonium measurement with a reasonable beam time, one can calculate the beam intensity necessary to collect a significant number of charmonia in a fixed number of days. The result, shown in Fig.~\ref{fig:na60psi4}(right), refers to the beam intensity needed to have $3\cdot10^4$ J/$\psi$ in 15 days. It assumes the burst structure, the target thickness, acceptance and suppression factors introduced in the previous paragraph. It is worth noting that in these conditions a few hundred $\psi(2\mathrm{S})$ mesons can also be collected, allowing a less accurate but still significant measurement. The result shows that in the energy range $60<E/A<160$~GeV a beam intensity smaller than $2\cdot 10^7$ ions/s is enough to collect the aforementioned statistics. In order to push the reach of the measurement further down in energy, a considerably larger beam intensity, or equivalently, beam time availability, will be necessary.

The use of high intensity beams may lead to significant pile-up effects in the detectors. While the probability of having muons from nearby ion-ion  interactions is very small 
at these beam energies, the pile-up of hadrons produced in those interactions may bias the centrality measurement. For example, if centrality is estimated from hadron multiplicity 
in the vertex detector, a pile-up of two interactions corresponding to semi-peripheral collisions may simulate a central event. To estimate the occurrence of such a situation, a 
calculation based on a Poissonian distribution of the ions inside the burst has been carried out, with the conservative hypothesis that the vertex detector has a 100 ns strobe. 
Up to $\sim 3\cdot 10^7$ ions/s the pile-
up probability is smaller than 10-15\%. Provided that such events are identified, for example using a fast beam hodoscope (as in NA60), their rejection would therefore only slightly 
affect the total integrated luminosity available for the measurement.

\begin{figure}[t]
\centering
\resizebox{0.47\textwidth}{!}
{\includegraphics{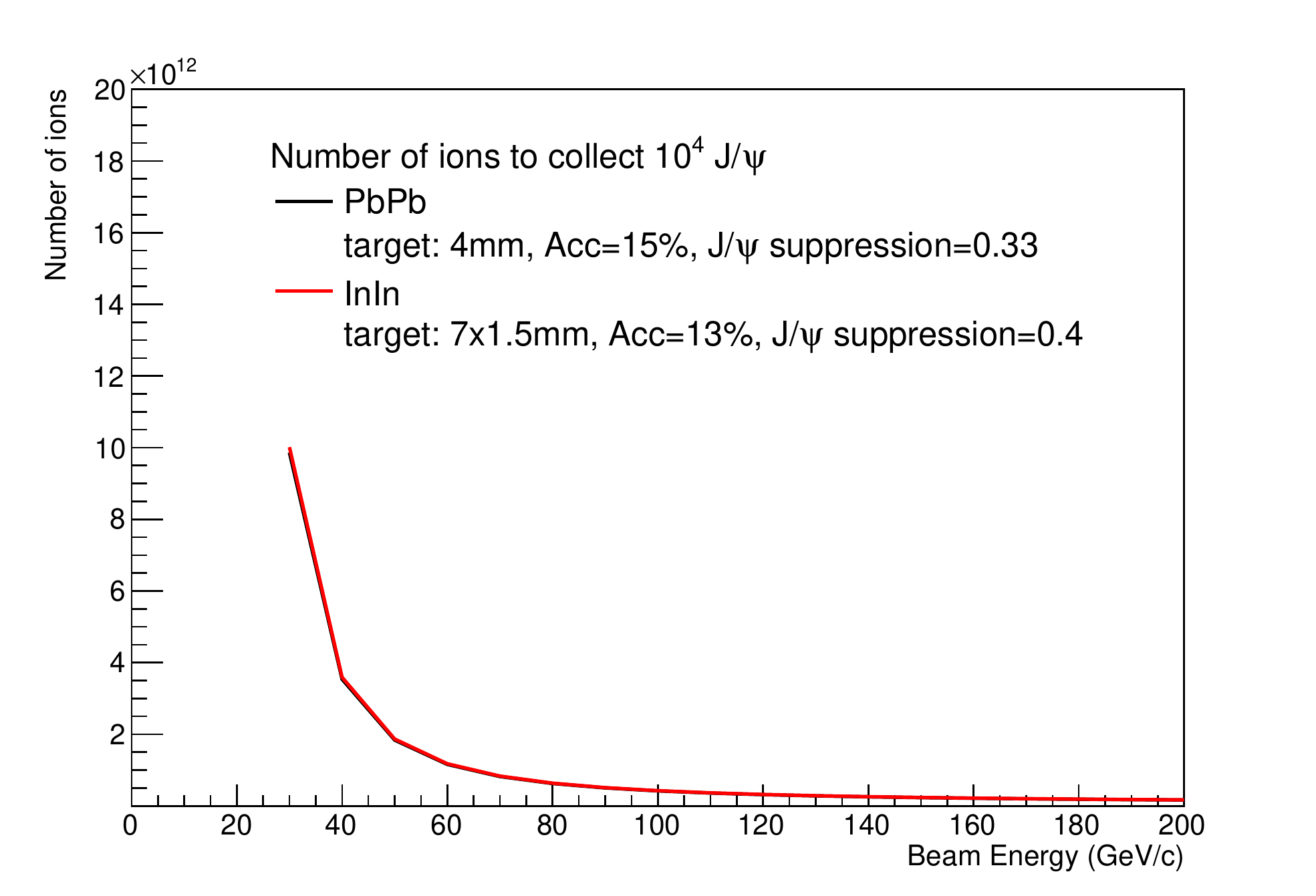}}
\resizebox{0.47\textwidth}{!}
{\includegraphics{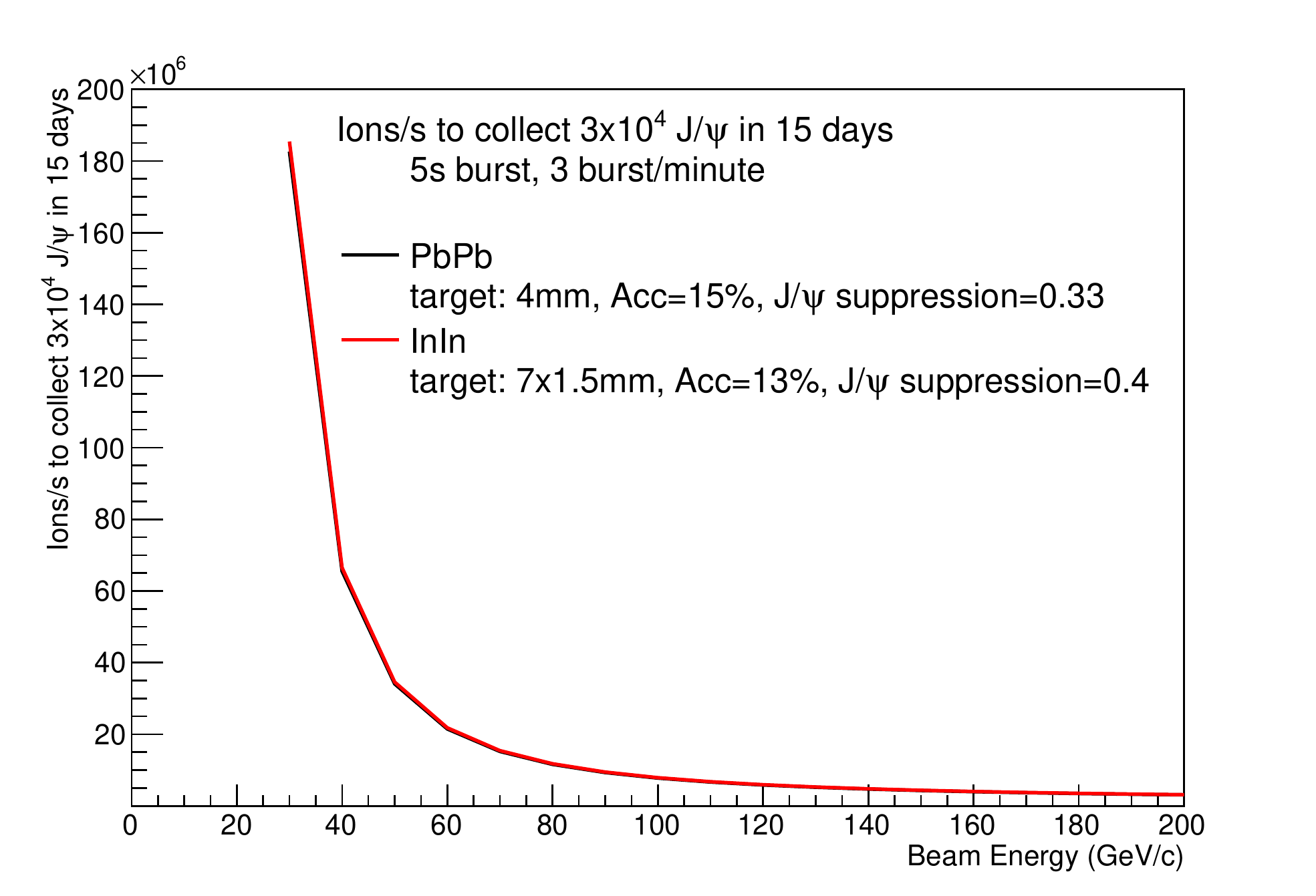}}
\caption{Left: number of Pb or In ions needed to collect 10$^4$ J/$\psi$, in a realistic experimental condition. Right: beam intensity necessary to collect $3\cdot 10^4$ J/$\psi$ in 15 days, in the same conditions.}
\label{fig:na60psi4}
\end{figure}


Having verified that a statistically significant charmonium measurement can be performed at low SPS energy, in the following a preliminary study of the acceptance of the experiment, assuming the set-up detailed in Section~\ref{sec:NA60applayout}, is described. Since the J/$\psi$ kinematic distributions (rapidity and transverse momentum) are not precisely known, the PYTHIA6 generator~\cite{Sjostrand:2006za}  has been used for this study. For the $y$-distribution, as an alternative, the empirical function of~\cite{Schuler:1994hy,Vogt:1999cu} has also been tested. As it can be seen in Fig.~\ref{fig:na60psi6}(left), both choices are in qualitative agreement with a parameterisation directly fitted on existing data for p-A collisions at 158~GeV (corresponding to top SPS energy for nucleus-nucleus collisions). Muons from the J/$\psi$ decays are propagated through the experimental apparatus using a GEANT3~\cite{Brun:1987ma} based MonteCarlo simulation, and the recorded detector hits are then fed into a code based on the NA60 reconstruction algorithm. In order to simulate the hadronic background in the vertex spectrometer, hits are added according to the expected pion and kaon multiplicity as measured by the NA49 experiment~\cite{Afanasiev:2002mx} and assuming a thermal $p_{\rm T}$ distribution and a gaussian rapidity shape. In Fig.~\ref{fig:na60psi6}(right) a reconstructed invariant mass distributions of muon pairs, for a simulation corresponding to $E=50$~GeV, is shown. The filled region represents events where the reconstructed tracks include at least one  background hit in the vertex spectrometer. It can be seen that the amount of events where such a ``fake'' match has occurred remains below 1\% and can therefore be considered as negligible. As the particle multiplicity only increases logarithmically with the collision energy, the ``fake'' match  contribution is expected to remain modest over all the SPS energy range.

\begin{figure}[!ht]
\centering
\resizebox{0.45\textwidth}{!}
{\includegraphics{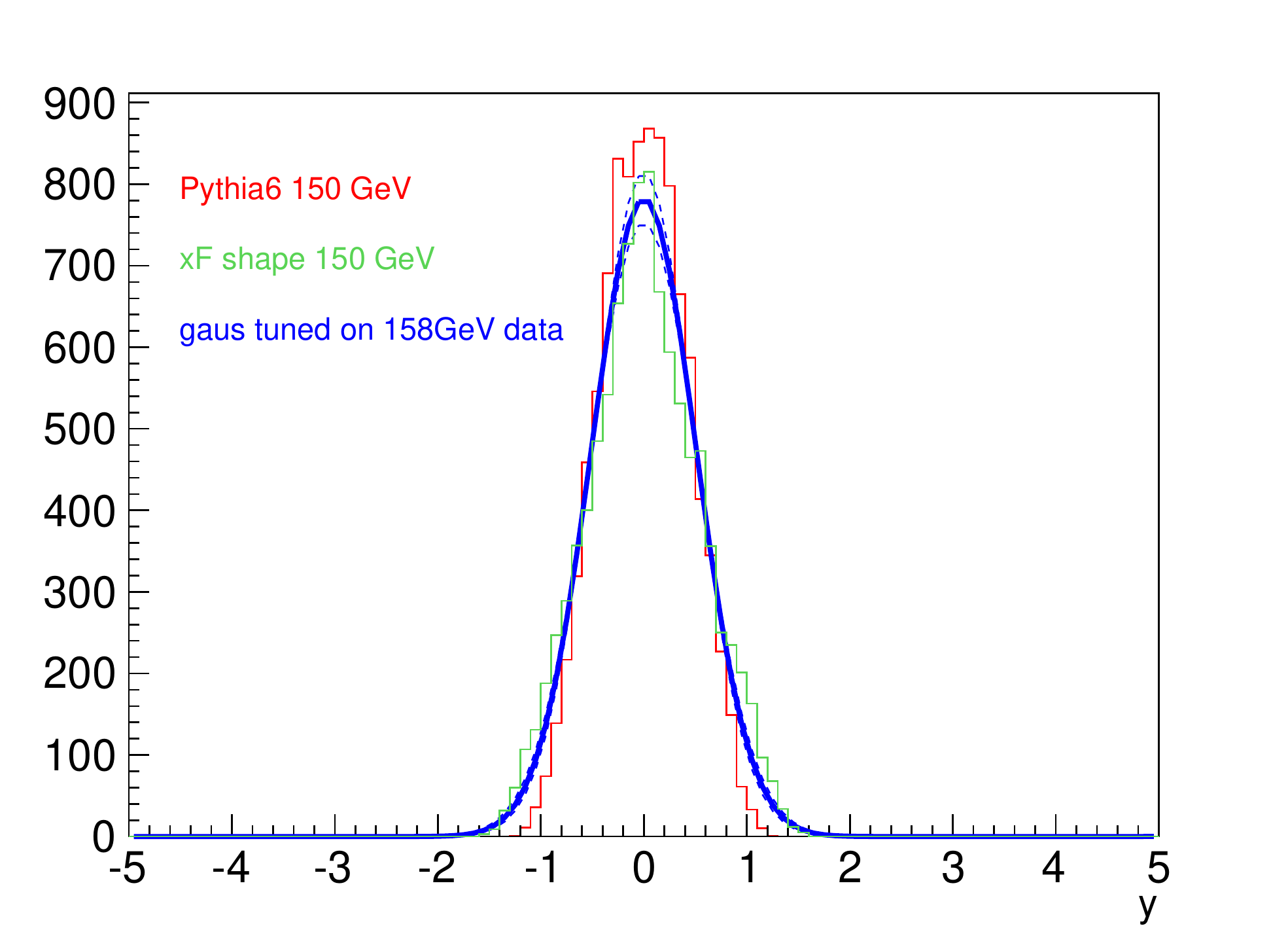}}
\resizebox{0.47\textwidth}{!}
{\includegraphics{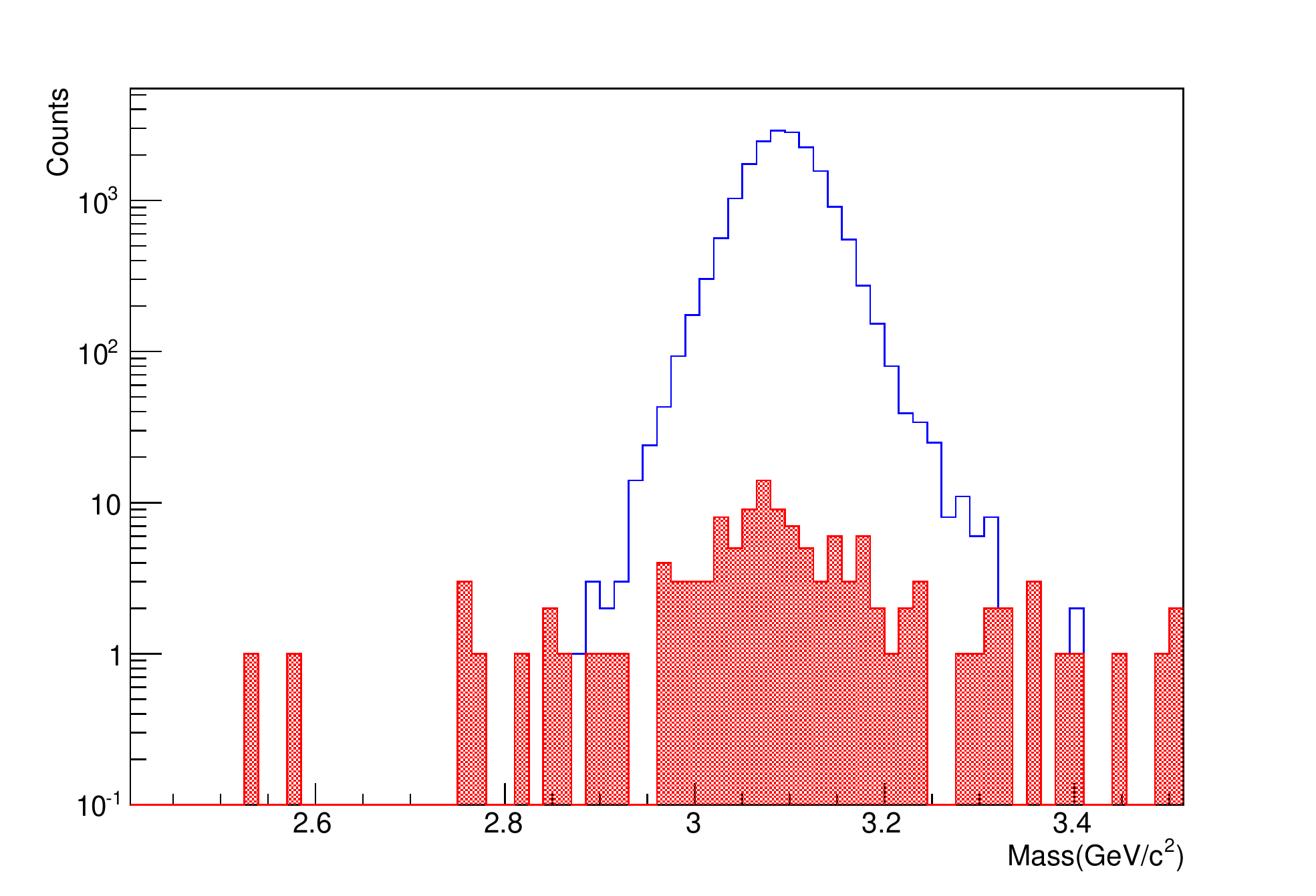}}
\caption{Left: a parametrisation of the J/$\psi$ rapidity distribution for p-A collisions at 158~GeV, compared with the corresponding shapes from PYTHIA6 and from~\cite{Schuler:1994hy,Vogt:1999cu}. Right: the invariant mass distribution of muon pairs from J/$\psi$ decays, reconstructed in the NA60 apparatus. Events where ``fake'' matches have occurred are also shown.}
\label{fig:na60psi6}
\centering
\resizebox{0.75\textwidth}{!}
{\includegraphics{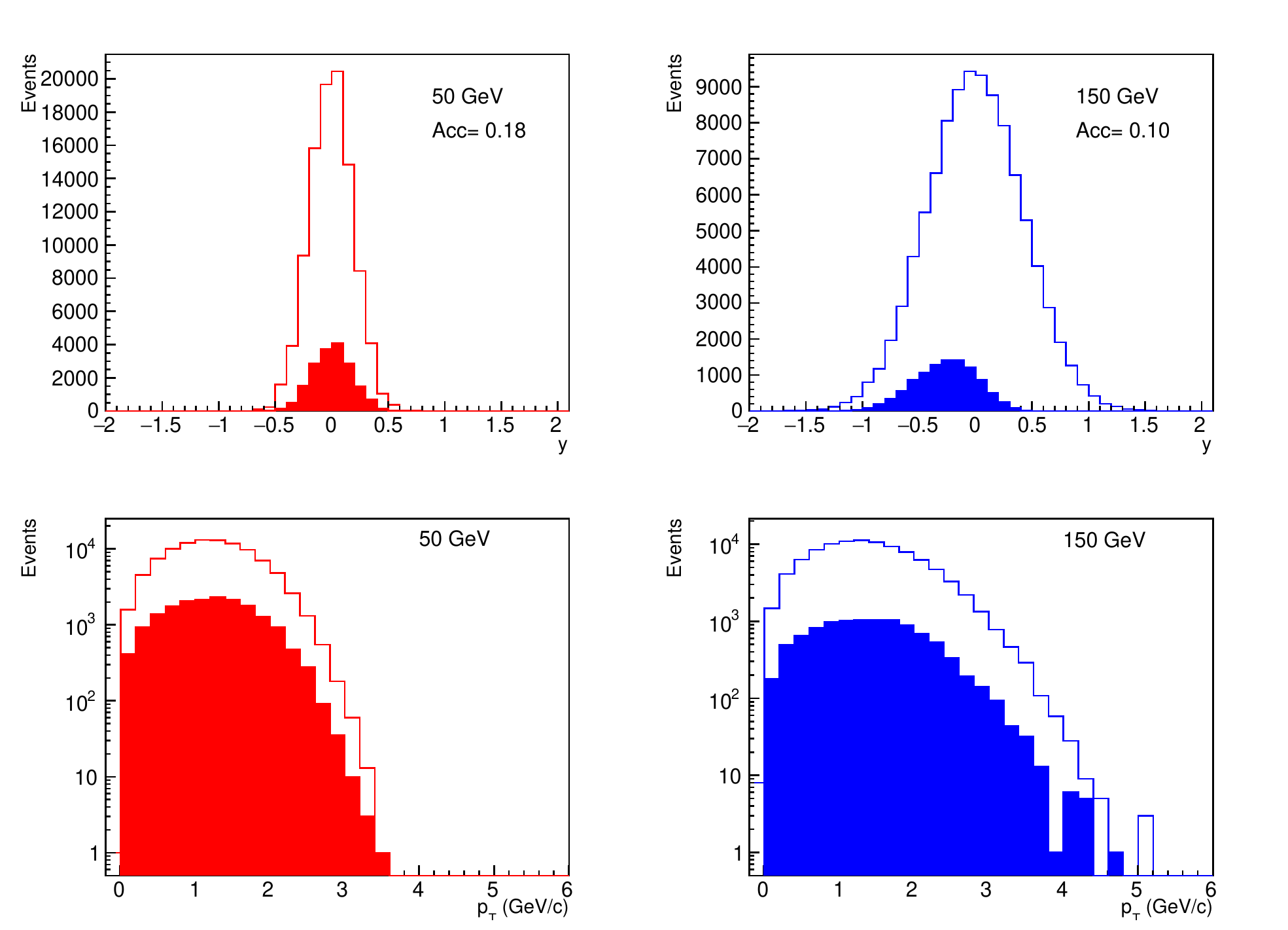}}
\caption{Generated (PYTHIA6) and reconstructed $p_{\rm T}$ and $y$ distributions for J/$\psi$, for 50 and 150~GeV incident energy.}
\label{fig:na60psi7}
\end{figure}

In Fig.~\ref{fig:na60psi7} the transverse momentum and rapidity distributions for generated and reconstructed J/$\psi$ are compared, for 50 and 150~GeV incident energy. For the moment, the same setup, optimised for the low end of the SPS energy range in such a way to cover the region around $y_{\rm cms}=0$ with maximum acceptance, has been used at both energies. Consequently, when the collision energy increases, the maximum of the acceptance shifts towards negative rapidity, due to the increased boost of the center-of-mass in the laboratory frame. As explained in Section~\ref{sec:NA60applayout}, the experimental set-up will be adapted to the varying beam energy by  contracting/extending the elements of the muon spectrometer in such a way to keep the maximum of the acceptance close to mid-rapidity. The acceptances, given by the ratio between the number of reconstructed and generated events, range between 10\% and 18\%, these values being similar to the ones of the NA50/NA60 experiments. As a function of $p_{\rm T}$, the acceptance shows a slight increase from low to high transverse momenta, the approximate $p_{\rm T}$ reach being of the order of 2-3~GeV/$c$ with the integrated luminosity described above.

In summary, the preliminary estimate of the achievable integrated luminosity and of the differential acceptances show that the measurement of J/$\psi$ production is feasible from top SPS energy down to a beam energy between 40 and 60~GeV, depending on the available beam time. A sample of $\sim 3\cdot 10^4$ J/$\psi$ can be collected with beam intensities similar to those already available in the NA50/NA60 experiments, running the experiment for $\sim$2 weeks at each energy (up to one month at the lower end of the energy range) and keeping the interaction pile-up level at about 10\%. With such a statistics, the measurement of the centrality dependence of J/$\psi$ production in both \mbox{In--In} and \mbox{Pb--Pb} collisions, as well as differential studies in $y$ and $p_{\rm T}$, should be feasible. A significant statistics for $\psi(2\mathrm{S})$, of a few hundred counts at each energy, allowing an integrated cross section estimate, is also within reach.

\subsection{Dimuon experiment proposal at the SPS: summary and timeline}

In order to address quantitatively the issue of chiral symmetry restoration and the first order phase transition in the region of moderate-large $\mu_B$, we have studied in detail novel high precision measurements
of thermal muon pair and J$/\psi$ production to be performed with a fixed target experiment operated at different energies below 160~GeV/nucleon at the  CERN SPS (low beam energy  scan). 

The pillars of the strategy are:
\begin{itemize}
\item chiral symmetry restoration: first measurement of the $\rho$--$a_1$ chiral mixing; measurement of the open charm yield.
\item onset of deconfinement:
\begin{itemize}
\item first measurement of the strongly interacting matter caloric curve: temperature vs. energy density; 
\item tagging of onset of deconfinement by measurement of QGP yield at different collision energies:
 measurement of the effective temperature extracted from dimuon transverse momentum spectra;
\item measurement of the fireball lifetime;
\item tagging of onset of deconfinement by measurement of J$/\psi$  anomalous suppression.
\end{itemize}
\end{itemize}

In order to 
to reach a break-through in physics, the thermal spectrum from the hadronic phase should be isolated for
 masses up to 2~GeV and the measurement of temperature should be performed with an accuracy at the level
 of few~MeV. In addition, J$/\psi$ production must be studied accurately as a function of centrality. This requires to collect:
\begin{itemize}
\item
$\sim5\cdot10^7$ reconstructed thermal muon pairs per energy point. The statistics increase at each energy point is a factor 100 over NA60  and $\sim10^5$ over RHIC and LHC experiments; 
\item
 $\sim2$--$3\cdot 10^4$ reconstructed J$/\psi$ mesons per energy point.
\end{itemize}

Tentatively,  data should be collected at 20-30-40-80-120-160~GeV/nucleon with Pb ions. 
Further energy points might be required depending on the first results. 
The goal is to collect the required statistics at a given energy in a run having 
a duration of 10--15 days. The experimental program requires also to collect p-Pb data at a few energy points.
This beam scan program might be accomplished in $\sim5$ years of data-taking. 

The measurements should be performed with a muon spectrometer complemented by a silicon vertex spectrometer with significantly improved performance with respect to the state-of-the-art experiment NA60.
The apparatus is designed to operate at interaction rates exceeding the MHz level, with a geometric acceptance able to cover the forward rapidity hemisphere for collisions in the foreseen energy range.

The physics performance were studied in detail for thermal radiation in central Pb--Pb collisions at 40~GeV/nucleon. For this observable,  the physics goals can be reached running at each energy for 10-15 days. The J$/\psi$ measurement is possible with a comparable beam-time period down to energies 40--60~GeV/nucleon.

It is envisaged to undertake the steps towards the formation of an international collaboration and the preparation of a Letter of Intent to be submitted to the SPS Committee within 2018. This would be timely in view of the update of the European Strategy for Particle Physics scheduled for that year. 

The construction and running of the experiment can be envisaged for the following decade assuming:
\begin{itemize}
\item 2--3 years devoted to R\&D for detectors and toroid magnet design;
\item 2 years for construction;
\item 5 years of data-taking.
\end{itemize}


\section{Further perspectives}
\label{sec:further}


\subsection{Fixed-target collisions with LHC beams: AFTER experiment}
\label{sec:after}


At the LHC collider, the highest center-of-mass energies available today for hadronic collisions can be reached. Correspondingly, if one could be able to extract the LHC beam and direct it towards a target, the highest conceivable energy in a fixed-target experiment could be obtained, as large as $\sqrt{s}\approx 115$~GeV for pp collisions, and $\sqrtsNN\approx 72$~GeV for \mbox{Pb--Pb} interactions. The extraction of the LHC beam, in order not to induce any perturbation on the beam colliding operation, could be done at the level of the halo of the beam itself, using bent crystals as a tool. 
This technique is currently under investigation at CERN~\cite{Uggerhoj:2005xz,Scandale:2010zzb,Scandale:2011za}, and it appears reasonable to extract in this way about $5\cdot 10^8$ protons/s and $\sim 2\cdot 10^5$ Pb ions/s. It can be shown that integrated luminosities up to 0.5~fb$^{-1}$ in pp,  and 10--20~nb$^{-1}$ with nuclear beams, can be reached using a target thickness of $\sim 1$~cm~\cite{Brodsky:2012vg}. Thanks to the strong boosting of the center-of-mass system in the laboratory (4.8 $y$-units for a 7~TeV beam) a typical fixed-target experimental set-up (AFTER, A Fixed Target ExpeRiment at LHC) could easily access all the backward hemisphere and push particle detection up to around mid-rapidity (corresponding to $\theta_{\rm lab}=0.9$~degrees)~\cite{Massacrier:2015qba}.

For QGP physics, one could explore the region corresponding to intermediate RHIC energies, with luminosities per year larger by a factor $\sim$100 with respect to same energy at RHIC, and similar to those expected for Pb--Pb at the LHC in Run-3~\cite{Brodsky:2012vg}. Therefore, precision measurements in the hard probe sectors, including quarkonia and jet quenching would become accessible, over an extended rapidity range. Thanks to such coverage, precise studies of long-range near-side angular correlations (the so-called ''ridge'') could be performed, as well investigations of the extended longitudinal scaling observed by PHOBOS~\cite{Back:2005hs}. Finally, detailed studies of cold nuclear matter effects, using proton (nuclear) beams on a nuclear (hydrogen) target, would be feasible, with the possibility of reaching the full kinematic domain and in particular the regions close to  $x_{\rm F}=\pm 1$~\cite{Vogt:2015dva}.

The beam extraction via bent crystals was already demonstrated at SPS energy by the UA9 Collaboration~\cite{Uggerhoj:2005xz,Scandale:2010zzb,Scandale:2011za}, in the frame of studies for beam collimation techniques, and is being extended to LHC beams, with encouraging preliminary results, by the LUA9 project~\cite{Scandale:1357606}. Clearly, bringing to reality the AFTER project, would possibly imply, apart from the construction of a new experiment, significant interventions at the LHC level (use of a dedicated experimental hall, creation of necessary infrastructures,...) and remains as of today an ambitious goal. Recently, a possible alternative implementation of this concept has been tested, with encouraging results, in the frame of the LHCb-SMOG project, which has implemented an internal gas target in the LHCb experiment~\cite{LHCbSMOG}. During the Run-1 LHC \mbox{p--Pb} run~\cite{LHCb:2012aka}, a short data taking corresponding to \mbox{Pb--Ne} interactions has been performed and a few J/$\psi$ counts have been observed. Finally, an expression of interest for the AFTER project is being prepared, in view of a submission to the CERN LHCC during 2016.


\subsection{Heavy-ion physics at the Future Circular Collider}
\label{sec:fcc}
A five-year international design study called Future Circular Collider
(FCC) was initiated by CERN in $2014$~\cite{FCCkickoff}. 
The main goal is to assess the feasibility and physics potential of a hadron collider with a centre-of-mass energy $\sqrt s$ of the order of $100$~TeV for pp collisions in a new $80$--$100$~km tunnel near Geneva. 
The operation starting date is targeted for $2035$--$40$. Operating such machine with heavy ions is part of the accelerator design studies.
First ideas on the physics opportunities with heavy ions at the FCC centre-of-mass energies and luminosities
are discussed in this section, covering the physics of the
quark--gluon plasma, high-density QCD and gluon saturation in the
initial state of the collisions 
and the possibility to use photon-induced interactions. More details can be found in~\cite{DaineseKickoff,Armesto:2014iaa,Armesto:2016qyo}.

For a centre-of-mass energy $\sqrt{s}= 100$~TeV for pp collisions, the relation $\sqrtsNN= \sqrt {s} \sqrt{Z_1 Z_2 / A_1 A_2}$ 
gives the energy in centre-of-mass per nucleon--nucleon collision, for
nuclear interactions. The beam parameters and luminosities expected~\cite{Schaumann:2015fsa} at FCC
when operating with lead--lead or proton--lead beams are reported in
Table~\ref{tab:beams}. The expected Pb--Pb integrated luminosity
per month is approximately $8$ times the current projections for the future LHC runs~\cite{rliup}.

\begin{table}[t]
\caption{Beam parameters and luminosities for Pb--Pb and p--Pb collisions at FCC. Reference values shown also for pp.  $L_\text{int,run}$ is intended
for a $30$ days run, typical for heavy-ion operations during a year. All values taken from~\cite{Schaumann:2015fsa}.}
\small
\begin{center}
\begin{tabular}{lccc}
\hline
Quantity & pp & Pb--Pb & p--Pb \\
\hline
Beam energy [TeV/A] & $50$ & $19.5$ & $50/19.5$ \\
$\sqrtsNN$ [TeV] & $100$ & $39$ & $63$ \\
$\mathcal{L}_\text{peak}$ [$10^{27}$ cm$^{-2}$s$^{-1}$] & $5.6\times 10^7$ & $7.3$ &  $1192$ \\
$L_\text{int,run}$ [nb$^{-1}$] & -- & $8.3$  & $1784$ \\
\hline
\end{tabular}
\end{center}
\label{tab:beams}
\end{table}

The QGP phase in Pb--Pb collisions at $\sqrtsNN=39$~TeV is expected to have larger volume, lifetime, energy density and temperature 
than at the top LHC energy. These properties can be estimated by extrapolating the measurements of global event characteristics
at lower energies ---namely: the charged particle multiplicity, the transverse energy and the parameters extracted from femptoscopic correlations.
The results were found consistent with results from a hydrodynamical
calculation ~\cite{Armesto:2014iaa}.
They are reported in Table~\ref{tab:PbPb} for central Pb--Pb collisions ($0$--$5$\%). The charged-particle multiplicity and the volume of the system
are expected to increase by  about a factor of two from top LHC to FCC energy, and its lifetime by $20$\%.  
The larger final volume and stronger flow field will result in an enhancement of collective effects. The twice larger multiplicity
may open the possibility of measurements of high-order flow harmonics
($v_n$ with $n\ge 5$) in individual events.
In addition, it has recently been suggested~\cite{Denicol} that the measurement of
the elliptic flow extracted from multi-particle correlations in Pb--Pb
events with the multiplicities reached at FCC energy could be sensitive
to the temperature-dependence of $\eta/s$ (shear viscosity over
entropy density), which is one of the most relevant properties of the
QGP (see Section~\ref{sec:evolution}).

\begin{table}[t]
\caption{Global properties measured in central Pb--Pb collisions at $\sqrtsNN=2.76$~TeV and extrapolated to $5.5$ and $39$~TeV.
The measurements at $2.76$~TeV are quoted for comparison only and without experimental uncertainties.}
\small
\begin{center}
\begin{tabular}{lccc}
\hline
Quantity & Pb--Pb $2.76$~TeV & Pb--Pb $5.5$~TeV & Pb--Pb $39$~TeV \\
\hline
${\mathrm d}N_{\mathrm ch}/{\mathrm d}\eta$ at $\eta=0$ & $1600$ & $2000$ & $3600$ \\
Total $N_{\mathrm ch}$ & $17000$ & $23000$ & $50000$ \\
${\mathrm d}E_{\mathrm T}/{\mathrm d}\eta$ at $\eta=0$ & $2$~TeV & $2.6$~TeV & $5.8$~TeV \\
Homogeneity volume $$ & $5000$~fm$^3$  & $6200$~fm$^3$ & $11000$~fm$^3$ \\
Decoupling time & $10$~fm/$c$ &  $11$~fm/$c$ & $13$~fm/$c$ \\
\hline
\end{tabular}
\end{center}
\label{tab:PbPb}
\end{table}


The Bjorken relation $\varepsilon(t) = \frac{1}{c\, t} \frac{1}{\pi R_A^2} {\mathrm d}E_{\mathrm T}/{\mathrm d}\eta$ allows to estimate the time dependence of the 
energy density for a system with given final state transverse energy profile.
Together with the Stefan-Boltzmann equation it provides an estimate of the 
temperature evolution of the system: 
$T(t)= [\varepsilon(t)\,(30/\pi^2)/n_{\rm d.o.f.}]^{1/4}$, where $n_{\rm d.o.f.}=47.5$ is the number of degrees of freedom for a system with gluons and three quark flavours.
The energy density increases by a factor of two from LHC to FCC, reaching a value of $35$~GeV/fm$^3$ at the time of $1$~fm/$c$. 
While the increase at a given time is modest, the thermalization time of the system 
(QGP formation time) is expected to be significantly smaller at FCC than at LHC, where it is usually assumed to be $\tau_0=0.1$~fm/$c$. If the thermalization time 
is significantly lower than $0.1$~fm/$c$, the initial temperature could be as large as $T_0\approx800$~MeV. 

A consequence of the increase of the system temperature could be a sizeable production of secondary charm and anti-charm quark ($c\overline c$) pairs from partonic
interactions during the hydrodynamical evolution of the system.
Calculations for top LHC energy indicate that this secondary production can be of the same order as the initial production in hard scattering processes
and is very sensitive to the initial temperature and its evolution during the QGP phase~\cite{Uphoff:2010sh,Zhang:2007dm,Zhou:2016wbo}. Figure~\ref{fig:charmandq2} (left) shows the time evolution of the 
number of $c\overline c$ pairs per unit of rapidity at mid-rapidity with the value at $\tau=0$ corresponding to the production yield from hard scattering. 
The secondary charm production results in an enhancement of charmed hadron 
production at very low $p_{\mathrm T}$, with respect to the expectation from binary scaling of the production in pp collisions. This enhancement potentially provides a handle on the
temperature of the QGP and it is expected to vary from $20\%$ to $50\%$~\cite{ko2} at FCC energies depending on medium properties such as initial temperature and thermalization time.
The abundance of charm quarks also has an effect on the QGP equation of state, which includes a dependence on the number of degrees of freedom.
Inclusion of the charm quark in the lattice QCD calculations results in a sizeable 
increase of $P/T^4$ for temperatures larger than about $400$~MeV~\cite{Borsanyi:2012vn}.

The higher energy and larger luminosities will make new rare hard probes available. The cross section of the 
top, Z$^0$+$1\,{\rm Jet}\,(p_{\mathrm T}>50~{\rm GeV}/c)$, bottom and Z$^0$ production are estimated to increase by factors $80$, $20$, $8$ and $7$, respectively~\cite{mcfm}. 
This opens the possibility to obtain a high-statistics sample for currently unexplored observables, such as pairs of top quarks.
Using as a basis the projections made by the CMS Collaboration for
future LHC runs~\cite{CMStop}, it was estimated that an experiment at FCC could record about $4\cdot 10^4$ fully-reconstructed 
events with $t\overline t\to b\overline b + \ell^+\ell^- + E_{\mathrm T}^{\mathrm{missing}}$ topology  for a sample of Pb--Pb collisions with 
integrated luminosity of $8~{\rm nb^{-1}}$~\cite{Armesto:2014iaa}.

\begin{figure}[!t]
\begin{center}
\includegraphics[width=0.52\textwidth]{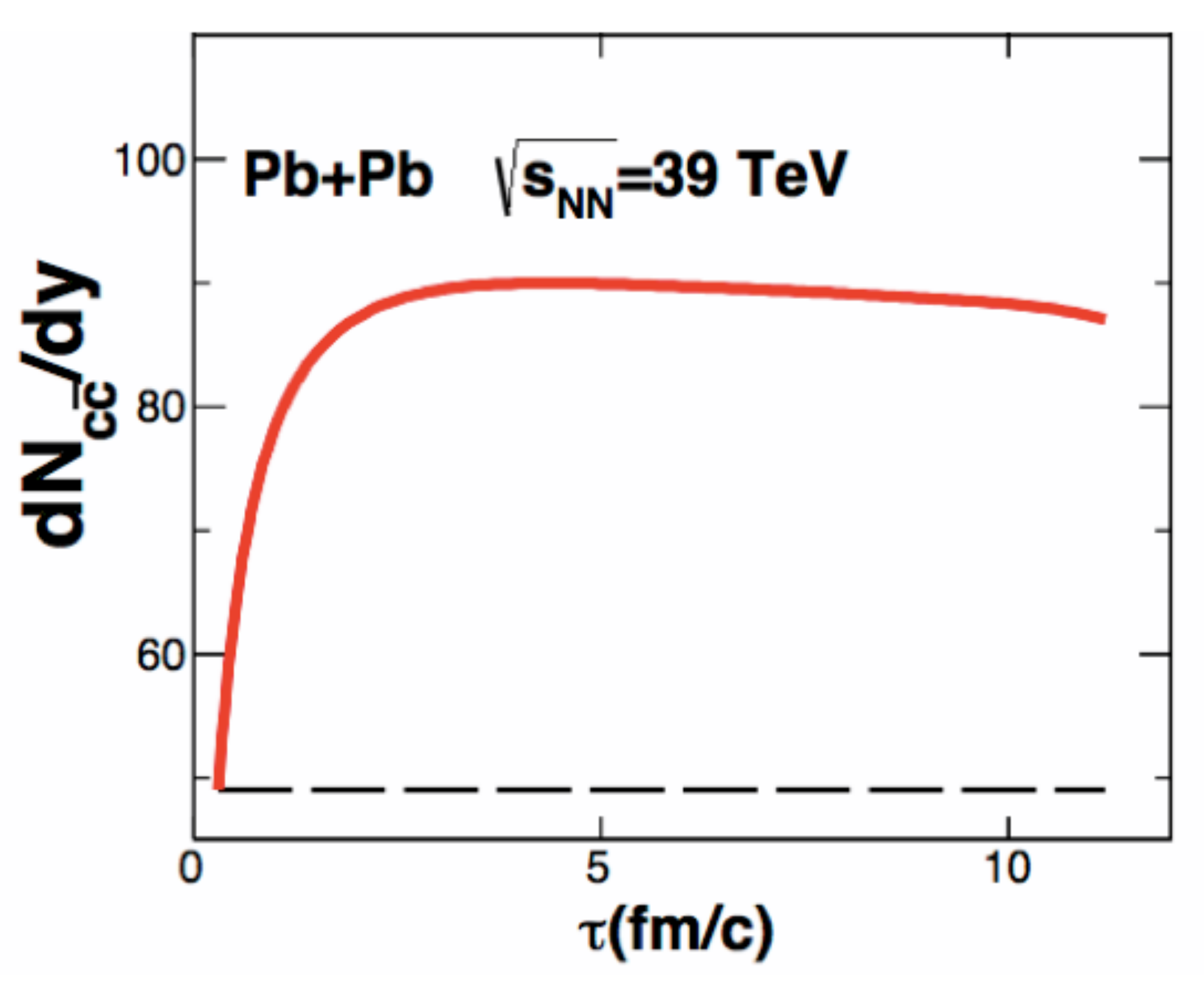}
\hfill
\includegraphics[width=0.47\textwidth]{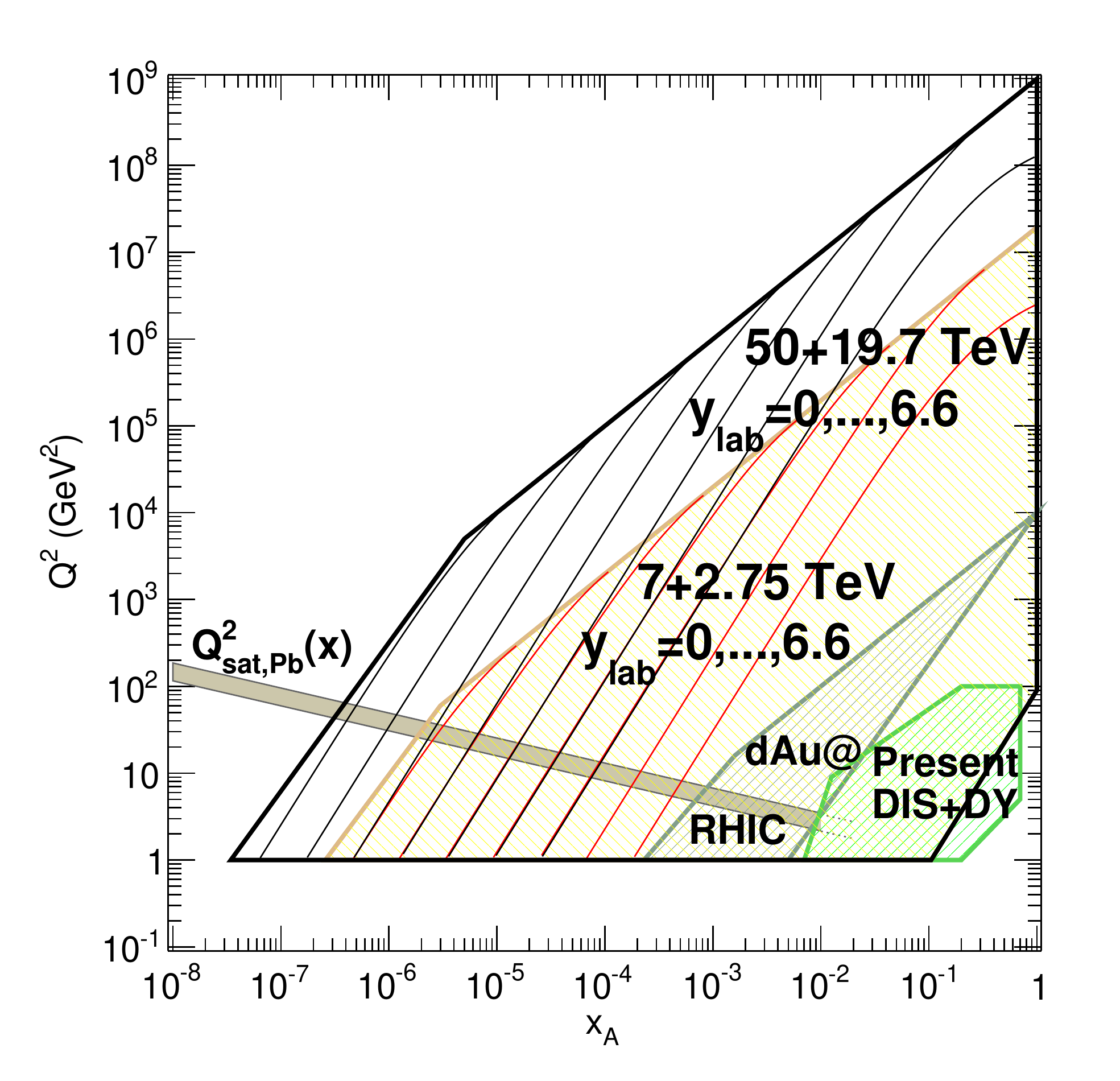}
\caption{Left: time-evolution of the charm and anti-charm quark pair yield (per unit of rapidity at mid-rapidity) for central Pb--Pb collisions 
at FCC energy~\cite{Zhou:2016wbo}. Right: kinematic coverage in the $(x,Q^2)$ plane for p--Pb collisions at RHIC, LHC and FCC energies, computed as  $x(y,Q^2) =  Q\cdot {\mathrm e}^{-y} / \sqrtsNN$~\cite{Armesto:2014iaa};
constant-rapidity lines and an estimate of the saturation scale for Pb nuclei are shown.}
\label{fig:charmandq2}
\end{center}
\end{figure}

Proton--nucleus, nucleus--nucleus and electron--nucleus collisions at very high energy provide a unique opportunity to study the dynamics of high-density gluon distributions in the initial 
nuclear state. The increase of the gluon density at small values of virtuality $Q^2$ and momentum fraction $x$ is expected to be at some point tamed by 
a saturation effect (see e.g.\,the recent review~\cite{Albacete:2014fwa}). 
When density becomes large, the dilute linear evolution of parton density breaks down and non-linear terms have to be considered.
Such non-linear 
terms describe merging processes of the type $gg\to g$ that balance the growth of the gluon density given by the splitting processes of the type $g\to gg$.
Saturation effects become sizeable in processes with virtualities smaller than a few times the saturation scale $Q^2_{\mathrm sat}$,
which is estimated to scale as $Q^2_{\mathrm sat}\sim A^{1/3}/x^{1/3}\sim A^{1/3}(\sqrt s\, {\mathrm e}^{+y})^{1/3}$ ($A$ is the mass number of the nucleus). 
The saturation scale can be increased (thus making its effects more prominent) by increasing $\sqrt s$ or by changing the system size, e.g.\,colliding large nuclei instead of protons.
The saturation effects are also stronger for particle production at forward rapidity, $y$. 
Figure~\ref{fig:charmandq2} (right) shows the kinematic coverage in the $(x,Q^2)$ plane for p--Pb collisions at top LHC ($\sqrtsNN=8.8$~TeV) and FCC ($63$~TeV) energies. 
The constant-rapidity lines from $0$ (right) to $6.6$ (left) and
an estimate of the $x$-dependence of the saturation scale for Pb nuclei are also shown.
The FCC extends the coverage by almost one order of magnitude down to $x\sim 10^{-7}$ at small $Q^2$ and down to $10^{-6}$ around $10~$GeV$^2$.
This region can be explored with probes, such as heavy quarks and charmonium, which are theoretically under good control and can be calculated perturbatively.
Furthermore, the nuclear modification of the valence and sea quark
distribution functions (PDFs) at large $Q^2$ values of the order of $10^4$~GeV$^2$ 
can be accessed with high precision, also in the central rapidity region, by the measuring the production of W and Z bosons, 
as well as of top quarks~\cite{d'Enterria:2015jna}.


Heavy ions accelerated to very high energies generate strong electromagnetic fields,
equivalent to a flux of quasi-real photons, which can be used to study high-energy $\gamma$--$\gamma$,
$\gamma$--p and $\gamma$--A processes in ultra-peripheral collisions (UPCs) where the colliding nuclei
pass close to each other without interacting hadronically. 
The coverage is significantly extended to low $x$ for ultra-peripheral Pb--Pb and p--Pb collisions at FCC energies.
At these energies quarkonium production is driven by gluon densities at $x$ values down to $10^{-7}$, which is 
more than two orders of magnitude lower than at LHC energy. Heavy-ion beams at the FCC
would therefore provide a unique intense photon source able to test the kinematic space in a completely
unexplored region.

From an experimental viewpoint, an important  topic to be investigated is the possibility to design and operate
a general purpose detector for pp collisions also for all observables to be studied within a heavy-ion physics programme. Some of 
the requirements may be conflicting (high-$p_{\mathrm T}$ vs. low-$p_{\mathrm T}$ sensitivity): the possibility
to operate a general-purpose detector with lower magnetic field for
heavy-ion collisions may be 
considered. The need to curb the material budget as well as to have some particle identification sub-systems are other
requirements that pose challenges to the design of a pp detector for high luminosity, with hermetic calorimeters. 
On the other hand, the need to have an extended coverage up to $\eta=5$--$6$ is a common requirement for
both the pp and heavy-ion programmes. 

The Italian heavy-ion community is represented in the INFN FCC group~\cite{TonelliCSN1} as
well as in the FCC CERN-based group working on heavy-ion physics~\cite{TwikiFCC-HeavyIons}.
Even if this is clearly a long-term project, it will be important to continue bringing a contribution of ideas (and a corresponding
simulation effort) related to heavy-ion physics and QGP studies during
the ongoing conceptual design phase ($2014$--$2018$) 
to pursue these important physics opportunities.


%
%
%


\section{Conclusions}
\label{sec:summary}

We conclude this document by summarising the main development lines in the 
study of the phase diagram of strongly-interacting matter with heavy-ion collisions
and by presenting our view of the possible contributions by the Italian community.

The experimental exploration of the phase diagram will continue in two parallel directions:
\begin{description}
\item[\uppercase{H}igh-energy experiments.]  At LHC and at top RHIC energy, where the high-temperature/low-baryon-density region of the phase diagram is covered, the experiments will move towards high-precision measurements, in order to constrain the properties of the QGP and determine its equation of state and characteristic parameters ---namely, the temperature, the shear-viscosity-to-entropy-density ratio and the transport coefficients, as well as their time-dependence during the collision evolution. 
Heavy-ion collisions at the FCC could provide several opportunities to extend these studies to the highest energies. 
\item[\uppercase{L}ow-energy experiments.]  A second beam-energy scan at RHIC, the continuation of the SPS programme 
and the new experiments at the future low-energy facilities NICA and FAIR will explore the region of the phase diagram with moderate-to-high baryonic density, in order to address the search for the onset of deconfinement and for the critical endpoint. In this scope, the SPS is a unique facility, because it offers at the same time a very high interaction rate and, with a beam energy scan, a coverage of a large portion of the phase diagram.
\end{description}

Our present view on the future involvement of the Italian heavy-ion community in these studies is summarised in the following.

\begin{itemize}
\item The Italian community is strongly involved in the upgrade of the ALICE experiment, in particular with the construction of the new Inner Tracking System and the upgrade of the readout electronics of the Time-Of-Flight, Zero-Degree-Calorimeter and Muon Spectrometer detectors. The experiment will be upgraded during the LHC second long shutdown (LS2, $2019$--$2020$) and will then collect data at LHC during Run-$3$ ($2021$--$2023$) and Run-$4$ ($2026$--$2029$), with the goal of integrating a luminosity of about $13$~nb$^{-1}$ in Pb--Pb collisions.
The largely-improved tracking resolution and the foreseen increase by two orders of magnitude of the sample of minimum-bias collisions will enable a detailed characterisation of the QGP properties with new and high-precision measurements, in particular in the sectors of heavy flavour and quarkonium production.
The Italian ALICE groups will contribute to the commissioning and operation of the experiment and to the physics analysis.
\item There is interest for a proposal of a new fixed-target experiment at the SPS, currently denoted as NA$60$+. This experiment would focus on novel high-precision measurements of thermal radiation, light vector mesons and charmonia via the detection of muon pairs, in order to search for the onset of deconfinement and the restoration of chiral symmetry.
In view of the required interaction rates and beam energy range, such an experiment can only be carried out, among the existing or future facilities, at the SPS.
 A first conceptual study of the experimental layout and physics performance is reported in this document. 
The Italian community recommends undertaking the steps towards the formation of an international collaboration and the preparation of a Letter of Intent to be submitted to the SPS Committee before 2018.
This would be timely in view of the update of the European Strategy for Particle Physics scheduled for that year.
The construction and running of the experiment can be envisaged for the following decade.
\item Studies for an experiment using  the LHC proton and Pb beams in a fixed-target mode (AFTER) 
after LS3 are currently in progress. An Expression of Interest will be submitted in $2016$. 
The Italian community looks with interest at the progress, although a definite commitment cannot be envisaged at the moment.
 \item Heavy-ion collisions at the FCC-hh are regarded as a interesting long-term opportunity ---the FCC would start operation after $2035$. INFN is part of the international collaboration that aims at the preparation of the Conceptual Design Study by $2018$, in view of the update of the European Strategy for Particle Physics.
 The Italian community supports the studies on the physics potential and detector requirements for heavy ions at the FCC.
\item The Italian theory community is involved in several aspects of the study of the QCD phase diagram. In the scope of lattice-QCD simulations, the activity encompasses the main hot topics, namely the study of the phase transition at finite baryon chemical potential, the search for the critical endpoint and the determination of the QGP transport coefficients. Concerning the fluid-like behaviour displayed by the QGP, advanced numerical tools, such as ECHO-QGP, will enable an extensive campaign of studies of soft observables and their relation with collectivity in large and small collision systems. In order to enable a quantitative extraction of the QGP properties from high-precision heavy-flavour and quarkonium production measurements, such as those planned with the ALICE upgraded detector, the Italian theory groups are called to a common effort to achieve a highly-realistic description of the various stages of heavy-quark production, transport and hadronization, and of the corresponding medium-induced modifications. All these studies will profit from the fruitful interaction with the experimental heavy-ion community, which we plan to further strengthen in the future. 
\end{itemize}


\renewcommand{\refname}{\spacedlowsmallcaps{References}} 

\bibliographystyle{utphys}

\footnotesize

\bibliography{biblio/intro,biblio/theory,biblio/nuclei,biblio/ohf,biblio/upgrade,biblio/jets,biblio/onia,biblio/soft,biblio/dilept,biblio/fcc,biblio/expintro,biblio/na60plus,biblio/after}


\end{document}